\documentclass[a4paper,fleqn,usenatbib]{mnras}

\usepackage{newtxtext,newtxmath}

 \UseRawInputEncoding

\usepackage[T1]{fontenc}
\usepackage{ae,aecompl}
\usepackage{rotating}
\usepackage{booktabs}
\usepackage{siunitx}
\usepackage{adjustbox}
\usepackage{threeparttable}
\usepackage{color}

\usepackage{graphicx}	
\usepackage{amsmath}	
\usepackage{amssymb}	
\usepackage{rotating}
\usepackage{pdflscape}

\title[Accretion and outflow activity in proto-brown dwarfs]{Accretion and outflow activity in proto-brown dwarfs}

\author[Riaz \& Bally]{B. Riaz$^{1}$\thanks{E-mail: briaz@usm.lmu.de }
\& J. Bally$^{2}$ 
\\
$^{1}$ Universit\"{a}ts-Sternwarte M\"{u}nchen, Ludwig Maximilians Universit\"{a}t, Scheinerstra$\beta$e 1, D-81679 M\"{u}nchen, Germany  \\
$^{2}$ Department of Astrophysical and Planetary Sciences, University of Colorado, Boulder, Colorado 80389, USA  \\
}

\date{Accepted XXX. Received YYY; in original form ZZZ}

\pubyear{2020}

\begin{document}
\label{firstpage}
\pagerange{\pageref{firstpage}--\pageref{lastpage}}
\maketitle

\begin{abstract}

We present a near-infrared study of accretion and outflow activity in 6 Class 0/I proto-brown dwarfs (proto-BDs) using VLT/SINFONI spectroscopy and spectro-imaging observations. The spectra show emission in several [Fe~II] and H$_{2}$ lines associated with jet/outflow activity, and in the accretion diagnostics of Pa~$\beta$ and Br~$\gamma$ lines. The peak velocities of the [Fe~II] lines ($>$100 km s$^{-1}$) are higher than the H$_{2}$ lines. The Class 0 proto-BDs show strong emission in the H$_{2}$ lines but the [Fe~II] lines are undetected, while the Class I objects show emission in both [Fe~II] and H$_{2}$ lines, suggesting an evolutionary trend in the jets from a molecular to an ionic composition. Extended emission with knots is seen in the [Fe~II] and H$_{2}$ spectro-images for 3 proto-BDs, while the rest show compact morphologies with a peak on-source. The accretion rates for the proto-BDs span the range of (2$\times$10$^{-6}$ -- 2$\times$10$^{-8}$) M$_{\sun}$ yr$^{-1}$, while the mass loss rates are in the range of (4$\times$10$^{-8}$ -- 5$\times$10$^{-9}$) M$_{\sun}$ yr$^{-1}$. These rates are within the range measured for low-mass protostars and higher than Class II brown dwarfs. We find a similar range in the jet efficiency for proto-BDs as measured in protostars. We have performed a study of the Brackett decrement from the Br~7-19 lines detected in the proto-BDs. The upper Brackett lines of Br~13--19 are only detected in the earlier stage systems. The ratios of the different Brackett lines with respect to the Br$\gamma$ line intensity are consistent with the ratios expected from Case B recombination.





\end{abstract}

\begin{keywords}

brown dwarfs -- stars: jets -- stars: winds, outflows -- ISM: individual objects: Mayrit 1701117 -- ISM: individual objects: HH 1165 -- ISM: individual objects: MHO 2156 -- ISM: individual objects: MHO 3256 -- ISM: individual objects: MHO 3257

\end{keywords}



\section{Introduction}

Mass accretion and ejection are the fundamental processes during the early stages of star formation. Outflows carry away the excess angular momentum that would otherwise prevent accretion onto the central object. Understanding the physical conditions underlying this coupled accretion-ejection mechanism is thus important at the earliest Class 0/I stages of evolution, when accretion dominates the energetics of the system and the mechanism to extract angular momentum is expected to be more efficient. Investigation of accretion and outflow activity in Class 0/I protostars has been challenging due to the high circumstellar extinction pertaining to these objects that limits the measurement of their stellar and accretion properties and a direct view of the jet base. However, constraints on the jet launching mechanism can be inferred from observations of outflows within a few arcsecond of the driving source where the jet is still expected to be largely unaffected by the ambient gas.

While spatially extended optical jets and bipolar CO molecular outflows have been observed in numerous Class 0/I protostars (e.g., Reipurth \& Bally 2001; Bally 2016; {\it and references therein}), near-infrared high-resolution spectroscopy and spectro-imaging observations in the past two decades have made it possible to study the kinematics of the outflowing gas and physical properties at the base of the jet within a few hundred au of the driving source in Class 0/I protostars (e.g., Davis et al. 2001; 2003; 2011; Caratti o Garatti et al. 2006; Antoniucci et al. 2008; 2011; 2017; Garcia Lopez et al. 2008; 2013; Takami et al. 2006; Nisini et al. 2005; 2016). These micro-jets are bright in [Fe~II] forbidden and H$_{2}$ ro-vibrational emission lines, hence showing the presence of forbidden emission line (FEL) regions and molecular hydrogen emission line (MHEL) regions in low-mass Class  0/I protostars. While multiple low and high velocity components are observed in both MHELs and FELs, the higher velocity gas is slightly further offset from the driving source than the slower gas, and the kinematics of the H$_{2}$ emission differs from [Fe~II] emission, revealing complicated kinematic structures. Evidence of H$_{2}$ emission from cavity walls is also seen in some protostars, suggesting the presence of a wide-angled wind. Strong emission in the well known accretion diagnostics of Paschen and Brackett hydrogen recombination lines is observed in protostars, with the ratio of the accretion luminosity to bolometric luminosity spanning from $\sim$0.1 to $\sim$1. The mass accretion and loss rates for Class 0/I low-mass protostars span the range of 10$^{-6}$ -- 10$^{-8}$ M$_{\sun}$ yr$^{-1}$, and the derived jet efficiencies (ratio between mass ejection and accretion rates) range between $\sim$1\% and 10\% (e.g., Davis et al. 2001; 2003; 2011; Caratti o Garatti et al. 2006; Antoniucci et al. 2008; 2011; 2017; Garcia Lopez et al. 2008; 2013; Takami et al. 2006; Nisini et al. 2005; 2016). These measurements are within the range predicted by the MHD jet launching models (e.g., Frank et al. 2014).

Among sub-stellar objects (M$_{*} \leq$ 0.08 M$_{\sun}$; L$_{bol} <$ 0.1 L$_{\sun}$), similar detailed studies remain to be conducted for brown dwarfs in their early Class 0/I evolutionary stages, or proto-brown dwarfs (hereafter, proto-BDs). We have reported the discovery of the HH~1165 jet, the first spatially extended ($\sim$0.26 pc) jet driven by a Class I proto-BD (Riaz et al. 2017), and comparatively more compact ($<$ 0.01 pc) jets/outflows in two other Class 0/I proto-BDs (Riaz \& Whelan 2015; Whelan et al. 2018). To build up on these sparse results, we have performed an extensive study of the MHEL and FEL regions and the accretion and outflow properties associated with a comparatively larger sample of 6 Class 0/I proto-BDs, using VLT SINFONI spectroscopy and spectro-imaging observations. Our aim is to understand how the jet/outflow physical properties and morphologies, the kinematics of the outflow and accretion tracers, the accretion and outflow activity rates and the jet efficiencies in Class 0/I proto-BDs compare with low-mass Class 0/I protostars and Class II brown dwarfs. This will allow us to investigate any possible trends with the evolutionary stage among brown dwarfs as well as with decreasing bolometric luminosities. 

We describe the sample and observations in Sect.~\ref{targets} and Sect.~\ref{obs}, respectively. Section~\ref{analysis} explains the derivation of the physical properties of the MHEL and FEL regions and the accretion and outflow activity rates. Results on the individual targets are presented in Sect.~\ref{results}. A discussion on the general properties of the [Fe~II] and H$_{2}$ lines observed for the whole sample, the possible dependence of the accretion and outflow properties with the evolutionary stage of the brown dwarfs, and the trends in comparison with low-mass protostars is presented in Sect.~\ref{discuss}. Also discussed in Sect.~\ref{discuss} is decrement of the Brackett HI lines for the proto-BD sample.

\section{Targets}
\label{targets}

\begin{figure*}
  \centering              
     \includegraphics[width=1.7in]{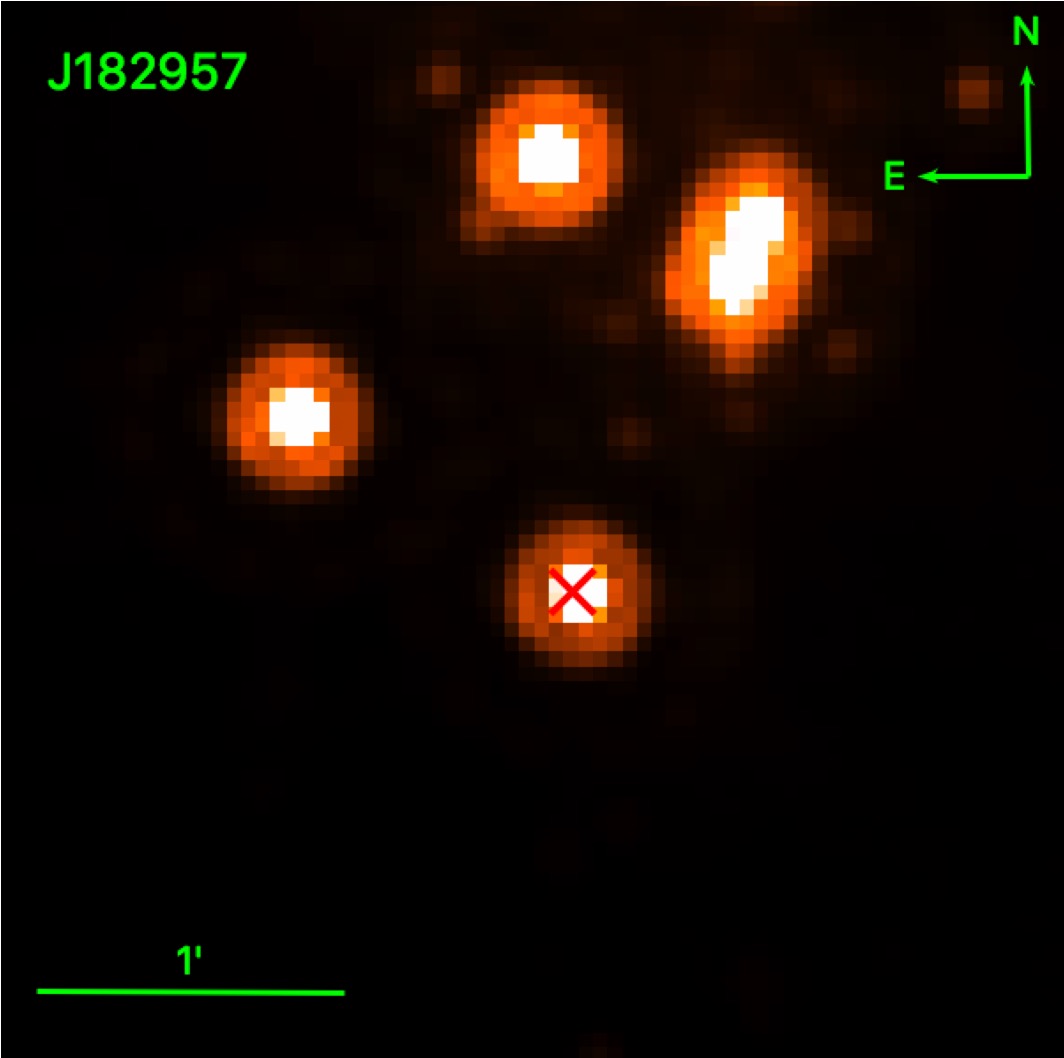} \hspace{0.1in}
     \includegraphics[width=1.7in]{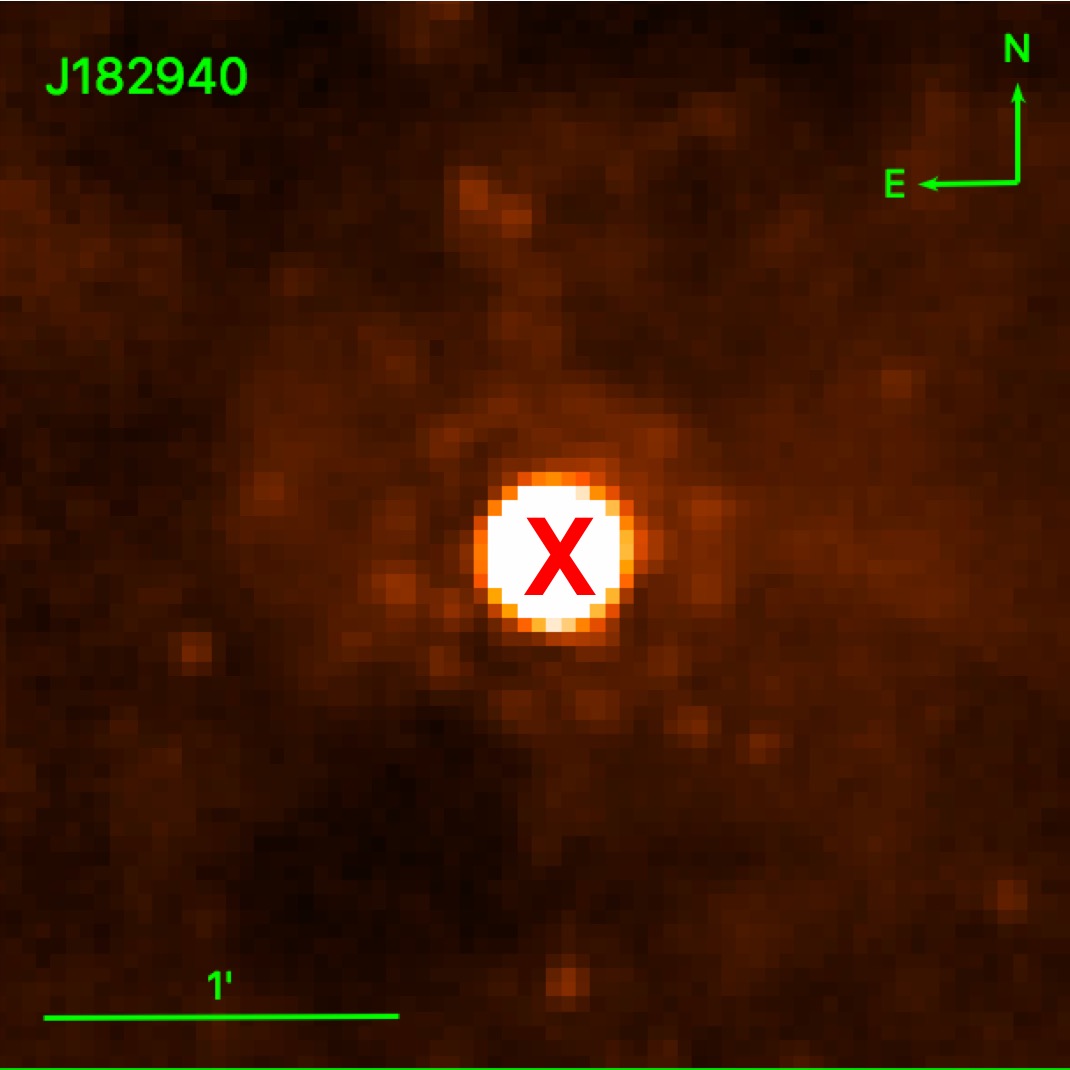} \hspace{0.1in}
     \includegraphics[width=1.7in]{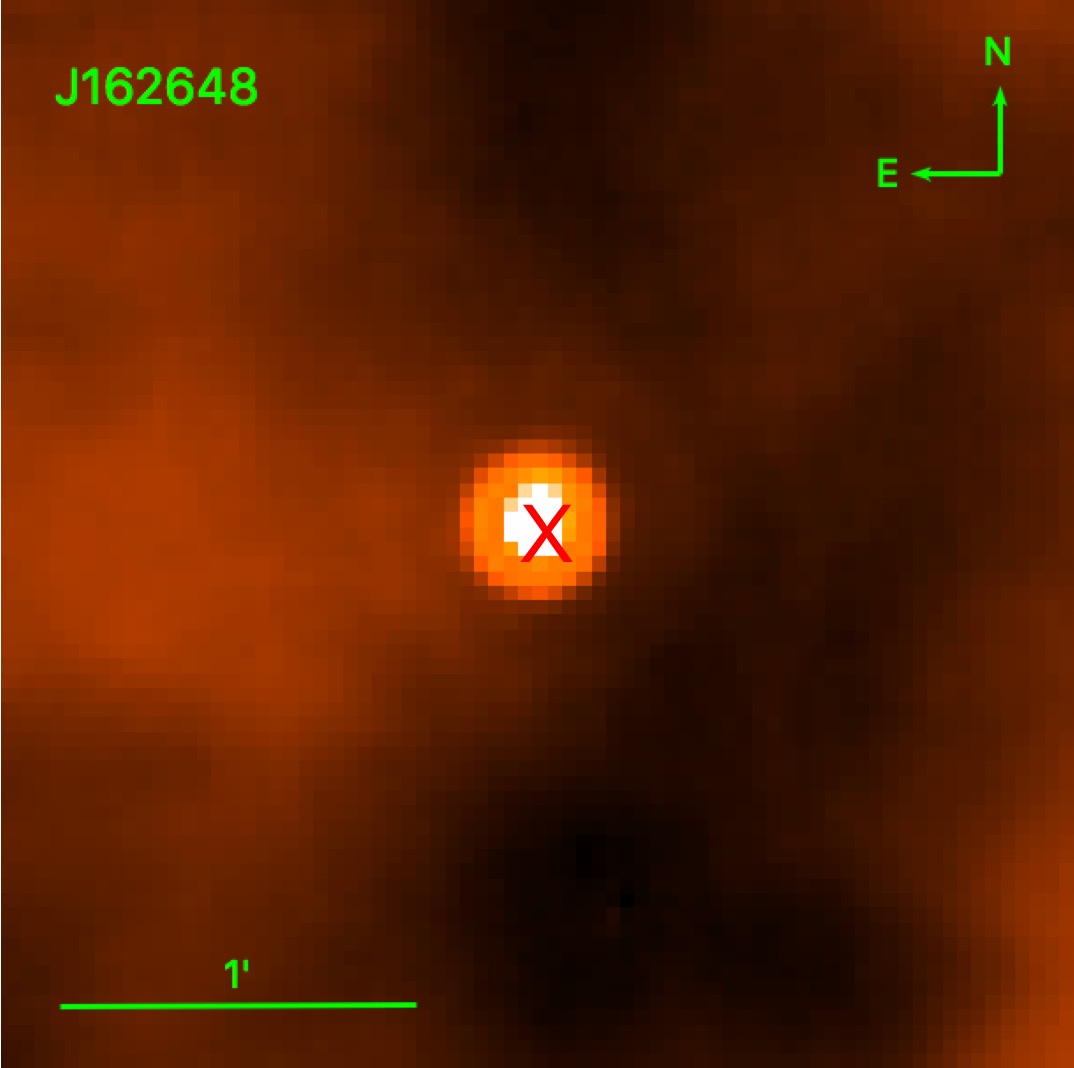} \\  \hspace{0.01in} 
     \vspace{0.5in}
     \includegraphics[width=1.7in]{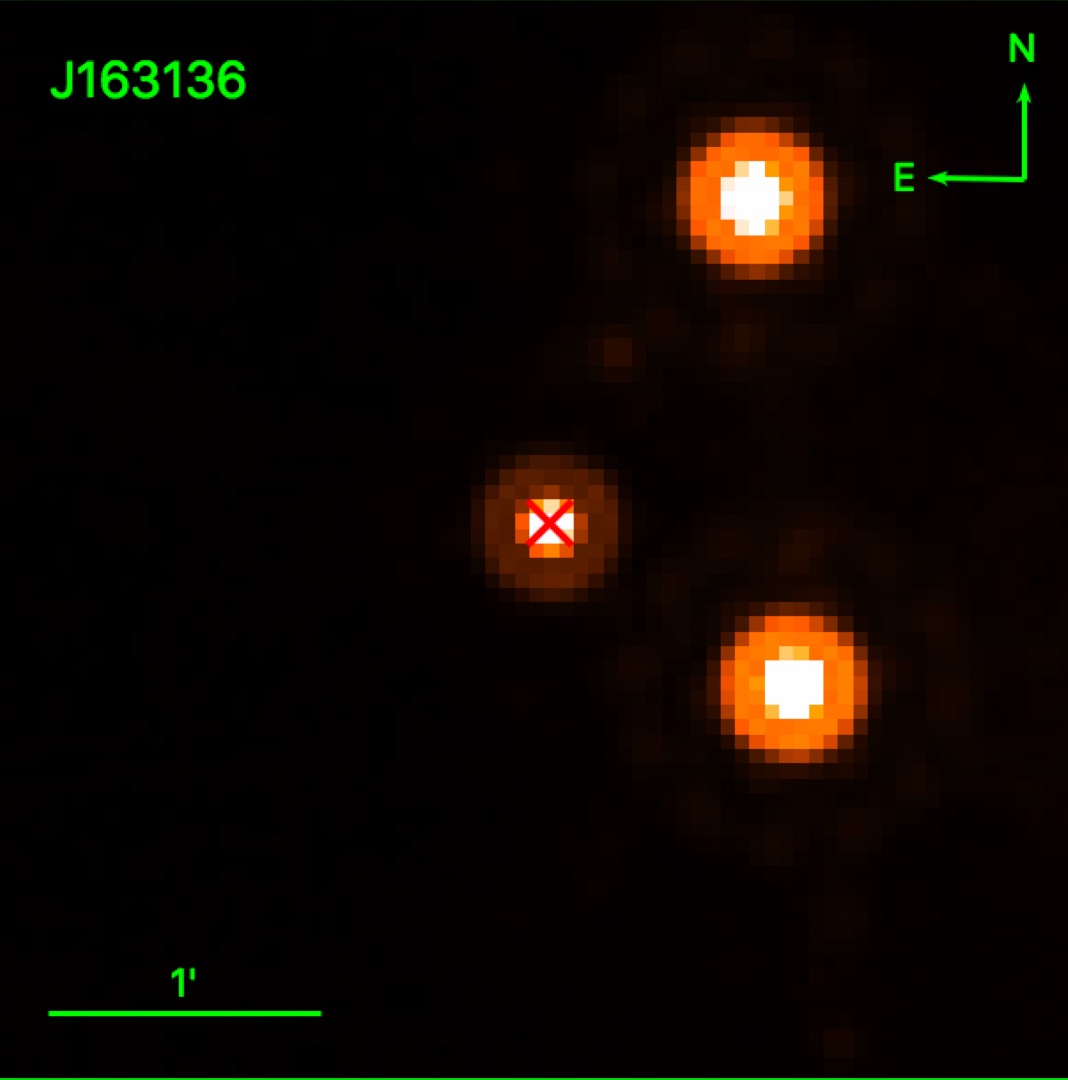} \hspace{0.1in}
     \includegraphics[width=1.7in]{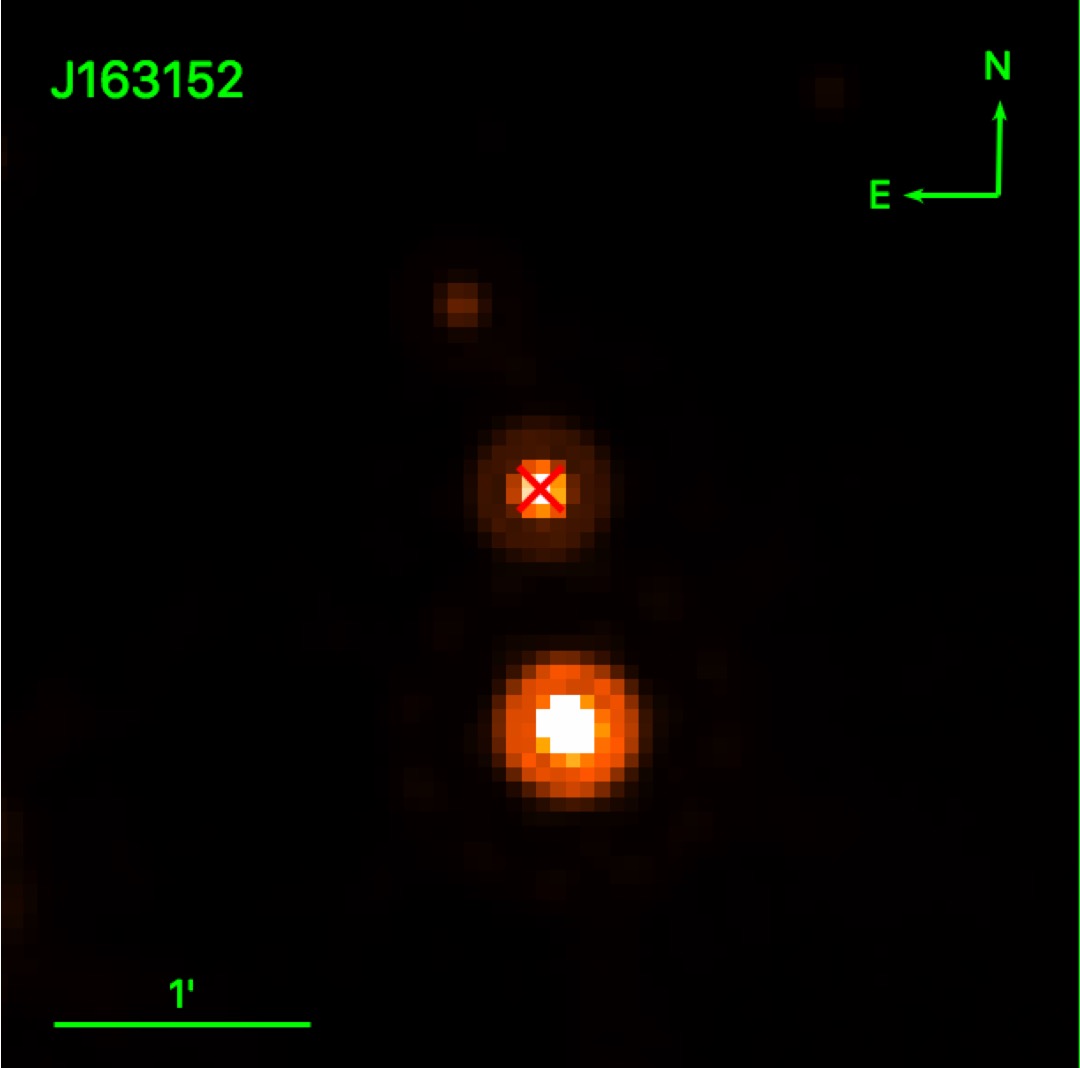} \hspace{0.1in}
     \includegraphics[width=1.7in]{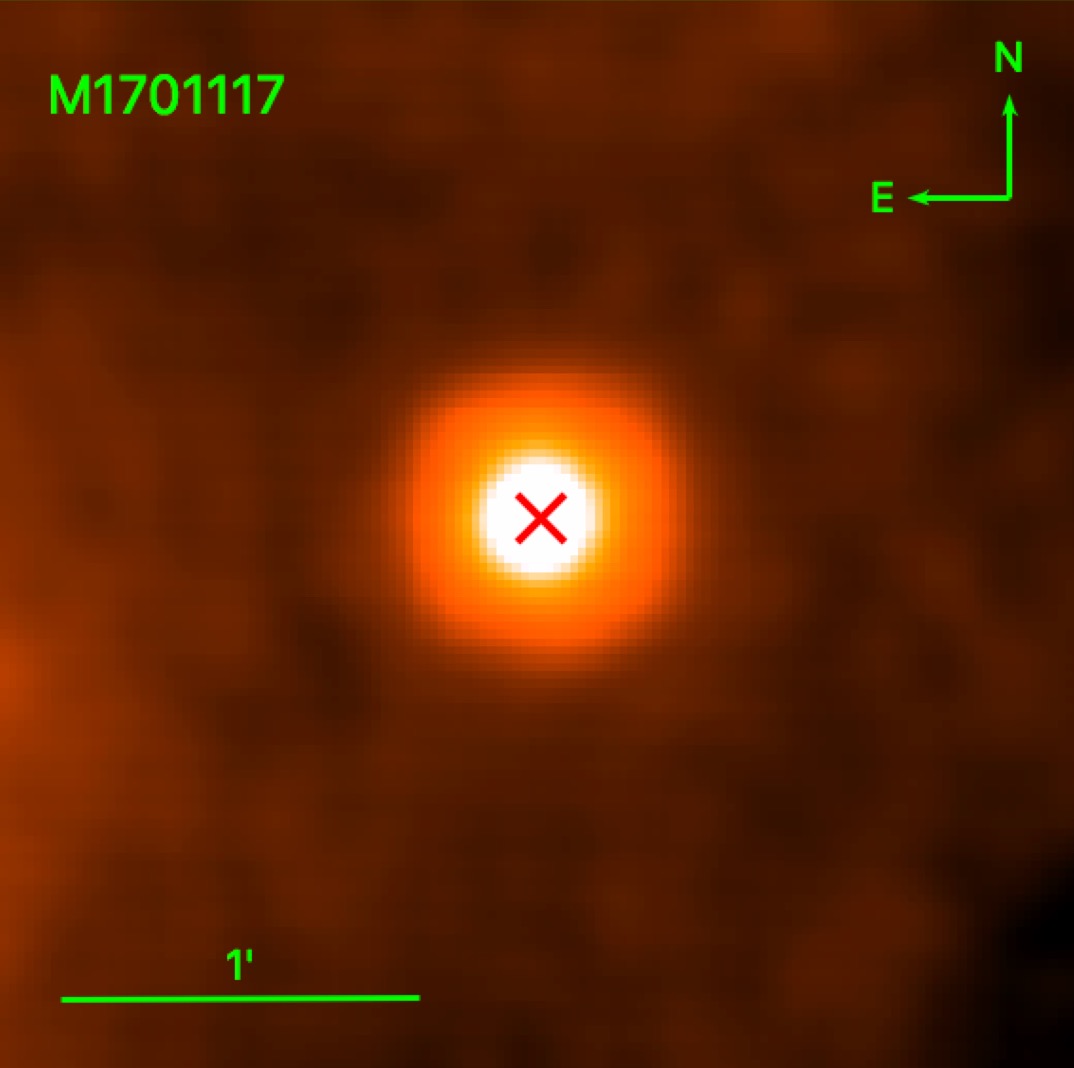}          
     \caption{The Spitzer MIPS 24$\micron$ images for the proto-BDs in the sample. Each image is about 3$\arcmin \times$3$\arcmin$ in size. The target is marked by a red cross.  }
     \label{images}
  \end{figure*}

Our sample consists of 6 proto-BDs, 2 of which were identified in Serpens, and 3 in the Ophiuchus region. Also included in the sample is the proto-BD Mayrit 1701117 (M1701117), which is the driving source of the large-scale HH~1165 jet and is located in the $\sigma$ Orionis region (Riaz et al. 2017). Table~\ref{sample} lists the properties for the targets in the sample. We applied the same target selection and identification criteria as described in Riaz et al. (2015; 2016), which is based on a cross-correlation of infrared data from the UKIDSS and/or 2MASS, Spitzer, and Herschel archives, followed by deep sub-millimeter (sub-mm) 850$\mu$m continuum observations obtained with the JCMT/SCUBA-2 bolometer array. Figure~\ref{images} shows the Spitzer MIPS 24$\micron$ images for the targets. None of the objects lie in a confused region and are at least an arcminute away from the nearest star. 



We have derived the total (dust+gas) circumstellar mass, M$^{d+g}_{\textrm{total}}$, arising from the (envelope+disk) for the proto-BDs using the sub-mm/mm flux density, assuming a dust temperature of 10 K, a gas to dust mass ratio of 100, and a dust mass opacity coefficient of 0.0175 cm$^{2}$ gm$^{-1}$ at 850 $\mu$m (Ossenkopf \& Henning 1994). The masses thus derived are in the range of $\sim$0.02--0.06 M$_{\sun}$. We can set constraints on the mass of the central object using the L$_{int}$-M$_{obj}$ tracks for Class 0/I objects from Vorobyov et al. (2017), where L$_{int}$ is the internal luminosity which is estimated to be $\sim$70\%-80\% of L$_{bol}$ (e.g., Riaz et al. 2016). The central object masses thus estimated are $\sim$0.01--0.02 M$_{\sun}$ for our targets. 


The bolometric luminosity, L$_{\textrm{bol}}$, for the targets is in the range of $\sim$0.03--0.09 L$_{\sun}$. We have measured L$_{\textrm{bol}}$ from integrating the observed infrared to sub-mm spectral energy distribution (SED). The flux densities were corrected for extinction using the source extinction values listed in Table~\ref{Av}. We assumed a distance of 436$\pm$9 pc to the Serpens region, 144$\pm$6 pc to Ophiuchus, and 410$\pm$60 pc to $\sigma$ Orionis, based on astrometric observations (Gaia DR2; Gaia Collaboration 2018; Ortiz-Leon et al. 2017; Dzib et al. 2010; Mamajek 2008; Schlafly et al. 2014). 

We note that since these objects are still accreting, the total mass may not represent the final mass for these systems. However, based on the accretion models by Baraffe et al. (2017) and Vorobyov et al. (2017), and as discussed in detail in Riaz et al. (2016), the L$_{\textrm{bol}}$ for these YSOs is below the luminosity threshold considered between very low-mass stars and brown dwarfs ($<$0.2 L$_{\sun}$). Considering the sub-stellar mass reservoir in the (envelope+disk) for these systems, and the prediction that about 30\%-50\% of the mass is expected to be expelled by the outflow (e.g., Machida et al. 2009), the final mass is also expected to stay within the sub-stellar limit.


The 2-24 $\micron$ spectral slope for the proto-BD candidate is $>$0.3, consistent with a Class 0/I classification. We have also determined the evolutionary stage of the candidates using the Stage 0+I/II criteria based on the strength in the HCO$^{+}$ (3-2) line emission, and the Stage 0/I/I-T/II classification criteria based on the physical characteristics of the system. A detailed description of these criteria is presented in Riaz et al. (2016). Briefly, a young stellar object (YSO) with an integrated intensity of $>$0.4 K km s$^{-1}$ in the HCO$^{+}$ (3-2) line is classified as a Stage 0+I object (e.g., van Kempen et al. 2009; Riaz et al. 2016). Based on the physical characteristics estimated from the radiative transfer modelling of the observed spectral energy distribution (SED), a Stage 0 object is expected to have a disk-to-envelope mass ratio of $<<$1, while for a Stage I YSO, this mass ratio would be in the range of 0.1--2 (e.g., Whitney et al. 2003). A Stage 0 object is also expected to have a total circumstellar (disk+envelope) mass to stellar mass ratio of $\sim$1. The Stage thus determined has been compared to the Class 0/I/Flat classification based on the 2-24 $\mu$m slope of the observed SED. Stage I-T and Class Flat are considered to be intermediate between the Stage I and Stage II, or Class I and Class II evolutionary phases. 

As listed in Table~\ref{sample}, we find a good match between the Stage of the system based on the physical characteristics and the strength in the molecular line emission, whereas the classification based on the SED slope does not relate well to this evolutionary stage. An edge-on inclination could result in a flatter spectral slope in the SED compared to genuine Stage 0/I systems (e.g., Riaz et al. 2016). Thus, the proto-BDs J163152 and J182957 in our sample are in the early Stage 0 evolutionary phase, J163136, J162648, and M1701117 are in Stage I, while J182940 is in the intermediate Stage I-T/Stage II phase. 


Table~\ref{Av} lists the estimates on the interstellar extinction towards the target, A$_{V}$(source), and the extinction caused by the circumstellar envelope, A$_{V}$(envelope). The estimates on the source extinction were obtained following the methods described in Riaz et al. (2012). The circumstellar envelope extinction was estimated from radiative transfer modelling of the observed SED for the targets, following the methods described in Riaz et al. (2016). The estimates on the jet extinction are discussed in Sect.~\ref{temp_den}.




\begin{table*}
\centering
\caption{Sample}
\label{sample}
\begin{adjustbox}{scale=0.9,center}
\begin{threeparttable}
\begin{tabular}{lcccccccccc} 
\hline
Designation &  HH/MHO\tnote{a} & RA (J2000) & Dec (J2000) & L$_{bol}$ [L$_{\sun}$] \tnote{b} & Classification\tnote{c} & $JHK_{s} [mag] $\tnote{d} \\
		&				&			&			& L$_{int}$ [L$_{\sun}$] & 	&	\\
\hline

Mayrit 1701117 (M1701117) & HH~1165 & 05h40m25.79s & -02d48m55.41s & 0.09$\pm$0.05 & Class 0/I, Stage 0+I, Stage I & J=15.15, H=14.10, K$_{s}$=13.07  \\
	&	&	&	& 0.063$\pm$0.05	&	&	J=15.45, H=14.43, K$_{s}$=13.33 \\	\hline

SSTc2d J163136.8-240420 (J163136) & MHO~2156 & 16h31m36.77s & -24d04m19.77s & 0.1$\pm$0.05 & Class Flat, Stage 0+I, Stage I & J=15.74, H=13.85, K$_{s}$=12.50  \\	
	&	&	&	& 0.07$\pm$0.05	&	&	J=16.00, H=13.98, K$_{s}$=12.68 \\	\hline

SSTc2d J163152.5-245536 (J163152) & -- & 16h31m52.32s & -24d55m36.1s & 0.08$\pm$0.05 & Class 0/I, Stage 0+I, Stage 0  & H=14.83, K$_{s}$=12.72   \\ 		
	&	&	&	& 0.057$\pm$0.05	&	&	H=15.11, K$_{s}$=17.72, 17.44, 16.54 \\	\hline

SSTc2d J162648.5-242839 (J162648) & -- & 16h26m48.48s & -24d28m38.91s & 0.07$\pm$0.05 & Class Flat, Stage I-T, Stage I   & J=18.47, H=13.93, K$_{s}$=11.22	\\	
	&	&	&	& 0.052$\pm$0.05	&	&	J= 18.67, H=14.09, K$_{s}$=11.87 \\		\hline

SSTc2d J182957.6+011304 (J182957) & MHO~3256 & 18h29m57.66s & +01d13m04.6s & 0.09$\pm$0.05 & Class 0/I, Stage 0+1, Stage 0 &  J=16.05, H=13.77, K$_{s}$=12.02 	  \\ 
	&	&	&	& 0.065$\pm$0.05	&	&	\\ \hline

SSTc2d J182940.2+001513 (J182940) & MHO~3257 & 18h29m40.20s & 00d15m13.11s & 0.05$\pm$0.03 & Class 0/I, Stage 1-T, Stage II  &  J=15.39, H=14.78, K$_{s}$=14.16   \\	
	&	&	&	& 0.035$\pm$0.03	&	&	J=15.11, H=14.39, K$_{s}$=13.80 \\

\hline
\end{tabular}
\begin{tablenotes}
  \item[a] The Herbig-Haro (HH) and/or Molecular Hydrogen Object (MHO) number assigned to the target.
  \item[b] The top value is for the bolometric luminosity, bottom value is for the internal luminosity.  
  \item[c] The first, second, and third values are using the classification criteria based on the SED slope, the integrated intensity in the HCO$^{+}$ (3-2) line, and the physical characteristics, respectively.
  \item[d] Near-infrared photometry from the 2MASS (top) and UKIDSS (bottom) databases. There is no J-band detection for J163152. There is no match in the UKIDSS database for J182957. 
\end{tablenotes}
\end{threeparttable}
\end{adjustbox}
\end{table*}

\begin{table*}
\centering
\caption{Extinction}
\label{Av}
\begin{threeparttable}
\begin{tabular}{lcccccccccc} 
\hline
Designation & A$_{V}$ (source) [mag] \tnote{a} & A$_{V}$ (envelope) [mag] \tnote{a} & A$_{V}$ (jet) [mag] \tnote{a} \\
\hline

M1701117 & 0.8 & 14 & 10  \\

J163136 & 10 & 7 & 20  \\	

J163152 & 9 & 9 & 20--30    \\ 		

J162648 & 10 & 9 & --    \\	

J182957 & 10 & 9 & 15-20    \\	

J182940 & 9 & 5 & 15--20     \\	

\hline
\end{tabular}
\begin{tablenotes}
  \item[a] The uncertainty on A$_{V}$ is 15\%-20\%. 
\end{tablenotes}
\end{threeparttable}
\end{table*}


\section{Observations and Data Reduction}
\label{obs}

Data were obtained using the integral field spectrograph SINFONI at the ESO Very Large Telescope (VLT) in December, 2018, and June, 2019. We used the J and H+K gratings, which provide a spectral resolution of 2000 and 1500, respectively, covering a wavelength range of $\sim$1.1--2.45 $\micron$. Observations were taken in the noAO mode with a field of view on the sky of 8$\arcsec$$\times$8$\arcsec$ and a spatial scale of 125$\times$250 mas. The seeing was 0.7$\arcsec$ -- 0.8$\arcsec$ in all observations. Observations were obtained in the standard object-sky ABB$^{\prime}$A$^{\prime}$ mode. The total exposure time was 1200 s in the J-band and 600 s in the H+K band for each target. 

The SINFONI data reduction pipeline was used to apply dark and bad pixel masks, flat field correction, and to correct for the detector's geometric distortions. The object and standard star frames were then wavelength calibrated using Xenon and argon arc lamp frames, and a sky frame was subtracted from the object frames to correct for variations in the night sky emission lines. We used the standard stars Feige110 and LTT3218 to flux calibrate the spectra in the J and H+K bands, respectively. The telluric standards of Hip026694, Hip081214, Hip089384, Hip088201, and Hip098641 were used to apply telluric correction and remove the atmospheric transmission features. To inter-calibrate the fluxes for the three J, H, K spectral segments, we normalized the spectra to the 2MASS or UKIDSS photometry (Table~\ref{sample}). There is some variability seen between the 2MASS and UKIDSS photometry, particularly for J163152 in the K$_{s}$-band (Table~\ref{sample}). The inter-calibration factors slightly differ for the segments possibly due to the different atmospheric conditions during the observations.

\section{Data Analysis}
\label{analysis}

The results on the spectra and spectro-images for the individual targets are discussed in Sect.~\ref{results}. The spectra were extracted in a 0.5$\arcsec\times$0.5$\arcsec$ aperture centered on the source. We have measured the parameters of the line center, line width, the peak and integrated line intensities for all emission lines detected at a signal-to-noise ratio (SNR) of $\geq$5. The line fluxes were measured after continuum-subtraction from a single-peaked fit for the typical Gaussian shaped profiles seen in all lines. The error on flux measurements due to the uncertainty in defining the pseudo-continuum is 20\%-30\%. The parameters thus derived are listed in Sect.~\ref{fluxes}. The brightest emission lines associated with jets in the near-infrared are H$_{2}$ and [Fe~II] observed throughout the wavelength range covered by the SINFONI spectra, while the strongest accretion diagnostics are the Paschen $\beta$ and the Brackett $\gamma$ lines.

A continuum-subtracted image in the brightest emission lines was produced by estimating the mean of the flux in the pseudo-continuum on both sides of the emission line, and then subtracting the continuum image from the line image. The central source position in the spectro-images shifts slightly with wavelength due to extinction. The position uncertainty is estimated to be $\leq$0.5$\arcsec$. We have measured the jet position angle (PA) using the convention from north of west.

\subsection{Extinction, Temperature, and Density in the Emission Line Regions}
\label{temp_den}

The extinction towards the forbidden emission line (FEL) and the molecular hydrogen emission line (MHEL) regions at the base of the jet can be measured using ratios of [Fe II], H$_{2}$, or H~I Brackett emission lines that are well separated in wavelength and share the same upper energy level. The H$_{2}$ and [Fe II] lines trace the extended jet emission while the H~I lines originate from very close to the driving source (e.g., Davis et al. 2011). Extinction measured using the various [Fe II], H$_{2}$, or H~I line ratios listed in the literature could result in a factor of two or larger differences, possibly due to the uncertainties in the transition probabilities of the lines involved (e.g., Nisini et al. 2005; Takami et al. 2006). Previous works have shown that the H$_{2}$ 2.122$\micron$/2.424$\micron$ ratio, the [Fe~II] 1.644$\micron$/1.25$\micron$ ratio, and the H$_{2}$ excitation diagram analysis are more reliable methods for estimating the extinction (e.g., Davis et al. 2011; Nisini et al. 2005; Takami et al. 2006). Another commonly used method is to equate the accretion luminosity for the Pa$\beta$ and Br$\gamma$ lines for different values of the extinction, A$_{V}$ (e.g., Alcal\'{a} et al. 2017). The H$_{2}$ excitation diagram analysis can also provide an estimate on the gas temperature, T$_{e}$, while the electron density, n$_{e}$, at the base of the jet can be derived from a comparison of the [Fe~II] line ratios with the models of Nisini et al. (2002) and Takami et al. (2006).

For the target J182957, the A$_{V}$ measured using the [Fe~II] 1.644$\micron$/1.25$\micron$ line ratio is 15$\pm$5 mag. This estimate is consistent within the uncertainty with A$_{V}\sim$20 mag estimated by equating the accretion luminosity for the Pa$\beta$ and Br$\gamma$ lines, and the estimate derived from the H$_{2}$ line ratio and the excitation diagram analysis. The electron temperature derived from the best-fit relation in the H$_{2}$ excitation diagram is 1650$\pm$300 K. We have estimated n$_{e} \geq$ 10$^{5}$ cm$^{-3}$ from the [Fe~II] 1.533$\micron$/1.644$\micron$ and the 1.600$\micron$/1.644$\micron$ line ratios.

For J163152, there are no [Fe~II] line detection in H or K bands (Fig.~\ref{oph2-spec}). This object shows several H$_{2}$ lines (Fig.~\ref{oph2-spec}) and therefore the excitation diagram technique can be used. Following the method described in Davis et al. (2011), the peak value for the correlation coefficient corresponds to an A$_{V}$ $\sim$ 20--30 mag, and the electron temperature derived from the best-fit relation is 1428$\pm$300 K. The J-band spectrum for this object is extremely noisy, due to which there is no definitive Pa$\beta$ detection and thus we cannot equate the accretion luminosity for the Pa$\beta$ and Br$\gamma$ lines to obtain another estimate on A$_{V}$. 



For J162648, there is detection in the H$_{2}$ 1.688$\micron$ and 2.122$\micron$ lines but none in the 2.424$\micron$ line (Fig.~\ref{oph3-spec}), due to which the H$_{2}$ 2.122$\micron$/2.424$\micron$ line ratio cannot be used for an A$_{V}$ estimate, and the H$_{2}$ excitation diagram technique results in a degenerate fit using just three data points. There is no definitive Pa$\beta$ detection due to which the accretion luminosity cannot be used. Therefore, there is no measurement on the jet extinction for J162648 and we have assumed the A$_{V}$ (source+envelope) estimate of 19 mag for this object. The lower limit on n$_{e} \geq$ 10$^{4}$ cm$^{-3}$ was estimated from the [Fe~II] 1.533$\micron$/1.644$\micron$ and the 1.600$\micron$/1.644$\micron$ line ratios.

For J163136, the A$_{V}$ estimate is 15$\pm$5 mag using the [Fe~II] 1.644$\micron$/1.25$\micron$ and the H$_{2}$ 2.122$\micron$/2.424$\micron$ line ratios. This object shows several H$_{2}$ line detections (Fig.~\ref{oph1-spec}). The A$_{V}$ estimate from the H$_{2}$ rotational diagram and from equating the Pa$\beta$ and Br$\gamma$ accretion luminosities is 20$\pm$5 mag. The electron temperature derived from the H$_{2}$ rotational diagram analysis is 2500$\pm$200 K. The electron density derived from the [Fe~II] 1.533$\micron$/1.644$\micron$ line ratio is (1--3)$\times$10$^{3}$ cm$^{-3}$. 

For J182940, the A$_{V}$ estimate from the H$_{2}$ 2.122$\micron$/2.424$\micron$ ratio and the H$_{2}$ rotational diagram is 15--20 mag. The A$_{V}$ (source+envelope) for J182940 is 14 mag, indicating that this is not a highly extincted source. The electron temperature derived from the H$_{2}$ rotational diagram analysis is 1428$\pm$300 K. The electron density derived from the [Fe~II] 1.533$\micron$/1.644$\micron$ line ratio is $\geq$10$^{5}$ cm$^{-3}$. 


For M1701117, the A$_{V}$ estimate from the [Fe~II] 1.644$\micron$/1.25$\micron$ line ratio and the H$_{2}$ rotational diagram is 10$\pm$5 mag. A similar estimate of $\sim$10 mag is obtained from the H$_{2}$ 2.424$\micron$ line ratio and by equating the Pa$\beta$ and Br$\gamma$ accretion luminosities. The electron temperature derived from the H$_{2}$ rotational diagram analysis is 3333$\pm$300 K. 

The jet extinction for the targets are comparable or lower than the A$_{V}$ (source+envelope) estimates (Table~\ref{Av}), with the exception of J182940 for which the jet A$_{V}$ estimate is higher than the source+envelope extinction. While jet extinction is expected to be the lower limit to the source extinction in typical T Tauri stars, the extinction towards the emission line regions is noted to be much higher than the source extinction in earlier stage protostellar systems (e.g., Davis et al. 2011). The discrepancy is likely due to the differences in the method of measuring extinction as well as the different wavelength regimes at which these measurements are made. The source A$_{V}$ is estimated using near-infrared photometry, the envelope A$_{V}$ using SED modelling over the full wavelength range from near-infrared up to sub-mm wavelengths, while the jet extinction from near-infrared emission lines. The envelope A$_{V}$ could be under estimated because the angular resolution of the sub-mm observations only allows observing the outer low density envelope region while the inner dense regions are beam diluted (e.g., Riaz et al. 2019). In comparison, the extinction near the jet launching region should be higher as it emanates from a much denser inner zone of the proto-brown dwarf system. Matching the resolution at all wavelengths can provide a better estimate on the extinction in the different jet/envelope regions.

\subsection{Accretion and Outflow Activity Rates}

The mass accretion rate, $\dot{M}_{acc}$, was calculated using the extinction-corrected Pa$\beta$ and Br$\gamma$ line luminosities and the L$_{line}$-L$_{acc}$ relations from Alcal\'{a} et al. (2017). To estimate the intrinsic stellar properties, we have used the measured bolometric and internal luminosity, and numerical simulations of stellar evolution described in the accretion models of Baraffe et al. (2017) and Vorobyov et al. (2017). A detailed discussion on the uncertainties in estimating the stellar parameters for such early-stage brown dwarfs using various evolutionary models is presented in Riaz et al. (2015; 2016; 2020). The $\dot{M}_{acc}$ measurements are listed in Table~\ref{macc}.

The mass outflow rate, $\dot{M}_{out}$, was calculated using the extinction-corrected H$_{2}$ and [Fe~II] line fluxes and the relations from Davis et al. (2011) and Garcia-Lopez et al. (2013). The main parameters required to calculate $\dot{M}_{out}$ are the H$_{2}$ and H column densities, the area of the H$_{2}$ or [Fe~II] emitting region, the projected length of the jet and the tangential velocity. The jet dimensions (length and width) were measured from the spectro-image of the particular H$_{2}$ or [Fe~II] emission lines. For cases such as J163152 where the jet emission is unresolved and peaks on-source, the jet dimensions are comparable to the seeing ($\sim$0.7$\arcsec$-0.8$\arcsec$). The tangential velocity requires information about the inclination of the outflow with respect to the line of sight. Since this is difficult to measure, we have used the approximation considered in Davis et al. (2011) and used the peak radial velocity measured from the observation. The outflow rates thus derived are listed in Table~\ref{mout}.




The uncertainty in the $\dot{M}_{acc}$ and $\dot{M}_{out}$ measurements is due to the range in the extinction estimated using different methods, the stellar parameters that are derived using the evolutionary models, the error on flux measurements due to the uncertainty in defining the pseudo-continuum, as well as the uncertainty on the velocity and jet parameters measurements. From propagating these errors, we estimate a 20\%-30\% uncertainty on the derived values for the activity rates. 



\begin{figure*}
  \centering              
     \includegraphics[width=2.5in]{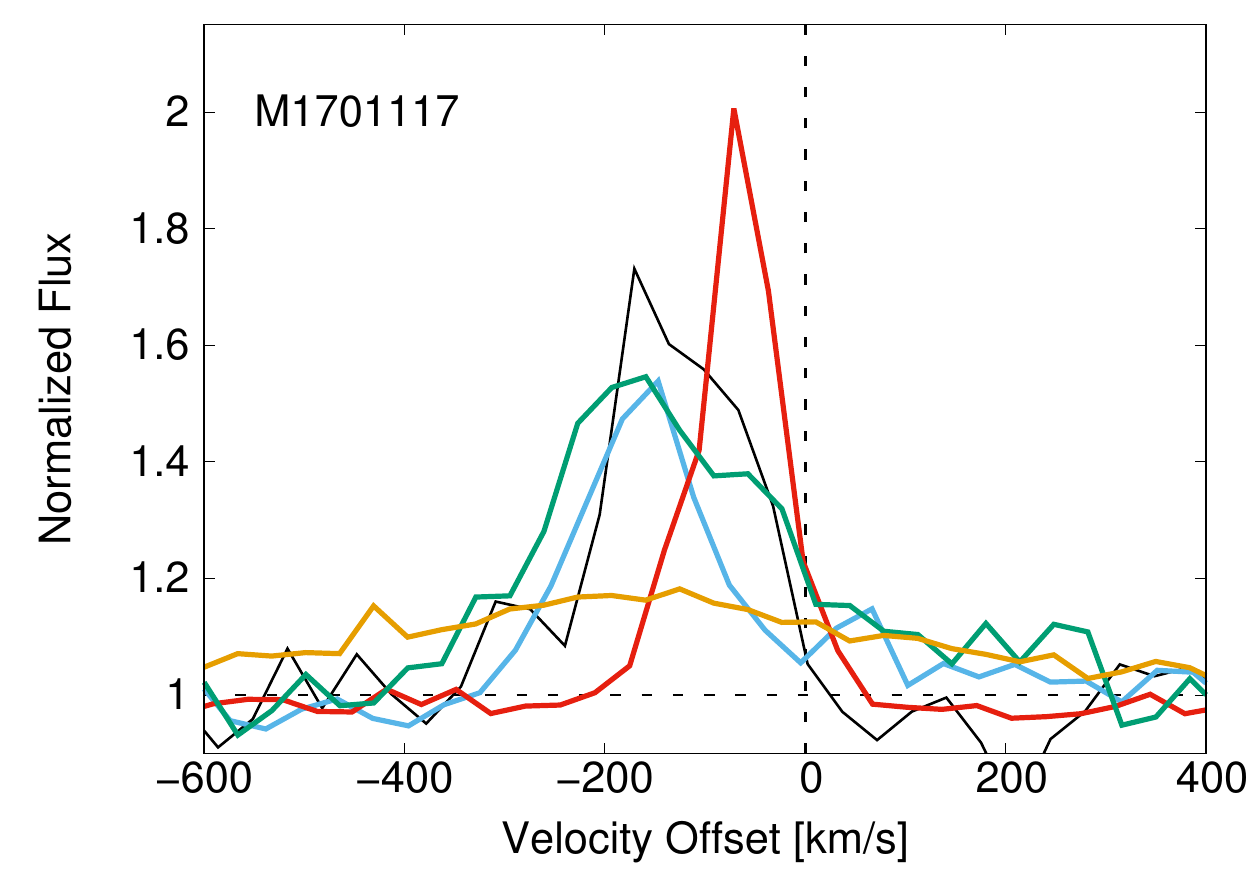}
     \includegraphics[width=2.5in]{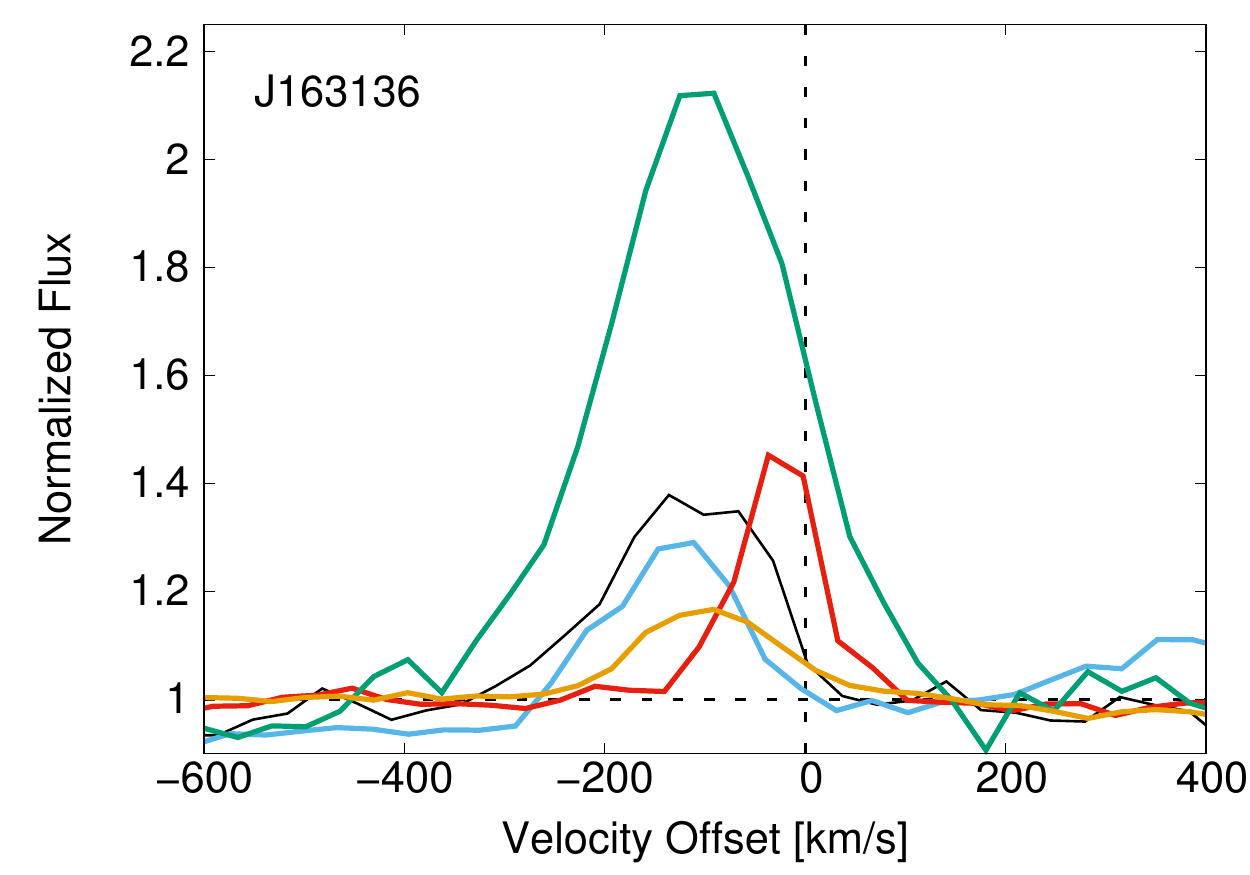}
     \includegraphics[width=2.5in]{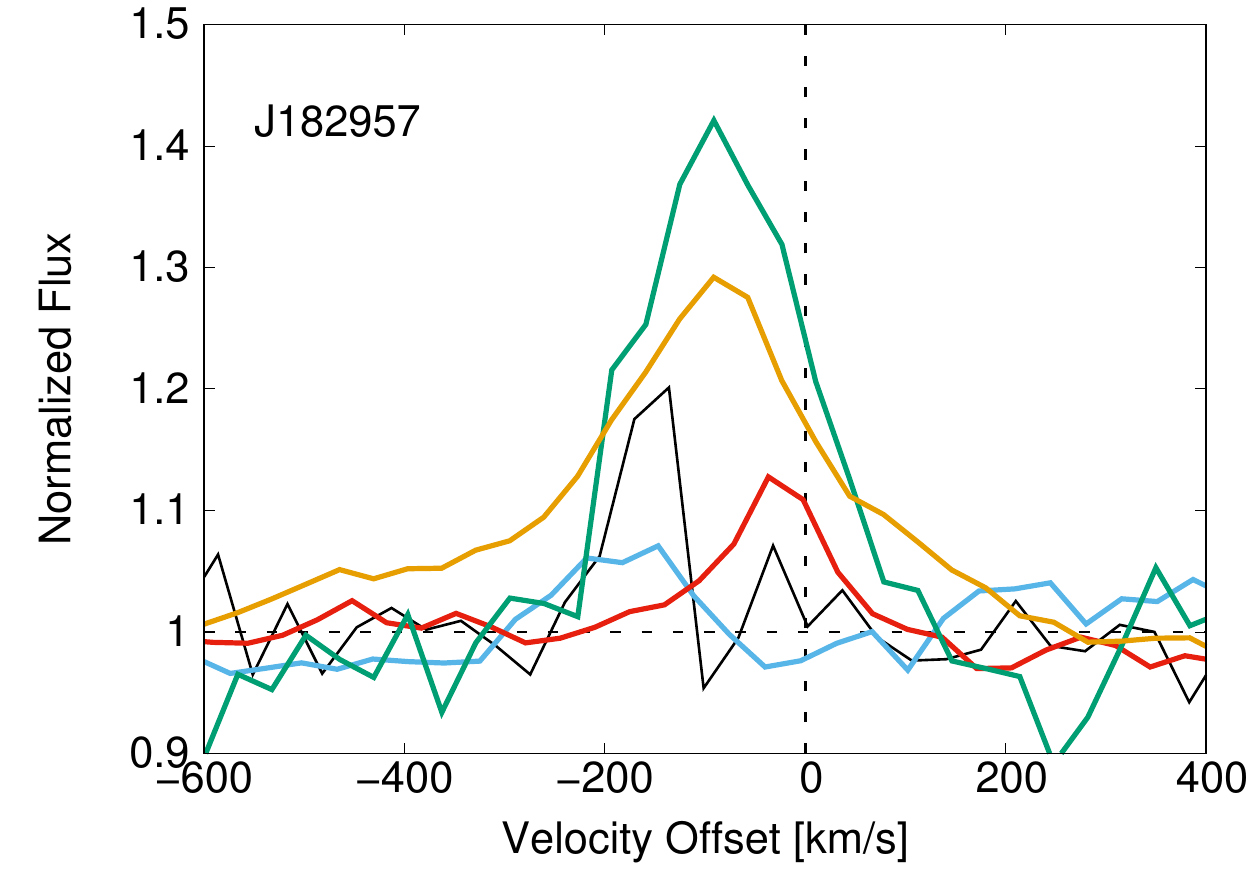}
     \includegraphics[width=2.5in]{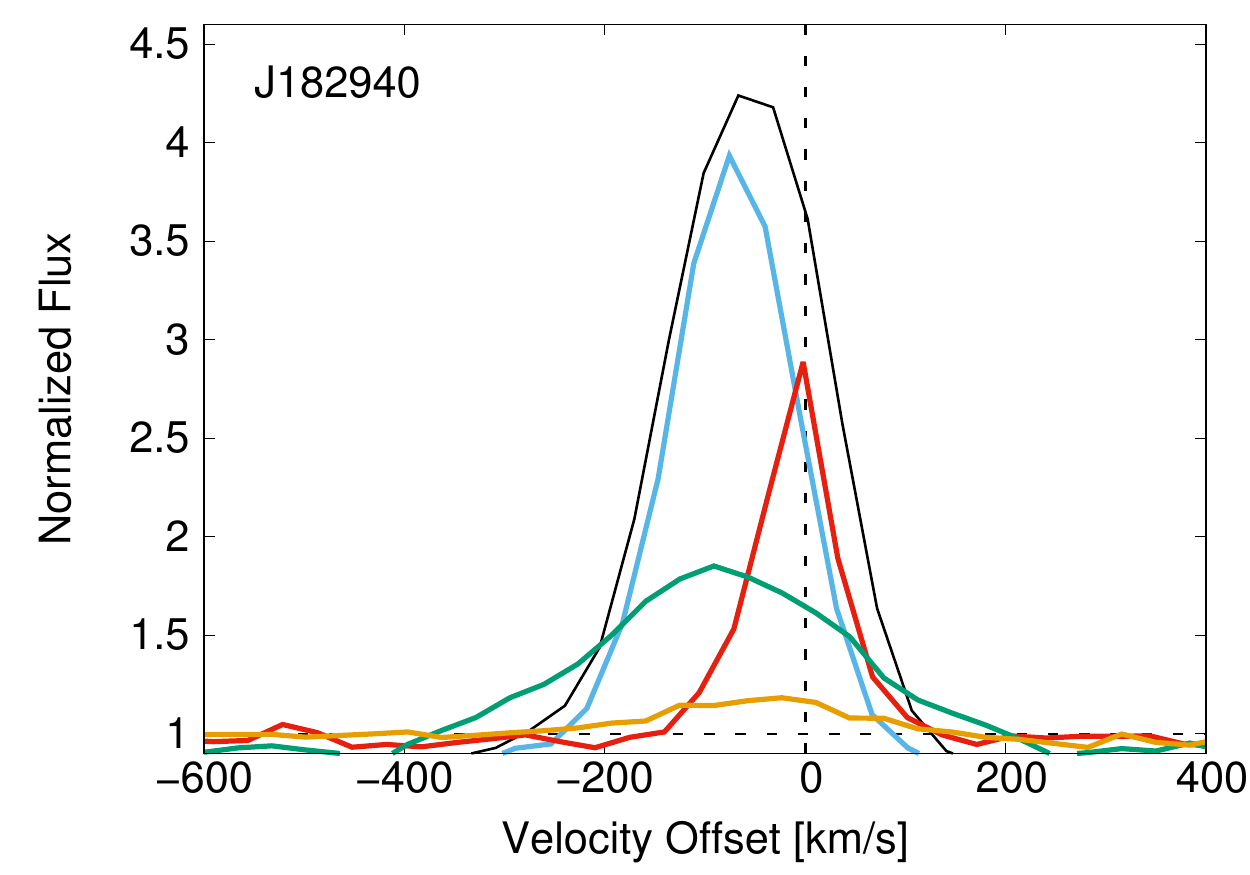}
     \includegraphics[width=2.5in]{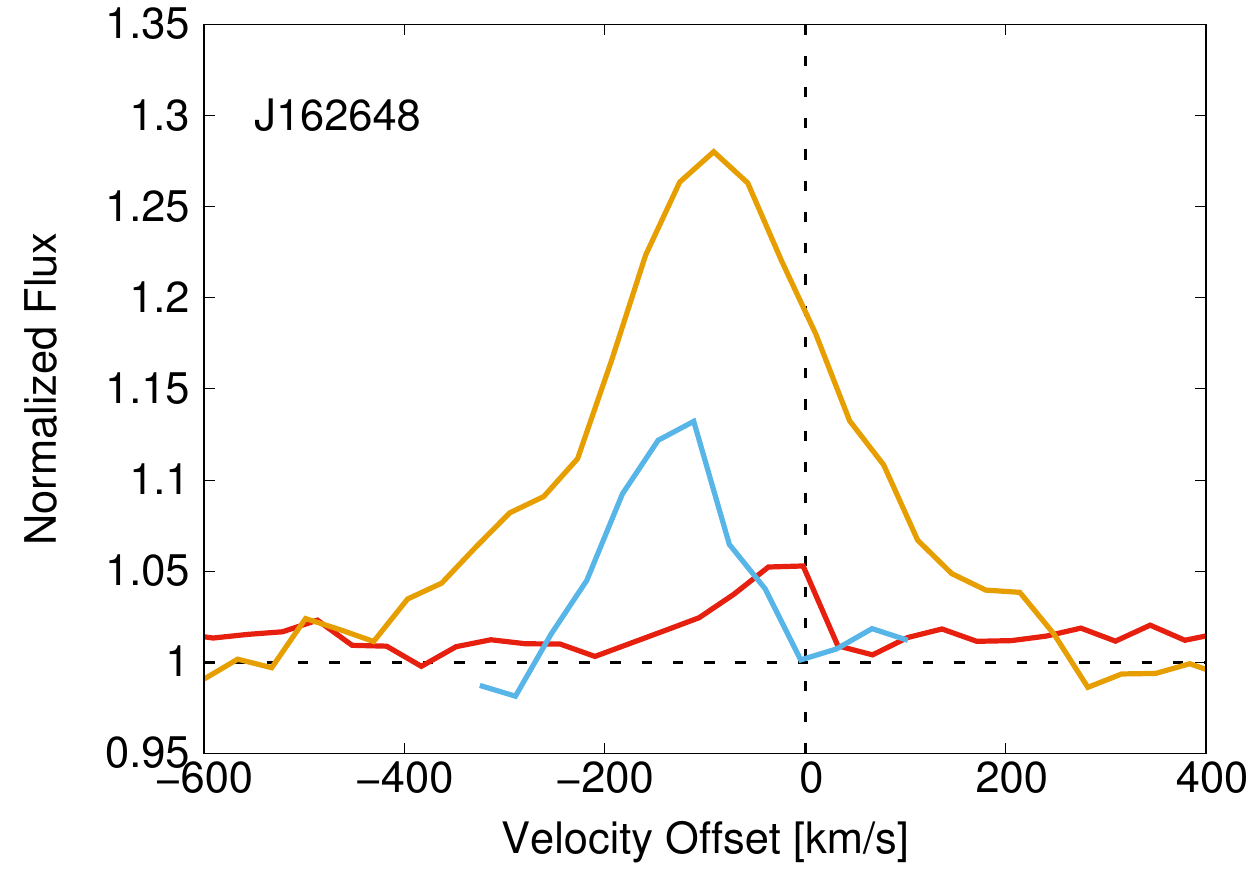}
     \includegraphics[width=2.5in]{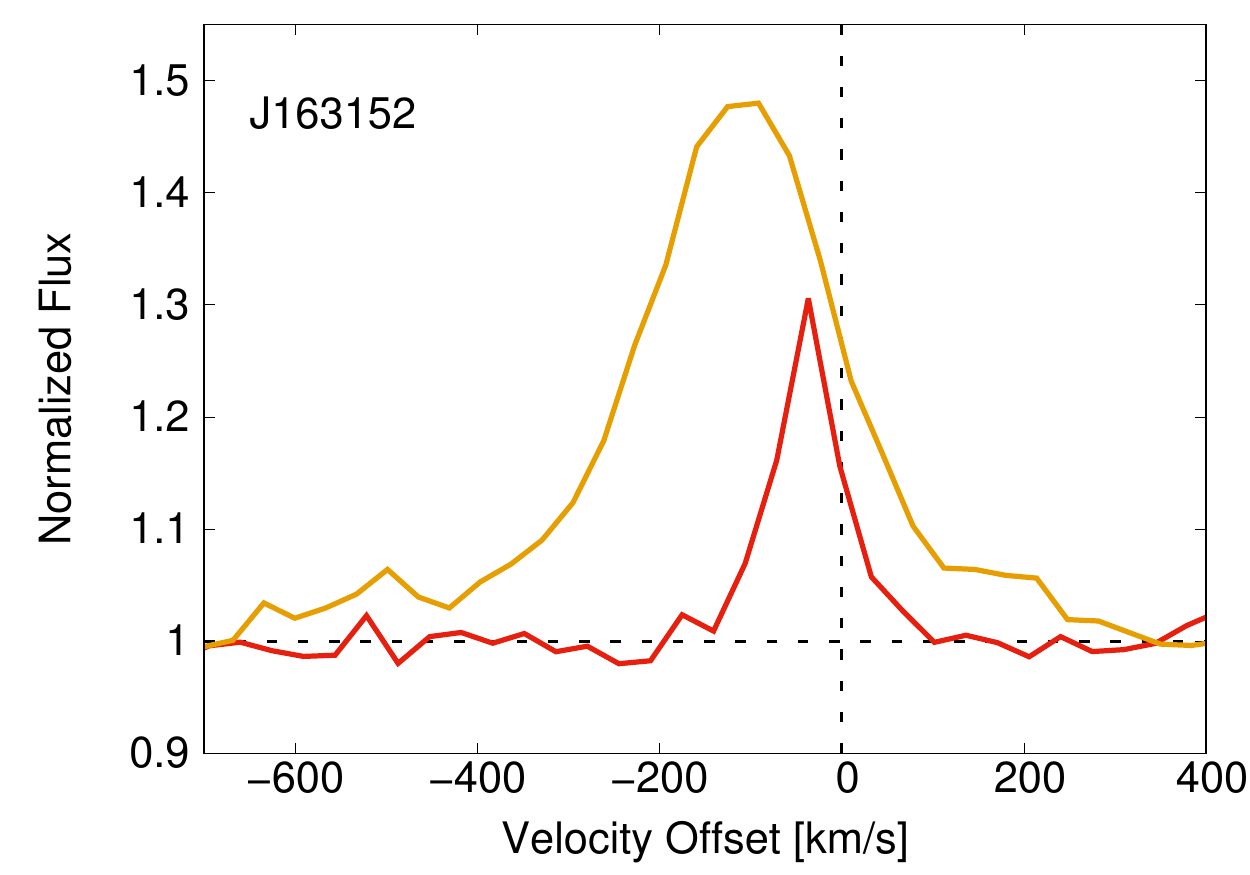}
     \caption{Continuum-normalized profiles of the brightest lines observed for the targets. Lines from [Fe~II] 1.257$\micron$ (black), [Fe~II] 1.644$\micron$ (blue), H$_{2}$ 2.12$\micron$ (red), Br$\gamma$ (orange), and Pa~$\beta$ (green) are plotted. All velocities are systemic. }
     \label{lines}
  \end{figure*}

Note that the accretion tracers may not sample well the accretion zone due to the obscuring and scattering effects of the envelope/disk around the source, as well as possible contamination from the outflow component. Figure~\ref{lines} shows a comparison of the accretion and outflow line diagnostics. The overlap in the line profiles suggests that the Pa$\beta$ and Br$\gamma$ profiles are likely made up of a suppressed accretion component plus an outflow component. While it is difficult to quantify the contribution from the outflow to the Pa$\beta$ and Br$\gamma$ line luminosities, previous works have noted the uncertainties due to such a contamination to be within the 20\%-30\% error estimated by propagating the error on the $\dot{M}_{acc}$ and $\dot{M}_{out}$ measurements, as discussed above. The large spread in the mass accretion rates for the proto-BD sample (Sect.~\ref{compare}) could be due to these factors. Additionally, no tracer is perfect. The outflow diagnostics of [Fe~II] and H$_{2}$ may only trace a portion of the outflow where shocks excite these lines while the quiescent gas in the outflow remains invisible. Thus, $\dot{M}_{out}$ based on any specific tracer is likely to be a lower-bound.


\begin{table}
\centering
\caption{Mass Accretion Rate}
\label{macc}
\begin{tabular}{lcccc} 
\hline
Designation 	& \multicolumn{2}{c}{L$_{acc}$ [L$_{\sun}$]}	& \multicolumn{2}{c}{$\dot{M}_{acc}$ [M$_{\sun}$ yr$^{-1}$]} 	 \\
\hline
			&     Pa$\beta$  	&  Br$\gamma$	&     Pa$\beta$  	&  Br$\gamma$				\\
\hline		    
M1701117 	& 0.1  &  0.6 		& (5$\pm$2)$\times$10$^{-8}$		& (3$\pm$1)$\times$10$^{-7}$  \\	

J163136 		& 1.25  	& 0.85	& (6$\pm$2)$\times$10$^{-7}$ 	& (4$\pm$1)$\times$10$^{-7}$ 	\\

J163152 		&	--	&	2.7	& --  & (2$\pm$0.4)$\times$10$^{-6}$	\\

J162648 		&	--	& 0.035 	& 	   	--     			     	& (2$\pm$0.5)$\times$10$^{-8}$ 	\\

J182957 		&  0.45 	&  0.55 	& (2$\pm$0.4)$\times$10$^{-7}$ 	& (3$\pm$0.6)$\times$10$^{-7}$	\\ 

J182940 		&	0.94	&  0.12	& (4$\pm$1)$\times$10$^{-7}$ & (6$\pm$2)$\times$10$^{-8}$ \\
\hline
\end{tabular}
\end{table}

\begin{table*}
\centering
\caption{Mass Outflow Rates}
\label{mout}
\begin{tabular}{lcccccc} 
\hline
			& M1701117 	& J163136 	& J163152 	& J162648 	& J182957 	& J182940 \\
\hline
Emission Line	& \multicolumn{6}{c}{$\dot{M}_{out}$ [M$_{\sun}$ yr$^{-1}$]} \\
\hline

\relax
[Fe~II]~$^{4}D_{7/2}~{\textemdash}~^{6}D_{9/2}$ (1.257$\micron$) 	& (5$\pm$1)$\times$10$^{-8}$ & (4$\pm$1)$\times$10$^{-9}$ & 	--		&	--		& (8$\pm$2)$\times$10$^{-9}$	&	(4$\pm$1)$\times$10$^{-8}$	\\
\relax
[Fe~II]~$^{4}D_{5/2}~{\textemdash}~^{4}F_{9/2}$ (1.533$\micron$) 	&	--		&	(2$\pm$1)$\times$10$^{-8}$	&	--		&	(1$\pm$0.2)$\times$10$^{-9}$	&	(1$\pm$0.2)$\times$10$^{-8}$	&	(2$\pm$0.5)$\times$10$^{-8}$	\\

\relax
[Fe~II]~$^{4}D_{3/2}~{\textemdash}~^{4}F_{7/2}$ (1.603$\micron$) 	&	--		&	(2$\pm$1)$\times$10$^{-8}$		&	--		&	(6$\pm$1.5)$\times$10$^{-8}$	&	(1$\pm$0.2)$\times$10$^{-7}$		&	--	\\
\relax
[Fe~II]~$^{4}D_{7/2}~{\textemdash}~^{4}F_{9/2}$ (1.644$\micron$) 	&	(7$\pm$2)$\times$10$^{-8}$	&	(4$\pm$1)$\times$10$^{-9}$		&	--		&	(2$\pm$0.5)$\times$10$^{-9}$	&	(8$\pm$2)$\times$10$^{-9}$	&	(3$\pm$0.5)$\times$10$^{-8}$	\\
\relax
[Fe~II]~$^{4}D_{3/2}~{\textemdash}~^{4}F_{5/2}$ (1.712$\micron$) 	&	--		&	(2$\pm$1)$\times$10$^{-8}$		&	--		&	--	&	(2$\pm$0.6)$\times$10$^{-8}$	&	--	\\

\hline

H$_{2}$ (1-0) $S$(2)	 (2.034$\micron$) 	&	(4$\pm$1)$\times$10$^{-9}$	&	(4$\pm$1)$\times$10$^{-10}$	& (5$\pm$1)$\times$10$^{-10}$	&	--		&	--		&	(3$\pm$0.5)$\times$10$^{-10}$	\\

H$_{2}$ (1-0) $S$(2)	 (2.073$\micron$) 	&	(5$\pm$1.5)$\times$10$^{-9}$	&		--	&	--		&	--		&	--		&	--	\\	

H$_{2}$ (1-0) $S$(1)	 (2.12$\micron$) 	&	(1$\pm$0.3)$\times$10$^{-8}$	&	(1$\pm$0.2)$\times$10$^{-9}$	&	(5$\pm$1)$\times$10$^{-10}$	&	(8$\pm$2)$\times$10$^{-10}$	&	(1$\pm$0.3)$\times$10$^{-10}$	&	(6$\pm$2)$\times$10$^{-11}$	\\

H$_{2}$ (2-1) $S$(2)	 (2.154$\micron$) 	&	(2$\pm$0.5)$\times$10$^{-9}$	&	--		&	--		&		--	&		--	&	--	\\	

H$_{2}$ (1-0) $S$(0)	 (2.22$\micron$) 	&	(4$\pm$1)$\times$10$^{-9}$		&	(2$\pm$0.6)$\times$10$^{-9}$	&	(8$\pm$2)$\times$10$^{-10}$	&	--		&	--		&	(5$\pm$2)$\times$10$^{-11}$	\\

H$_{2}$ (2-1) $S$(1)	 (2.248$\micron$) 	&	(2$\pm$0.4)$\times$10$^{-10}$	&		--	&	--		&	--		&	--		&	(1$\pm$0.2)$\times$10$^{-11}$	\\	

H$_{2}$ (1-0) $Q$(1) (2.406$\micron$) 	&	(2$\pm$0.5)$\times$10$^{-9}$ 		&	(7$\pm$3)$\times$10$^{-10}$	&	(8$\pm$1)$\times$10$^{-10}$	&	--		&	(2$\pm$0.4)$\times$10$^{-10}$		&	(5$\pm$2)$\times$10$^{-11}$	\\

H$_{2}$ (1-0) $Q$(2) (2.413$\micron$) 	&	(2$\pm$1)$\times$10$^{-9}$		&	(6$\pm$2)$\times$10$^{-10}$	&	(5$\pm$1)$\times$10$^{-10}$	&	--		&	--		&	(3$\pm$1)$\times$10$^{-11}$	\\

H$_{2}$ (1-0) $Q$(3) (2.424$\micron$) 	&	(2$\pm$1)$\times$10$^{-9}$		&	(3$\pm$0.8)$\times$10$^{-10}$	&	(8$\pm$2)$\times$10$^{-10}$	&	--		&	(6$\pm$2)$\times$10$^{-11}$		&	--	\\

H$_{2}$ (1-0) $Q$(4) (2.437$\micron$) 	&	(1.5$\pm$0.3)$\times$10$^{-9}$	&	--						&	(1$\pm$0.2)$\times$10$^{-9}$		&	--		&	--		&	--	\\

H$_{2}$ (1-0) $Q$(5) (2.455$\micron$) 	&	(5$\pm$2)$\times$10$^{-10}$		&	--						&	--								&	--		&	--		&	(3$\pm$1)$\times$10$^{-11}$	\\

\hline
\end{tabular}
\end{table*}

\section{Results}
\label{results}

\subsection{M1701117}

\begin{figure*}
  \centering              
     \includegraphics[width=3in]{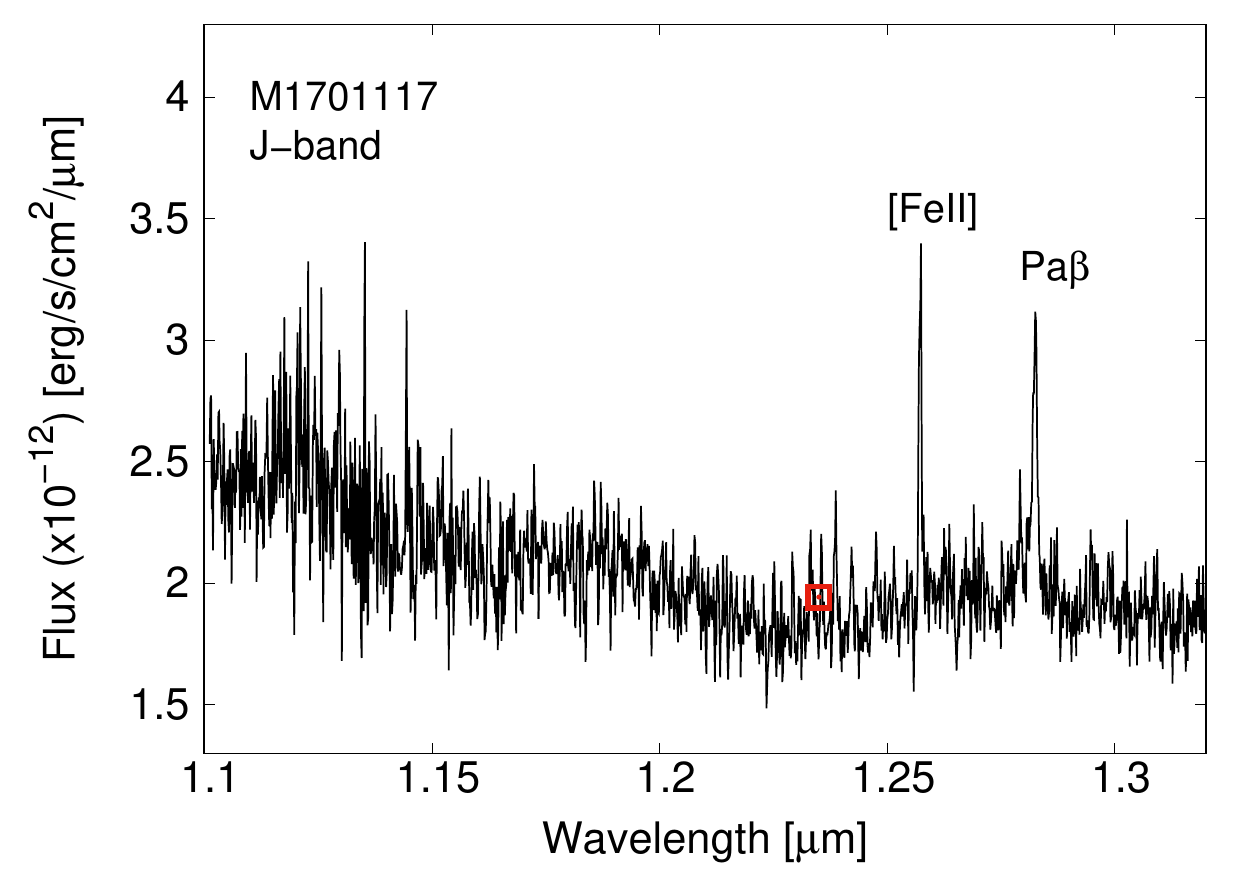}
     \includegraphics[width=3in]{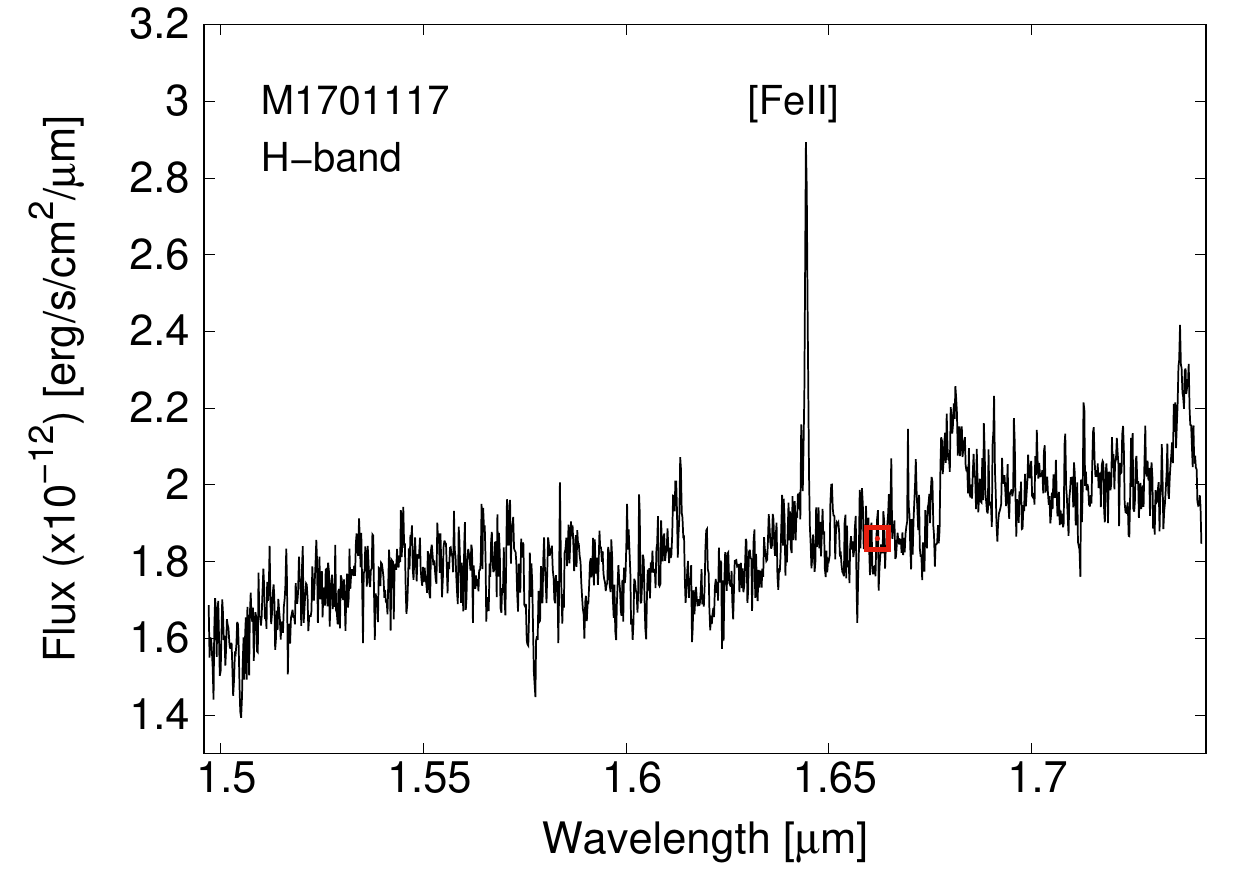}
     \includegraphics[width=3in]{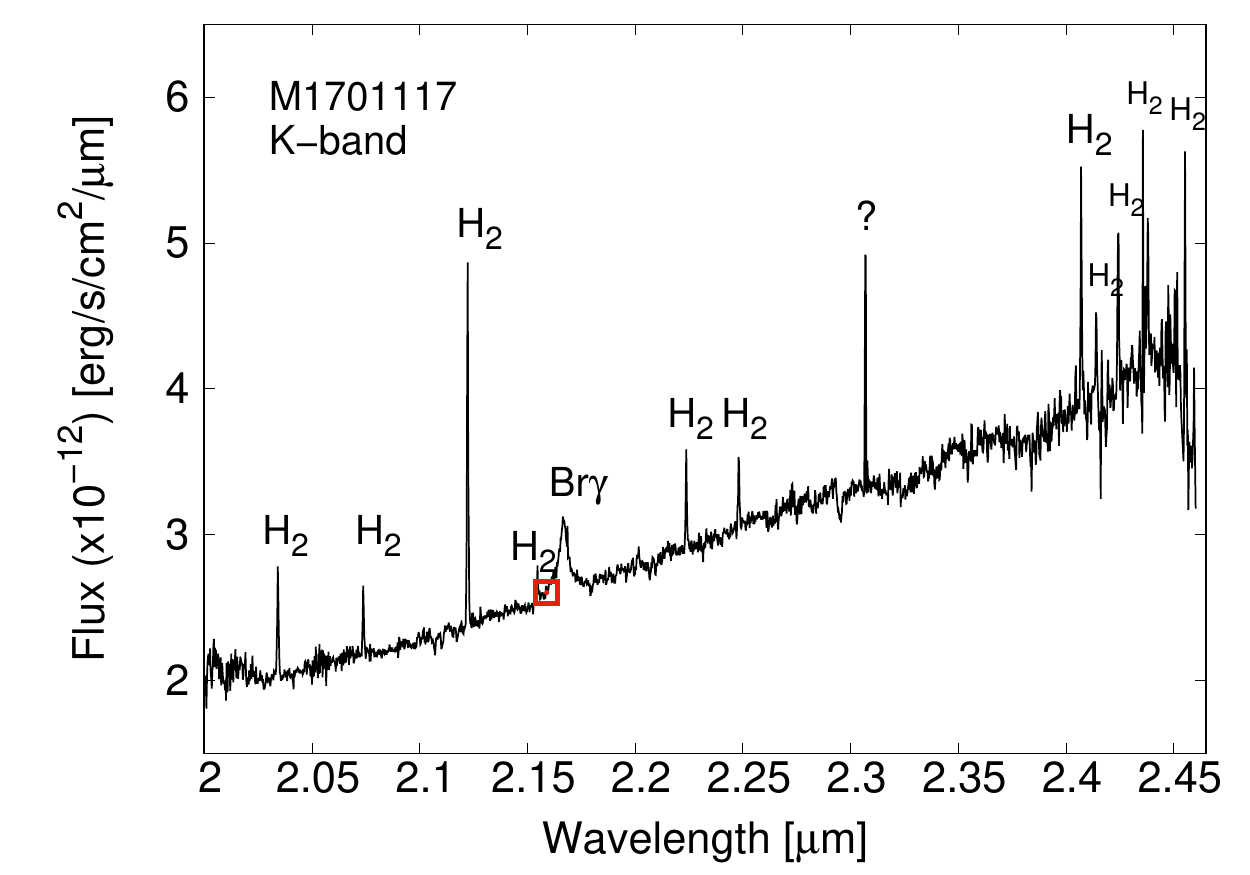}     
     \caption{The near-infrared spectra for M1701117. The prominent accretion and outflow tracers are labelled. The 2MASS/UKIDSS photometry is marked by the red open square.  }
     \label{M170-spec}
  \end{figure*}

M1701117 shows strong emission in the [Fe~II] 1.257$\micron$ and 1.644$\micron$ lines, and several H$_{2}$ line detections from different upper vibrational energy levels ranging from $\nu$ = 1 to $\nu$ = 2 (Fig.~\ref{M170-spec}; Table~\ref{M170-lines}). The flux in the higher vibrational lines are $\sim$2-10 times lower than the brightest H$_{2}$ (1-0) $S$(1) line at 2.12 $\micron$, implying that the H$_{2}$ emission is mainly tracing the cold gas component at E$_{(v,J)} <$ 6000 K. The gas temperature derived from the H$_{2}$ ($\nu$ = 1) rotational diagram is $\sim$3000 K. The weak detection in the $\nu$ = 2 H$_{2}$ lines indicates the presence of a hot gas component at E$_{(v,J)} \sim$ 6000-20,000 K. 

M1701117 shows emission in both the Pa$\beta$ and Br$\gamma$ lines (Fig.~\ref{lines}). The Br$\gamma$ line is much broader (FWHM $\sim$600 km s$^{-1}$) compared to Pa$\beta$ (FWHM $\sim$270 km s$^{-1}$). The mass accretion rate derived from the Pa$\beta$ line is 5$\times$10$^{-8}$ M$_{\sun}$ yr$^{-1}$, an order of magnitude lower than the accretion rate of 3$\times$10$^{-7}$ M$_{\sun}$ yr$^{-1}$ derived from Br$\gamma$. The average mass outflow rate derived for M1701117 using the [Fe~II] lines is 8$\times$10$^{-8}$ M$_{\sun}$ yr$^{-1}$, about an order of magnitude higher than the average rate of 3$\times$10$^{-9}$ M$_{\sun}$ yr$^{-1}$ derived using the H$_{2}$ lines (Table~\ref{mout}). This implies a jet efficiency of $\dot{M}_{out}$[Fe~II]/$\dot{M}_{acc}$ $\sim$0.5 and $\dot{M}_{out}$H$_{2}$/$\dot{M}_{acc}$ $\sim$0.02, using the mean accretion rate derived from the Pa$\beta$ and Br$\gamma$ lines. 

\begin{figure*}
  \centering              
     \includegraphics[width=2.95in]{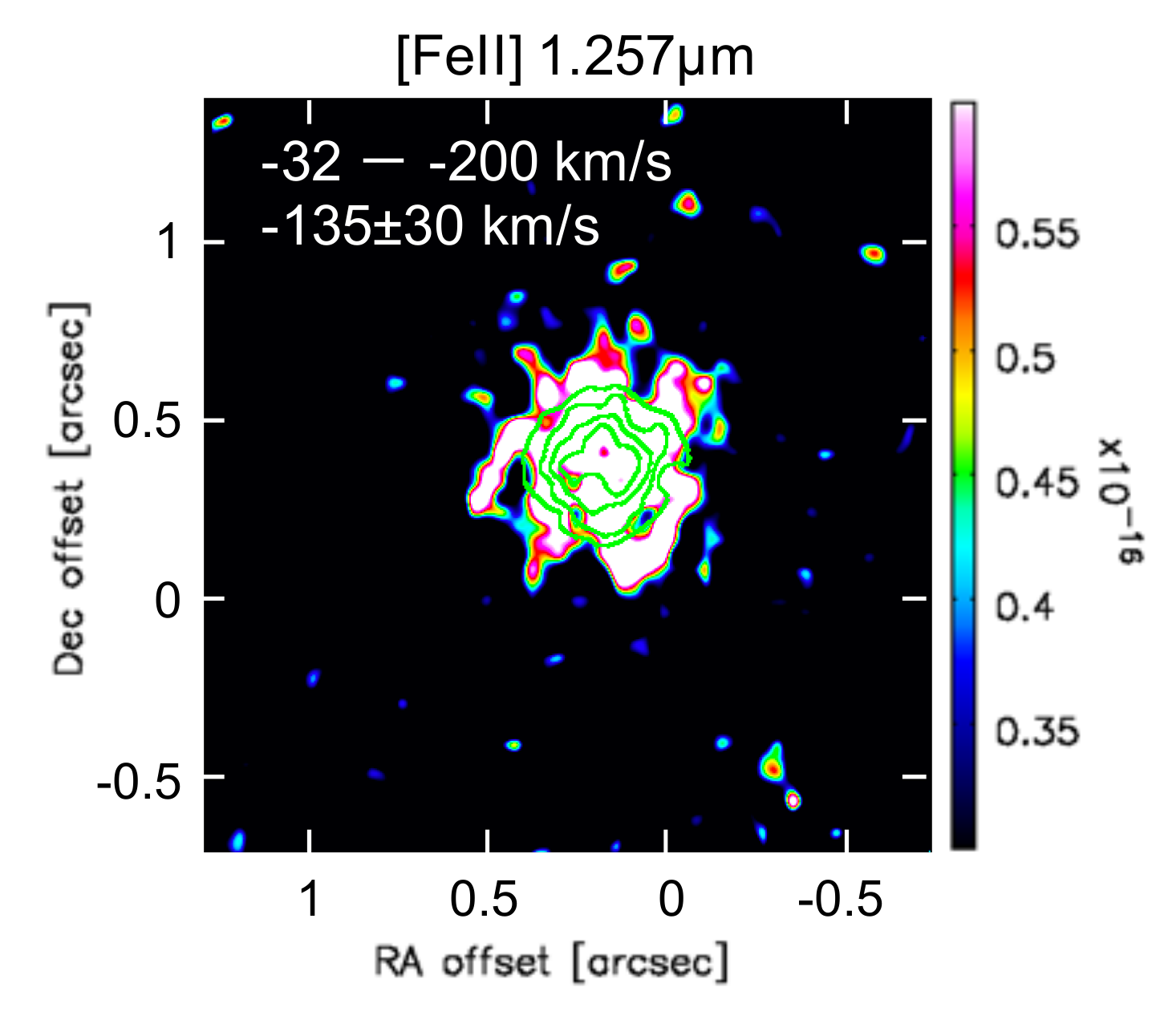}
     \includegraphics[width=3in]{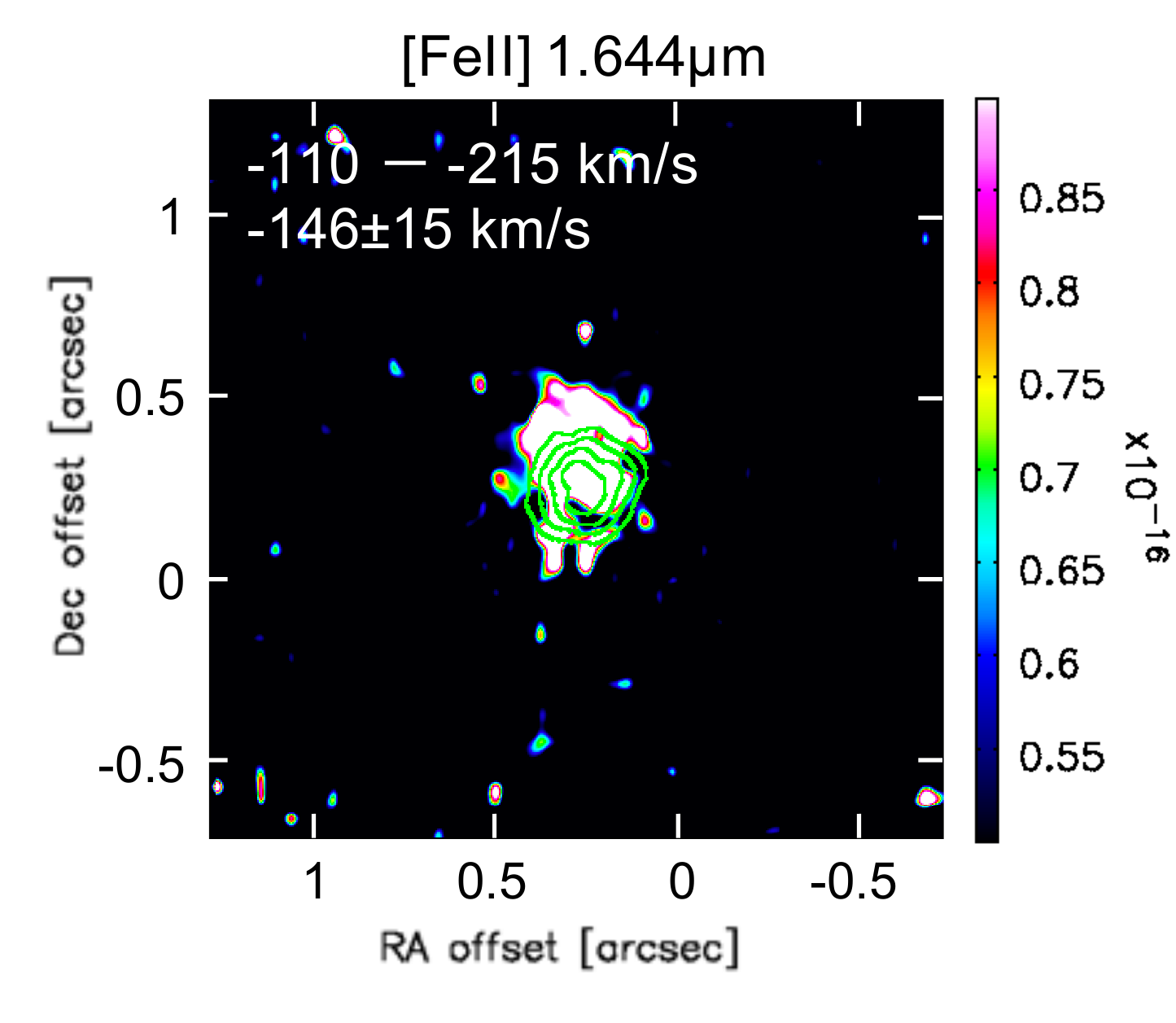}
     \includegraphics[width=3in]{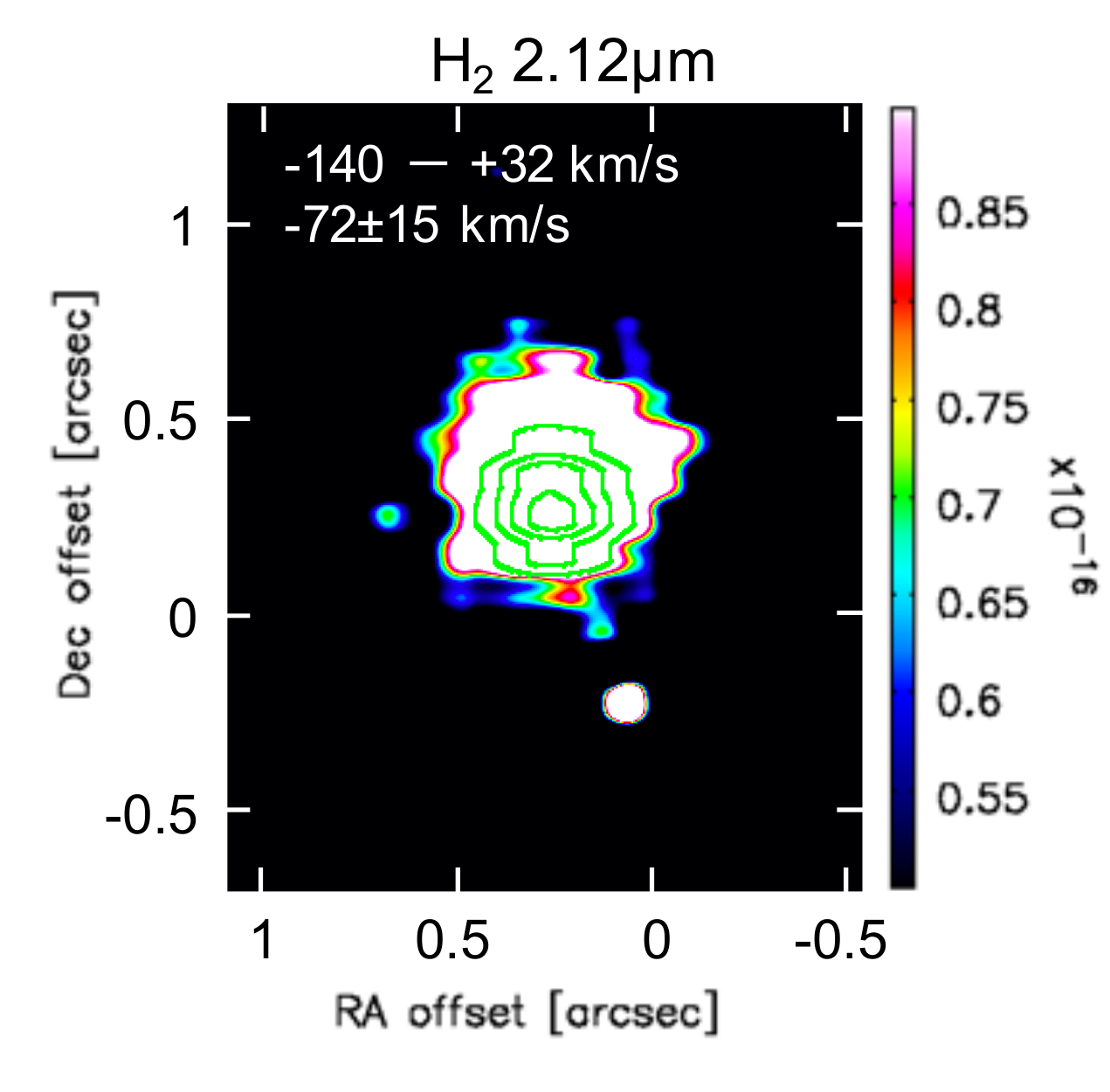}
     \caption{The continuum-subtracted integrated intensity line images for M1701117. The intensity is integrated over the range in velocities where line emission is seen. The velocity range and the peak velocity are shown at the top, left corner. The continuum emission is overplotted in green contours. The colour bar on the right shows the integrated flux in units of (erg s$^{-1}$ cm$^{-2}$). North is up, east is to the left.  }
     \label{M170-imgs}
  \end{figure*}

Figure~\ref{M170-imgs} shows the spectro-images in the [Fe~II] 1.257$\micron$ and 1.644$\micron$ lines, and the H$_{2}$ 2.12 $\micron$ lines, the brightest ones detected for this jet. No extended emission is seen in any of these line images. The most compact, point-like structure is seen in the [Fe~II] 1.644$\micron$ line image, while the brightest and broadest emission is seen in the H$_{2}$ 2.12 $\micron$ image. Figure~\ref{lines} shows a comparison of these line profiles. The [Fe~II] lines are broader (FWHM $\sim$ 170--190 km s$^{-1}$) than the H$_{2}$ line (FWHM $\sim$ 84 km s$^{-1}$). The peak in emission is blue-shifted in all of these lines. The peak velocity of the H$_{2}$ 2.12 $\micron$ line emission (-72 km s$^{-1}$) is comparatively lower than the [Fe~II] 1.257$\micron$ (-169 km s$^{-1}$) and the [Fe~II] 1.644$\micron$ (-146 km s$^{-1}$) lines (Fig.~\ref{M170-imgs}; Table~\ref{M170-lines}).




 \begin{figure*}
  \centering              
     \includegraphics[width=3in]{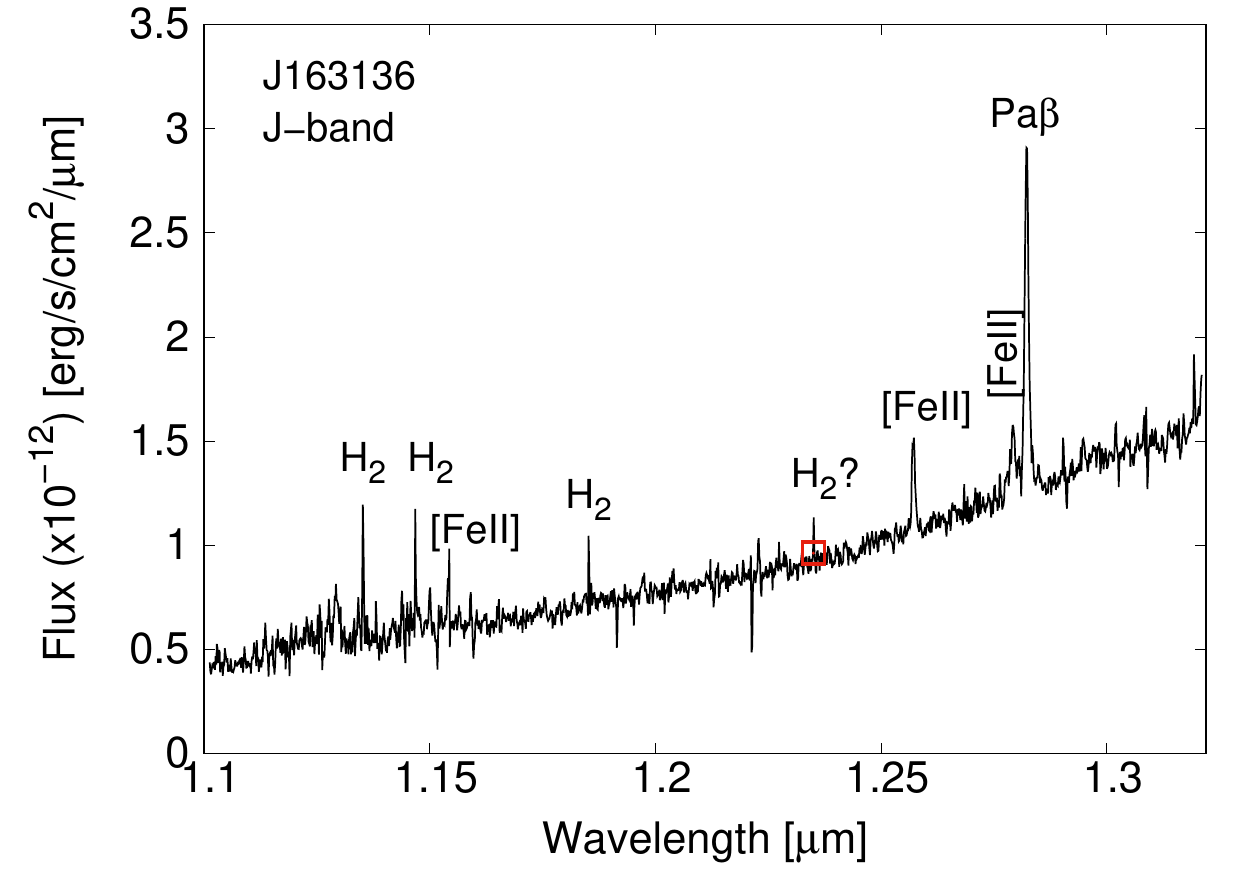}  
     \includegraphics[width=3in]{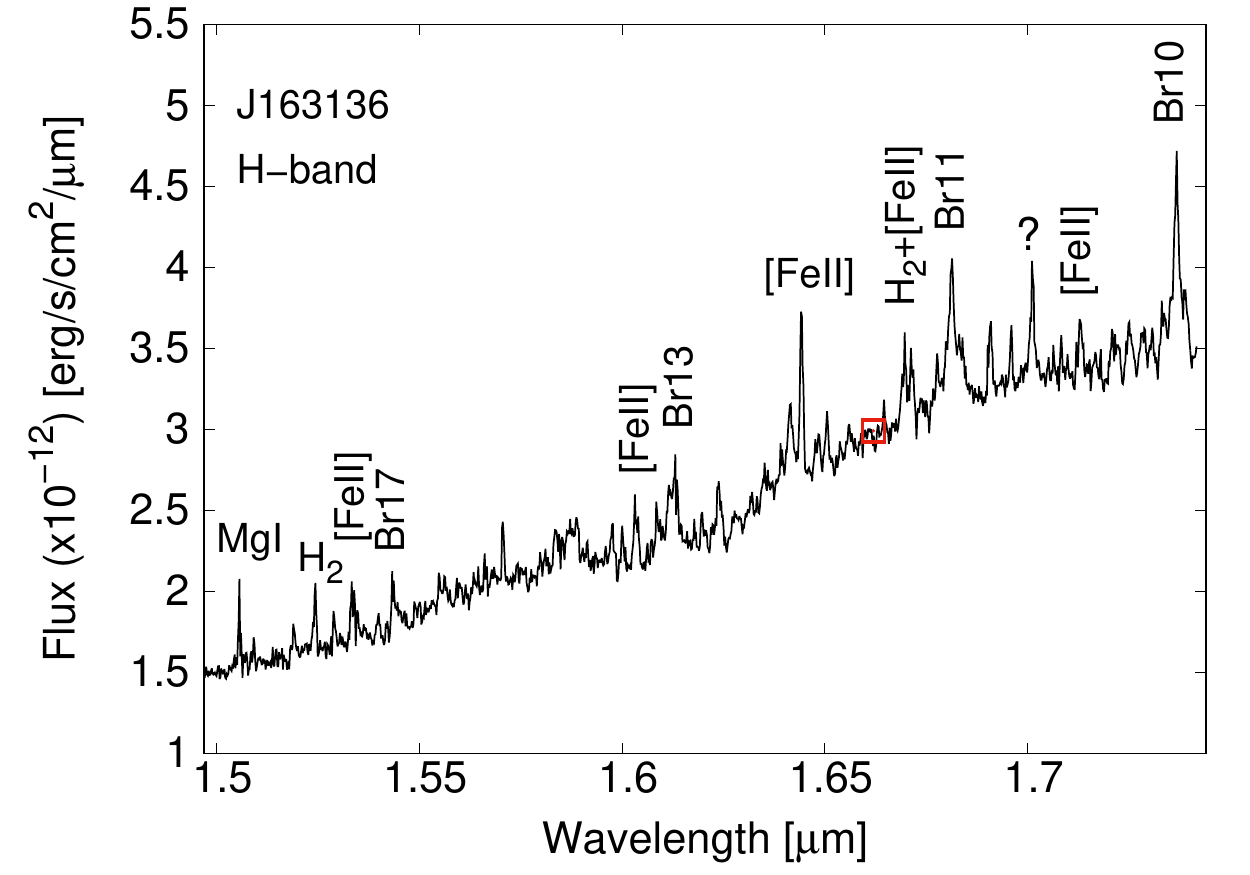}
     \includegraphics[width=3in]{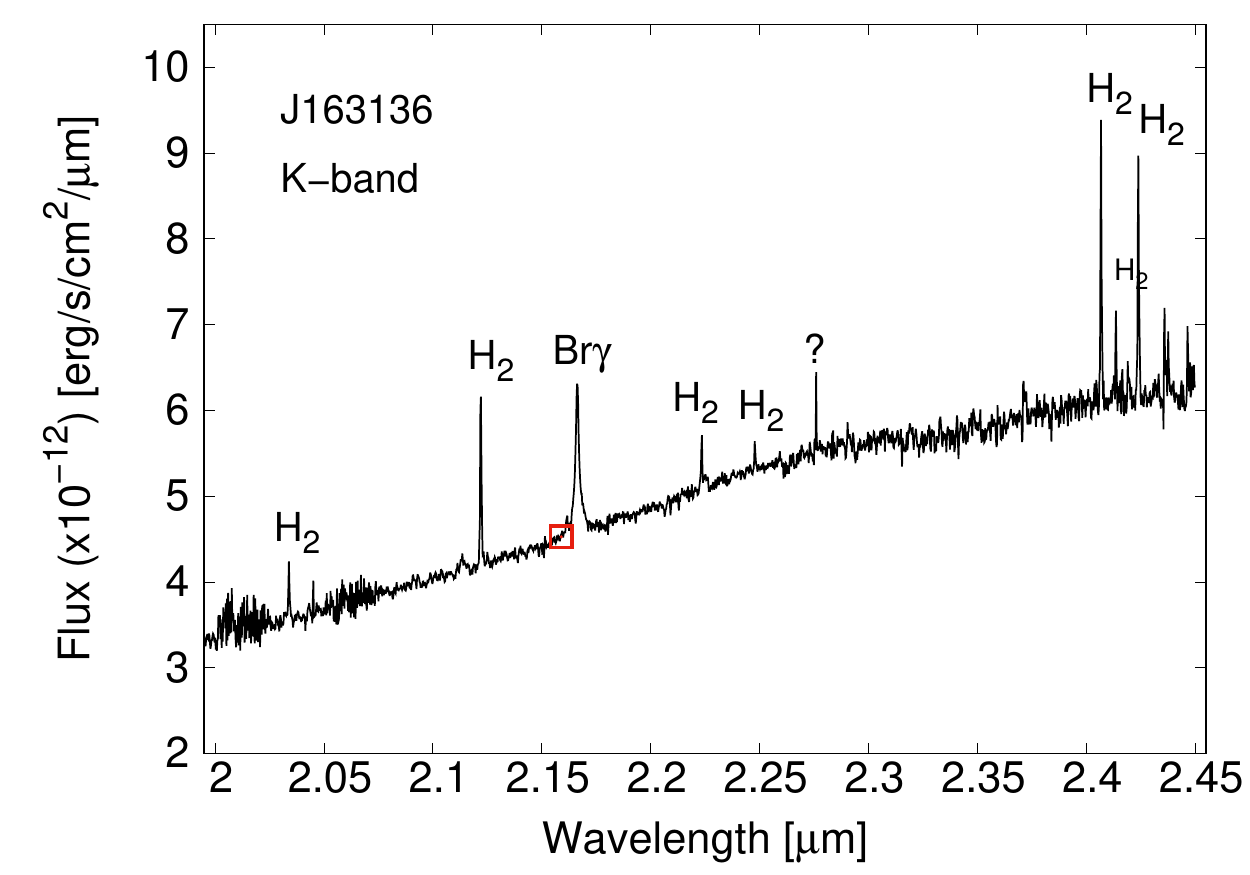}     
     \caption{The near-infrared spectra for J163136. The prominent accretion and outflow tracers are labelled. The 2MASS/UKIDSS photometry is marked by the red open square.   }
     \label{oph1-spec}
  \end{figure*}

 \begin{figure*}
  \centering              
     \includegraphics[width=3in]{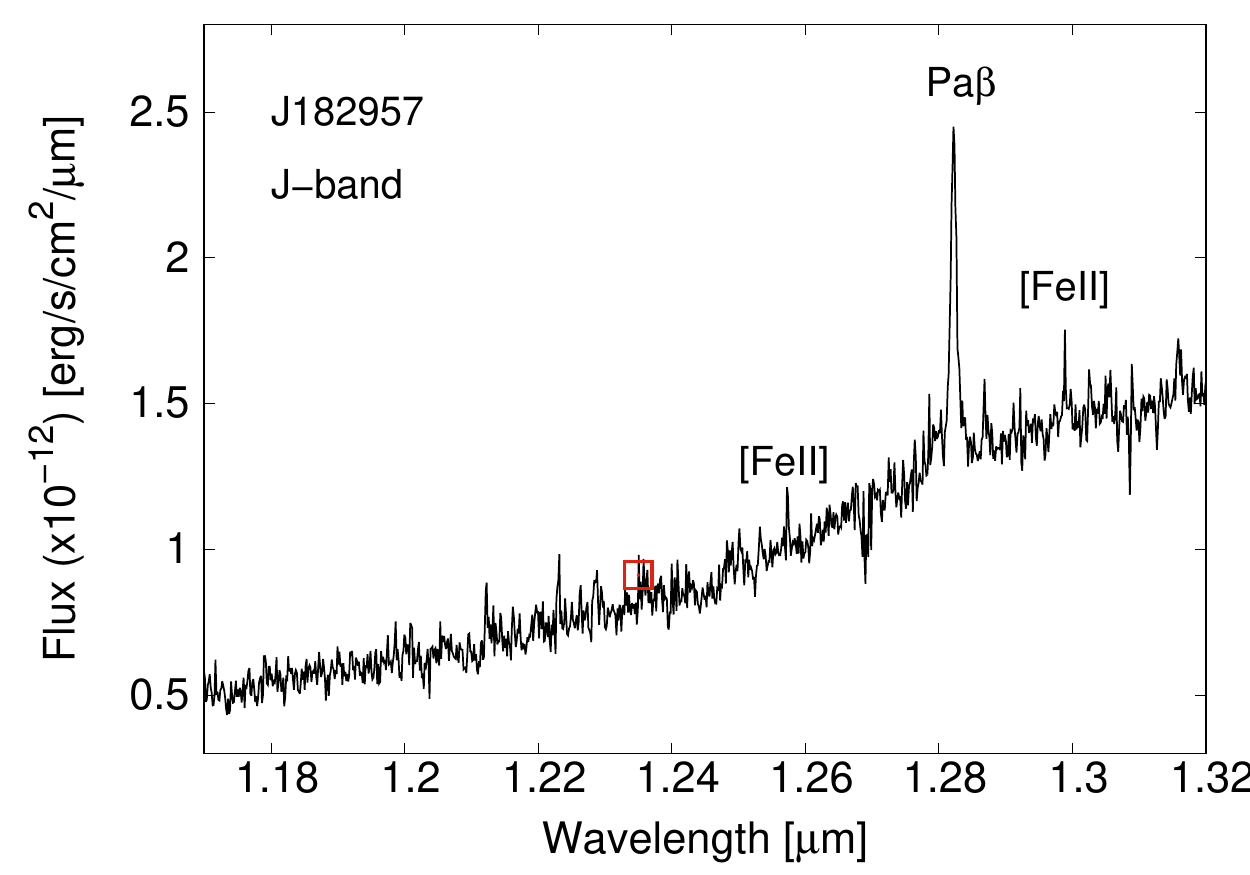}
     \includegraphics[width=3in]{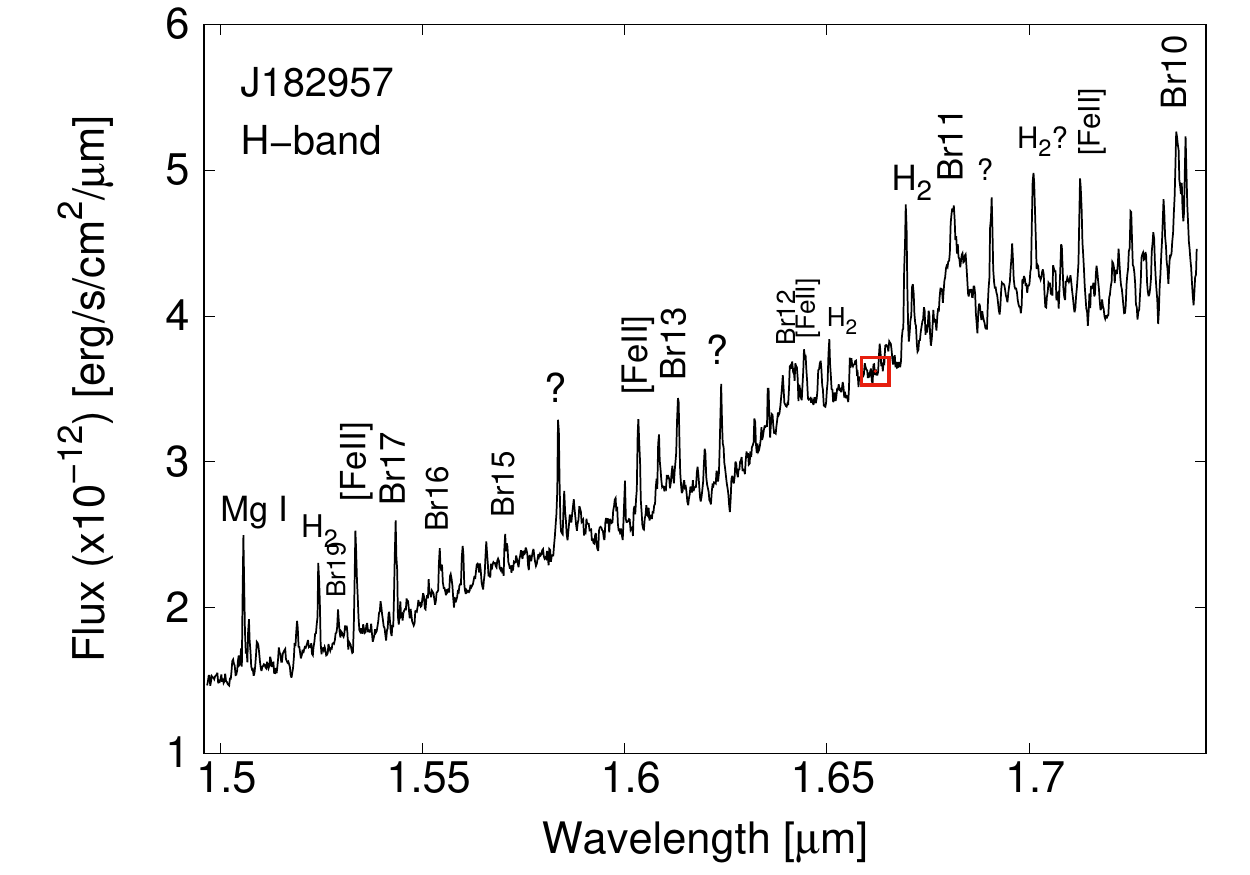}
     \includegraphics[width=3in]{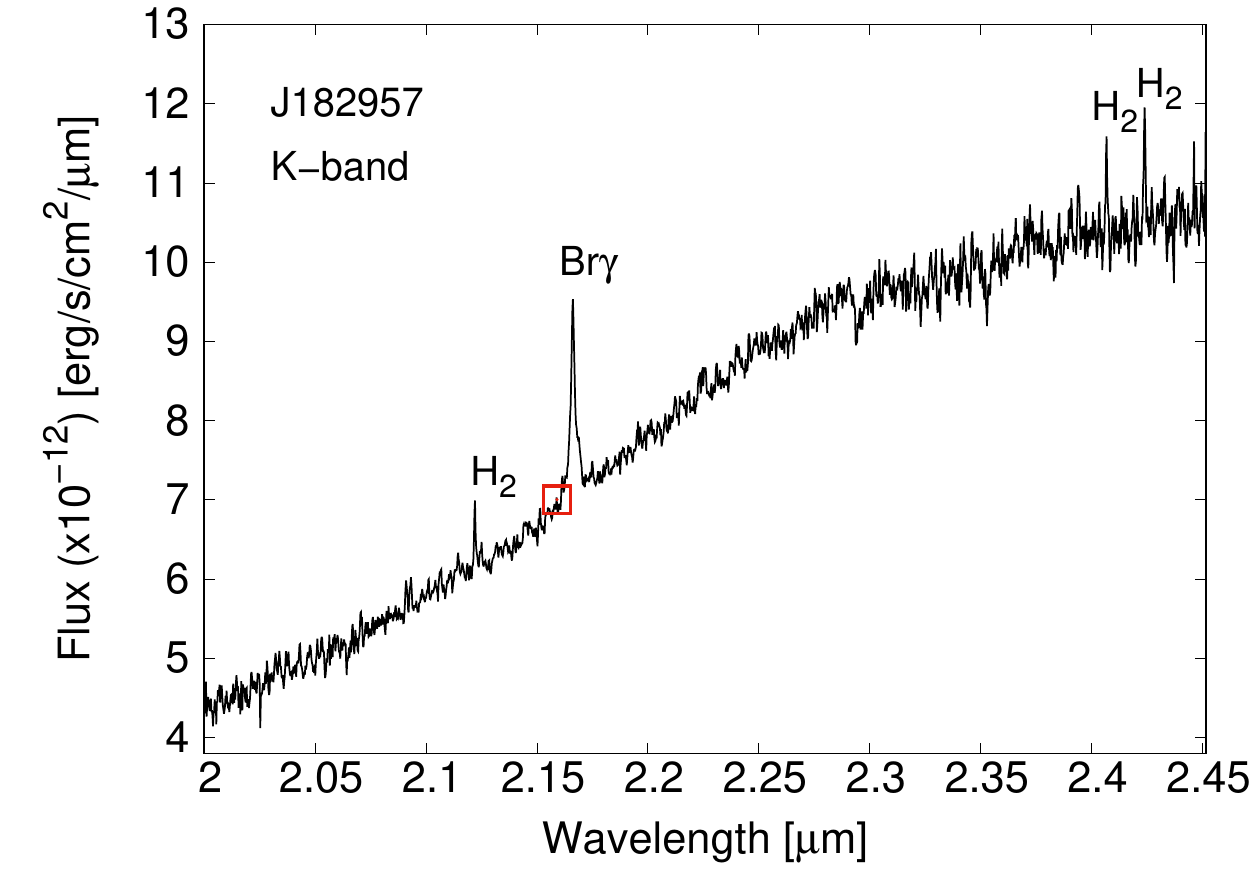}     
     \caption{The near-infrared spectra for J182957. The prominent accretion and outflow tracers are labelled. The 2MASS/UKIDSS photometry is marked by the red open square.   }
     \label{ser1-spec}
  \end{figure*}

 \begin{figure*}
  \centering              
     \includegraphics[width=3in]{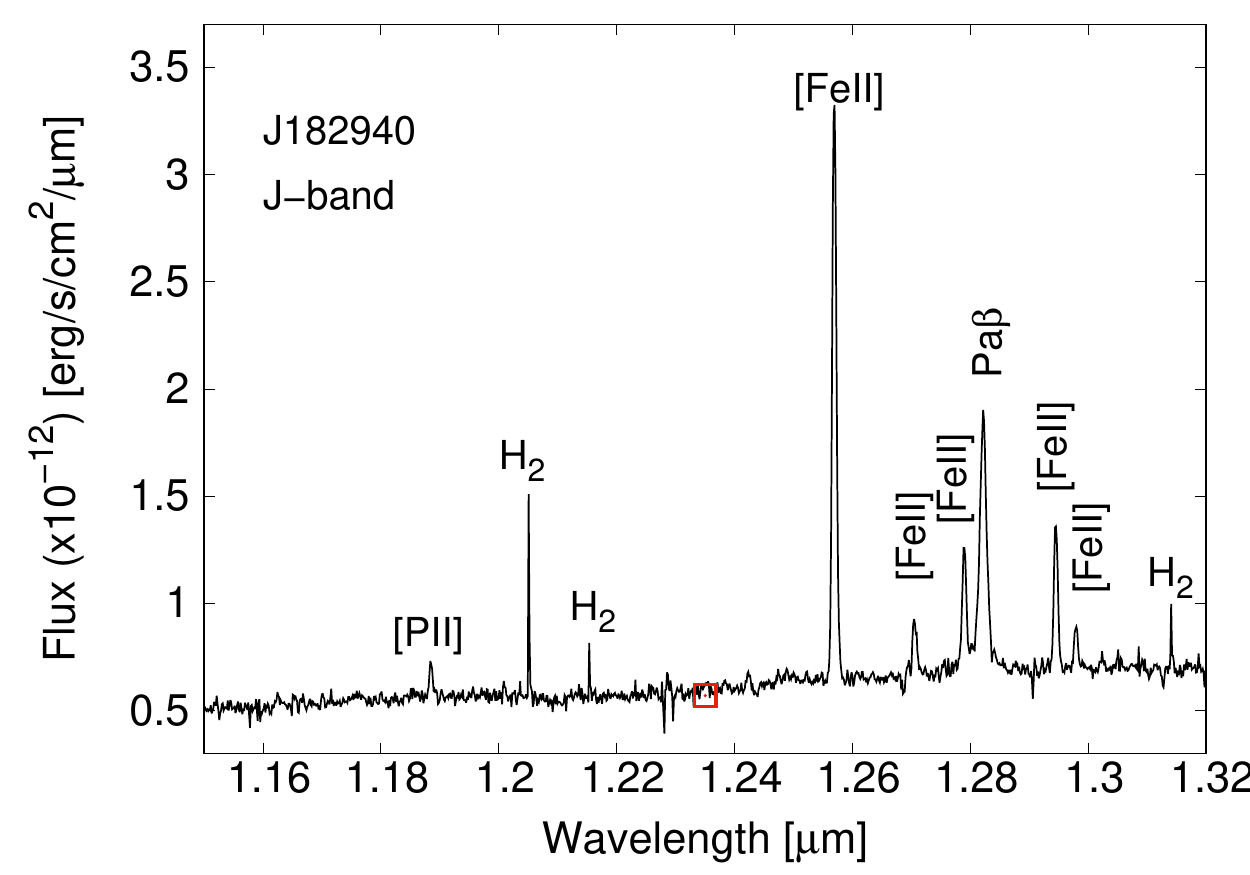}
     \includegraphics[width=3in]{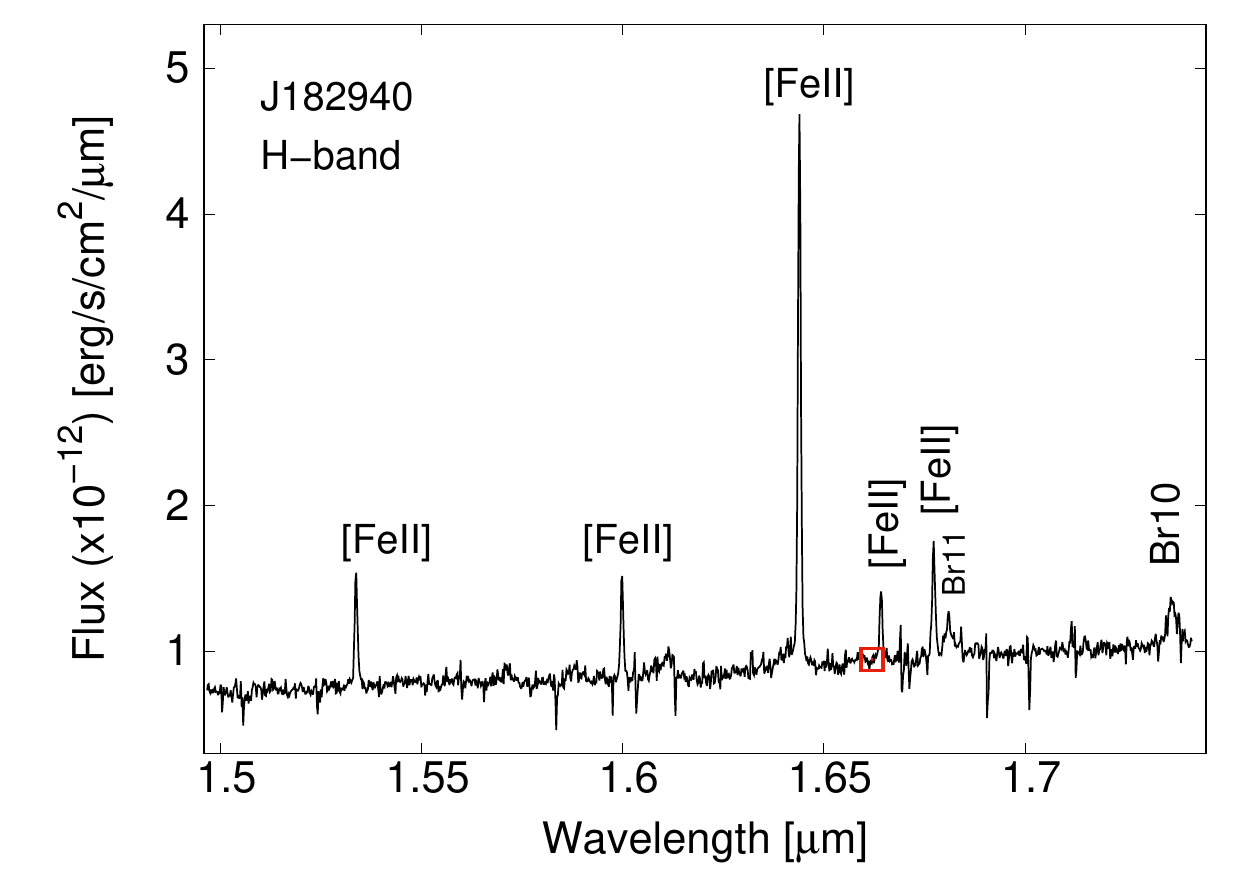}
     \includegraphics[width=3in]{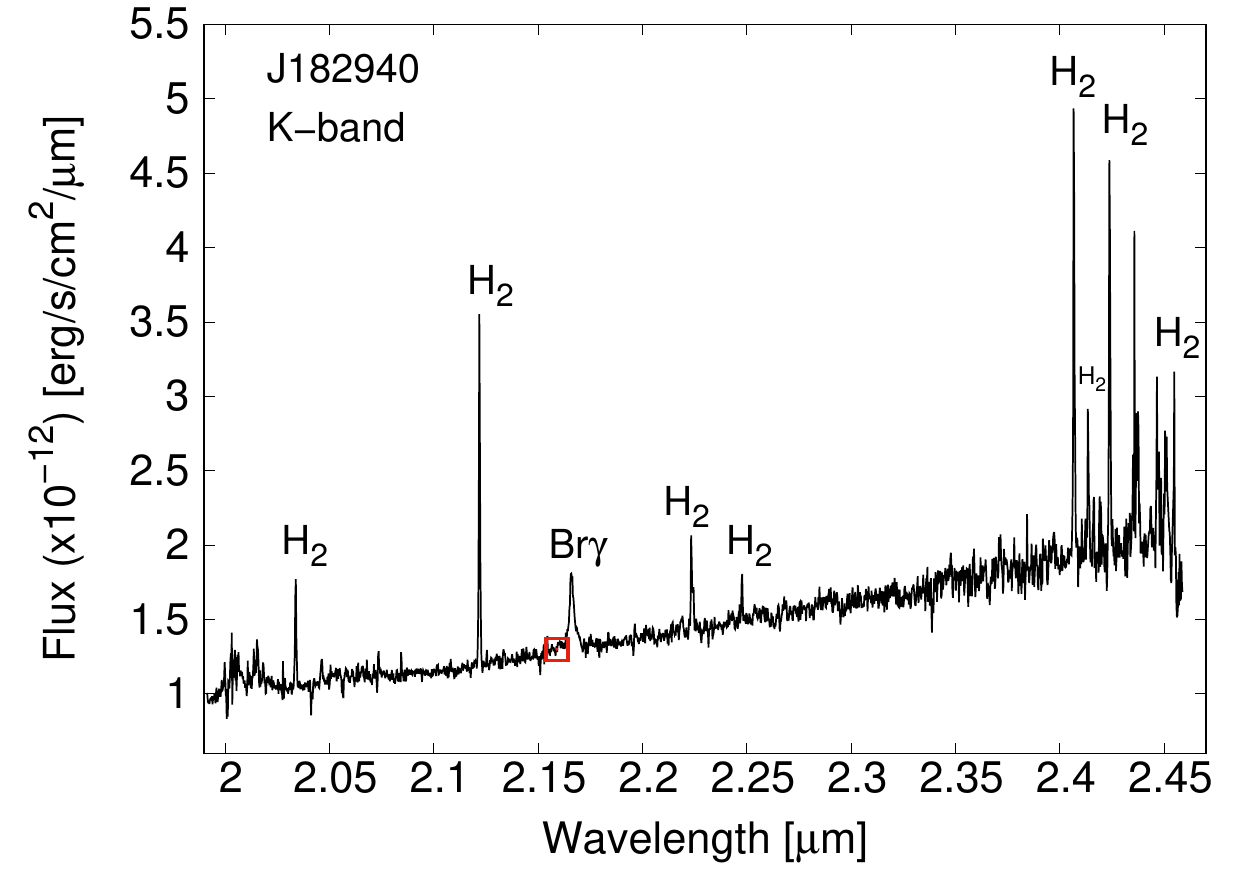}     
     \caption{The near-infrared spectra for J182940. The prominent accretion and outflow tracers are labelled. The 2MASS/UKIDSS photometry is marked by the red open square.   }
     \label{ser8-spec}
  \end{figure*}

 \begin{figure*}
  \centering              
     \includegraphics[width=3in]{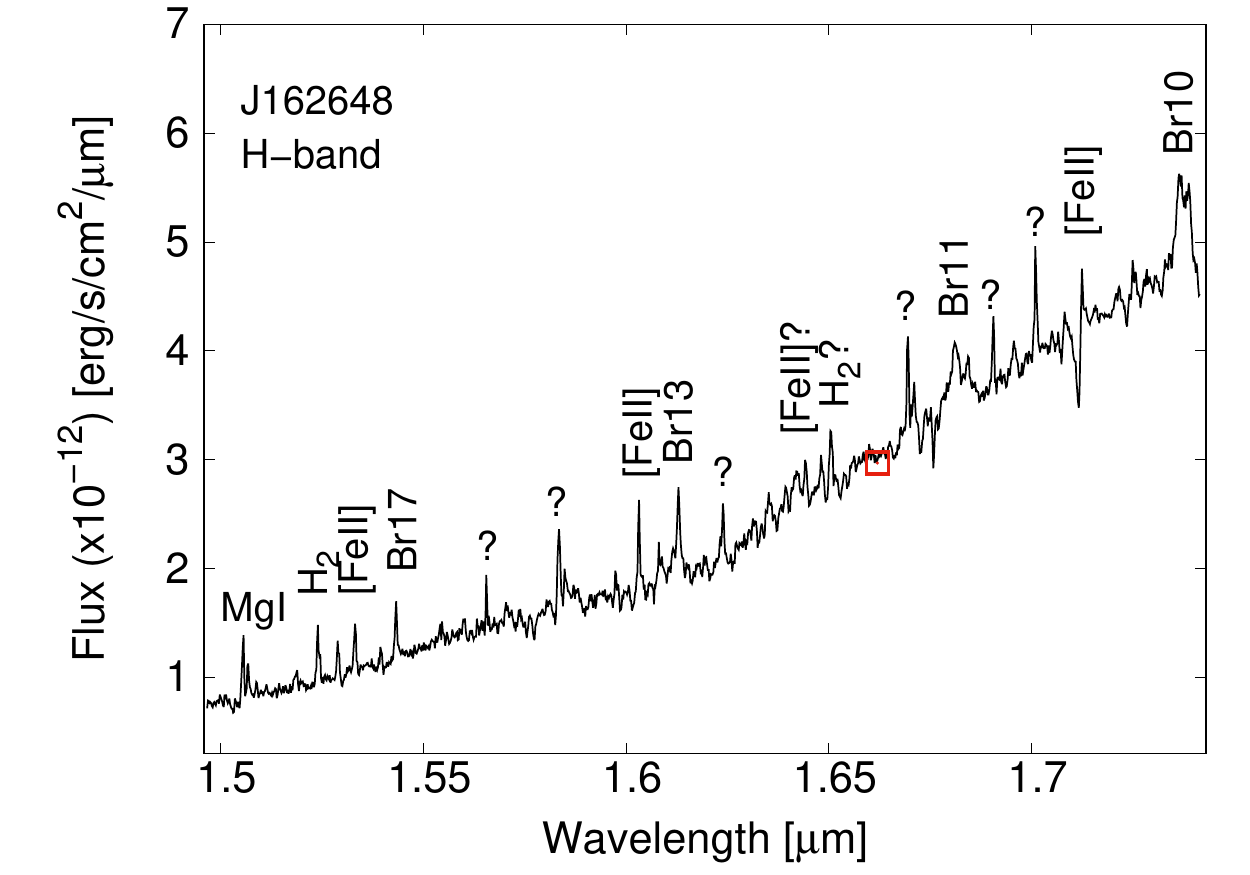}
     \includegraphics[width=3in]{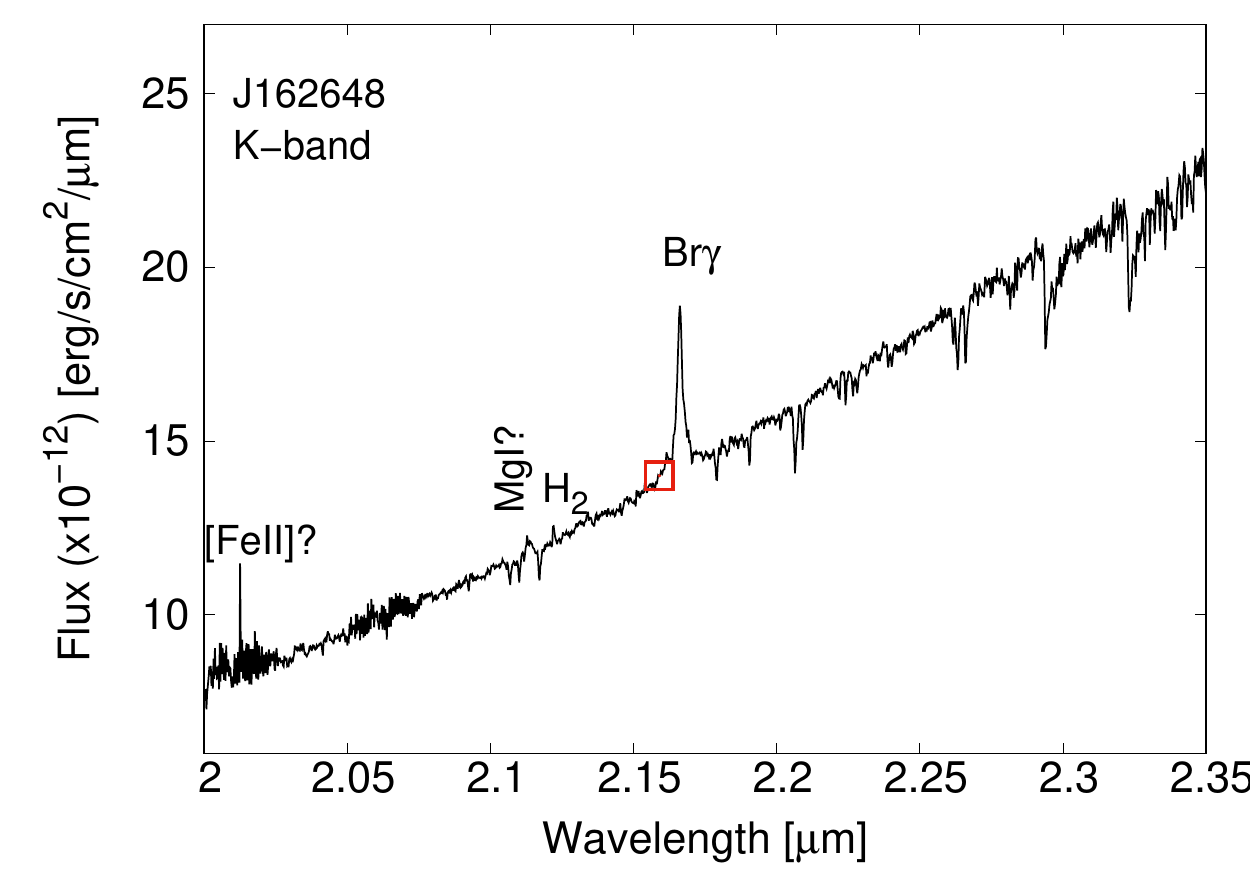}     
     \caption{The near-infrared spectra for J162648. The prominent accretion and outflow tracers are labelled. The 2MASS/UKIDSS photometry is marked by the red open square.   }
     \label{oph3-spec}
  \end{figure*}

 \begin{figure*}
  \centering              
     \includegraphics[width=3in]{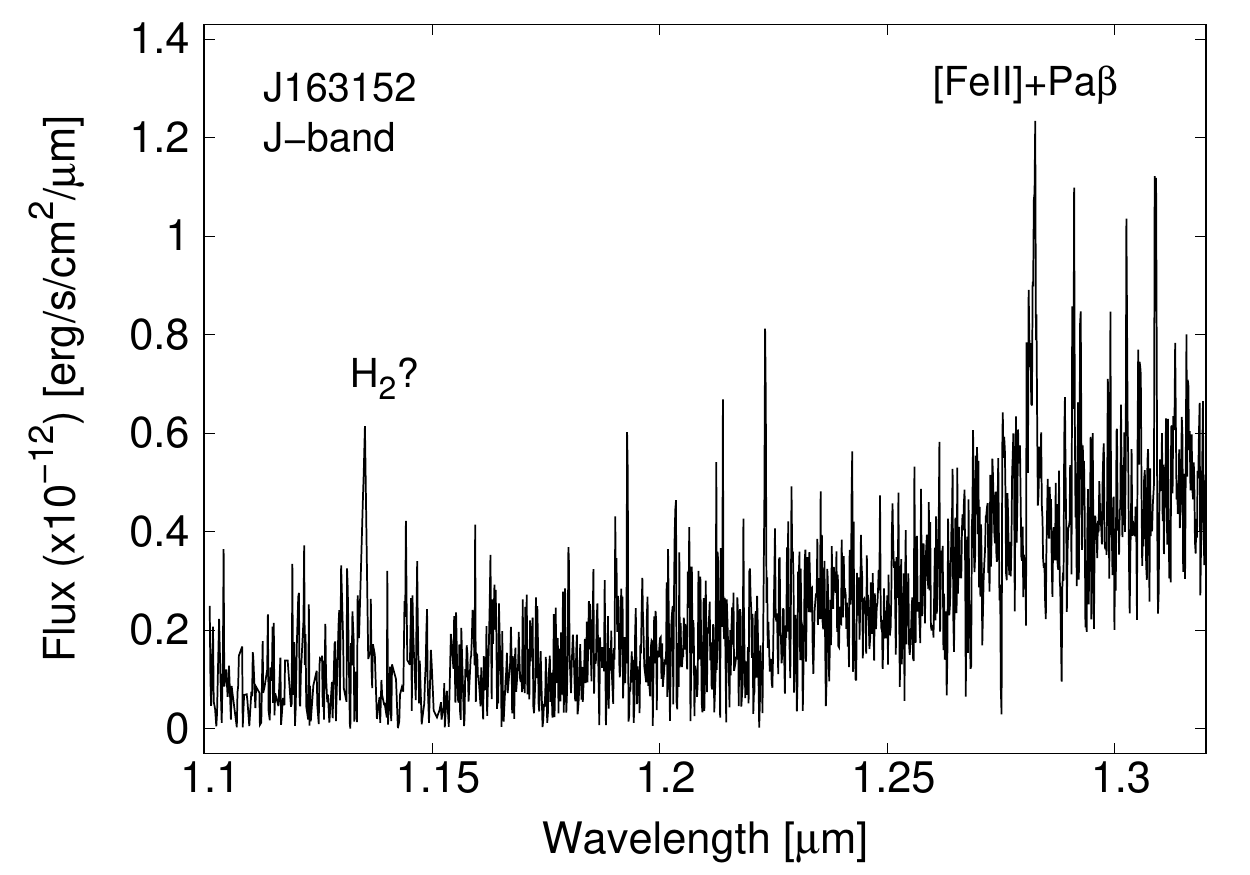}
     \includegraphics[width=3in]{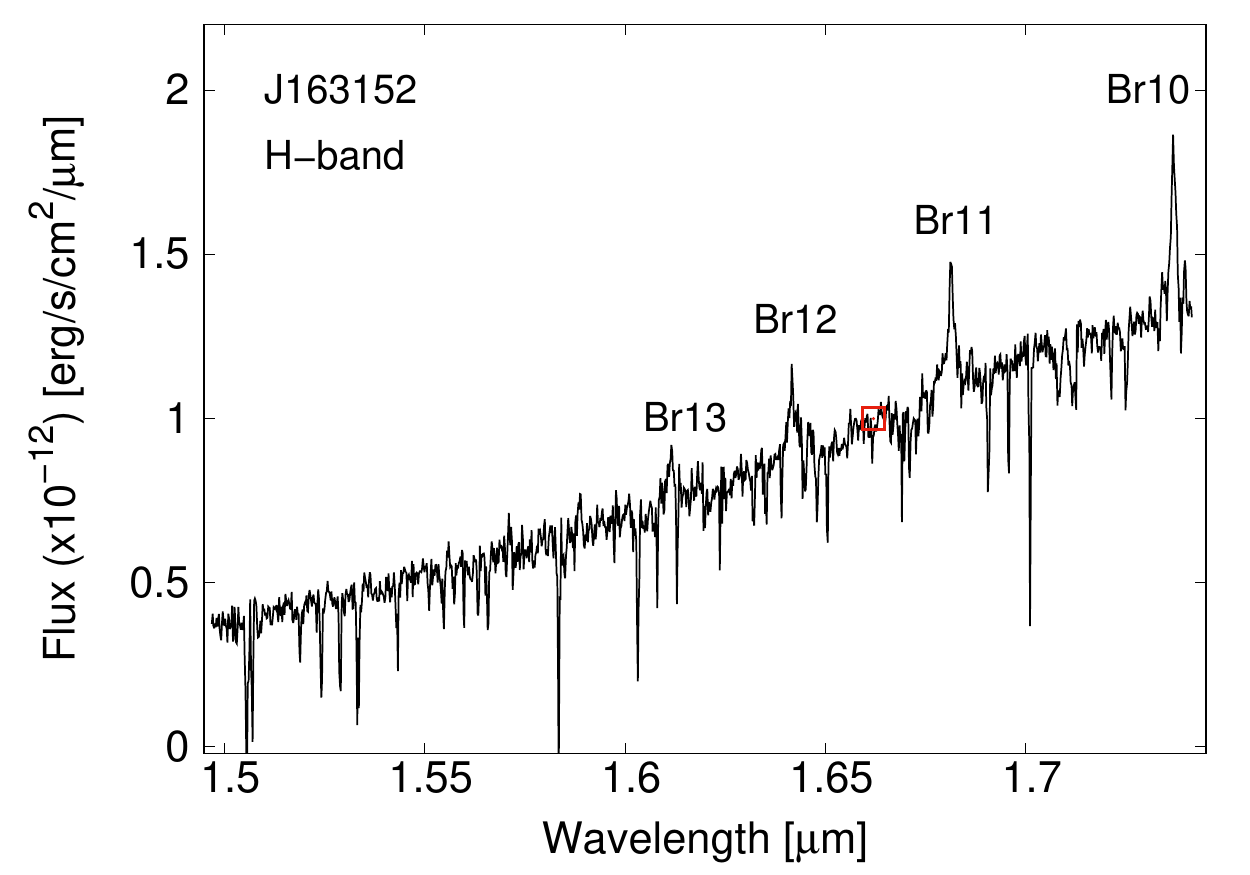}
     \includegraphics[width=3in]{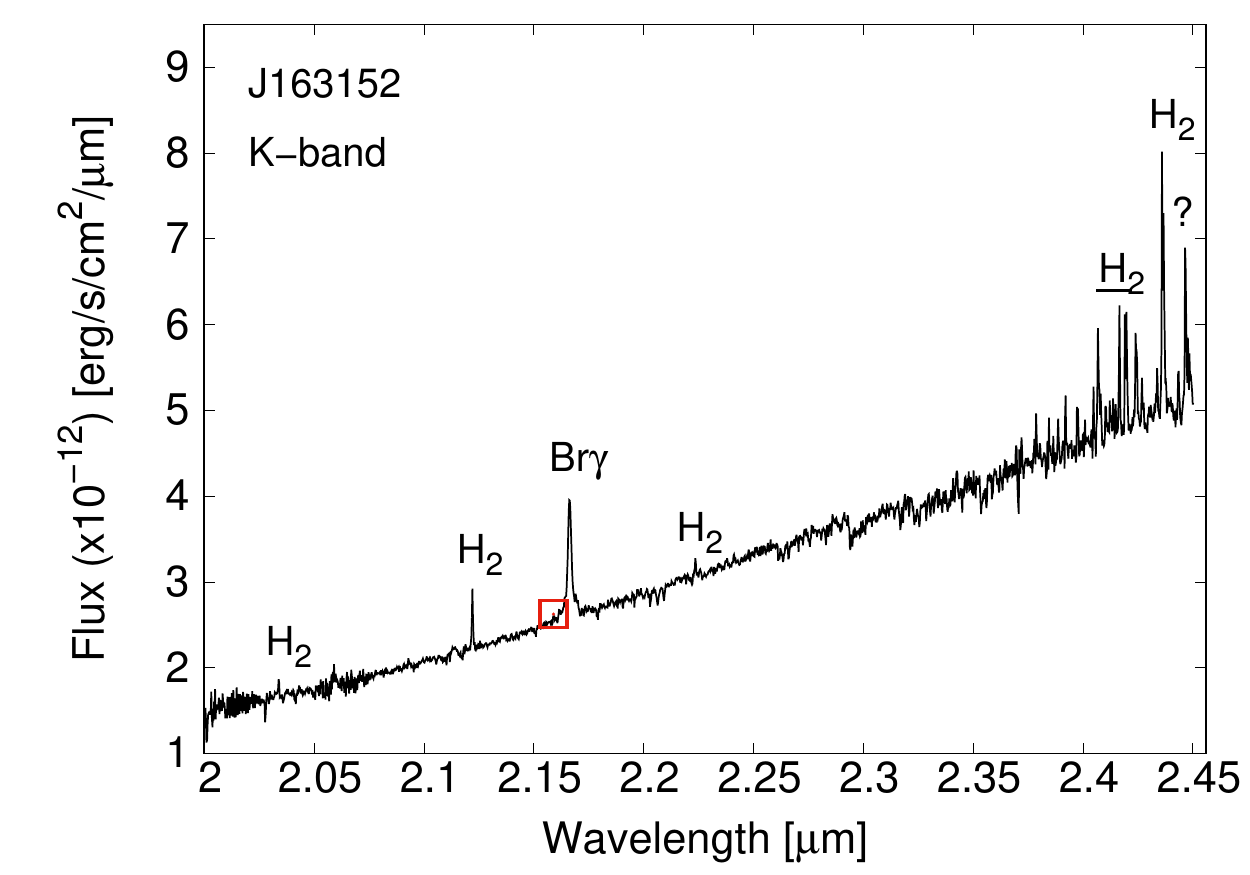}     
     \caption{The near-infrared spectra for J163152. The prominent accretion and outflow tracers are labelled. The 2MASS/UKIDSS photometry is marked by the red open square.    }
     \label{oph2-spec}
  \end{figure*}


 \begin{figure*}
  \centering             
     \includegraphics[width=3in]{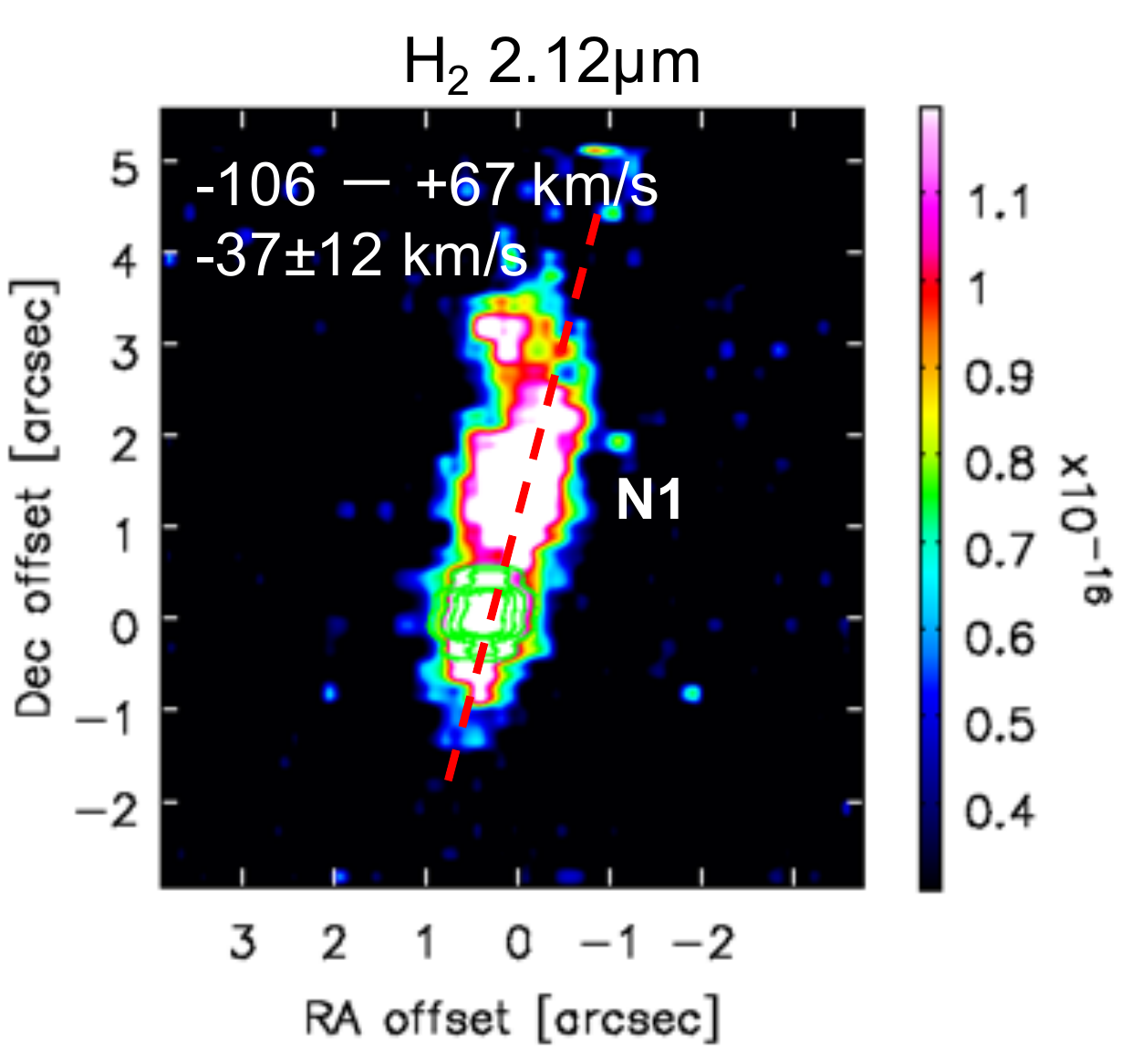}
     \includegraphics[width=3in]{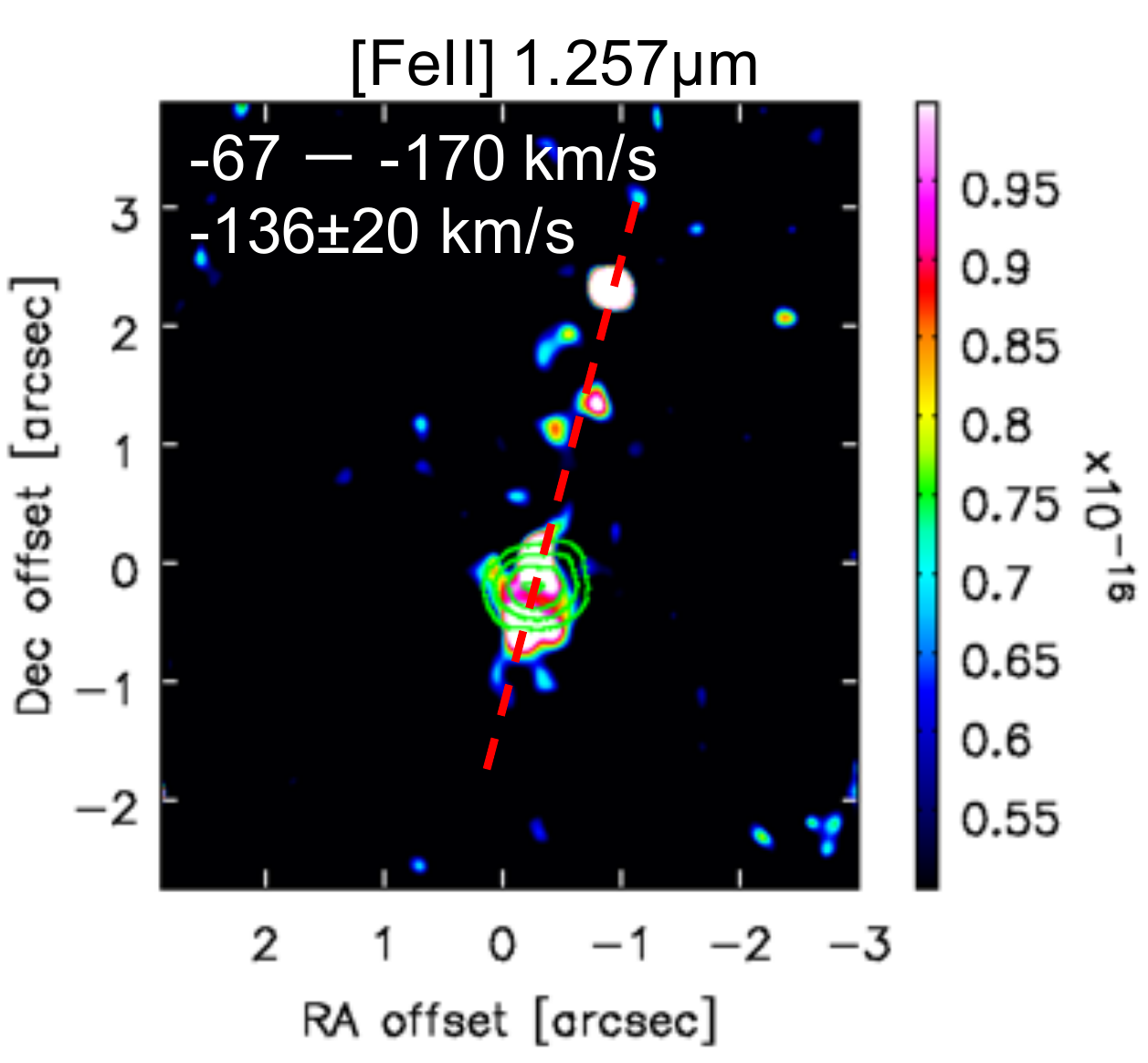}
     \includegraphics[width=3in]{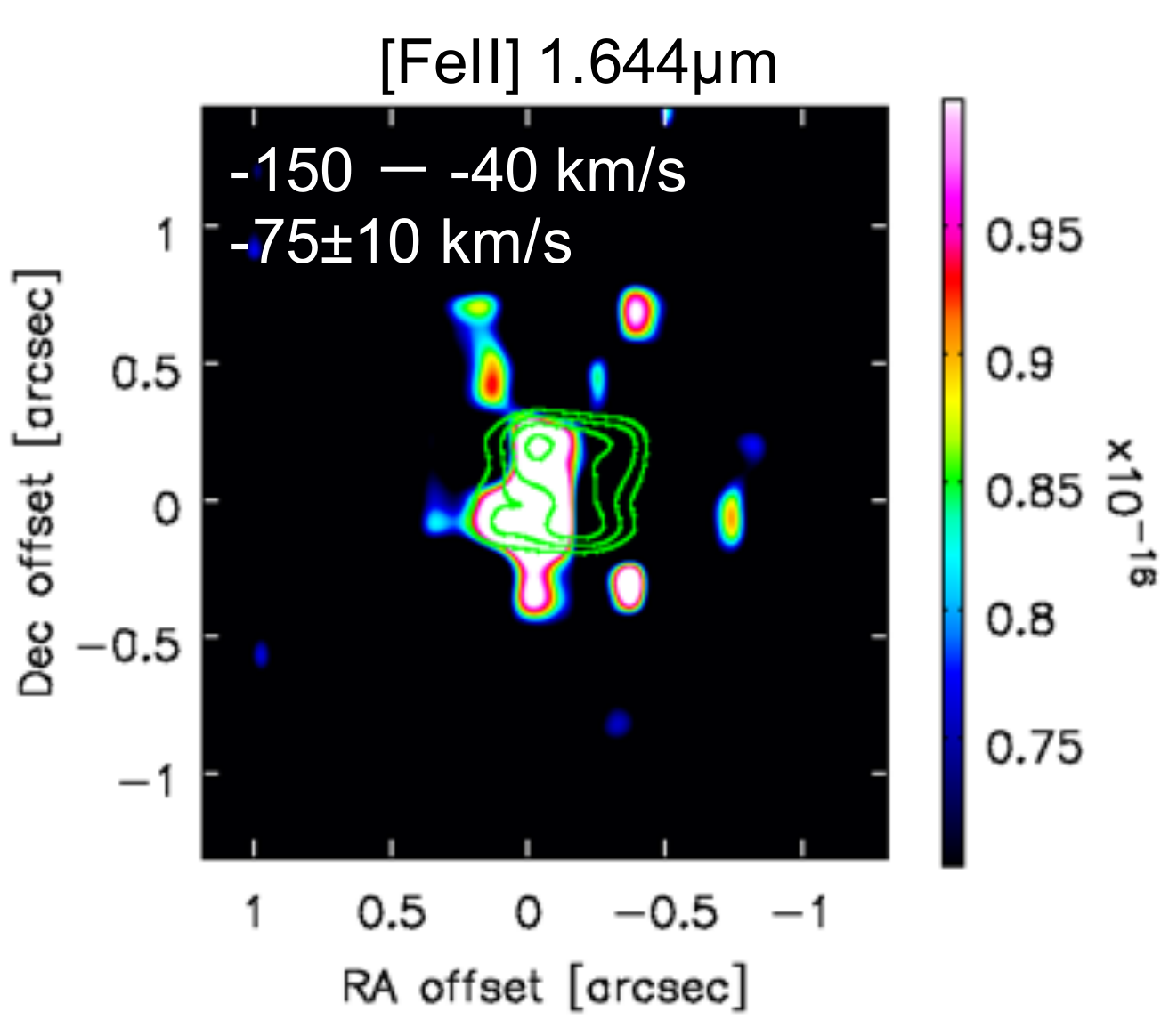}
     \includegraphics[width=3in]{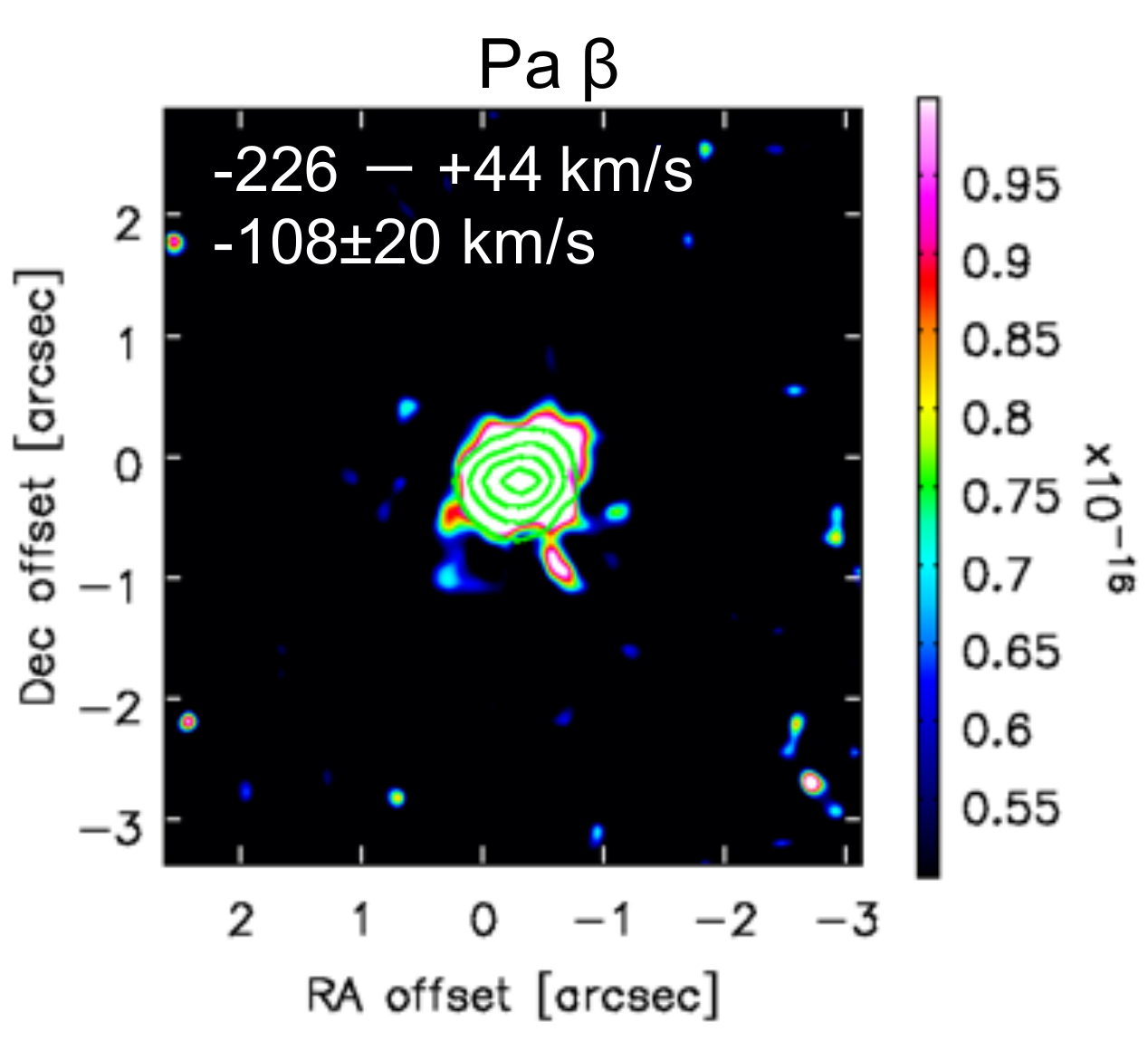}
     \includegraphics[width=3.3in]{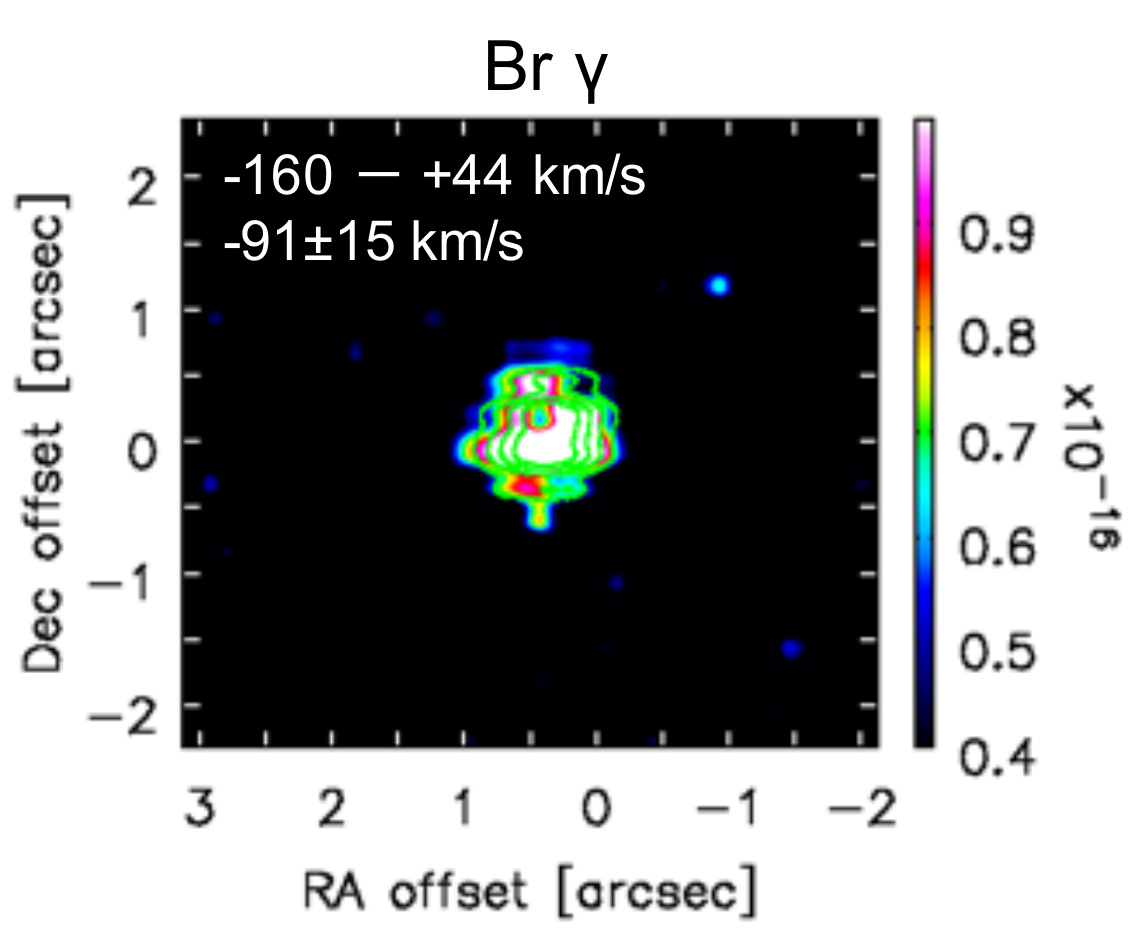}
     \caption{The continuum-subtracted integrated intensity line images for J163136. The intensity is integrated over the range in velocities where line emission is seen. The velocity range and the peak velocity are shown at the top, left corner. The continuum emission is overplotted in green contours. The colour bar on the right shows the integrated flux in units of (erg s$^{-1}$ cm$^{-2}$). North is up, east is to the left. }
     \label{oph1-imgs}
  \end{figure*}

 \begin{figure*}
  \centering              
     \includegraphics[width=3in]{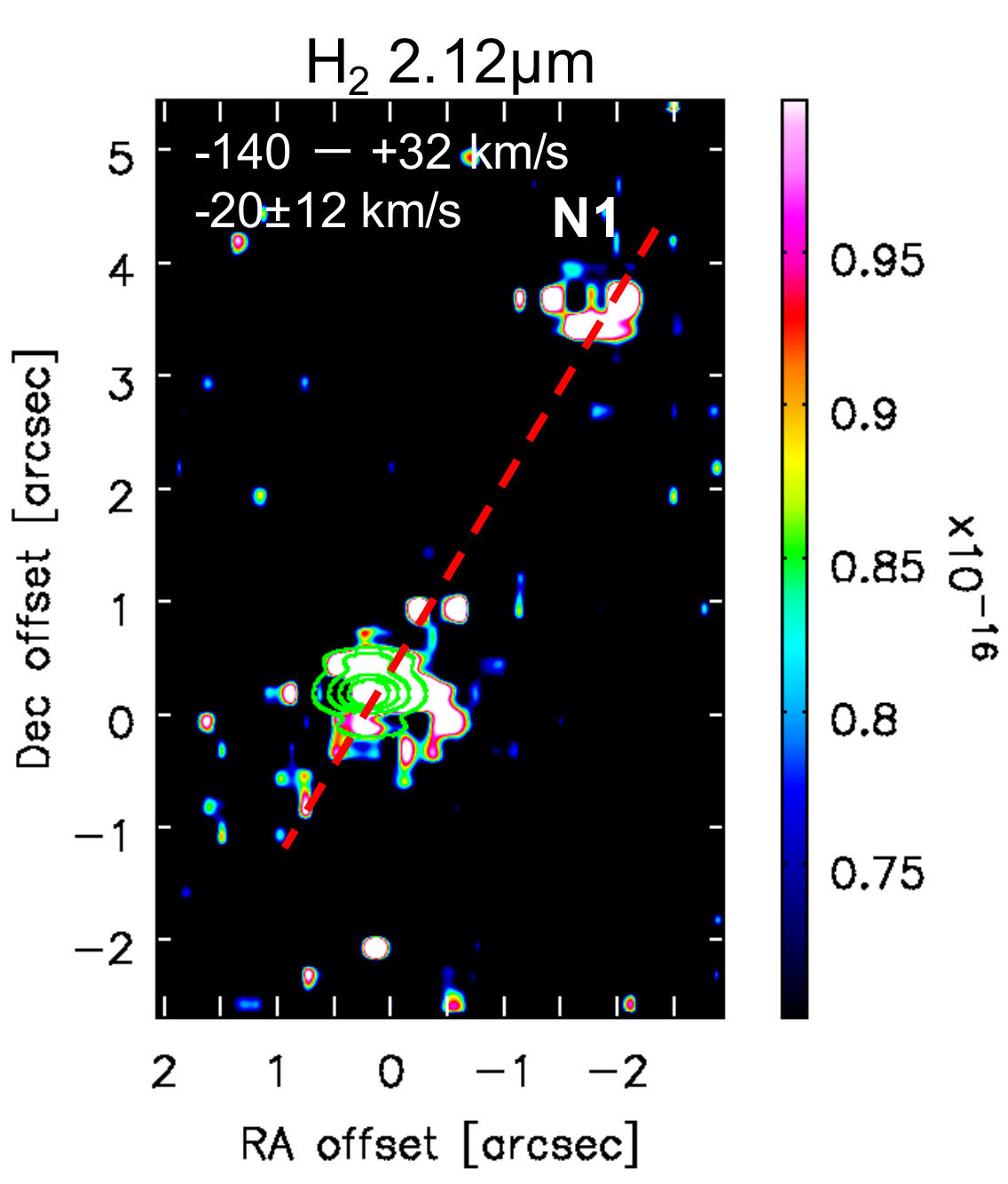} \\
     \includegraphics[width=2.9in]{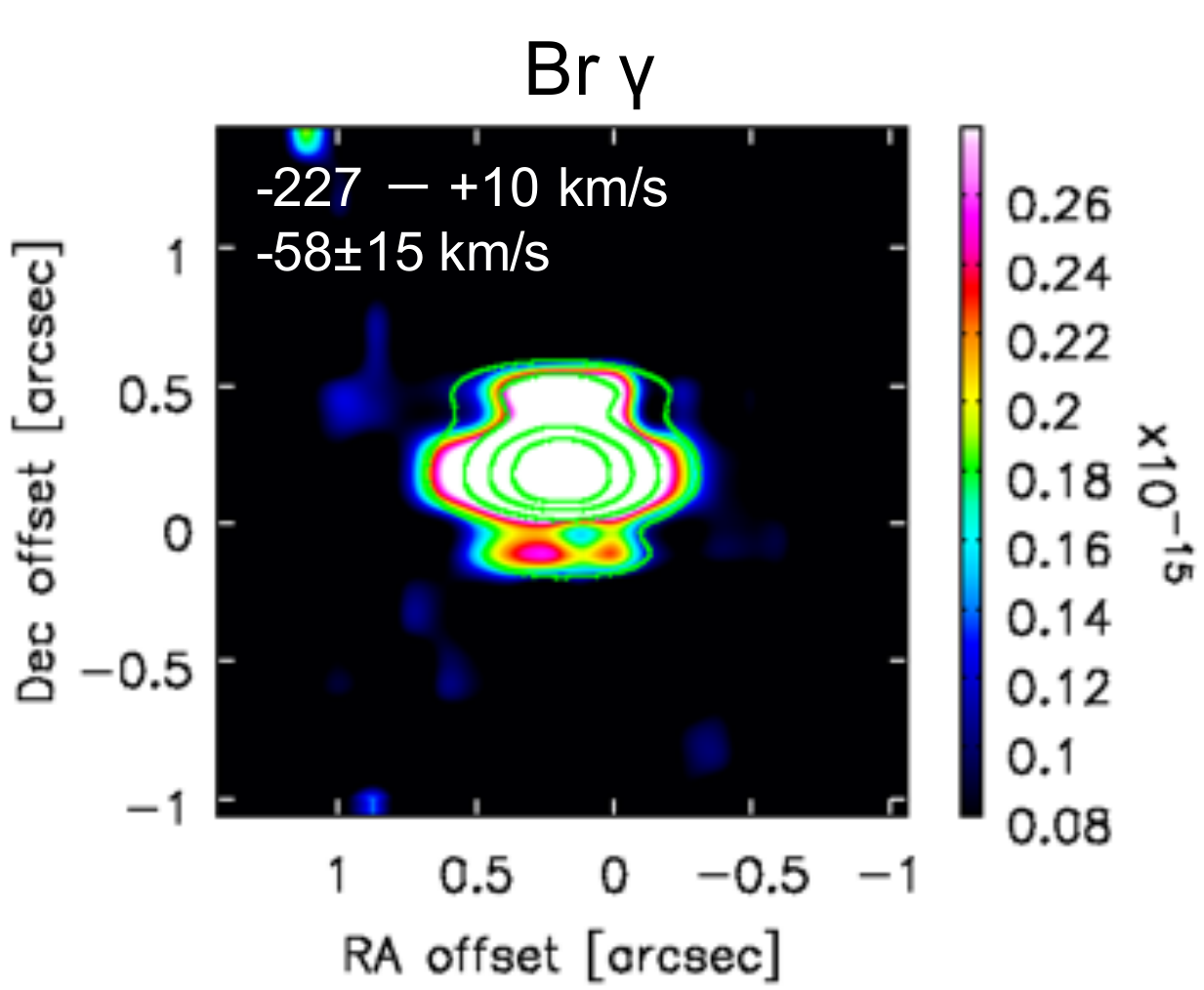}
     \includegraphics[width=3.2in]{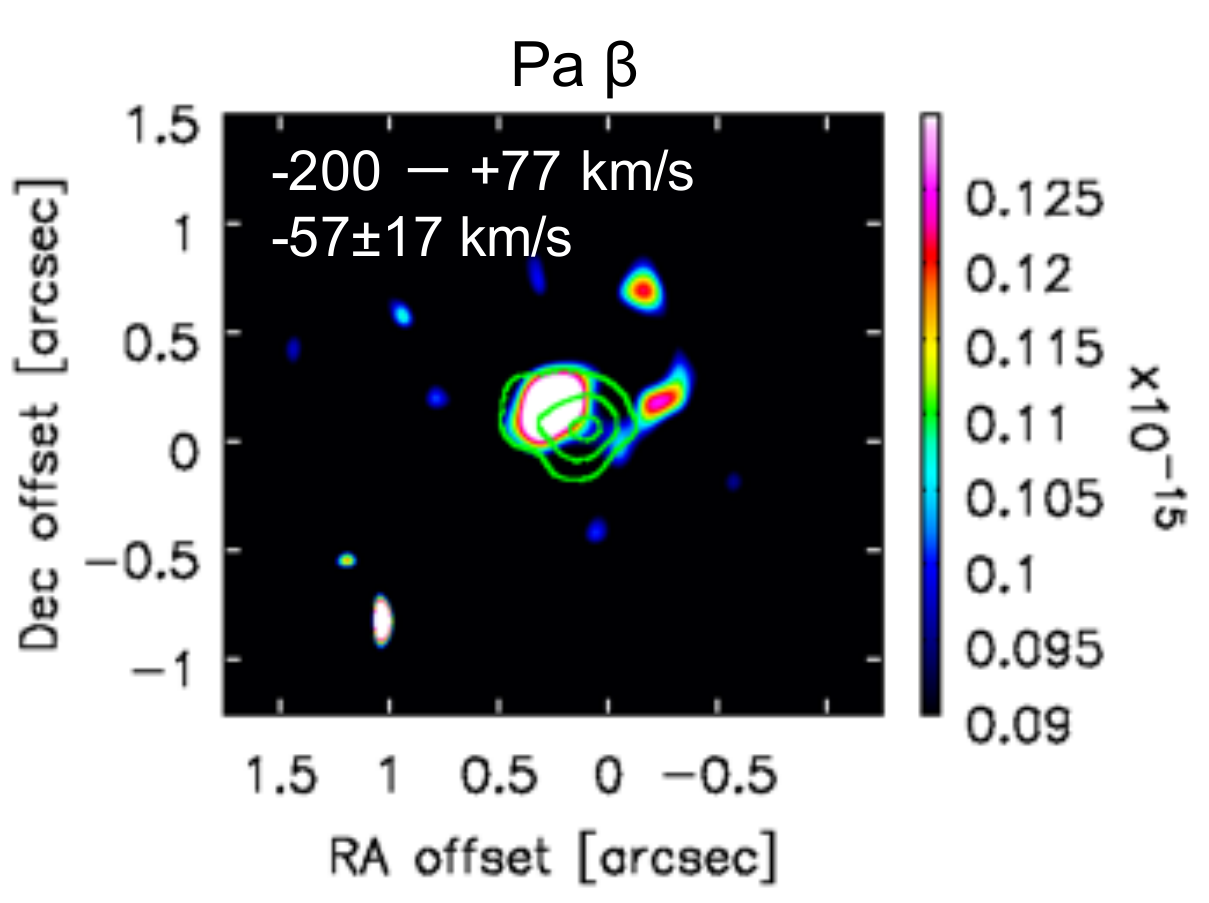}
     \caption{The continuum-subtracted integrated intensity line images for J182957. The intensity is integrated over the range in velocities where line emission is seen. The velocity range and the peak velocity are shown at the top, left corner. The continuum emission is overplotted in green contours. The colour bar on the right shows the integrated flux in units of (erg s$^{-1}$ cm$^{-2}$). North is up, east is to the left.  }
     \label{ser1-imgs}
  \end{figure*}

 \begin{figure*}
  \centering              
     \includegraphics[width=3.2in]{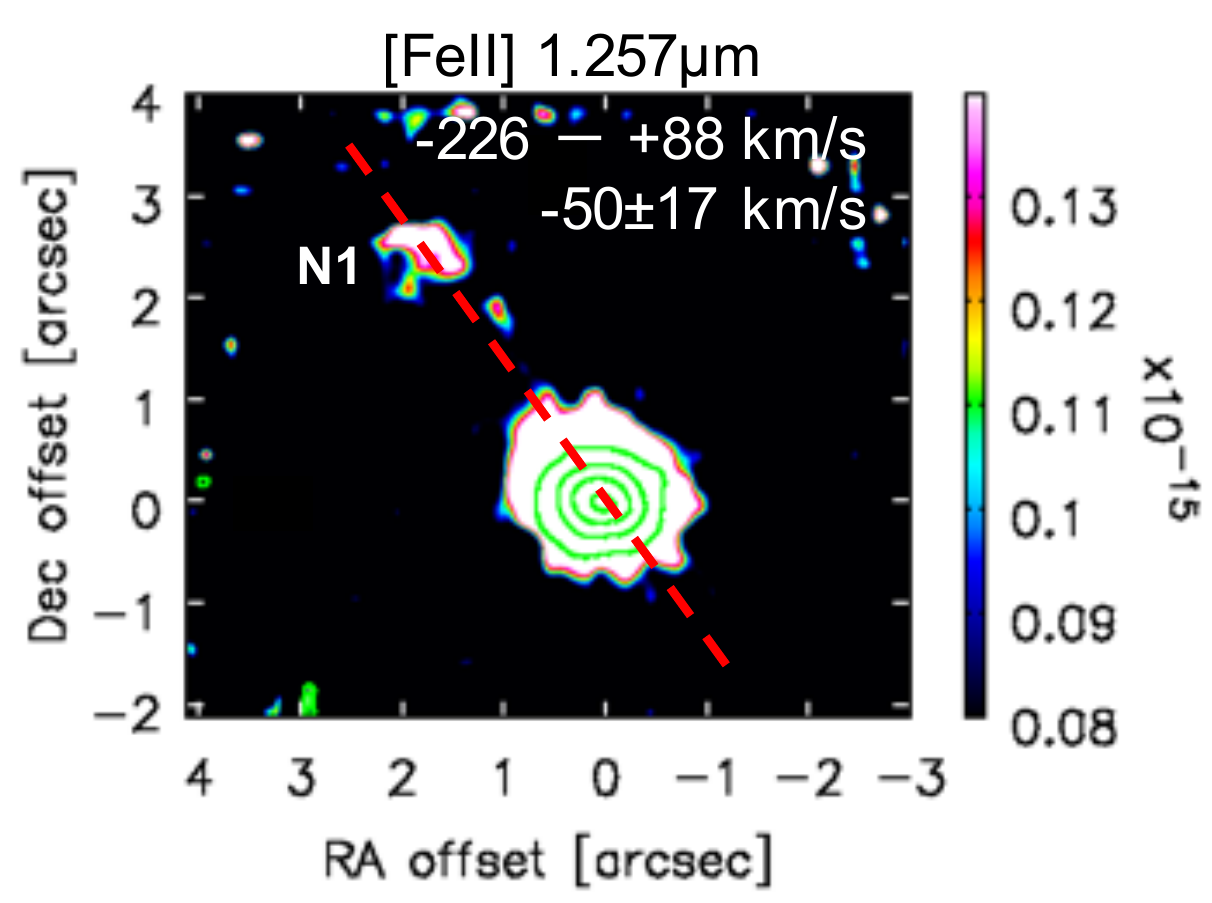}
     \includegraphics[width=3in]{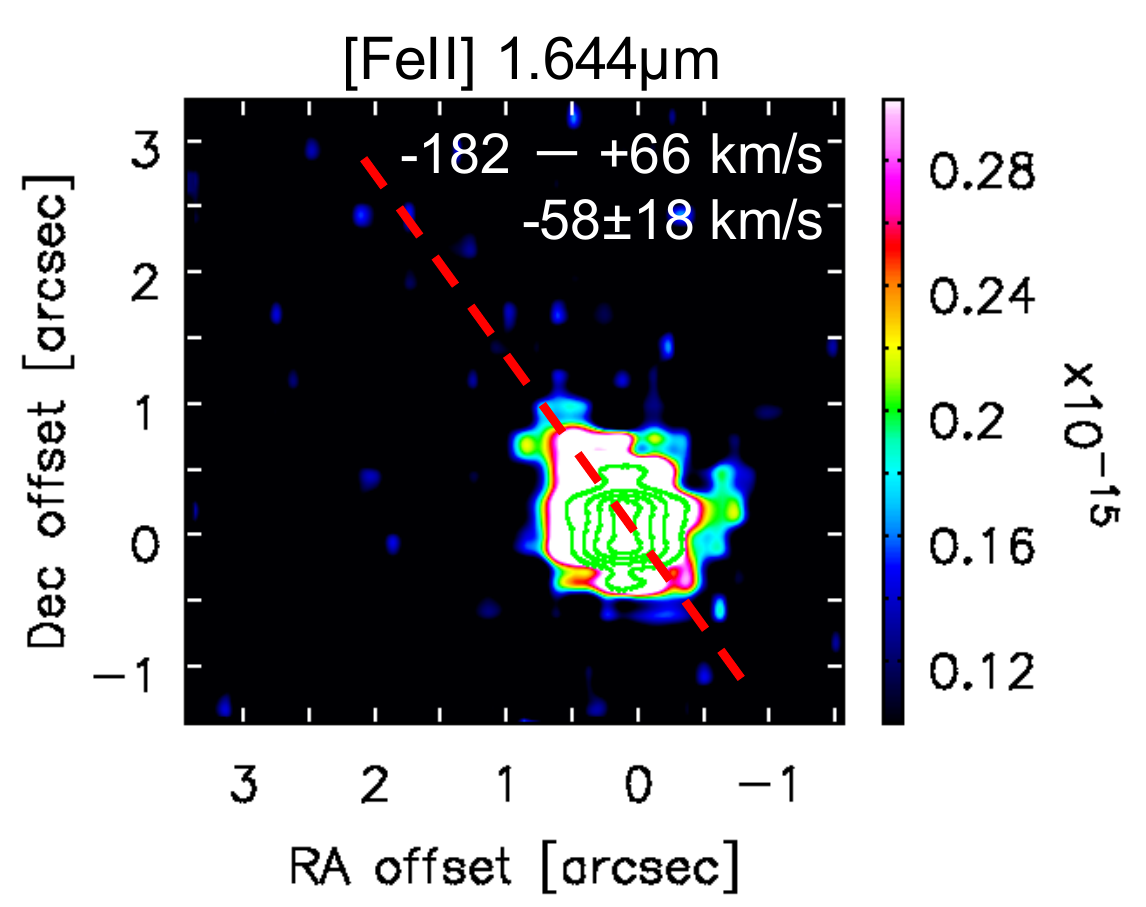}
     \includegraphics[width=3in]{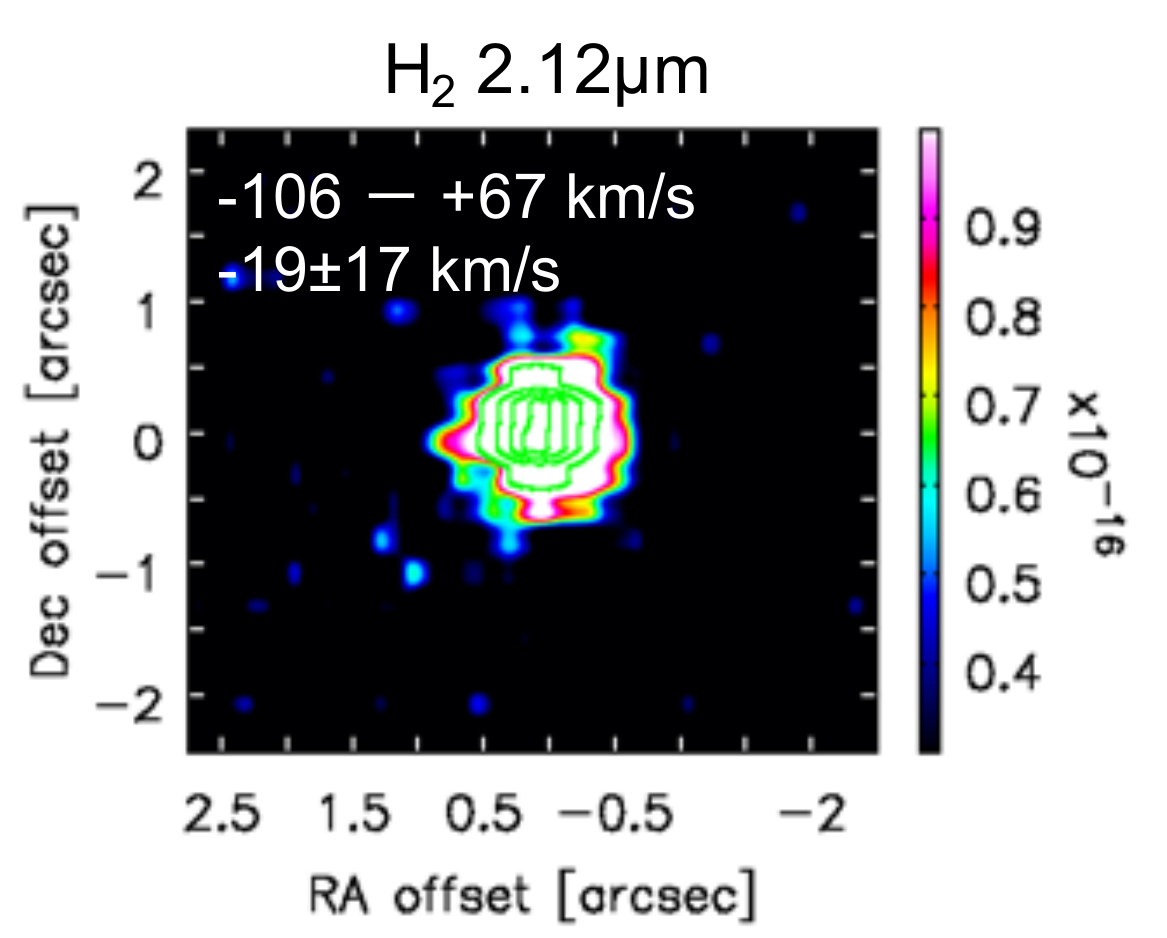}
     \includegraphics[width=3in]{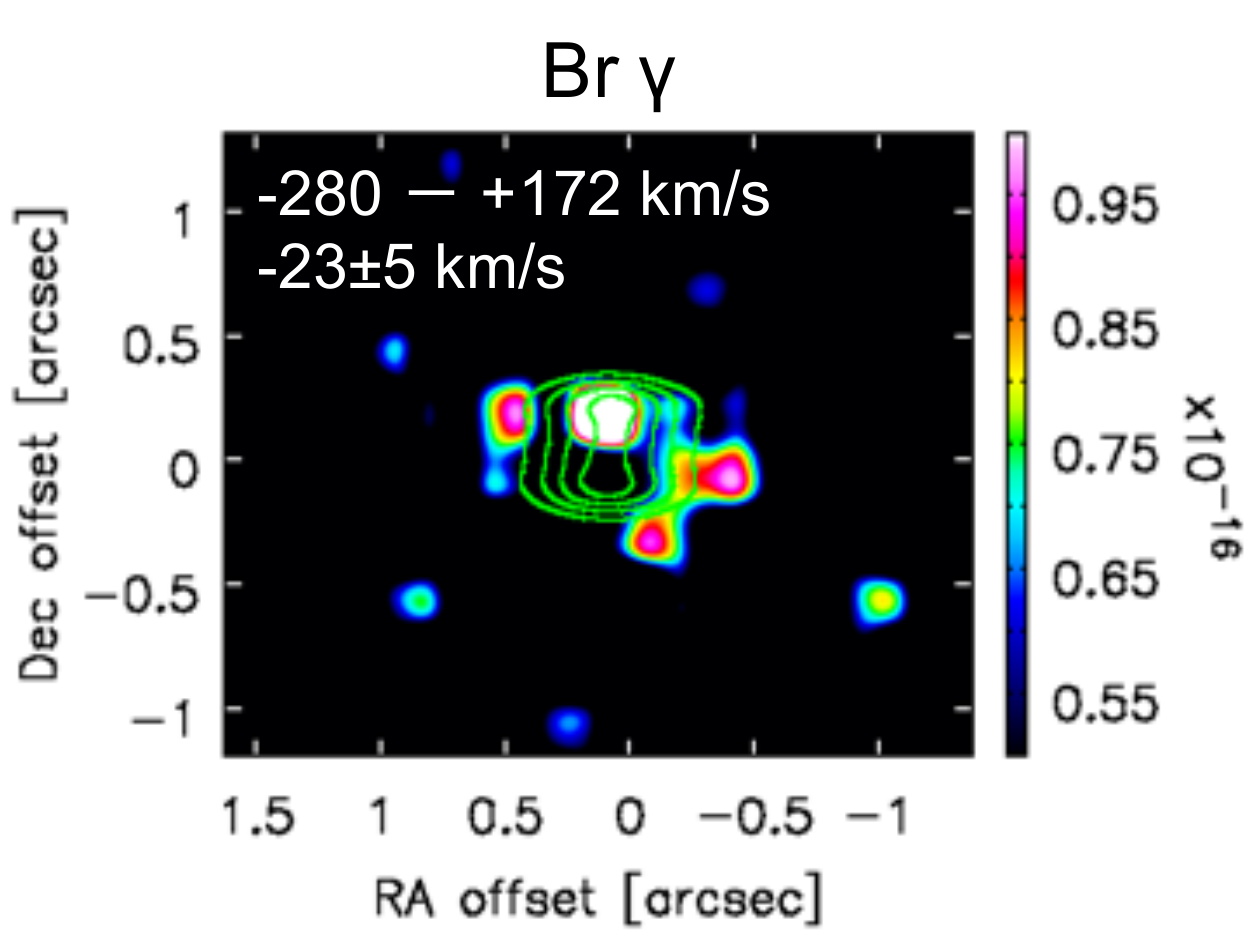}
     \includegraphics[width=2.7in]{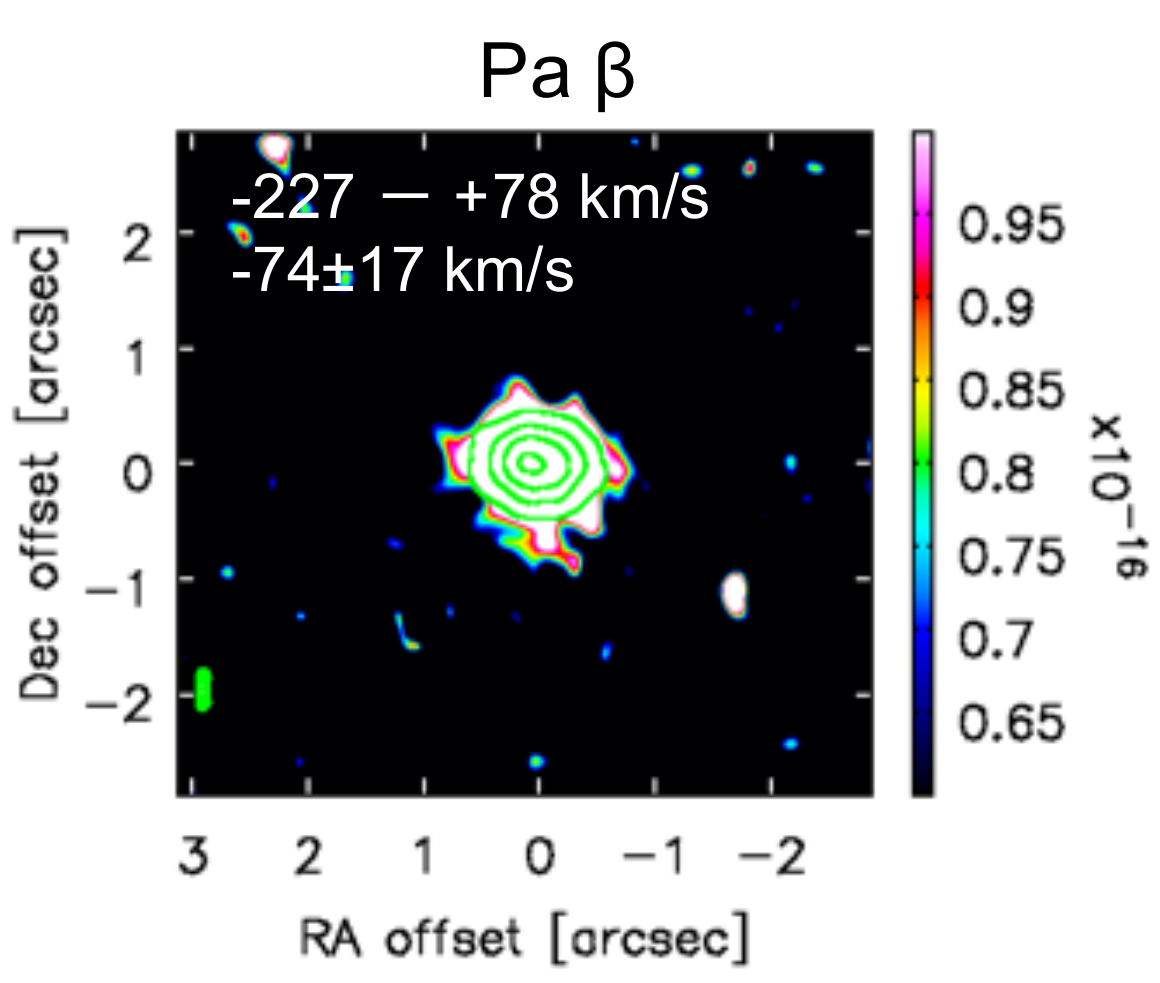}
     \caption{The continuum-subtracted integrated intensity line images for J182940. The intensity is integrated over the range in velocities where line emission is seen. The velocity range and the peak velocity are shown at the top, left corner. The continuum emission is overplotted in green contours. The colour bar on the right shows the integrated flux in units of (erg s$^{-1}$ cm$^{-2}$). North is up, east is to the left. }
     \label{ser8-imgs}
  \end{figure*}

 \begin{figure*}
  \centering              
     \includegraphics[width=3in]{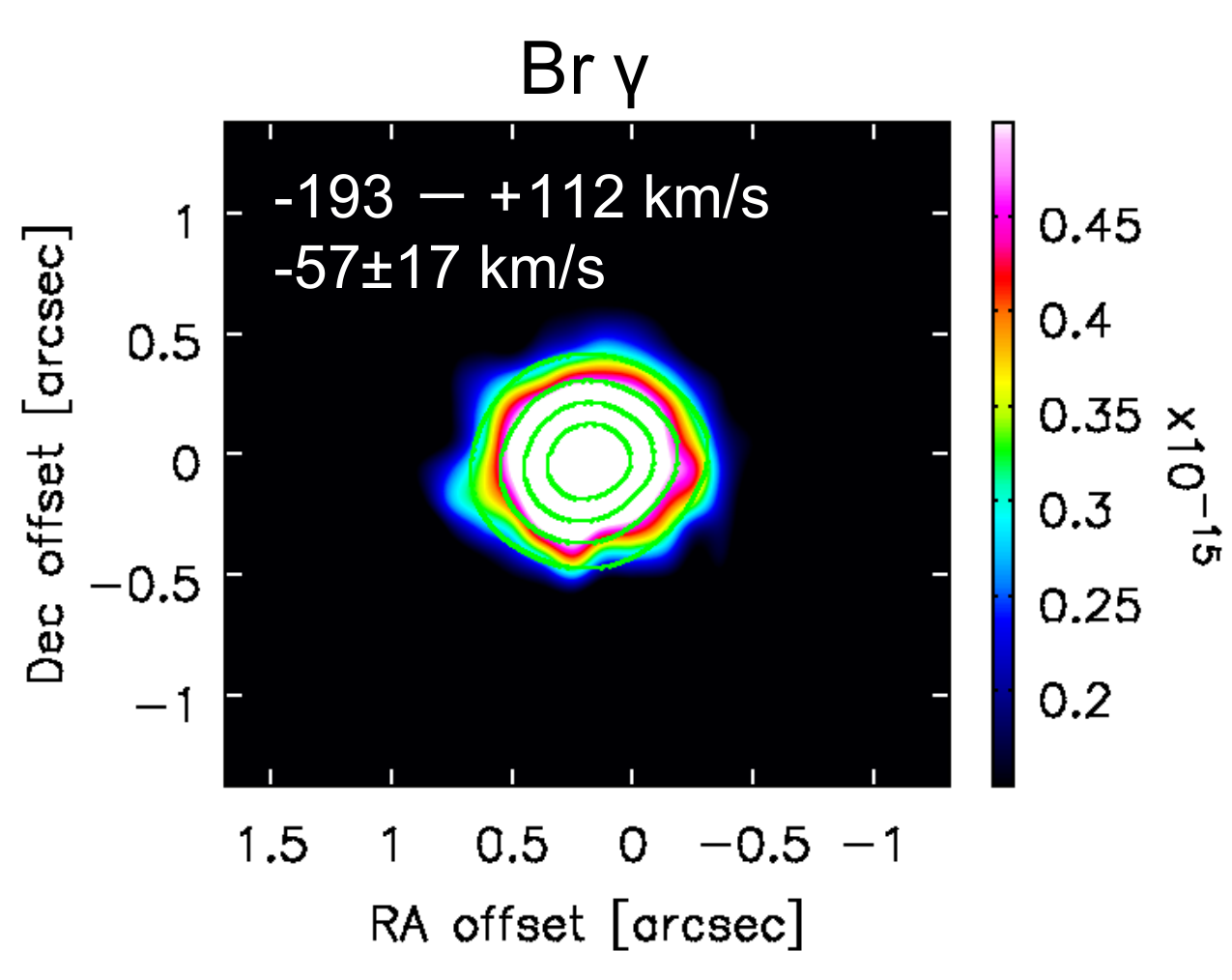}
     \caption{The continuum-subtracted integrated intensity line images for J162648. The intensity is integrated over the range in velocities where line emission is seen. The velocity range and the peak velocity are shown at the top, left corner. The continuum emission is overplotted in green contours. The colour bar on the right shows the integrated flux in units of (erg s$^{-1}$ cm$^{-2}$). North is up, east is to the left. }
     \label{oph3-imgs}
  \end{figure*}

 \begin{figure*}
  \centering              
     \includegraphics[width=2.6in]{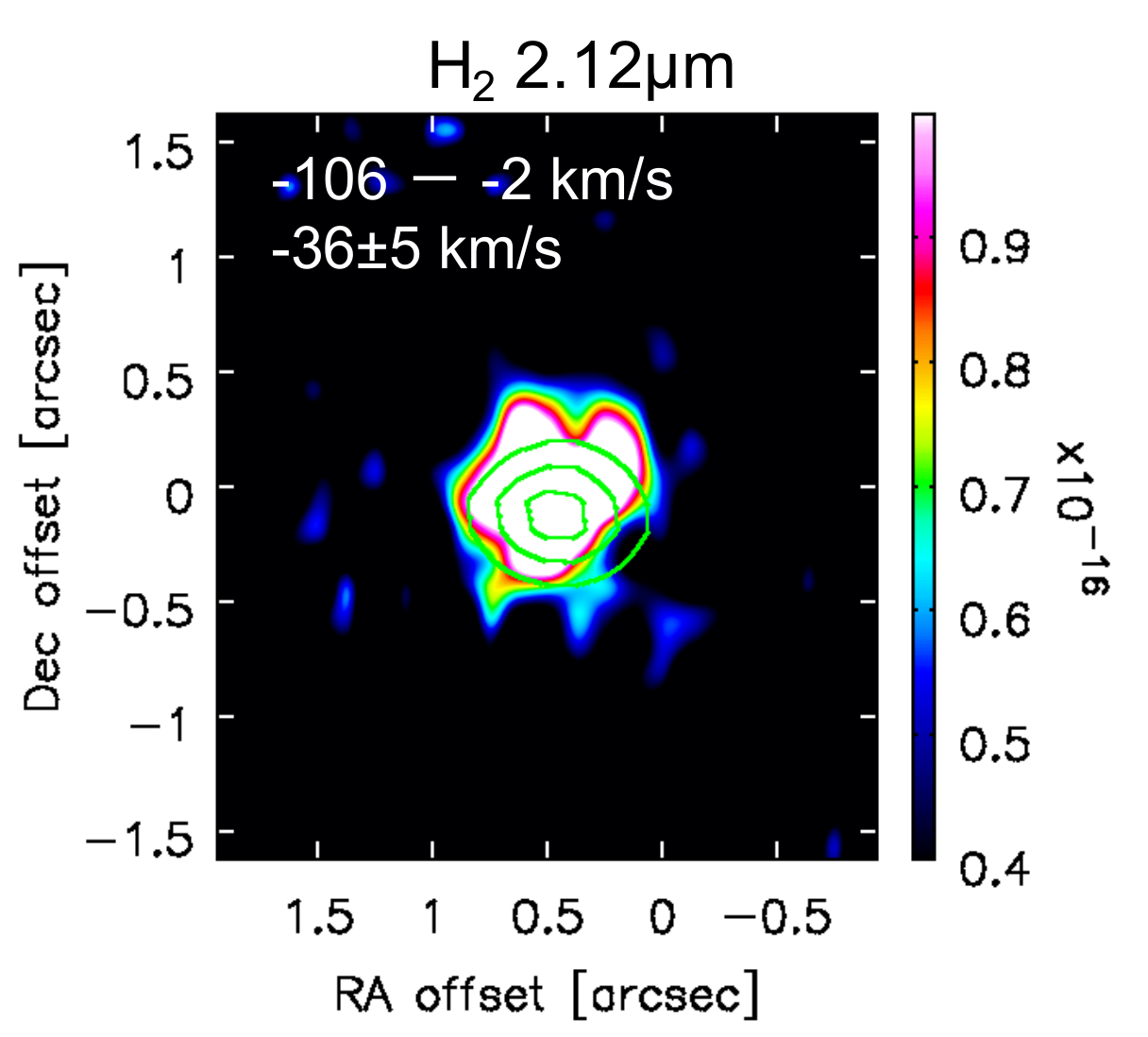}
     \includegraphics[width=3in]{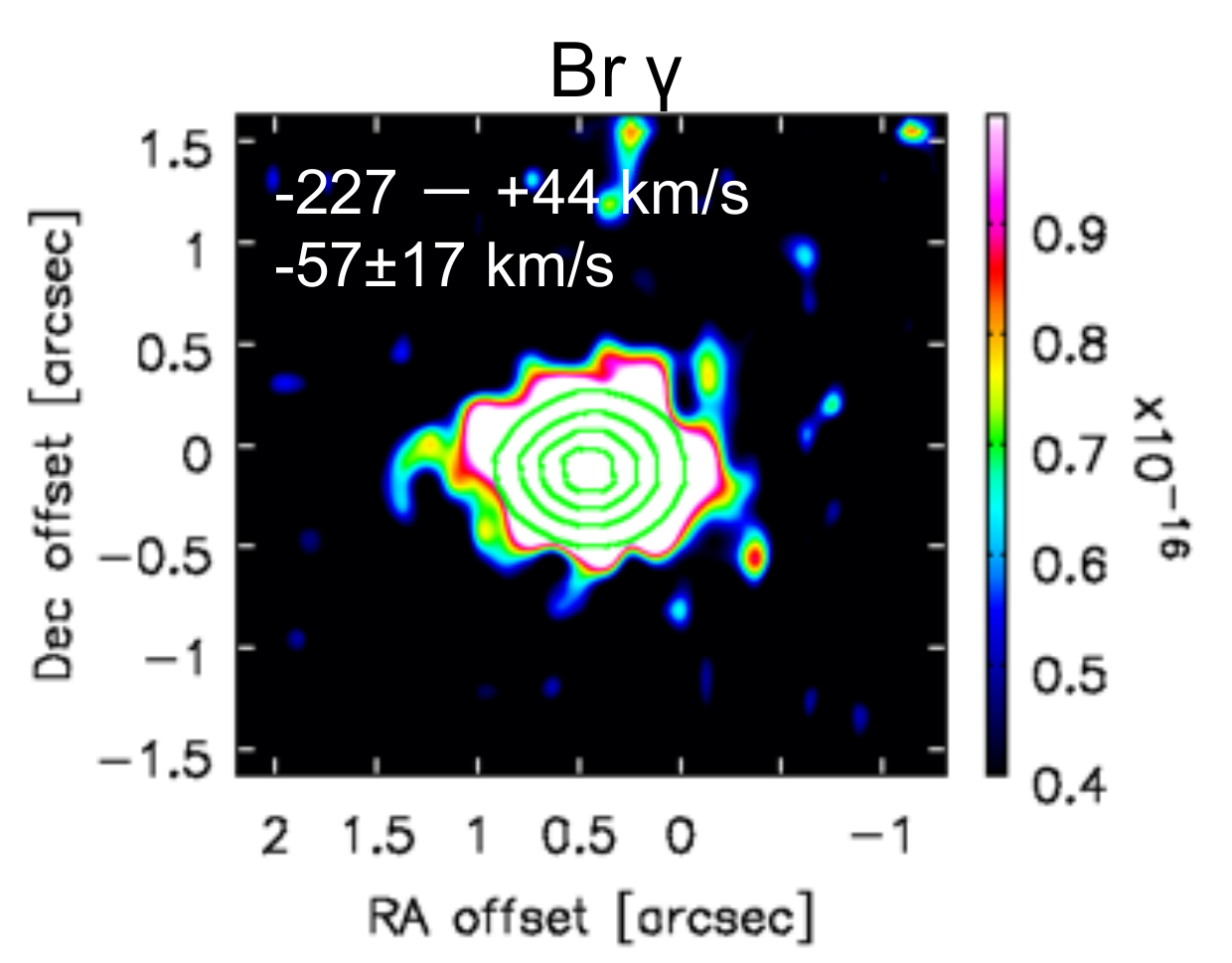}
     \caption{The continuum-subtracted integrated intensity line images for J163152. The intensity is integrated over the range in velocities where line emission is seen. The velocity range and the peak velocity are shown at the top, left corner. The continuum emission is overplotted in green contours. The colour bar on the right shows the integrated flux in units of (erg s$^{-1}$ cm$^{-2}$). North is up, east is to the left. }
     \label{oph2-imgs}
  \end{figure*}


\subsection{J163136}

The $JHK$ spectra for J163136 show emission in several H$_{2}$ and [Fe~II] lines (Fig.~\ref{oph1-spec}). The H$_{2}$ lines at higher vibrational levels of $\nu$ = 3 have fluxes more than order of magnitude lower than the brightest H$_{2}$ $\nu$ = 1 lines (Table~\ref{oph1-lines}). The electron temperature derived from the H$_{2}$ $\nu$ = 1 lines is 2500 K. There is strong emission seen in the Pa$\beta$ and the Br$\gamma$ lines, with weak detection in several upper Brackett lines (Br~10 -- Br~19). In addition to H~I, there is emission detected in the Mg~I permitted atomic line at 1.5$\micron$.

The mass accretion rate for J163136 derived from the Pa~$\beta$ line is 6$\times$10$^{-7}$ M$_{\sun}$ yr$^{-1}$, similar to the rate of 4$\times$10$^{-7}$ M$_{\sun}$ yr$^{-1}$ derived from the Br$\gamma$ line (Table~\ref{macc}). The mean $\dot{M}_{out}$ derived using the [Fe~II] lines is $\sim$1$\times$10$^{-8}$ M$_{\sun}$ yr$^{-1}$, while the mean $\dot{M}_{out}$ derived using the H$_{2}$ lines is $\sim$2 orders of magnitude lower at $\sim$8$\times$10$^{-10}$ M$_{\sun}$ yr$^{-1}$ (Table~\ref{mout}). This implies a jet efficiency of $\dot{M}_{out}$[Fe~II]/$\dot{M}_{acc}$ $\sim$0.02 and $\dot{M}_{out}$H$_{2}$/$\dot{M}_{acc}$ $\sim$0.002, using the mean accretion rate derived from the Pa$\beta$ and Br$\gamma$ lines.

\begin{figure*}
  \centering  
     \includegraphics[width=3.3in]{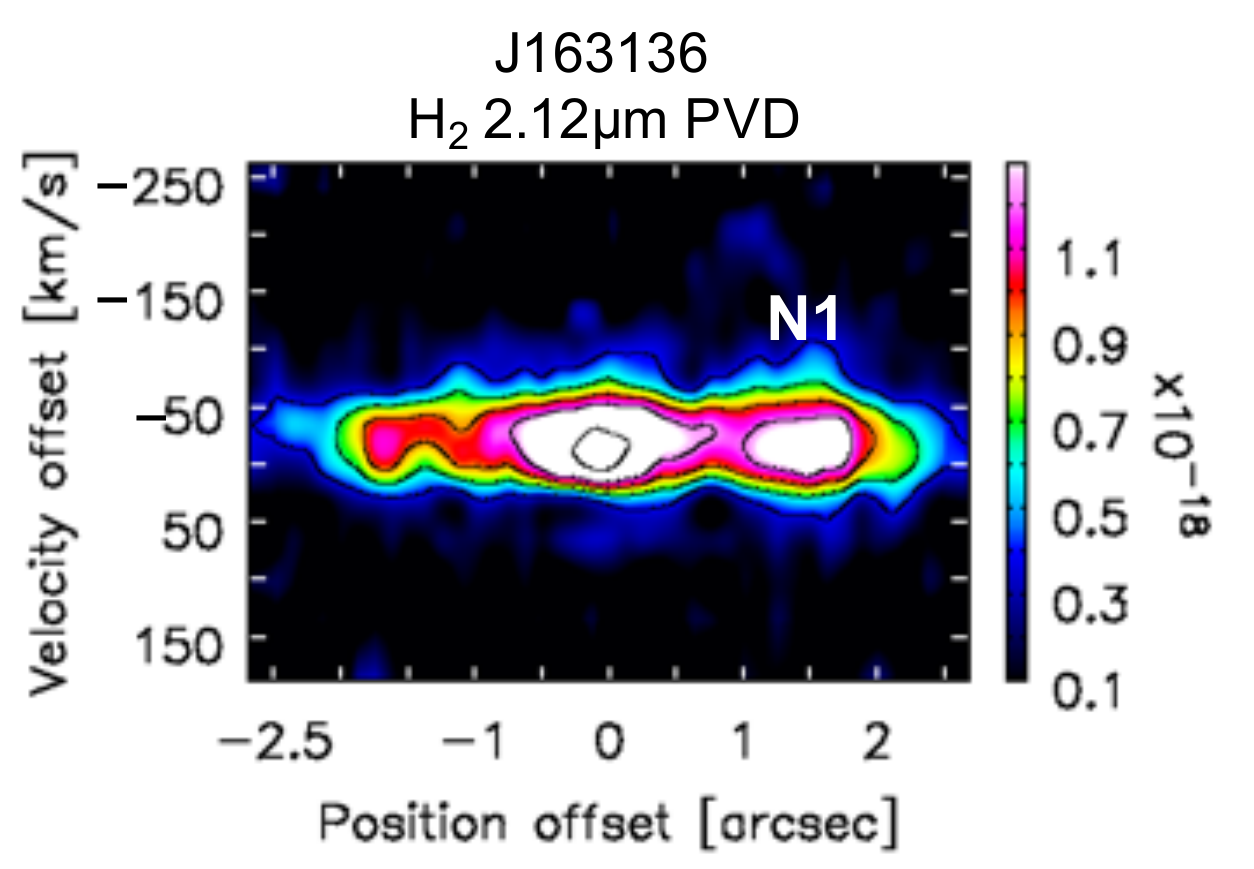}      
     \includegraphics[width=3in]{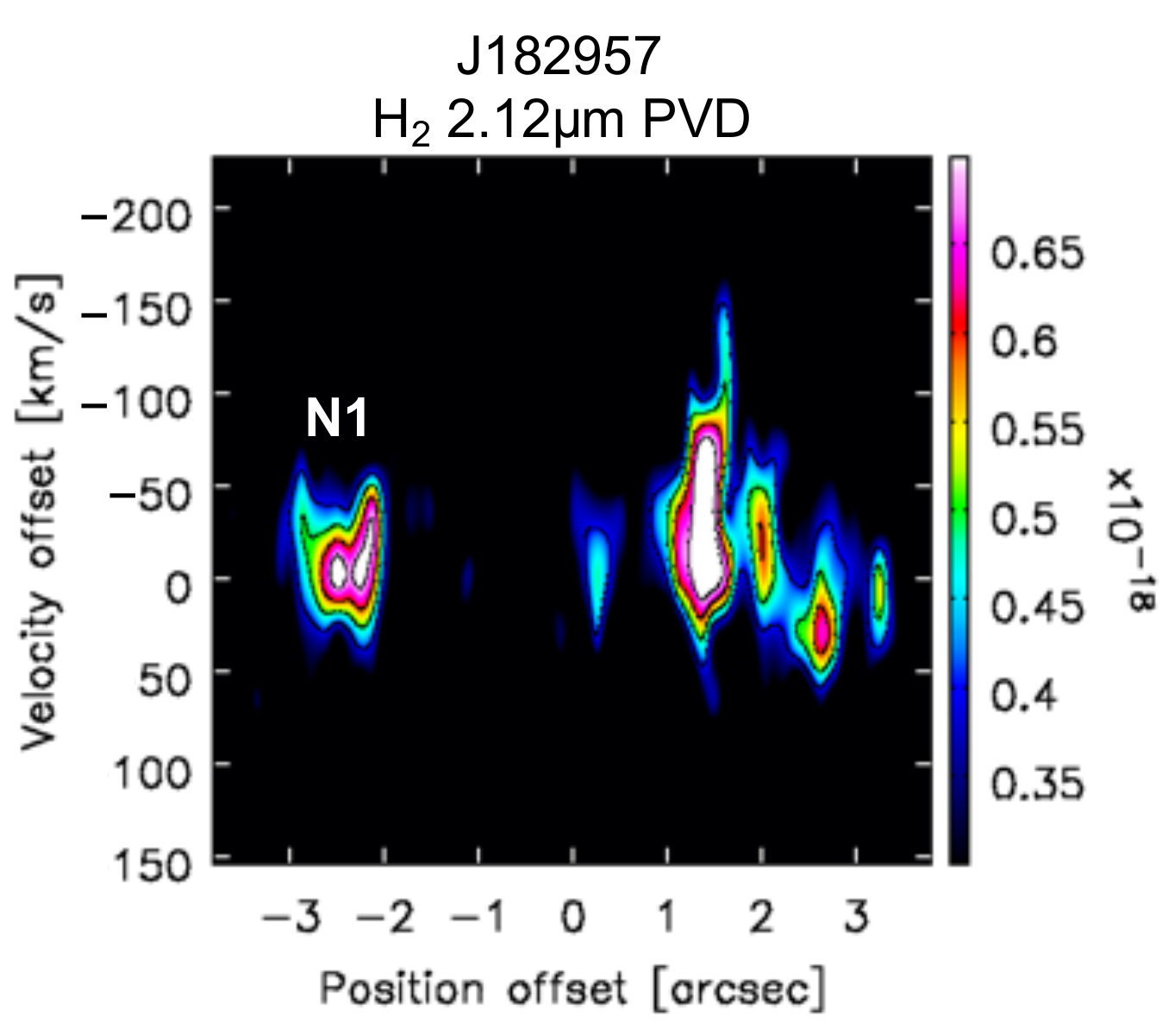}      
     \includegraphics[width=3in]{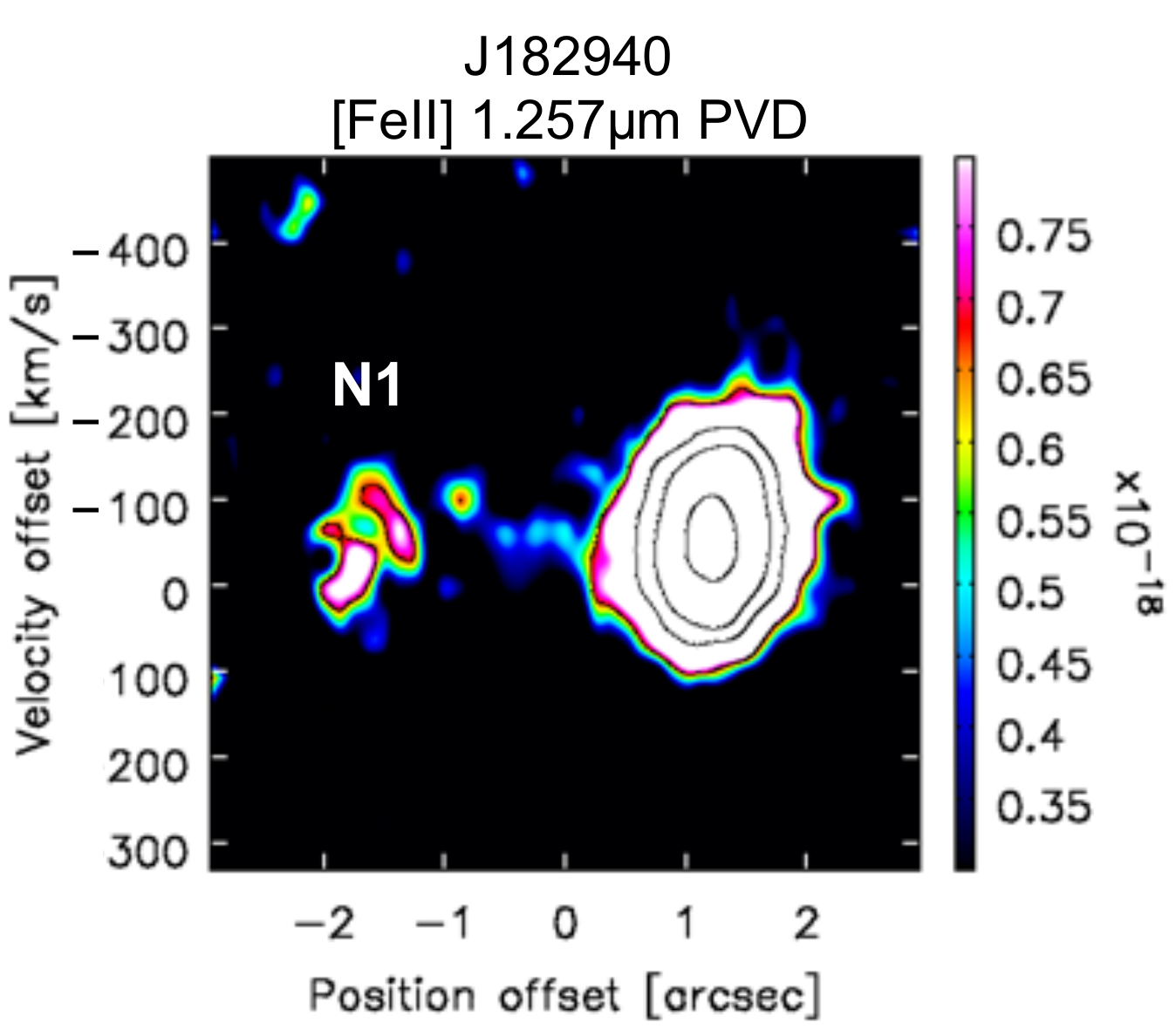}  \hspace{0.1in}
     \includegraphics[width=3in]{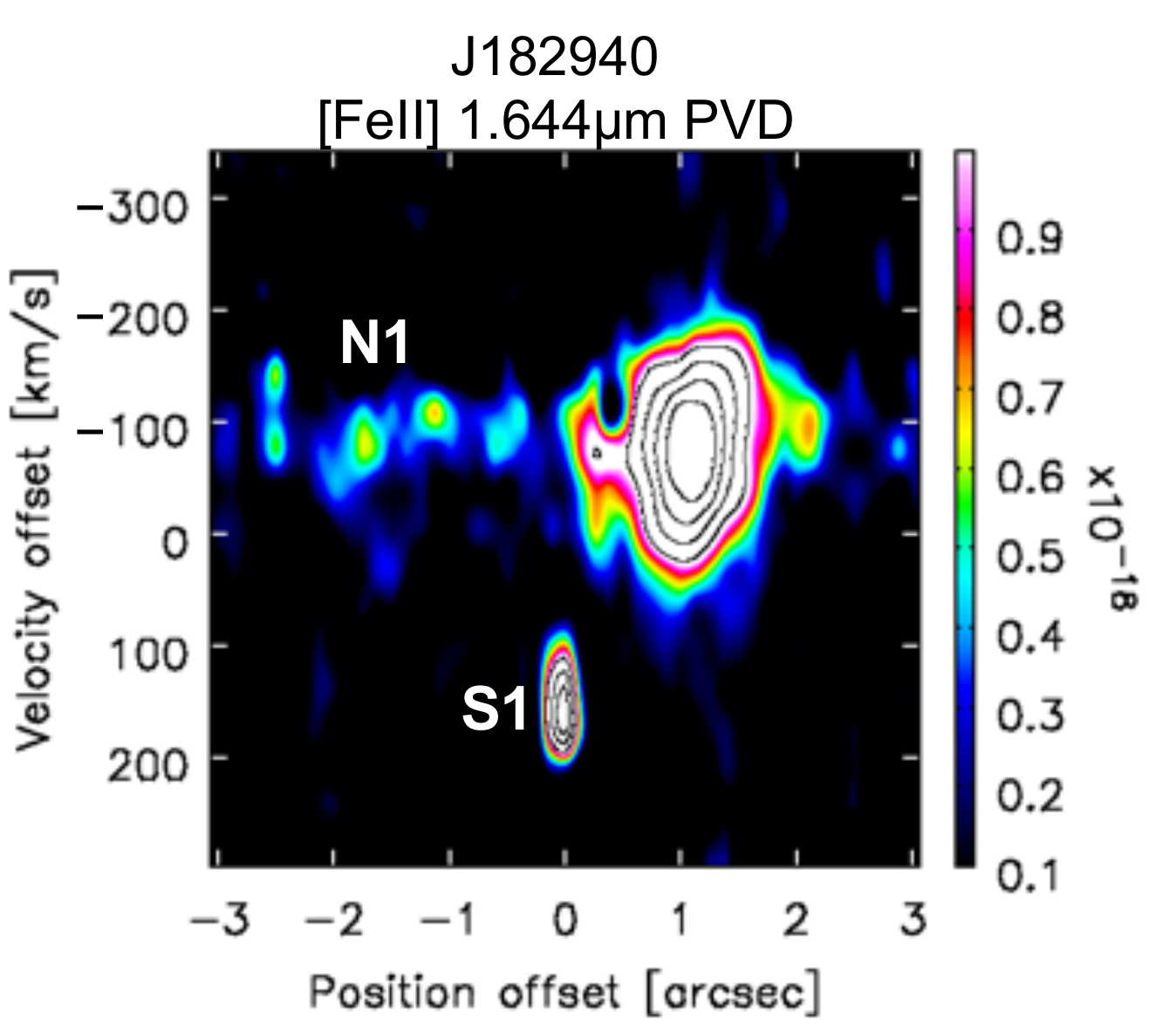}     
     \caption{The position-velocity diagrams for extended jets. The PVDs are produced after subtracting the continuum emission. The velocity offset is with respect to the vacuum wavelength of the emission line. The position offsets are with respect to the target position.The colour scale on the right shows the integrated flux in units of (erg s$^{-1}$ cm$^{-2}$).  }
     \label{PVDs}
  \end{figure*}

Figure~\ref{oph1-imgs} shows the spectro-images for J163136 in the brightest emission lines detected. J163136 shows extended jet emission in the H$_{2}$ 2.12$\micron$ line. The jet extends to $\sim$3.5$\arcsec$ ($\sim$500 au) north-west of the driving source, and $\sim$1$\arcsec$ ($\sim$144 au) south of the source. The PA of the jet is measured to be 74$\degr$$\pm$5$\degr$. The jet shows one extended knot, labelled N1, and a faint N2 knot. The width of the jet varies between $\sim$1--1.75$\arcsec$ ($\sim$144-252 au) across the length of the jet. Figure~\ref{PVDs} shows the position-velocity diagram (PVD) in the H$_{2}$ 2.12$\micron$ line. The PVD is produced along the jet PA = 74$\degr$ after continuum subtraction. The extended jet emission in the H$_{2}$ 2.12$\micron$ line is seen over the velocity range of -106--+67 km s$^{-1}$, with a peak velocity of -37 km s$^{-1}$. The full length of the jet is $\sim$4.8$\arcsec$ ($\sim$690 au). We also see weak extended knots in the [Fe~II] 1.257$\micron$ image at the same PA measured for the H$_{2}$ extended emission. On the other hand, the [Fe~II] 1.644$\micron$ shows compact emission at the source location. Likewise, Pa~$\beta$ and Br$\gamma$ line images also show compact structure concentrated at the source position.

Figure~\ref{lines} shows a comparison of the profiles in these spectral lines. The peak emission is blue-shifted in all lines. The narrowest profile is seen for the H$_{2}$ 2.12$\micron$ line (FWHM $\sim$ 50 km s$^{-1}$), while Pa~$\beta$ shows the broadest profile with FWHM $\sim$200 km s$^{-1}$. These two lines also show red-shifted emission at velocities of approximately +20 -- +100 km s$^{-1}$, though the blue- and red-shifted emission cannot be resolved from each other. The Br$\gamma$ line strength is weaker in comparison to Pa~$\beta$. The [Fe~II] lines are broader (FWHM$\sim$150-190 km s$^{-1}$) compared to H$_{2}$ and peak at a higher velocity of $\sim$ -100 km s$^{-1}$ (Fig.~\ref{lines}; Table~\ref{oph1-lines}).


\subsection{J182957}

J182957 is the only proto-BD in the sample that shows detection in the $\nu$ = 3,4,5,6 H$_{2}$ lines (Table~\ref{ser1-lines}). These lines indicate the presence of a hot gas component (E$_{(v,J)} >$ 20,000 K). Usually, the emission in vibrational excited H$_{2}$ ($\nu$ $\geq$3) is a sign of fluorescent excitation as opposed to shock excitation. It is possible that in the near-infrared, we are seeing H$_{2}$ that is pumped by UV emitted by the shocks where accretion flows from the inner accretion zone impact the central proto-BD surface.

J182957 shows strong emission in both the Pa~$\beta$ and Br$\gamma$ lines (Fig.~\ref{ser1-spec}). The FWHM is similar in both lines ($\sim$180 km s$^{-1}$). The H$_{2}$ 2.12$\micron$ and [Fe~II] 1.257$\mu$m lines are much narrower than the accretion tracers, with a FWHM of $\sim$70-80 km s$^{-1}$ (Fig.~\ref{lines}). All of these lines show a blue-shifted peak velocity, with a weak red-shifted wing seen in the H$_{2}$ 2.12$\micron$ line and the accretion tracers (Fig.~\ref{lines}). The peak velocity in the H$_{2}$ 2.12$\micron$ line (-37 km s$^{-1}$) is lower than the Pa~$\beta$ and Br$\gamma$ line emission (-91 km s$^{-1}$). In addition, several weaker [Fe~II], H$_{2}$, Brackett series lines and atomic Mg I line are detected in the H-band (Fig.~\ref{ser1-spec}; Table~\ref{ser1-lines}).

The mass accretion rate for J182957 derived from the Pa~$\beta$ line is 2$\times$10$^{-7}$ M$_{\sun}$ yr$^{-1}$, similar to the rate of 3$\times$10$^{-7}$ M$_{\sun}$ yr$^{-1}$ derived from the Br$\gamma$ line (Table~\ref{macc}). The mean outflow rate derived using the [Fe~II] lines is $\sim$3$\times$10$^{-8}$ M$_{\sun}$ yr$^{-1}$, which is $\sim$2 orders of magnitude higher than the $\dot{M}_{out}$ of 1$\times$10$^{-10}$ M$_{\sun}$ yr$^{-1}$ derived from the H$_{2}$ lines (Table~\ref{mout}). This implies a jet efficiency of $\dot{M}_{out}$[Fe~II]/$\dot{M}_{acc}$ $\sim$0.12 and $\dot{M}_{out}$H$_{2}$/$\dot{M}_{acc}$ $\sim$0.0004, using the mean accretion rate derived from the Pa$\beta$ and Br$\gamma$ lines.


The spectro-images in the brightest lines for J182957 are shown in Fig.~\ref{ser1-imgs}. There is an extended knot, N1, seen $\sim$4$\arcsec$ ($\sim$1744 au) north-west of the driving source at a PA = 32$\degr$$\pm$5$\degr$ in the H$_{2}$ 2.12$\micron$ line image. The H$_{2}$ 2.12$\micron$ line PVD constructed along this PA (Fig.~\ref{PVDs}) shows the peak emission at a $\sim$1.5$\arcsec$ offset from the source position. In addition to the bright knot N1 towards the north-west direction, a few faint knots are also seen towards the south-east of the driving source. The full length of the H$_{2}$ jet emission is $\sim$6$\arcsec$ ($\sim$2616 au). In comparison, the Pa~$\beta$ and Br$\gamma$ line images show a compact structure with a peak in emission within $\sim$0.5$\arcsec$ of the source position (Fig.~\ref{ser1-imgs}).

\subsection{J182940}


The strongest emission lines in several accretion and outflow tracers are seen in the spectra for J182940 (Fig.~\ref{ser8-spec}). The spectra show a wealth of [Fe~II] and H$_{2}$ line detections, with the strongest emission detected in the [Fe~II] 1.257$\micron$ and 1.644$\micron$ lines, and the H$_{2}$ 2.12$\micron$ line. The detection of H$_{2}$ lines at $\nu$ = 0,1,2 and $\nu \geq$3 (Table~\ref{ser8-lines}) indicates the presence of both a cold (E$_{(v,J)} <$ 6000 K) and a hot gas component (E$_{(v,J)} >$ 20,000 K). The gas temperature derived from the H$_{2}$ ($\nu$ = 1) rotational diagram is $\sim$1428 K. There is detection in both the Pa~$\beta$ and Br$\gamma$ accretion tracers for J182940, with similar strength in both lines. 


Figure~\ref{lines} shows a comparison of the profiles in the brightest lines. The lowest peak velocity is measured for the H$_{2}$ 2.12$\micron$ line (-3 km s$^{-1}$) compared to higher peak velocities of -67 -- -75 km s$^{-1}$ for the [Fe~II] 1.257$\micron$ and 1.644$\micron$ lines. The [Fe~II] lines are broader (FWHM $\sim$120--140 km s$^{-1}$) than the H$_{2}$ line (FWHM $\sim$75 km s$^{-1}$). The Br$\gamma$ line has a lower peak velocity (-23 km s$^{-1}$) compared to Pa~$\beta$ (-91 km s$^{-1}$). Pa~$\beta$ also shows a broad profile with a FWHM of $\sim$280 km s$^{-1}$.

The $\dot{M}_{acc}$ for J182940 derived from the Pa~$\beta$ and Br$\gamma$ lines is 4$\times$10$^{-7}$ M$_{\sun}$ yr$^{-1}$ and 6$\times$10$^{-8}$ M$_{\sun}$ yr$^{-1}$, respectively (Table~\ref{macc}). The mean mass outflow rate derived from the [Fe~II] lines is 3$\times$10$^{-8}$ M$_{\sun}$ yr$^{-1}$, while $\dot{M}_{out}$ derived from the H$_{2}$ lines is $\sim$8$\times$10$^{-11}$ M$_{\sun}$ yr$^{-1}$ (Table~\ref{mout}). This implies a jet efficiency of $\dot{M}_{out}$[Fe~II]/$\dot{M}_{acc}$ $\sim$0.13 and $\dot{M}_{out}$H$_{2}$/$\dot{M}_{acc}$ $\sim$0.0003, using the mean accretion rate derived from the Pa$\beta$ and Br$\gamma$ lines.

Figure~\ref{ser8-imgs} shows the spectro-images in the brightest lines. A small knot N1 is seen at a distance of $\sim$3$\arcsec$ ($\sim$1396 au) north-east of the driving source at a PA = 144$\degr$$\pm$5$\degr$ in the [Fe~II] 1.257$\micron$ line image. A slight extension along the same PA is also seen in the [Fe~II] 1.644$\micron$ image although there is no detection of the N1 knot. The [Fe~II] 1.644$\micron$ image instead shows a weak S1 knot south-east of the source at a distance of $\sim$1.7$\arcsec$ ($\sim$728 au) at a PA = 20$\degr$$\pm$3$\degr$. The H$_{2}$ 2.12$\micron$ image shows a broad structure centered at the source position with no extended emission or knots. The accretion tracers Pa~$\beta$ and Br$\gamma$ images show compact structures at the source position. The velocity offset in the knot N1 is within the range measured for the brightest emission at the source position, as seen in the [Fe~II] 1.257$\micron$ PVD in Fig.~\ref{PVDs}. In comparison, the faint S1 knot shows a higher velocity offset of $\sim$150 km s$^{-1}$ in the [Fe~II] 1.644$\micron$ PVD.

\subsection{J162648 and J163152}

Figures~\ref{oph3-spec};~\ref{oph2-spec} show the H and K band spectra for J162648 and J163152. The J-band spectra for these objects are at a low signal-to-noise ratio of $<$5, due to which no clear line detection can be made; as an example, we have shown the J-band spectrum for J163152 (Fig.~\ref{oph2-spec}). J162648 shows emission in several [Fe~II] and H$_{2}$ lines, along with upper Brackett lines and the Mg~I line in the H-band, while only the Br$\gamma$ line and weak H$_{2}$ 2.12$\micron$ line are detected in the K-band. Due to the weak [Fe~II] and H$_{2}$ line detection, there is no clear point source detection in any of these line images. In the K-band, there is a $>$10-$\sigma$ detection only in the Br$\gamma$ line, with a bright source detection in the spectro-image, as shown in Fig.~\ref{oph3-imgs}.

The H-band spectrum for J163152 shows detection in the upper Brackett lines of Br~10--13. Interestingly, J163152 does not show emission in any [Fe~II] lines. The K-band shows emission in several H$_{2}$ lines, however, only the H$_{2}$ 2.12$\micron$ and the Br$\gamma$ lines are detected at a $>$10-$\sigma$ level. The spectro-images in these two lines (Fig.~\ref{oph2-imgs}) show a compact structure with emission peaking at the source position.

Figure~\ref{lines} shows a comparison of the H$_{2}$ 2.12$\micron$ and Br$\gamma$ lines for J162648 and J163152. Also plotted is the [Fe~II] 1.644$\mu$m line profile for J162648. Both objects show a broad Br$\gamma$ profile with a FWHM of $\sim$250--300 km s$^{-1}$. In comparison, H$_{2}$ line shows a much narrower profile with a FWHM of $\sim$80--100 km s$^{-1}$. J162648 shows much weaker H$_{2}$ line emission compared to Br$\gamma$ or [Fe~II] 1.644$\mu$m line. For both objects, a red-shifted wing is seen in the H$_{2}$ 2.12$\micron$ and Br$\gamma$ lines. The peak velocity in H$_{2}$ 2.12$\micron$ and Br$\gamma$ lines is blue-shifted, and it is lower for H$_{2}$ (-37 -- -40 km s$^{-1}$) compared to Br$\gamma$ (-116 -- -118 km s$^{-1}$) (Tables~\ref{oph3-lines};~\ref{oph2-lines}).

The mass accretion rate derived from the Br$\gamma$ line is 2$\times$10$^{-6}$ M$_{\sun}$ yr$^{-1}$ for J163152 and 2$\times$10$^{-8}$ M$_{\sun}$ yr$^{-1}$ for J162648. The mean $\dot{M}_{out}$ derived from the H$_{2}$ lines for J163152 is 5$\times$10$^{-9}$ M$_{\sun}$ yr$^{-1}$. The mean $\dot{M}_{out}$ derived from the [Fe~II] lines for J162648 is 2$\times$10$^{-8}$ M$_{\sun}$ yr$^{-1}$, higher than the rate of 8$\times$10$^{-10}$ M$_{\sun}$ yr$^{-1}$ derived from the H$_{2}$ 2.12$\micron$ line (Table~\ref{mout}). This implies a jet efficiency of 0.002 for J163152 and 0.04--1 for J162648.



\section{Discussion}
\label{discuss}



\subsection{Two types of flows}


The blue-shifted emission seen in the [Fe~II] and H$_{2}$ lines for the proto-BDs in our sample is consistent with an origin in the outflow. Among the most prominent [Fe~II] and H$_{2}$ line detections (Fig.~\ref{lines}), the peak velocities of the [Fe~II] lines ($>$100 km s$^{-1}$) are higher than the H$_{2}$ lines ($<$100 km s$^{-1}$). The [Fe~II] forbidden emission lines and H$_{2}$ rotational-vibrational emission lines trace different flow components. The [Fe~II] atomic jet is associated with high-velocity (100-200 km s$^{-1}$), hot ($T_{e} \sim$10,000 K), high-density (n$_{e} \sim$10$^{5}$ cm$^{-3}$), partially-ionized gas, while the H$_{2}$ molecular outflow is associated with low-velocity ($\sim$10-50 km s$^{-1}$), cool ($T_{e} \sim$2000 K), low-density (n$_{e} \geq$10$^{3}$ cm$^{-3}$), low-excitation shocked gas.

Previous works on low-mass Class 0/I protostars have noted that the [Fe~II] lines trace the collimated jet while H$_{2}$ lines trace the wide-angled low-velocity outflow or the outflow cavity walls (e.g., Davis et al. 2011). We do not have the spatial resolution to measure the width of the observed structure at the source position and compare the width and opening angle of the [Fe~II] vs. H$_{2}$ emission. We see a peak in the [Fe~II] and H$_{2}$ emission within $\sim$1$\arcsec$ of the source, which for the nearest target implies a distance of $<$144 au. It is within these close distances that H$_{2}$ and [Fe~II] flows begin to separate and take the shape of a wide-angled outflow or a collimated jet. Beyond $\sim$100 au, the flows will interact with the ambient medium and produce shocked knots resulting in a mixture of the two emission line regions. The shock emission knots also indicate a variable outflow. 


In general, we do not see any notable offsets between the peak H$_{2}$ and [Fe~II] emission and the peak continuum emission in the spectro-images, whereas typical offsets of $\sim$10-200 au have been observed in protostars. This could be related to the compact spatial scales of proto-BDs compared to protostars. As shown in recent ALMA observations (Riaz et al. 2019), the pseudo-disk in a proto-BD is $\sim$150-200 au in size compared to $>$500 au in protostars. The inner Keplerian disk is predicted by models to be $\sim$10 au in proto-BDs compared to $\sim$50-100 au in protostars. 

Low-velocity, wide-angled outflows are expected to be driven near the outer pseudo-disk region, while high-velocity, collimated jets are driven near the inner Keplerian disk (e.g., Machida et al. 2009). The jet/outflow launching regions are therefore $\sim$10-150 au in proto-BDs, which at the distance to our targets would require $\sim$0.02$\arcsec$-0.07$\arcsec$ to resolve the [Fe~II] collimated jet and $\sim$0.3$\arcsec$-1$\arcsec$ for H$_{2}$ outflow spatial resolution to resolve the offsets. This could explain why any clear offsets of $\geq$0.5$\arcsec$ are only seen in H$_{2}$ but not [Fe~II] line images because the launching region of H$_{2}$ is expected to be in the outer pseudo-disk regions of $\sim$150-200 au. The H$_{2}$ and [Fe~II] emission requires high excitation and thus high densities such as those found at the base of the jet to be excited, while both the line intensities and densities are known to decline with distance from the driving source. Higher spatial resolution spectro-imaging observations can resolve the launching regions of the FEL and MHEL regions in proto-BDs.


\subsection{Dependence on evolutionary stage}
\label{evolution}

We find a range in the [Fe~II] and H$_{2}$ line strengths among the proto-BDs in the sample, which may be related to their evolutionary stage and/or the inclination of the system. If we compare the extreme cases of the Stage 0 objects J163152 and J182957 with the Stage I-T/II object J182940, then a clear difference is the lack of detection in the [Fe~II] lines and strong emission in the Br~$\gamma$ accretion tracer in the Stage 0 objects (Fig.~\ref{lines}). The Stage I objects show emission in both [Fe~II] and H$_{2}$ lines with varied strengths.


The absence of [Fe~II] emission could be due to low jet density of n$_{e} \sim$10$^{3}$ -- 10$^{5}$ cm$^{-3}$ estimated for the proto-BD targets (Sect.~\ref{temp_den}). If the gas excitation conditions at the base of the jet are modest, then such conditions will be favourable for H$_{2}$ emission but not [Fe~II]. The absence of [Fe~II] but presence of H$_{2}$ line detection indicates that the jet has low excitation conditions or it is composed mostly of molecular material. The absence of [Fe~II] emission could also be due to a density much above the critical, making collisional de-excitation the preferential mechanism of de-excitation of [Fe~II]. The presence of both [Fe~II] and H$_{2}$ lines in Stage I objects suggests that if [Fe~II] traces the collimated high-velocity jet, then the momentum in this jet is sufficient to drive, entrain, and accelerate the molecular outflow traced by H$_{2}$. The H$_{2}$ lines may be tracing the molecular gas entrained in a boundary layer between the jet and the ambient medium within a few hundred au of the driving source (e.g., Davis et al. 2001).



An evolutionary trend in the jets from a molecular to an ionic composition is expected due to changes in the density of the jet and the ambient medium where the jet propagates. We have three jets in Ophiuchus and two in Serpens. No particular change is seen in the jet emission between these two regions, although the sample size is too small. Jets from embedded Stage 0 sources travel in high-density gas where the protostar is embedded. In this embedded environment, non-dissociative C-type shocks that cool mainly through molecular lines such as H$_{2}$ should be favoured. In the relatively more evolved Stage I objects, the jet propagates in a medium at a lower density since previous mass loss events have swept out the ambient gas. In this case, dissociative J-type shocks are favoured due to the lower influence of the local magnetic fields whose strength is a function of density (e.g., Davis et al. 2011; Caratti o Garatti et al. 2006). Only two objects in our sample, M1701117 (Stage I) and J182940 (Stage I-T/II), are associated with optical HH jets while the rest are undetected in the optical, as expected from the more embedded systems. There may be optical HH jets at a small-scale very close to the embedded regions near the Stage 0/I systems that remain undetected due to high extinction in these seeing limited observations.


The first overtone of ro-vibrational CO bandheads are seen in emission at $\sim$2.3-2.5 $\micron$ in embedded protostars (e.g., Davis et al. 2011; Carrati o Garrati et al. 2006). The red-edge of the K-band spectra for our targets are either too noisy or do not show any clear CO bandheads in emission due to the strong H$_{2}$ line detection. The CO lines likely originate from the innermost regions of the accretion disk or in dense stellar winds or in accretion flows between the inner disk surface and the proto-BD. These innermost regions are expected to be at a distance of within 10 au from the central proto-BD and thus unresolved in the present observations. The non-detection in CO could be due to beam dilution of these innermost densest regions. We also note that there is detection in the Mg~I atomic line in J182957 (Stage 0), J162648 (Stage I), and J163136 (Stage I) objects. The Mg I line has a low ionisation potential of 7.6 eV compared to 13.6 eV for the H I lines, and likely originates from the outer disk regions (e.g., Nisini et al. 2004; Davis et al. 2011). 


No clear correlation is seen between the evolutionary stage and the presence of extended vs. compact jet emission. There is clear extended emission with knots seen in J163136 (Stage I), J182957 (Stage 0), and J182940 (Stage I-T/II). The extended emission is seen only in H$_{2}$ for J163136 and J182957, while J182940 shows extended emission only in [Fe~II]. The presence of shocked emission knots indicates multiple epochs or variable jet emission, while asymmetries in the extended H$_{2}$ emission as seen in J163136 could be due to inhomogeneities in the ambient medium. The brightest emission in the [Fe~II] and H$_{2}$ spectro-images is seen at the source position, indicating that the jet emission originates from close to the driving source. Any extended knots are much weaker in intensity. For the remaining targets, we do not see any notable offset from the continuum position in the [Fe~II] and H$_{2}$ line images, suggesting the presence of shock layers at different excitation behind the shock front that are not spatially resolved. 

None of the proto-BD jets show a definitive detection in the [Fe~II] lines in the K-band. The upper level energies of these transitions are $\sim$2.6$\times$10$^{4}$ K (e.g., Takami et al. 2006). Such high-level transitions have been observed in jets driven by high-mass protostars, such as SVS 13, B5 IRS 1, HH 34 IRS, with L$_{bol}$ $>$ 10 L$_{\sun}$. In contrast, all [Fe~II] lines in proto-BDs are detected for the 4D --> 4F transitions, implying upper level energies of $\sim$1.2$\times$10$^{4}$ K. 


Among the accretion tracers, J182957 (Stage 0), J163136 (Stage I) and J162648 (Stage I) show emission in the upper Brackett lines of Br~13--19, J163152 (Stage 0) shows emission in the Br~10--13 lines, whereas the more evolved object J182940 (Stage I-T/II) shows emission only in the Br~10-11 lines. J182940 also shows the weakest emission in Br$\gamma$ while J163152 (Stage 0) and J162648 (Stage I) show the strongest Br$\gamma$ emission (Fig.~\ref{lines}; Sect.~\ref{Brdec}). There is thus a possible trend of the detection of the upper Brackett lines of Br~13--19 in the earlier stage systems.

The Pa$\beta$ line is the strongest in J163136 (Stage I) and J182957 (Stage 0). There is no Pa$\beta$ detection in J163152 (Stage 0) that also does not show [Fe~II] emission, suggesting a possible trend of Pa$\beta$ detection if [Fe~II] is detected. Likewise, the fact that J162648 and J163152 show emission in only Br$\gamma$ and H$_{2}$ lines, while Pa$\beta$ is undetected and [Fe~II] is either weakly detected or undetected, suggests a possible correlation between the presence of MHEL and Br$\gamma$ emission line regions. 

The Br$\gamma$ emission derives from the inner accretion zone, i.e., the region linking the inner disk at its truncation radius and the surface of the central object, while the H$_{2}$ emission derives from the base of a large scale molecular outflow. Note that the lack of detection in the Pa$\beta$ could also be due to higher extinction in the $J$-band compared to $HK_{s}$ bands. J162648 and J163152 are the faintest objects in the $J$-band, with $J\sim$18-19 mag. This is close to the $J\sim$19-20 magnitude limit of the SINFONI instrument.

The Paschen and Brackett hydrogen recombination lines likely originate from shocks associated with magnetospheric accretion (e.g., Muzerolle et al. 1998). The Br$\gamma$ line has a critical density of $>$10$^{10}$ cm$^{-3}$ and originates from the inner, more ionised disk regions where Keplerian velocities are high. Emission in the upper Brackett lines is associated with the same high excitation regions observed in Br$\gamma$, namely the inner regions of the accretion disk, magnetospheric accretion flows, and/or the first few au of the jet. The Br$\gamma$ and Pa$\beta$ spectro-images for the proto-BDs show the peak line emission to be coincident with the peak continuum emission for all objects; however, it is difficult to resolve the emission from an accretion flow from the base of the jet or a less collimated ionized wind.

The highest mean accretion rate is measured for J163152 (Stage 0), while the lowest mean accretion rate is measured for J162648 (Stage I). The mass outflow rate derived using the [Fe~II] lines is at least an order of magnitude higher than H$_{2}$ lines. The highest mean mass outflow rate derived from the [Fe~II] lines is for M1701117 (Stage I), while the lowest $\dot{M}_{out}$ is for J163136 (Stage I). The highest mean outflow rate derived from the H$_{2}$ lines is for J163152 (Stage 0), while the lowest is for J182940 (Stage I-T/II). Overall, there is no clear relation between the evolutionary stage and activity rates for the proto-BDs. Both the evolutionary stage and the inclination of the system can play a role in enhancing or suppressing the activity levels; in the case of the inclination it is the tracers of activity that become enhanced or suppressed, but not the activity itself. The highest $\dot{M}_{acc}$ and $\dot{M}_{out}$ for J163152 is expected from its early Stage 0 phase, whereas the edge-on inclination of J182940 will allow a more direct view of the accretion disk as well as the extended region emitting the forbidden lines resulting in similar activity rate estimates compared with the more embedded systems.

The variable nature of accretion and outflow activity in young embedded systems and the influence of ambient medium on line intensities can also result in the non-correlation. Nevertheless, the similarities with the protostars in the accretion and outflow rates and other properties suggest that the MHEL and FEL regions are also common in the first evolutionary stages of brown dwarfs, suggesting similar launching mechanisms in low-mass stars and brown dwarfs.

On the other hand, the mean accretion and outflow rates for the more evolved Class II brown dwarfs are of the order of 10$^{-9}$ -- 10$^{-11}$ M$_{\sun}$ yr$^{-1}$ (e.g., Whelan et al. 2009), lower than our proto-BD sample, and is consistent with the activity levels being less intense at later evolutionary stages. The differences in the rates and other physical properties are more pronounced when comparing Class 0/I with Class II brown dwarfs, but not among Class 0 and I proto-BD systems. The scatter in the individual properties of the objects in our sample, added to the effects of geometry, inclination, extinction, sample size, etc., are large enough to mask any genuine differences between the properties of Class 0 and I proto-BD systems, which probably does not exclude the actual existence of such differences. 


\subsection{Comparison with low-mass protostars}
\label{compare}

\begin{figure*}
  \centering              
     \includegraphics[width=2.5in]{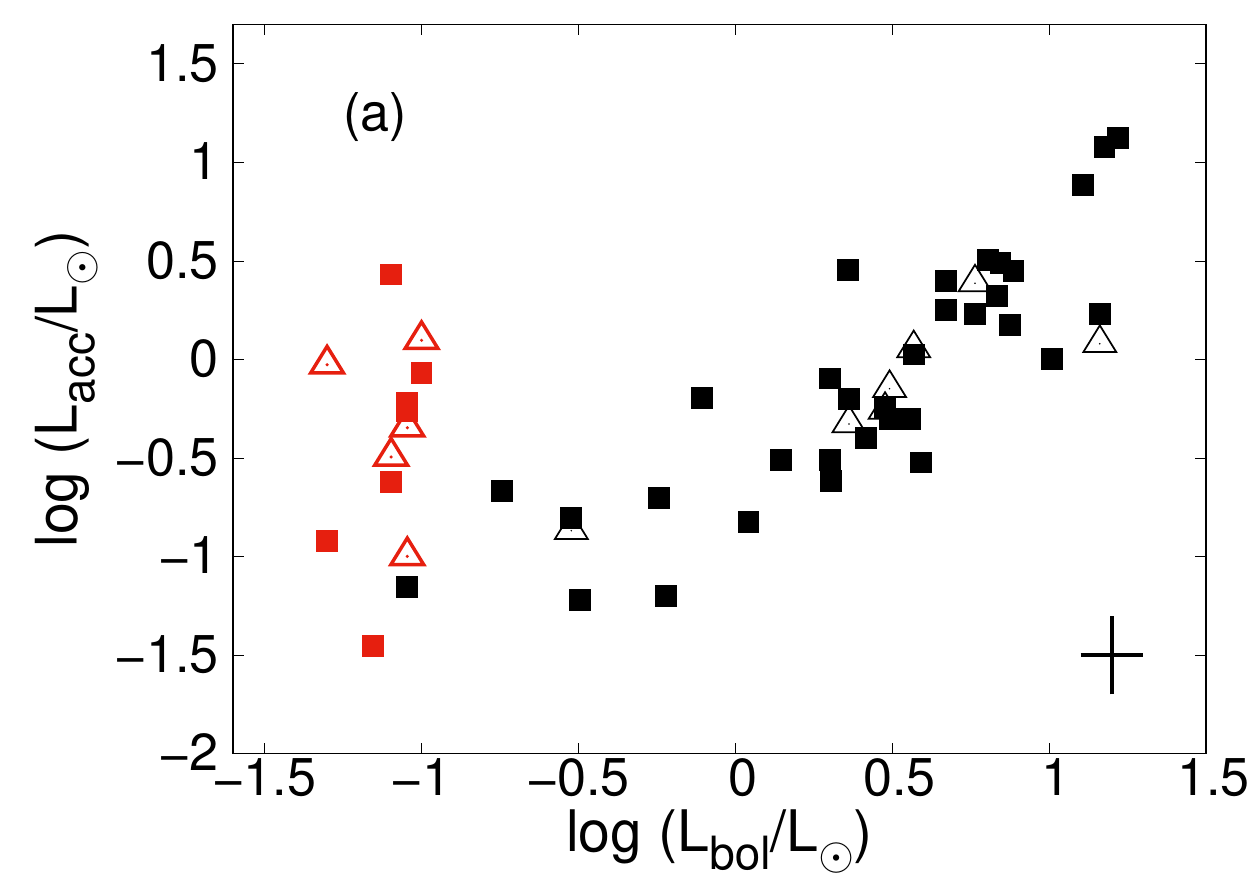}
     \includegraphics[width=2.5in]{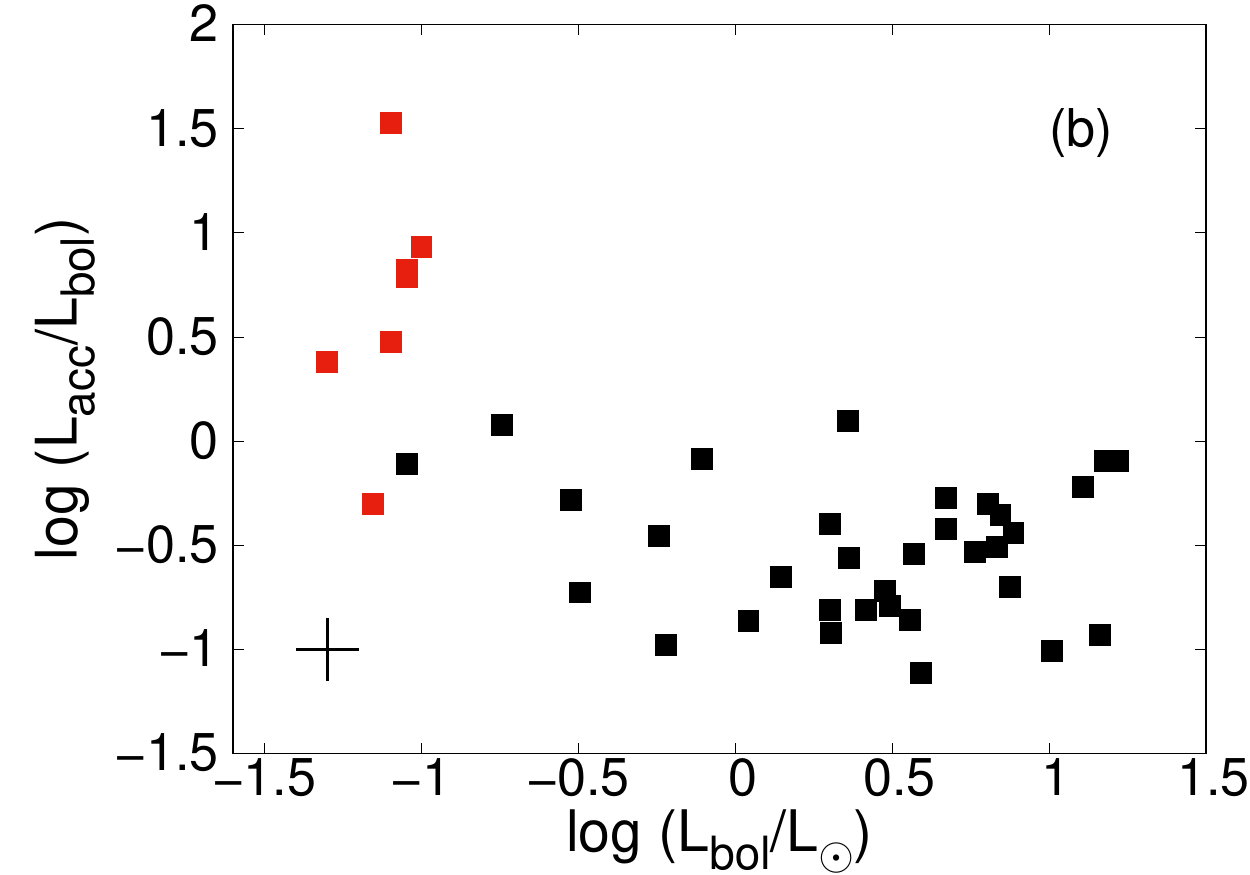}
     \includegraphics[width=2.5in]{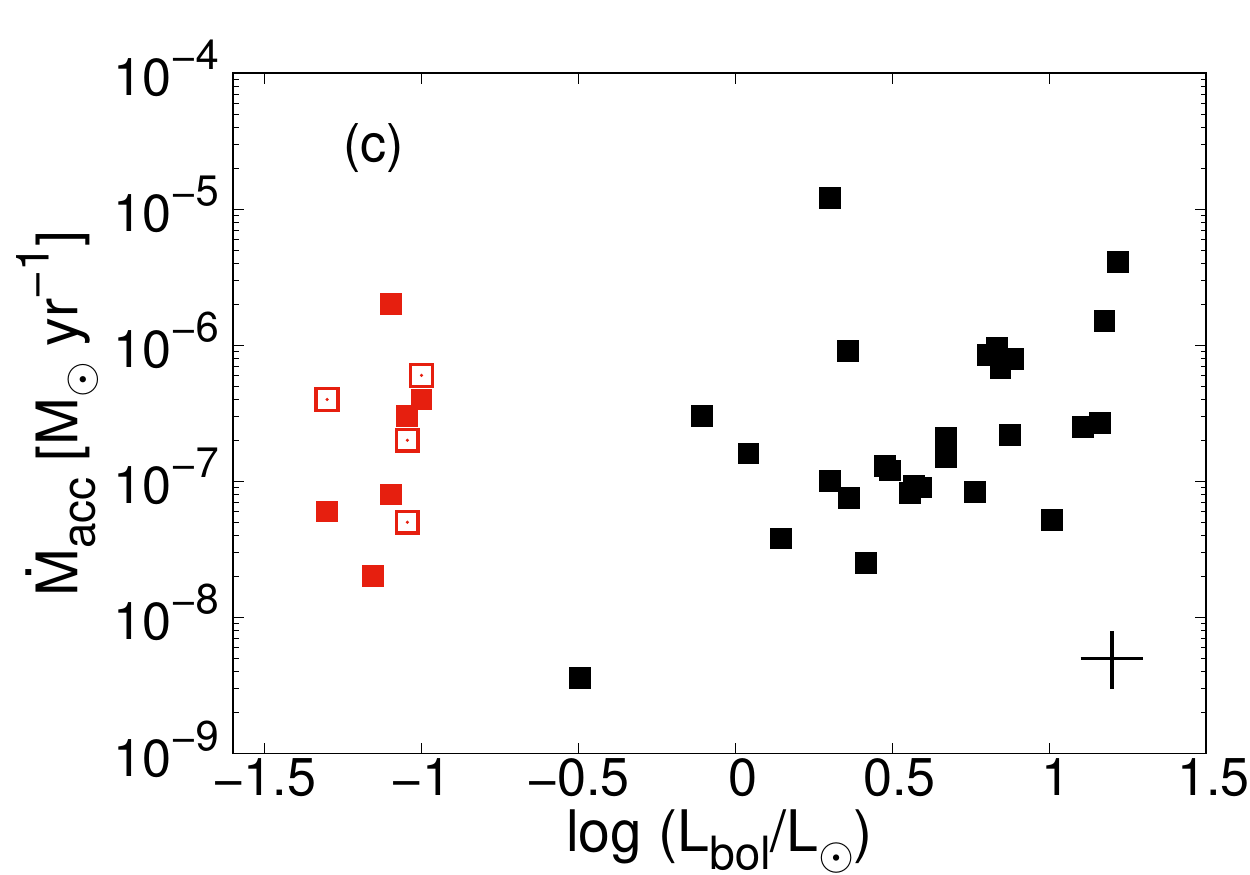}
     \includegraphics[width=2.5in]{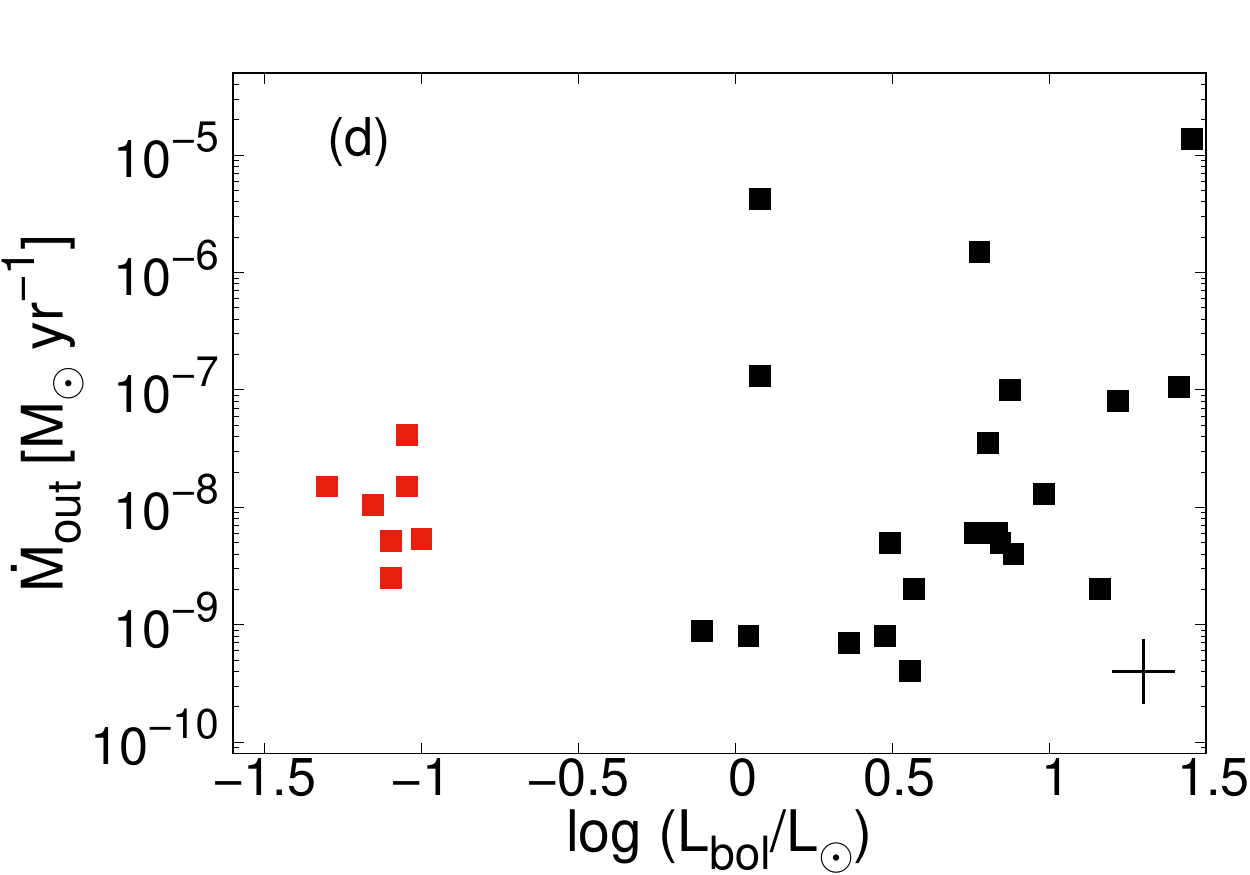}     
     \includegraphics[width=2.5in]{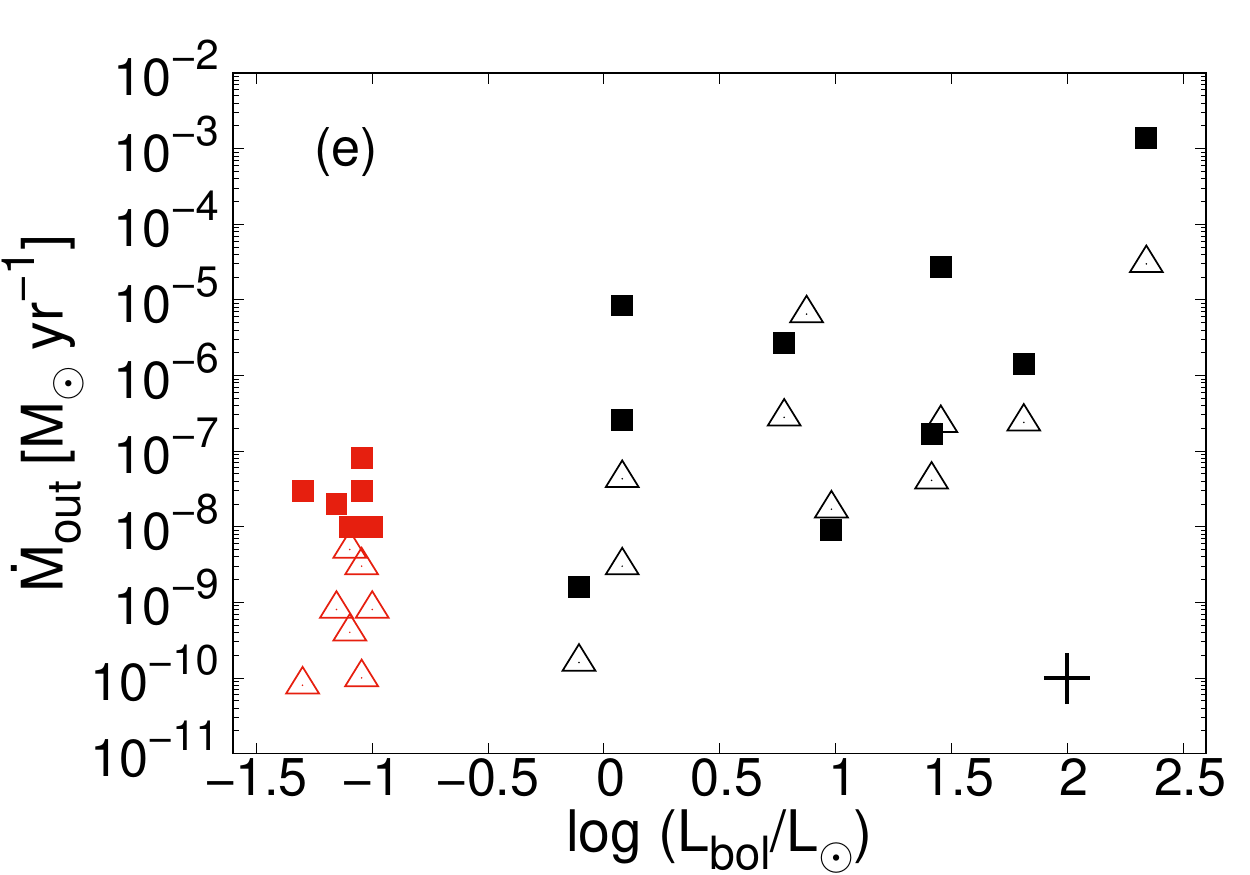}
     \includegraphics[width=2.5in]{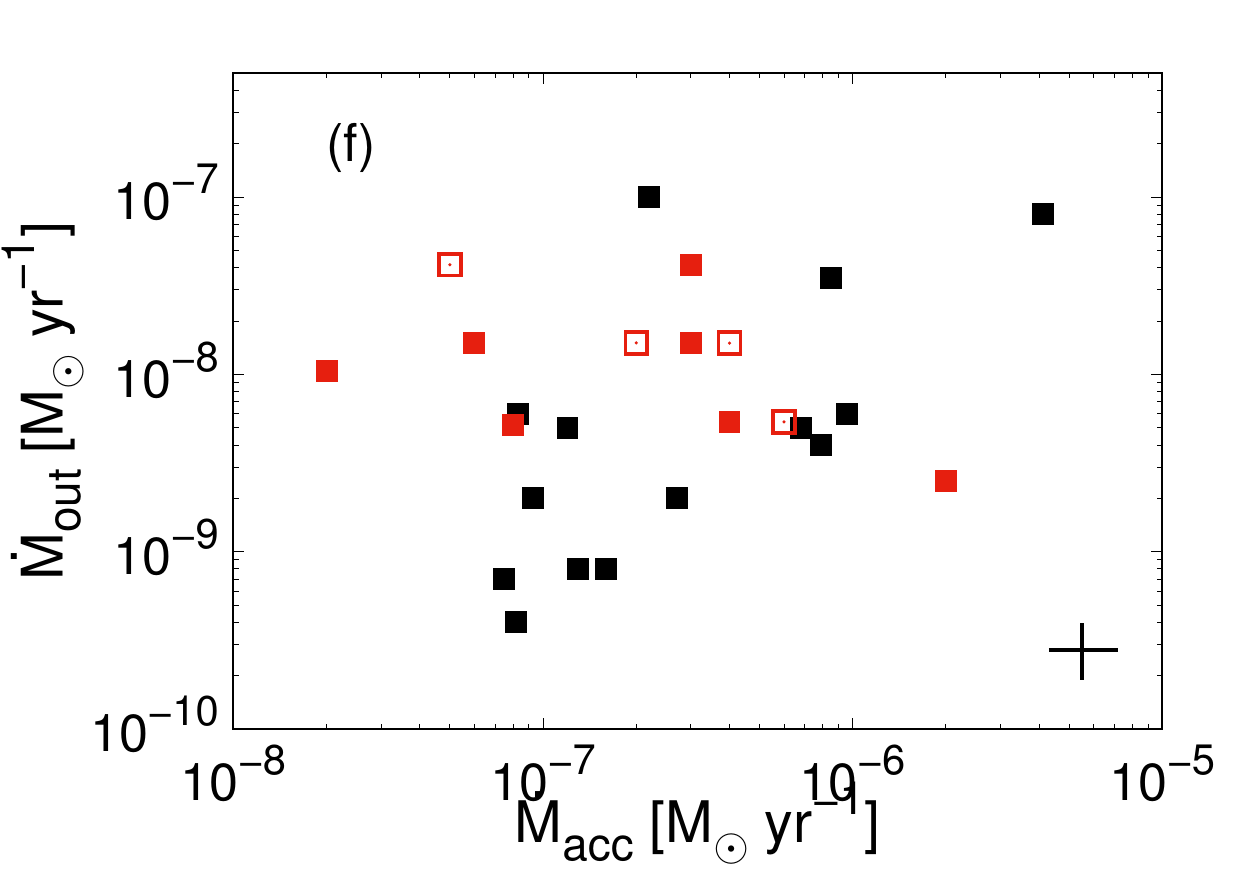}
     \includegraphics[width=2.5in]{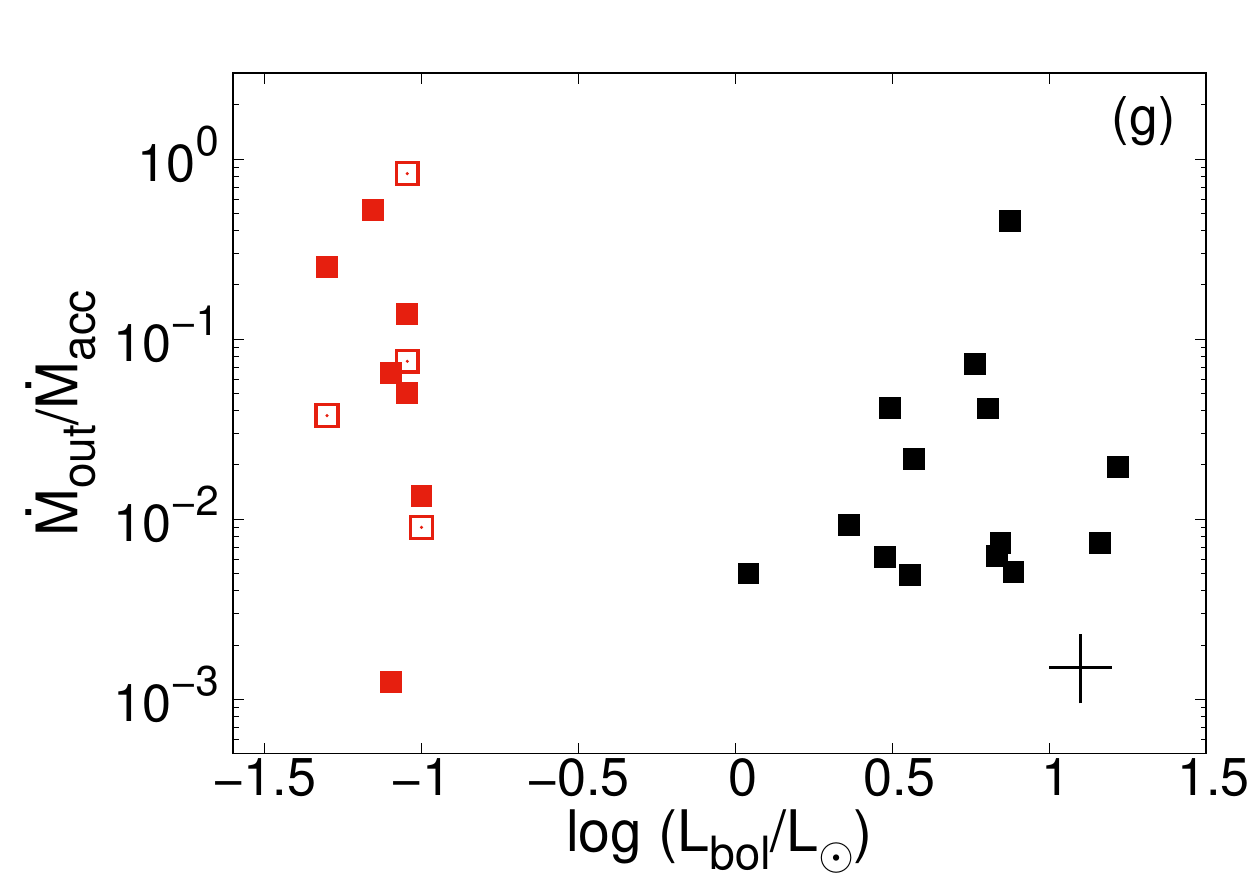}
     \caption{A comparison of accretion and outflow activity in proto-BDs (red) with low-mass protostars (black). (a) L$_{acc}$ vs. L$_{bol}$. The accretion luminosity derived using the Pa~$\beta$ line is plotted as triangles  and that derived using the Br$\gamma$ line is plotted as squares. (b) L$_{acc}$/L$_{bol}$ vs. L$_{bol}$. The L$_{acc}$ was measured using the Br$\gamma$ line. (c) $\dot{M}_{acc}$ measured using the Br$\gamma$ line (filled squares) and the Pa~$\beta$ line (open squares) vs. L$_{bol}$. (d) $\dot{M}_{out}$ vs. L$_{bol}$. The average values of $\dot{M}_{out}$ derived using the [Fe~II] and H$_{2}$ lines are plotted. (e) $\dot{M}_{out}$ vs. L$_{bol}$. The $\dot{M}_{out}$ measured using the [Fe~II] lines is plotted as squares and that derived using the H$_{2}$ lines is plotted as triangles. (f) $\dot{M}_{out}$ vs. $\dot{M}_{acc}$ measured using the Br$\gamma$ line is plotted as filled squares and using the Pa~$\beta$ line as open squares. (g) The jet efficiency ($\dot{M}_{out}$/$\dot{M}_{acc}$) vs. L$_{bol}$. The average values of $\dot{M}_{out}$ derived using the [Fe~II] and H$_{2}$ lines are plotted. $\dot{M}_{acc}$ measured using the Br$\gamma$ line is plotted as filled squares and using the Pa~$\beta$ line as open squares. }
     \label{relations}
  \end{figure*}

Figure~\ref{relations} shows a comparison of the accretion and outflow activity rates for the proto-BDs with low-mass protostars. The data for protostars is from Davis et al. (2011), Carrati o Garrati et al. (2006), Antoinucci et al. (2014), and Nisini et al. (2006). We have also included in these plots the activity rates for the proto-BD ISO-Oph 200 derived from similar SINFONI observations (Whelan et al. 2018). A large scatter in L$_{acc}$ is seen for both the proto-BDs and protostars (Fig.~\ref{relations}a), which could be explained by variable accretion and ejection rates of the sources. A similar range is seen for L$_{acc}$ derived using the Br$\gamma$ and Pa~$\beta$ lines.

In Fig.~\ref{relations}b, the proto-BDs show higher L$_{acc}$/L$_{bol}$ ratios compared to protostars, which suggests more intense accretion activity in proto-BDs. Considering the large scatter in L$_{acc}$ for the proto-BDs (Fig.~\ref{relations}a), the very low L$_{bol}$ could explain the high ratios. Sources with dense jets excited very close to the driving source are expected to show high L$_{acc}$/L$_{bol}$ ratios of $>$0.5 (e.g., Nisini et al. 2006). The high L$_{acc}$/L$_{bol}$ ratios of $\geq$1 indicates that there must be some extra source of H-recombination line emission and the Br~$\gamma$ and Pa~$\beta$ lines may not be tracing accretion alone but may have other contribution, such as, shocks in the outflows. This excess may be caused by hard shocks in the outflows which ionize H and add recombination line flux. The highest ratio is seen for J163152, a Stage 0 proto-BD, while lowest ratio is for J162648 (Stage I). The present trend suggests that Stage I objects appear to be in a phase of less intense accretion, however, the sample size is too small to make a definitive conclusion.

Figure~\ref{relations}cd show a slight decline in $\dot{M}_{acc}$ and $\dot{M}_{out}$ with decreasing L$_{bol}$, although the activity rates for the proto-BDs lie in a similar range as seen in protostars. Also notable are the nearly constant outflow rates for the proto-BDs over a range in L$_{bol}\sim$0.1--0.01 L$_{\sun}$. Figure~\ref{relations}e shows that the $\dot{M}_{out}$ measured using [Fe~II] lines is higher than H$_{2}$ lines, indicating that [Fe~II] traces a large fraction of the total outflow mass. No strong correlation is seen between the accretion and outflow activity rates for proto-BDs or protostars in Fig.~\ref{relations}f. The spread in the activity rate measurements is too large to definitively conclude on any trends.



Figure~\ref{relations}g shows a similar range in the jet efficiencies ($\dot{M}_{out}$/$\dot{M}_{acc}$) for the proto-BDs compared to protostars. There may be a possible trend of higher jet efficiency with decreasing luminosity. The highest ratio among proto-BDs is derived for J162648 (Stage I), while the lowest ratio is for J163152 (Stage 0). This suggests a possible trend of higher jet efficiencies among the more evolved cases. As noted in Sect.~\ref{evolution}, the mass loss rates derived using the [Fe~II] lines are at least an order of magnitude higher than H$_{2}$ lines (Table~\ref{mout}), and there is a tentative trend of non-detection in the [Fe~II] lines and strong emission in the accretion tracers among the Stage 0 proto-BDs. This can explain the lower jet efficiencies in the Stage 0 objects.

The overall picture from Figure~\ref{relations} suggests that there is no notable decline in the accretion and outflow activity levels or jet efficiencies for sub-stellar objects during the early evolutionary stages. A larger proto-BD sample is needed to further investigate these tentative trends.







\subsection{Brackett decrement}
\label{Brdec}

\begin{figure*}
  \centering              
     \includegraphics[width=2.in]{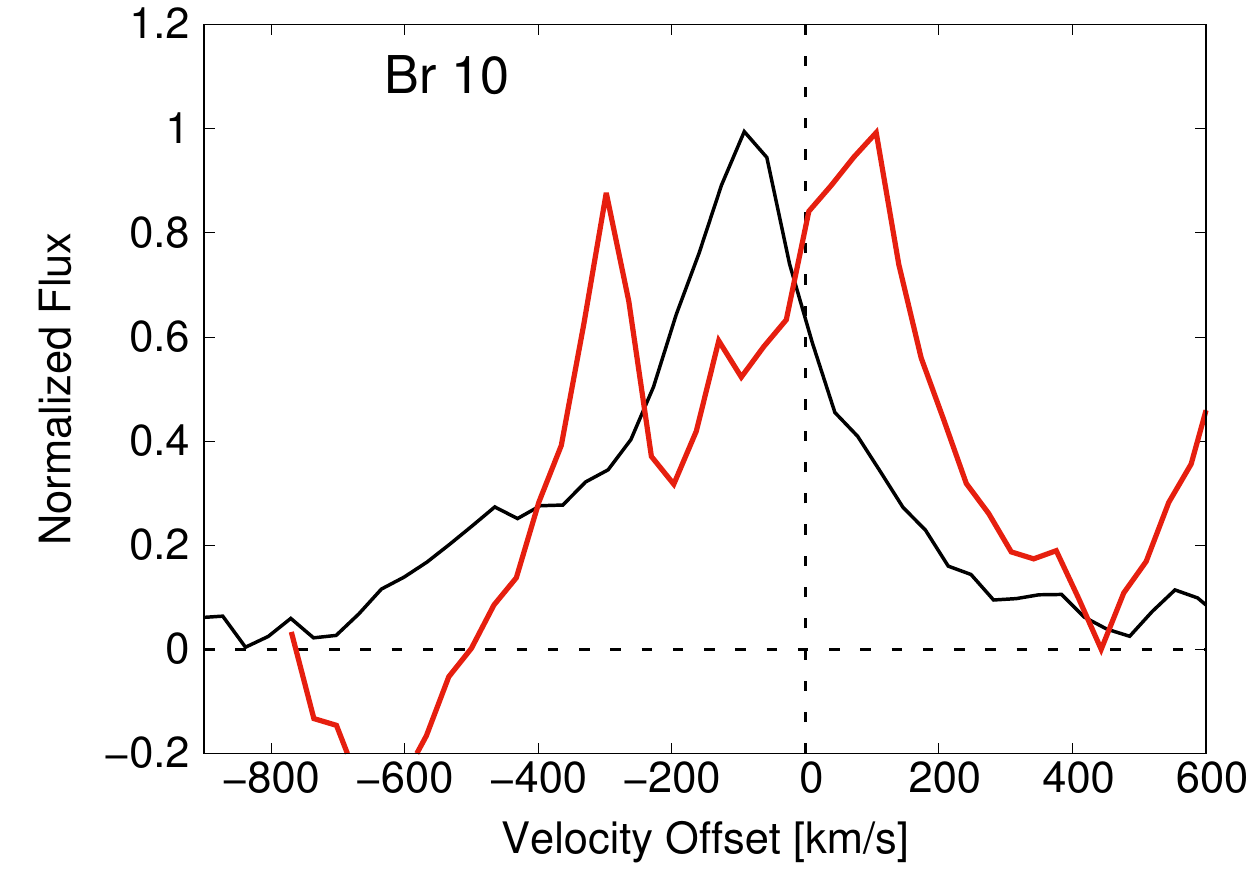}
     \includegraphics[width=2.in]{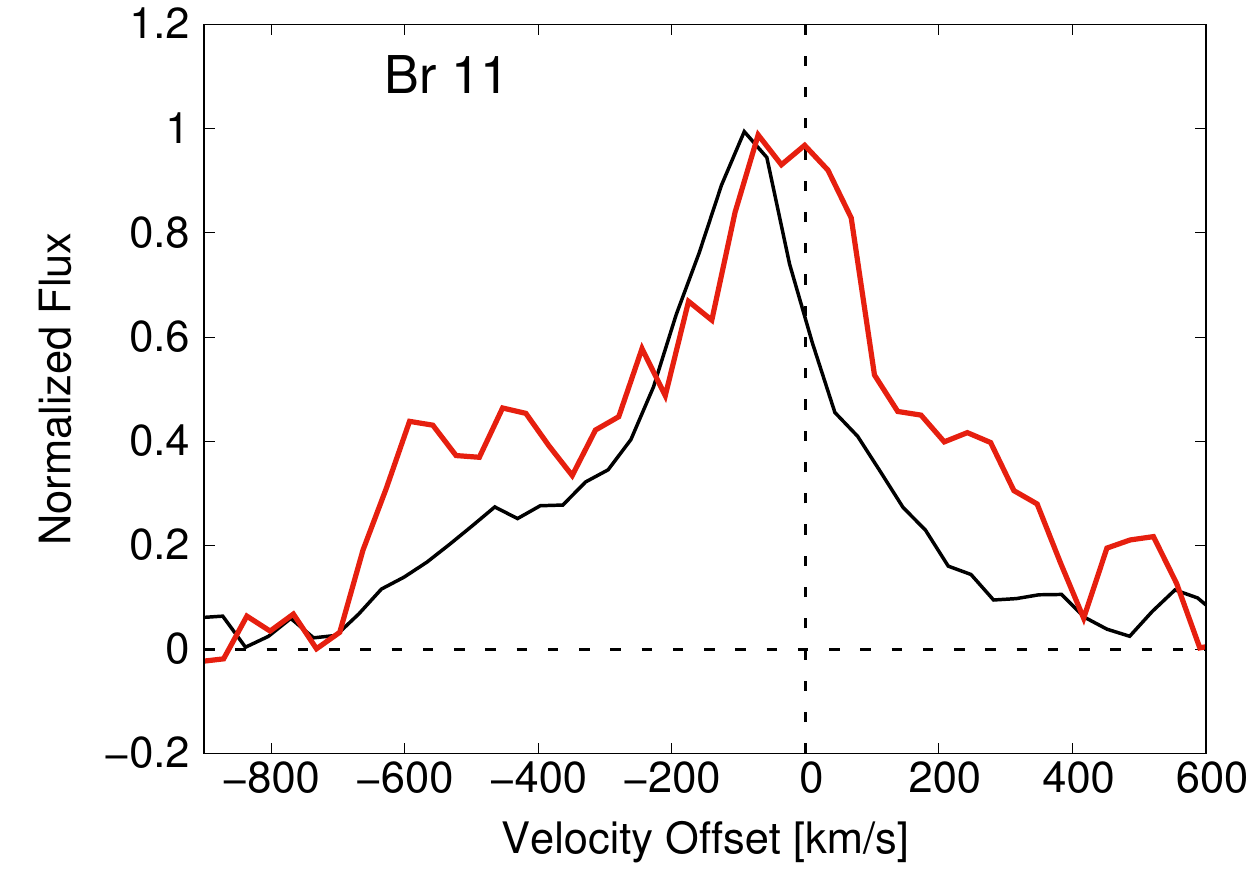}    
     \includegraphics[width=2.in]{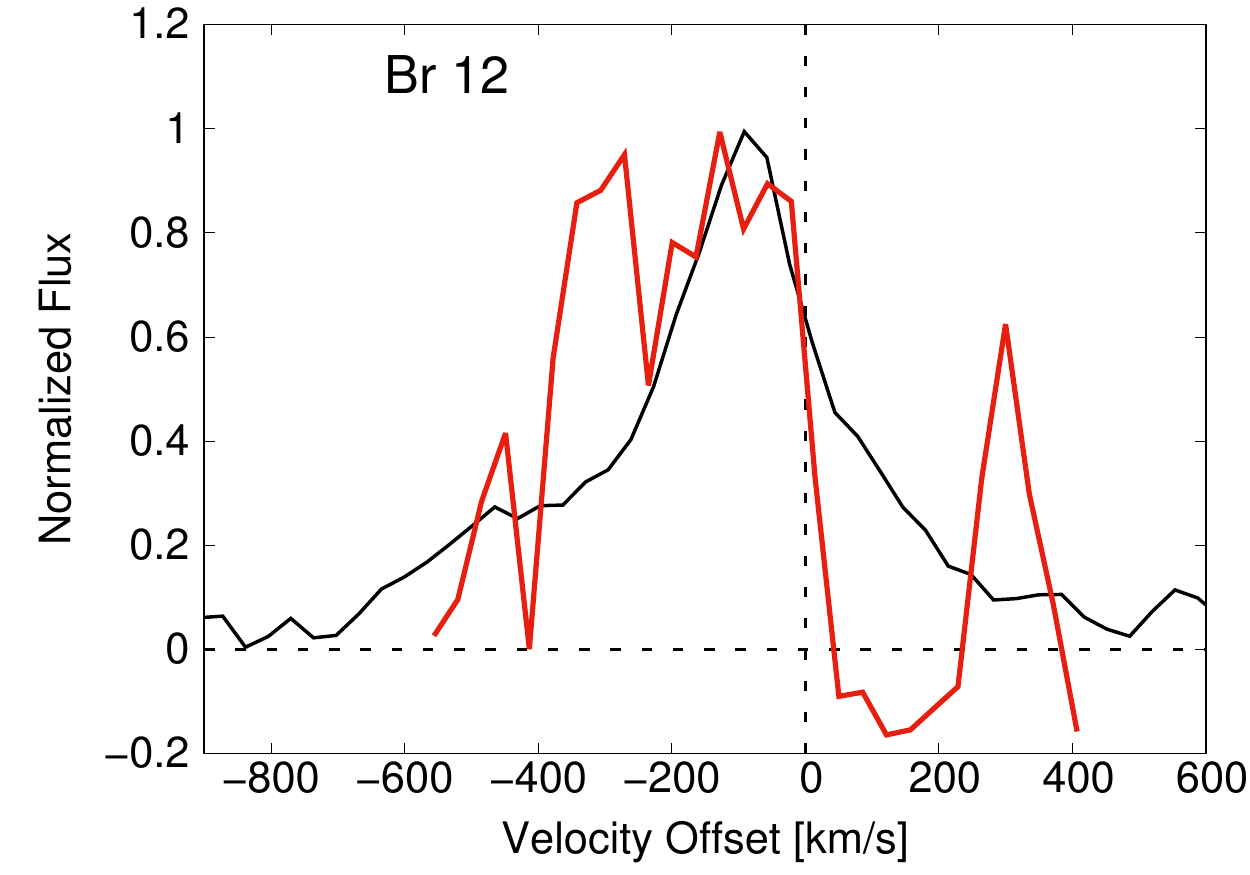}     
     \includegraphics[width=2.in]{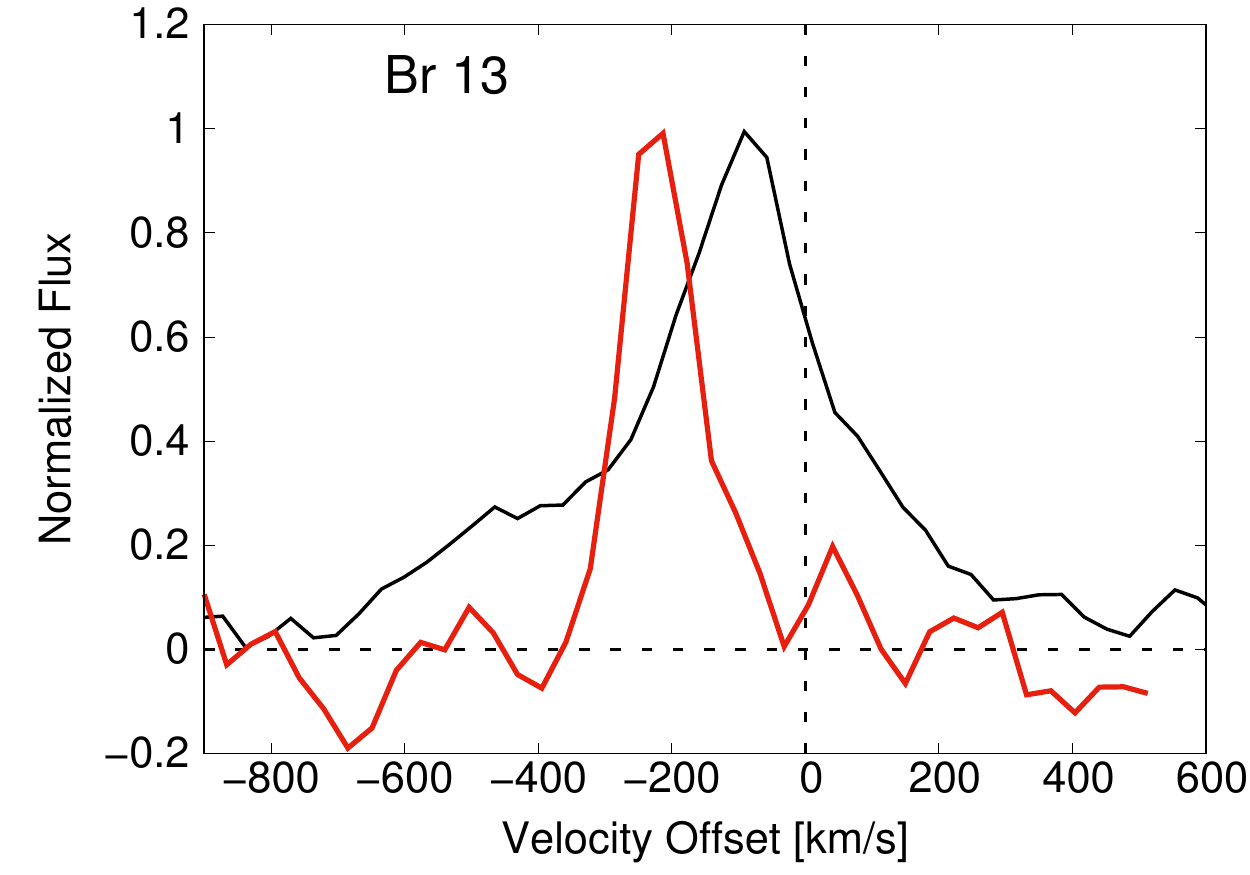}
     \includegraphics[width=2.in]{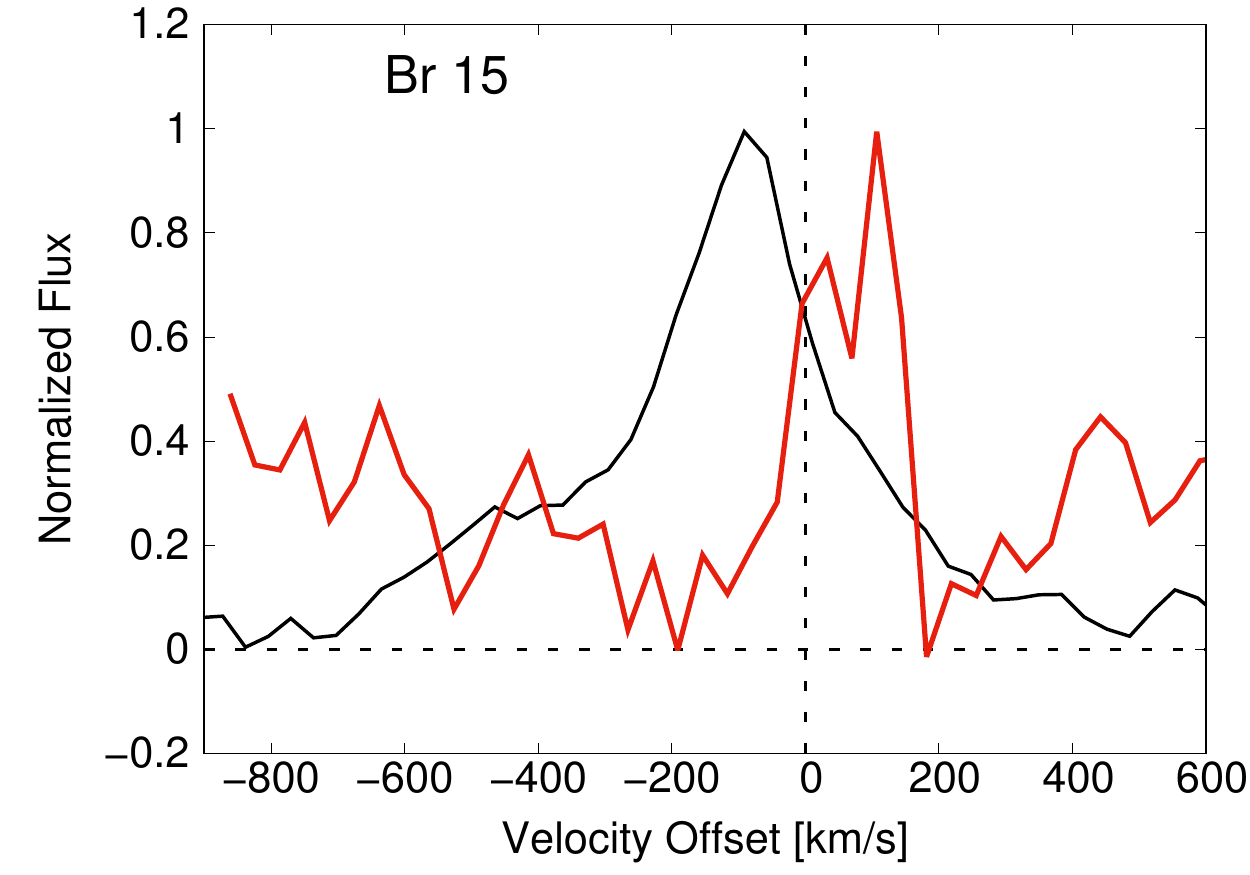}    
     \includegraphics[width=2.in]{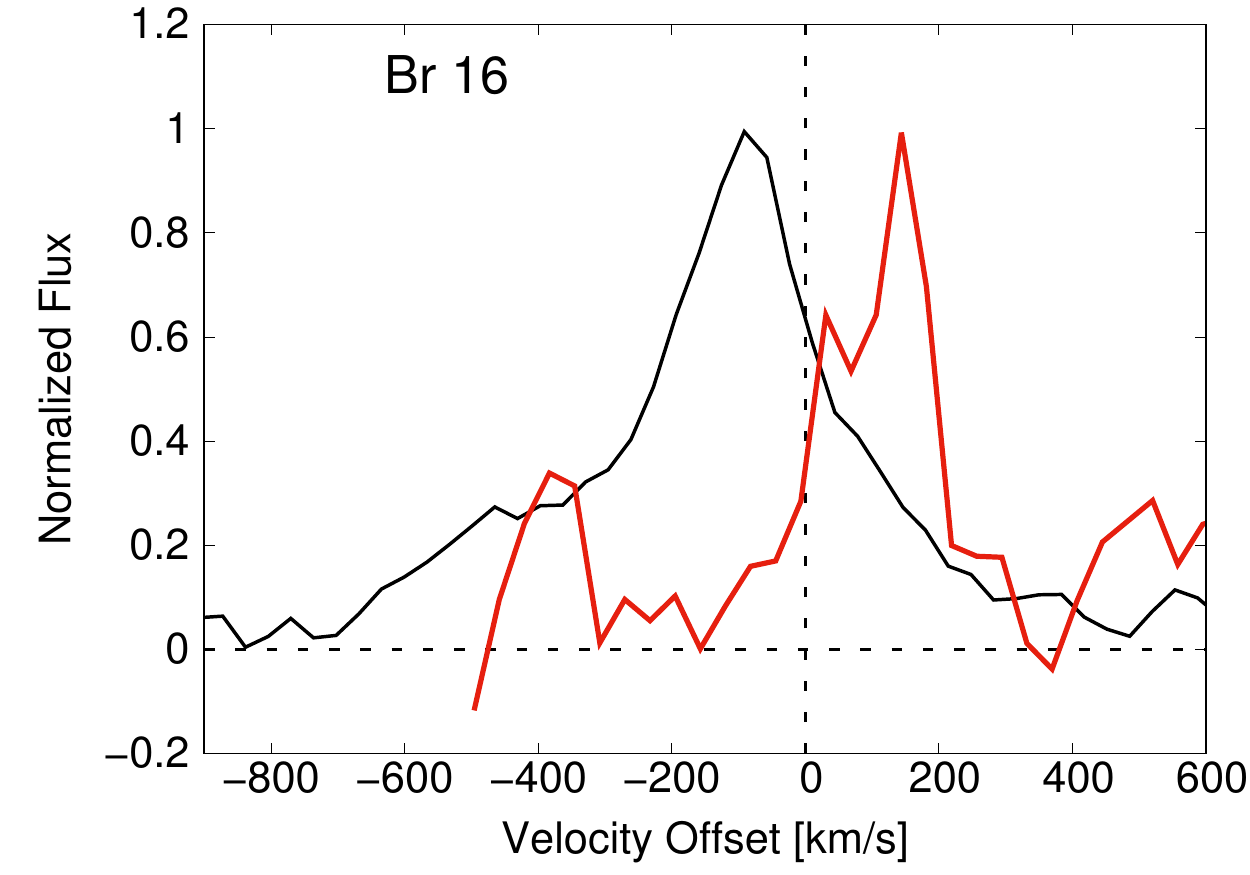}   
     \includegraphics[width=2.in]{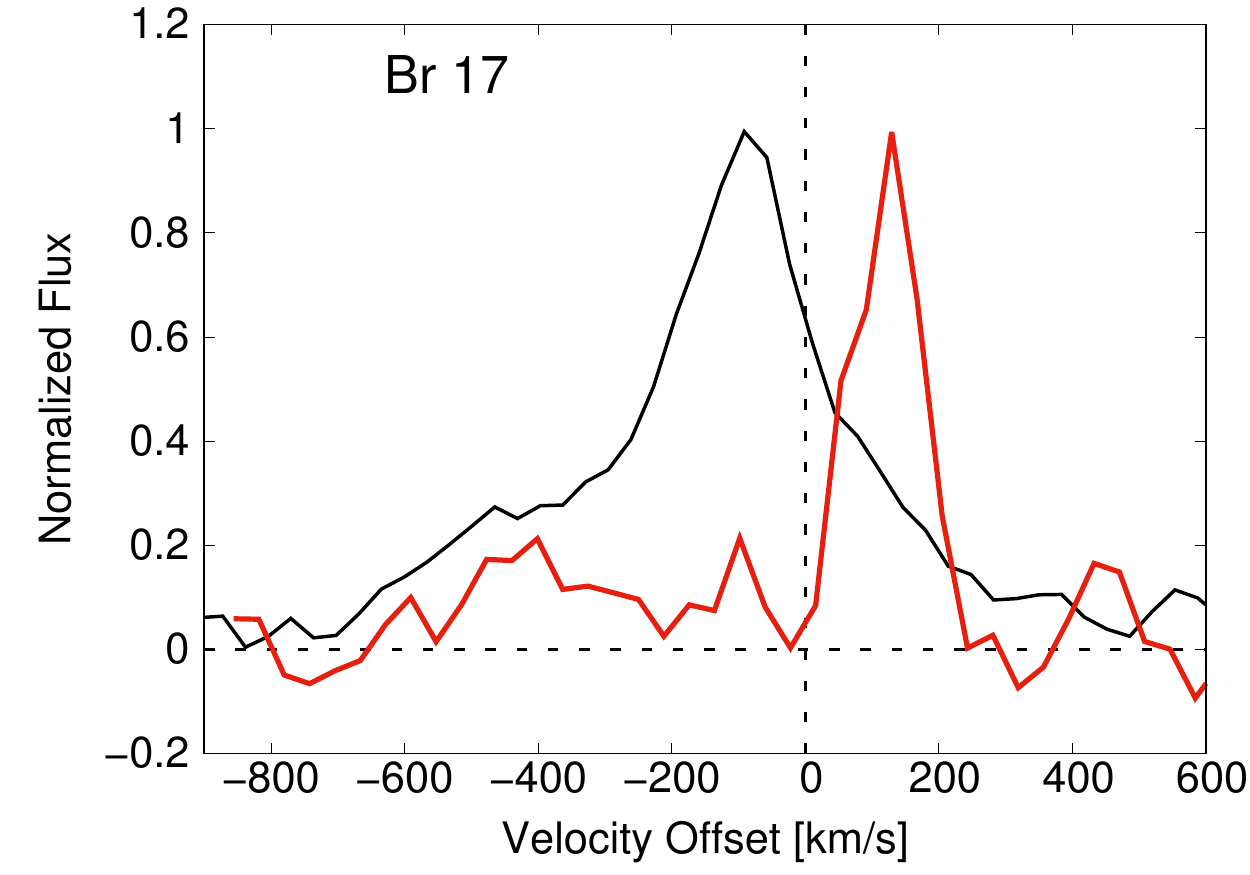}
     \includegraphics[width=2.in]{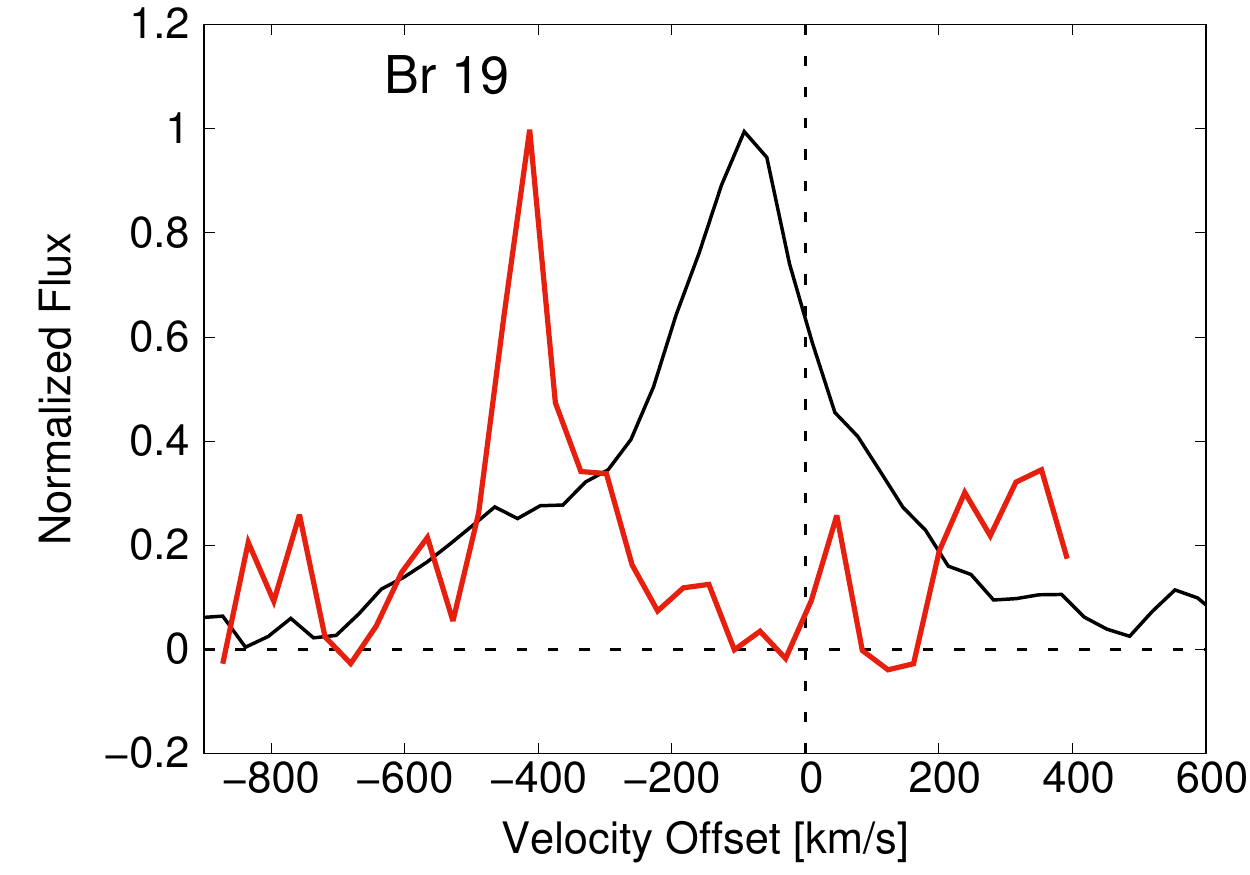}    
      \caption{Normalized spectra for the upper Brackett lines (red) compared with Br$\gamma$ (black) for J182957.  }
     \label{ser1-Br}
  \end{figure*}

\begin{figure*}
  \centering              
     \includegraphics[width=2.in]{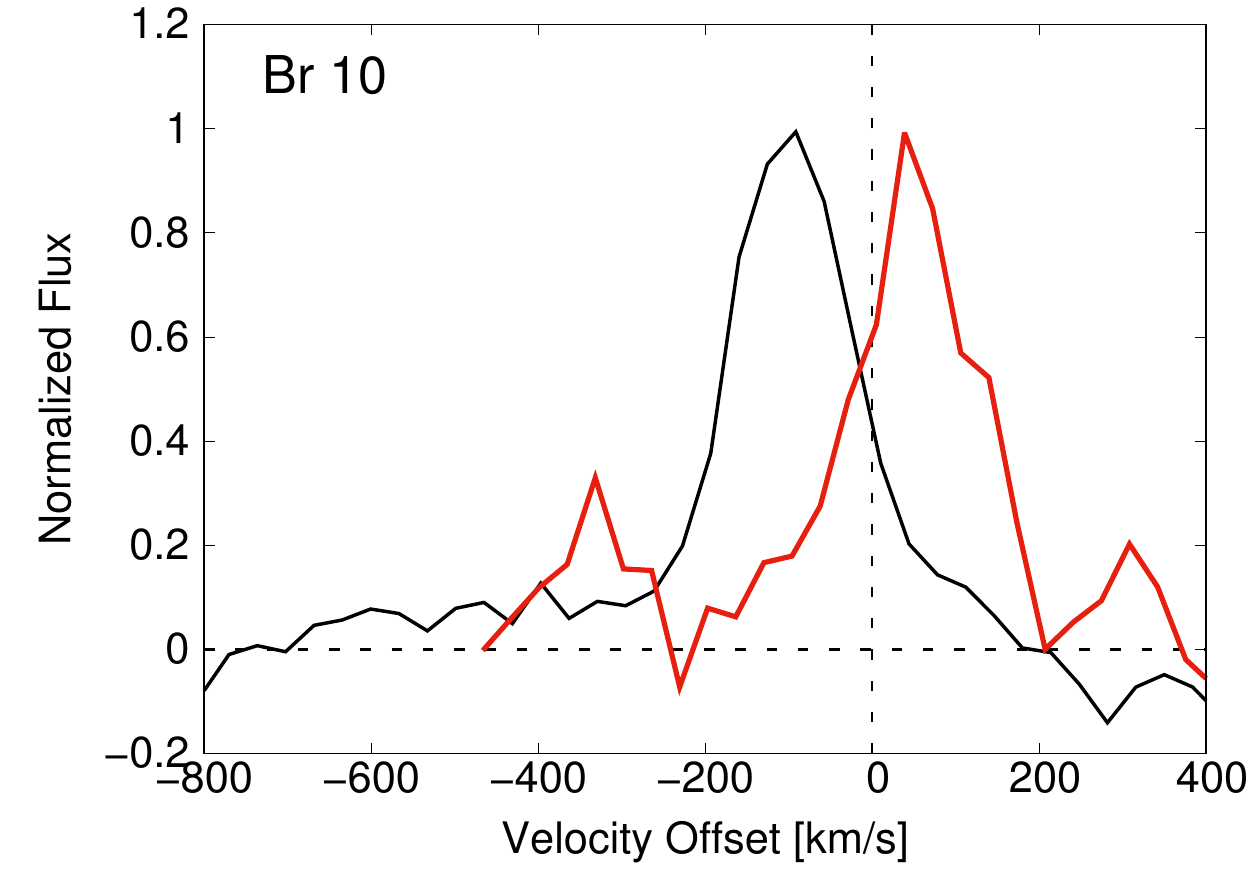}
     \includegraphics[width=2.in]{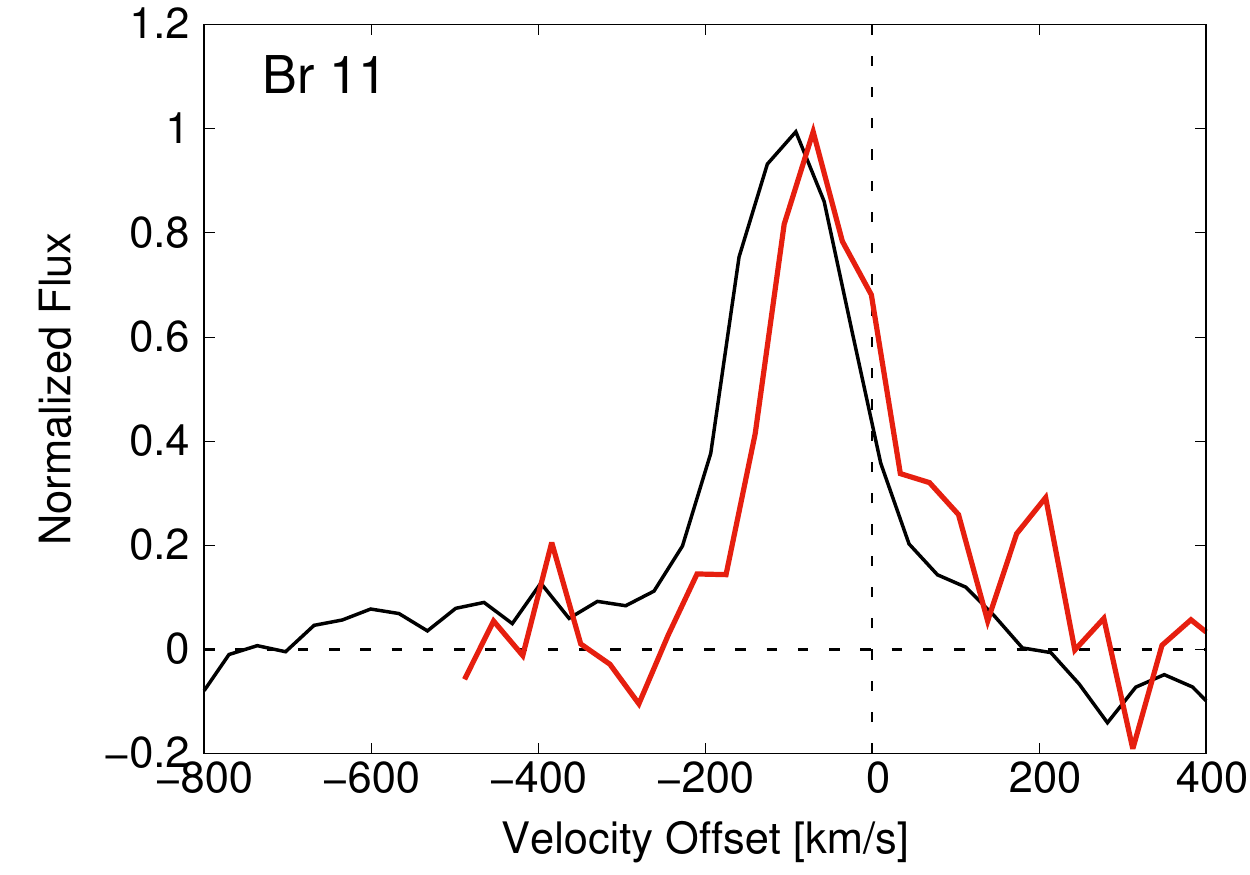}
     \includegraphics[width=2.in]{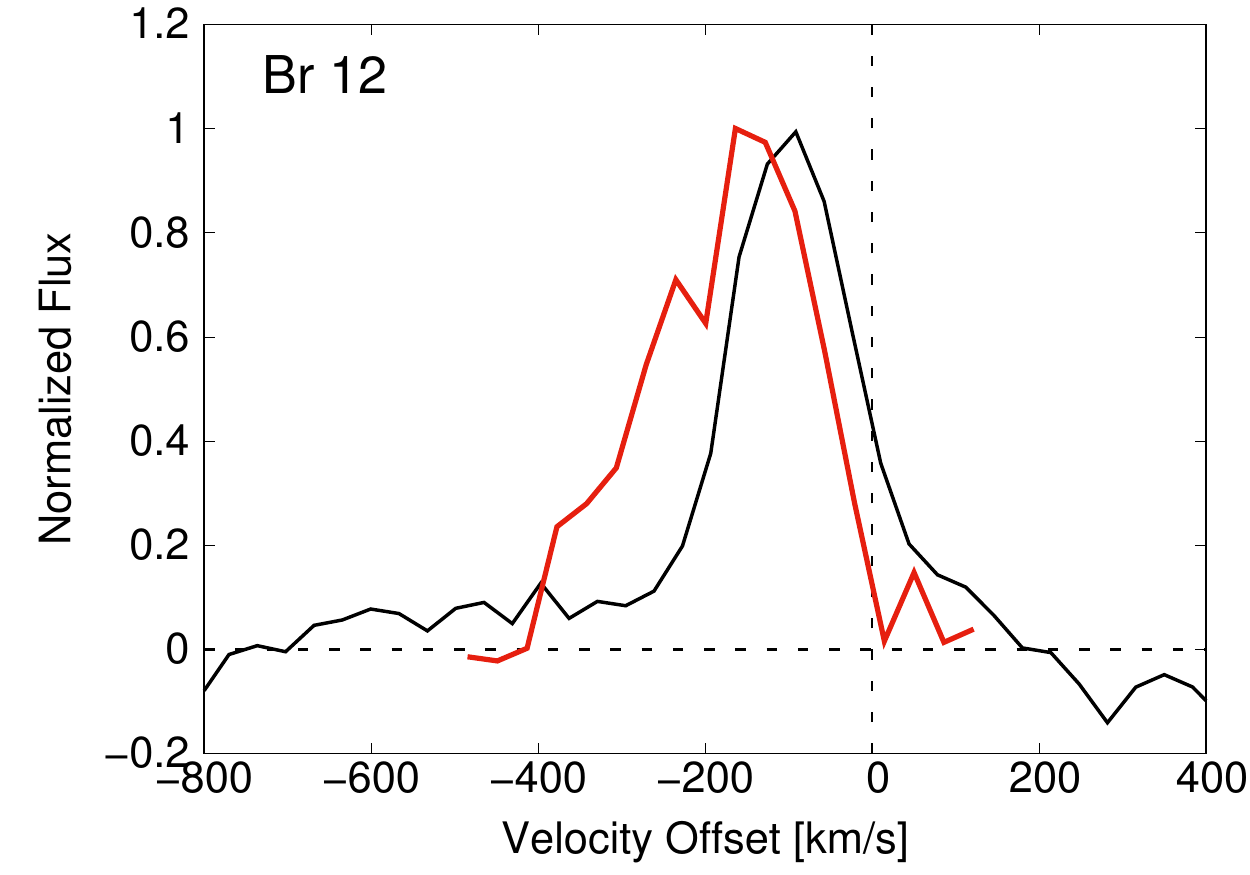}
     \includegraphics[width=2.in]{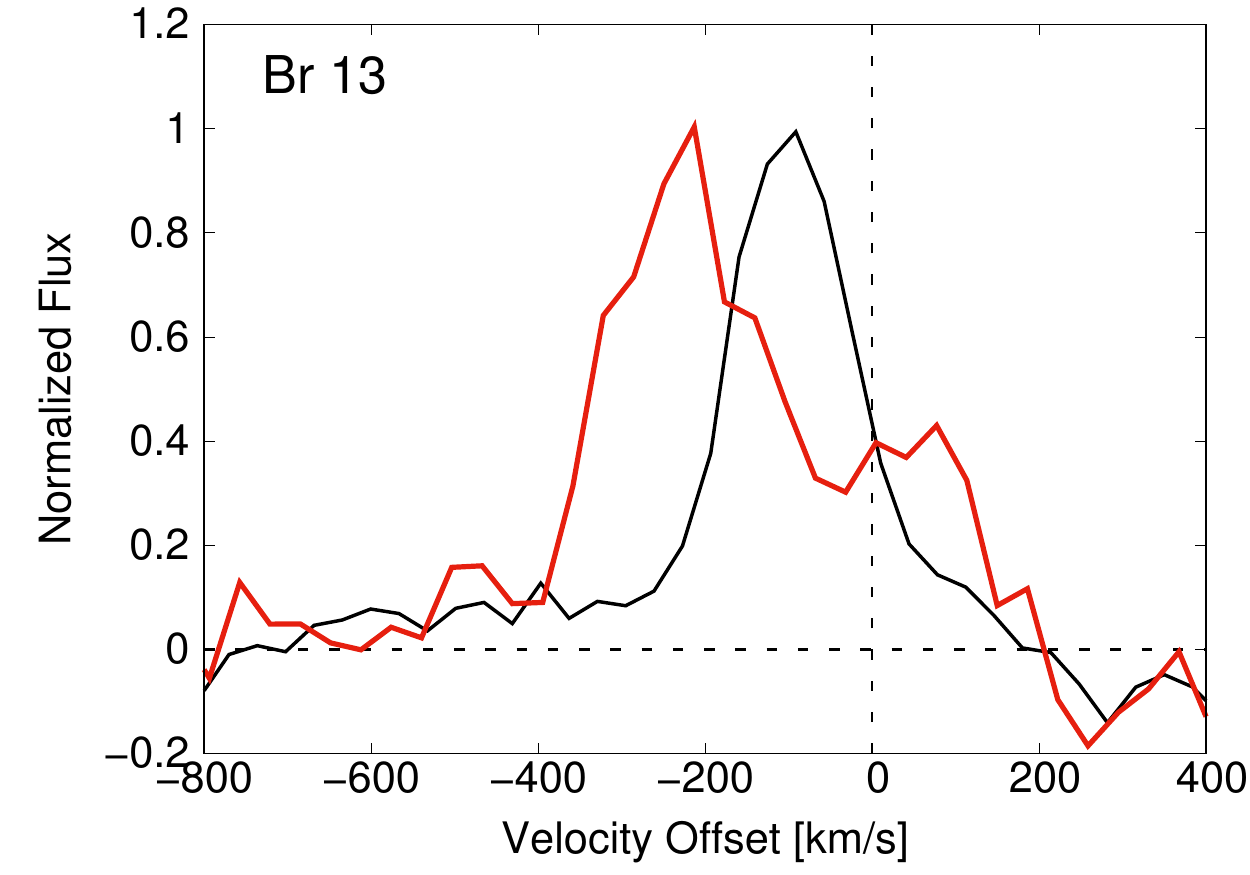}
     \includegraphics[width=2.in]{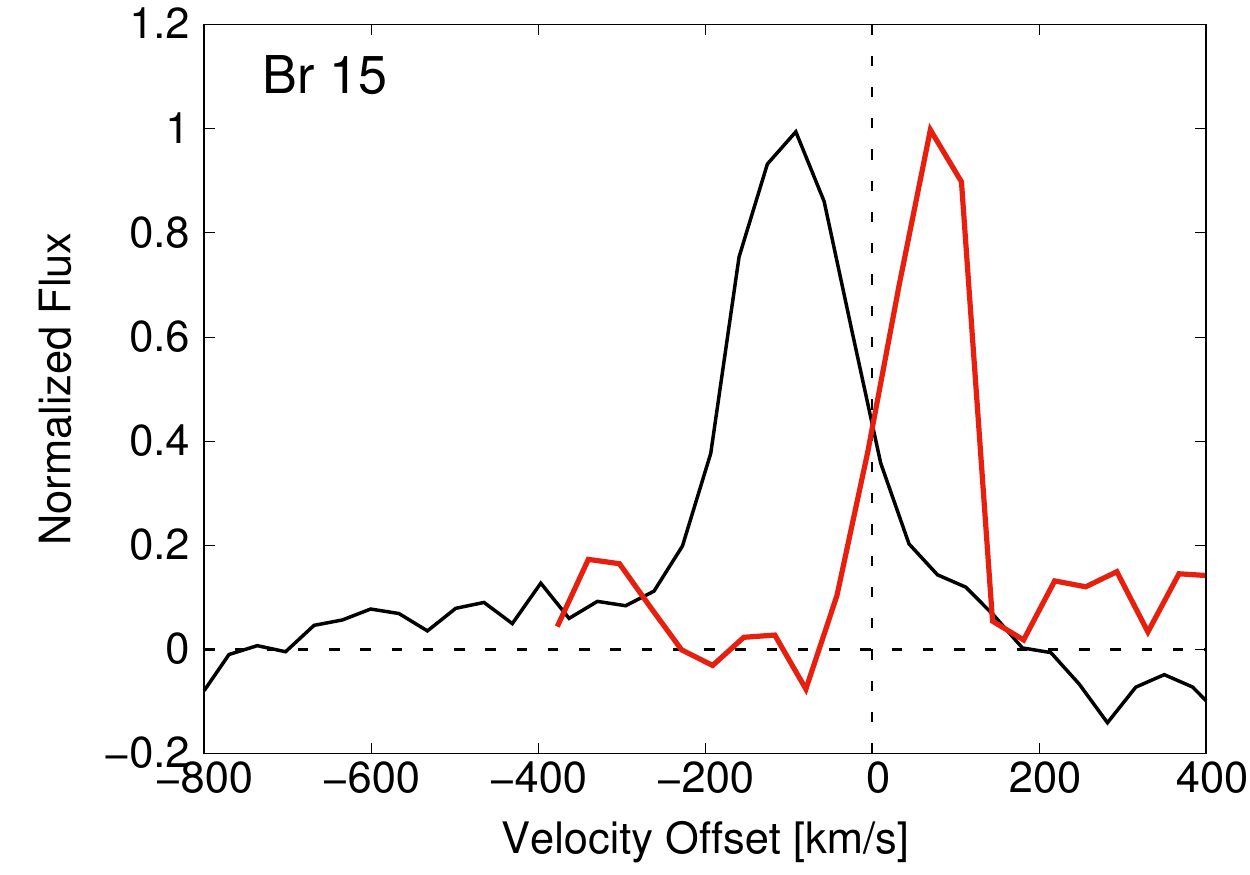}
     \includegraphics[width=2.in]{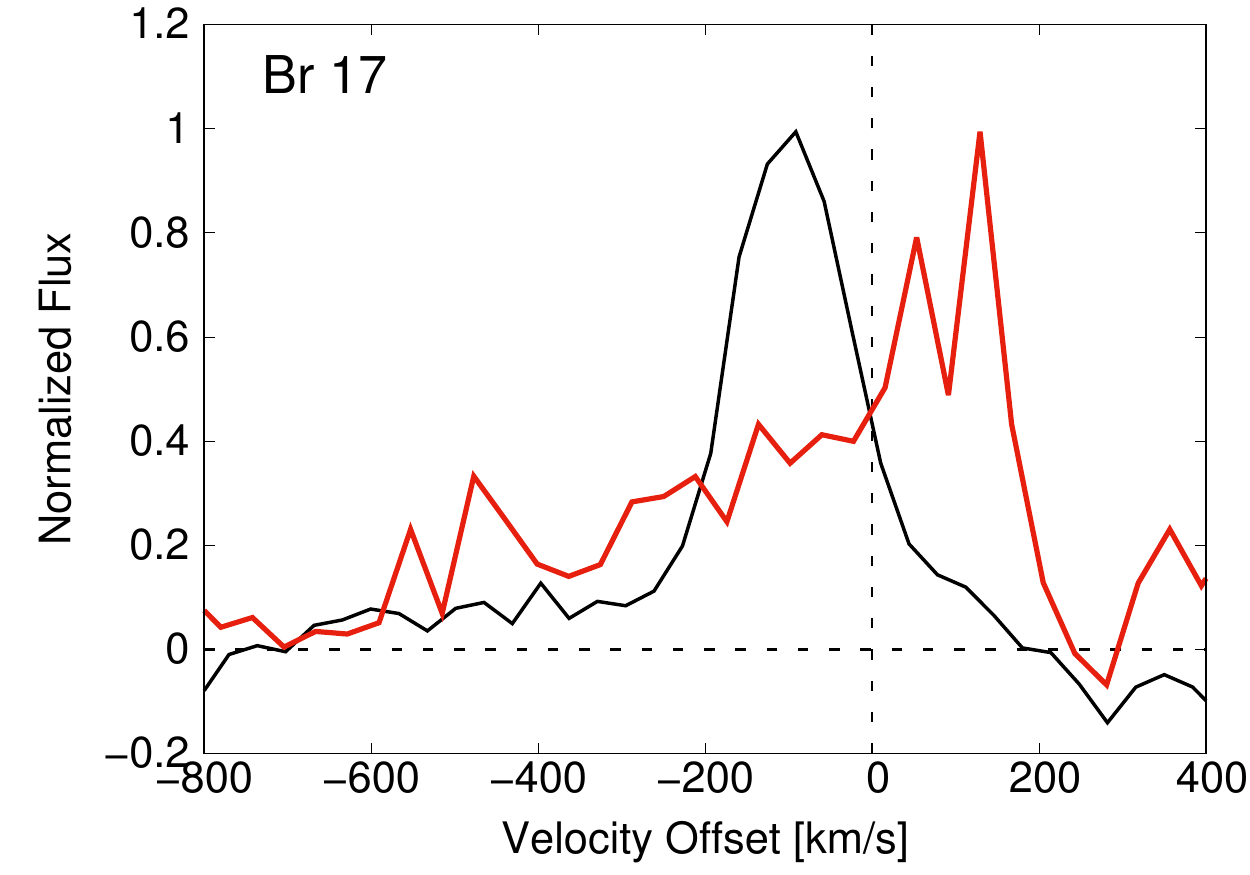}    
     \includegraphics[width=2.in]{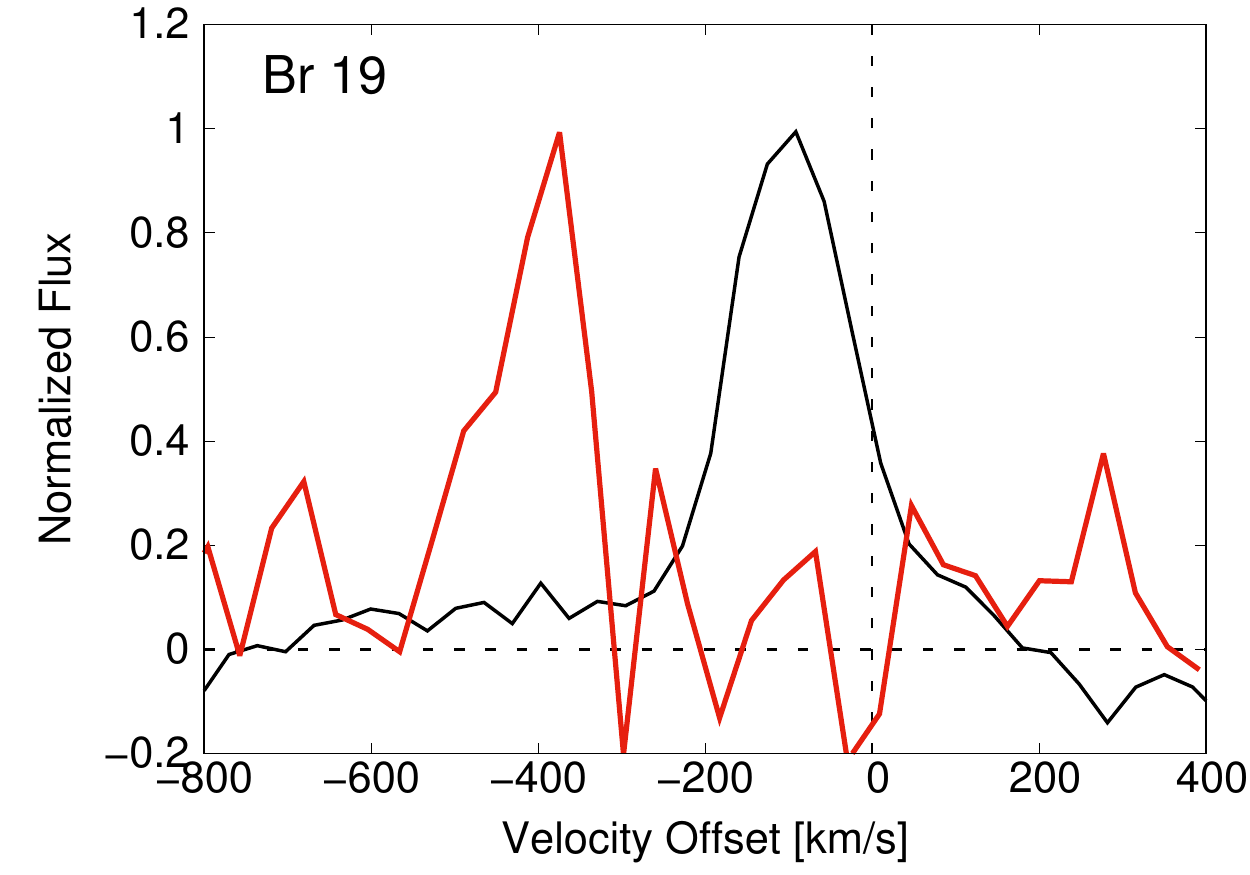}         
     \caption{Normalized spectra for the upper Brackett lines (red) compared with Br$\gamma$ (black) for J163136. }
     \label{oph1-Br}
  \end{figure*}

\begin{figure*}
  \centering              
     \includegraphics[width=2.in]{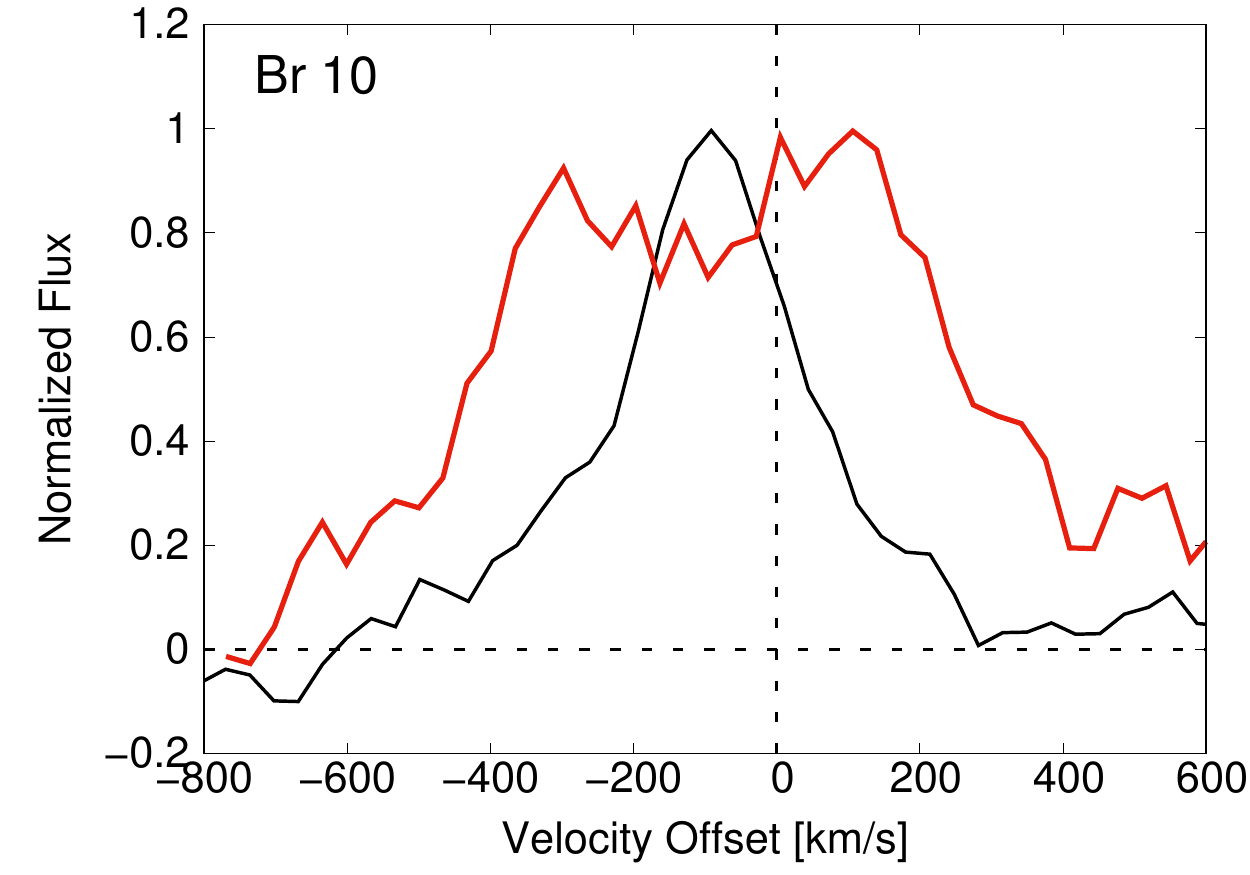}
     \includegraphics[width=2.in]{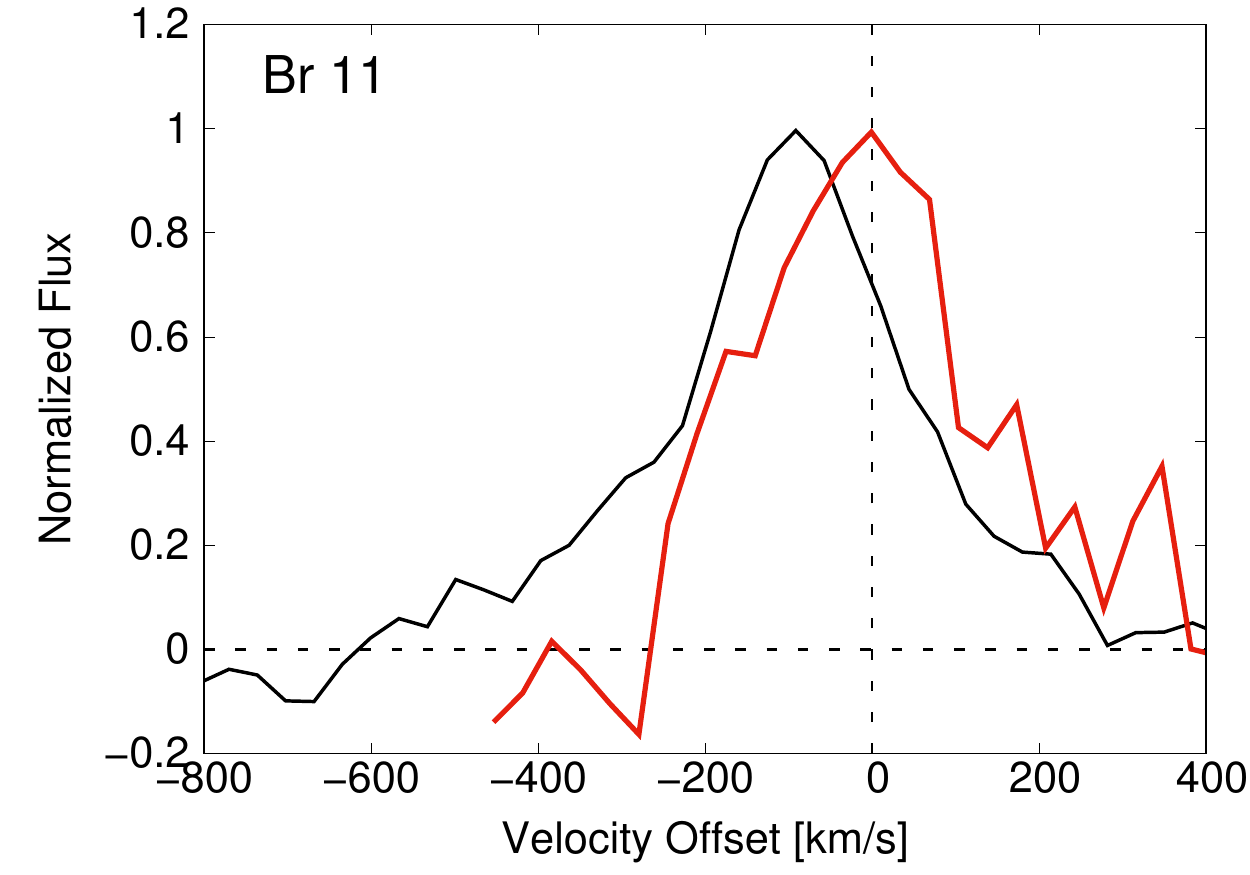}
     \includegraphics[width=2.in]{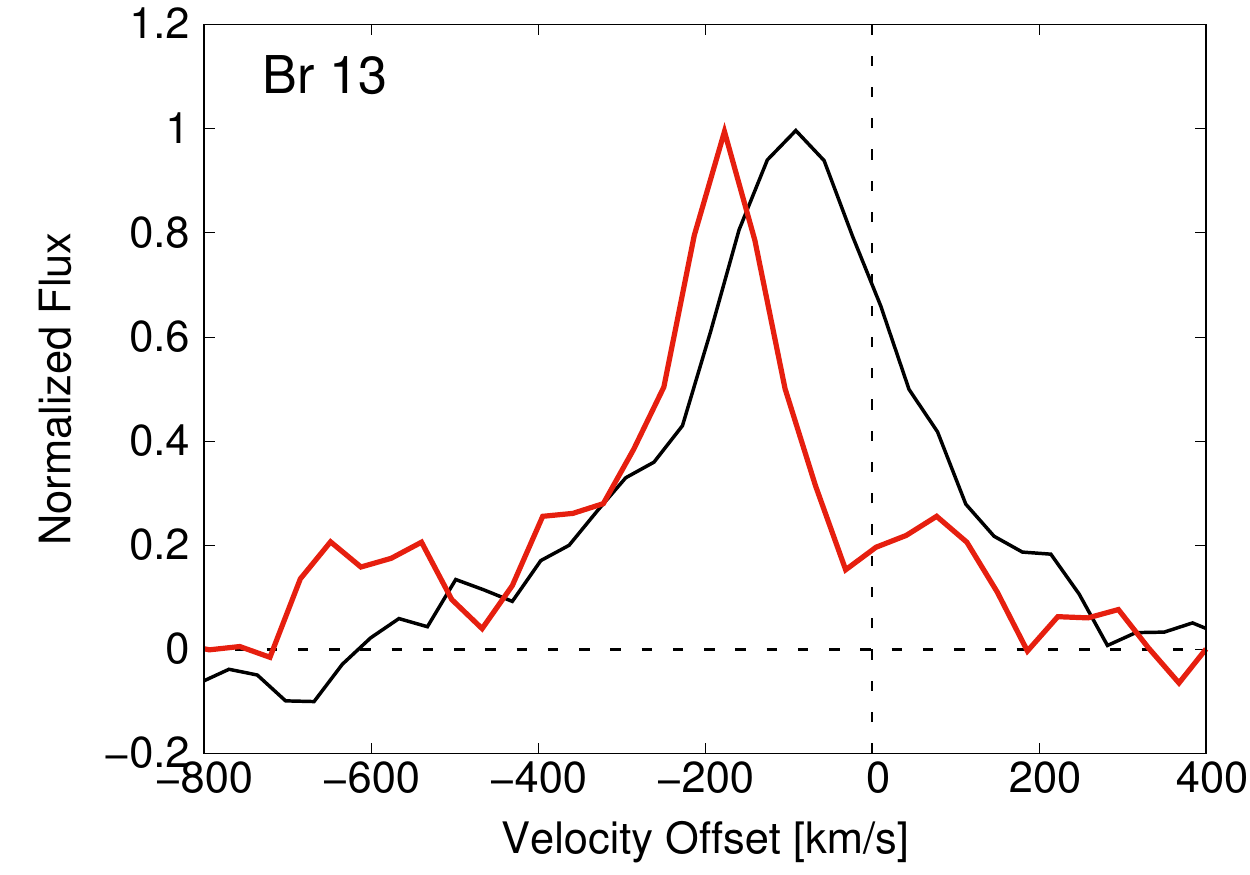}
     \includegraphics[width=2.in]{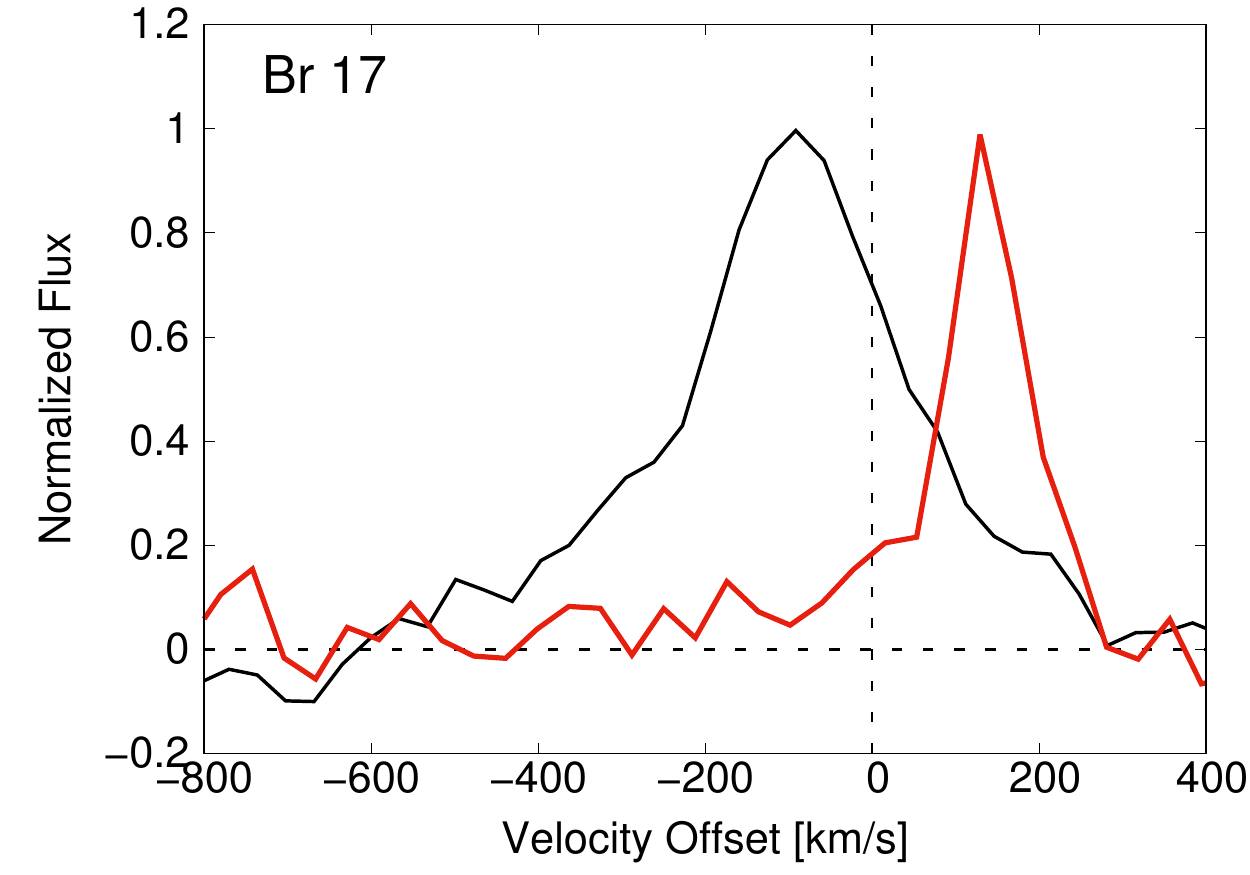}
     \includegraphics[width=2.in]{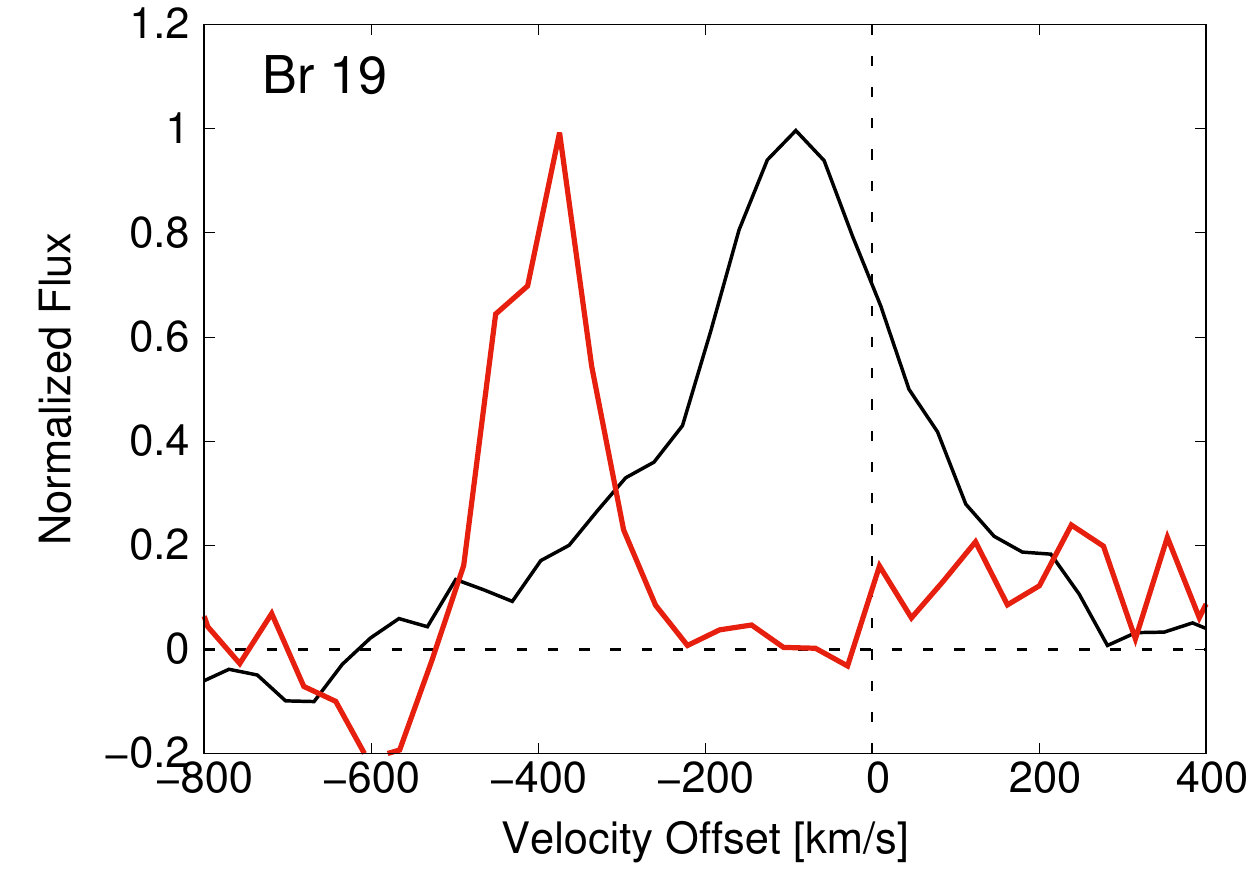}    
     \caption{Normalized spectra for the upper Brackett lines (red) compared with Br$\gamma$ (black) for J162648. }
     \label{oph3-Br}
  \end{figure*}

\begin{figure*}
  \centering              
     \includegraphics[width=2.in]{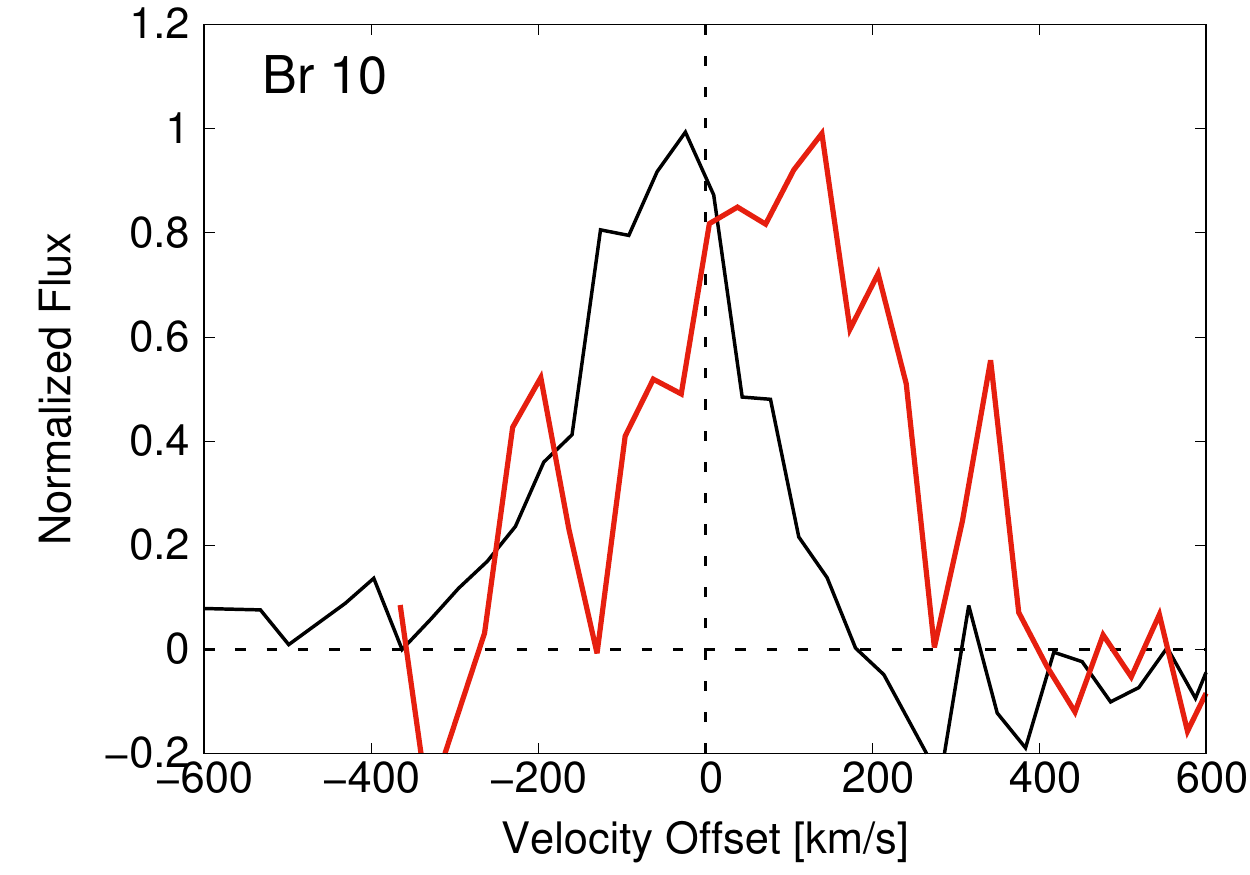}
     \includegraphics[width=2.in]{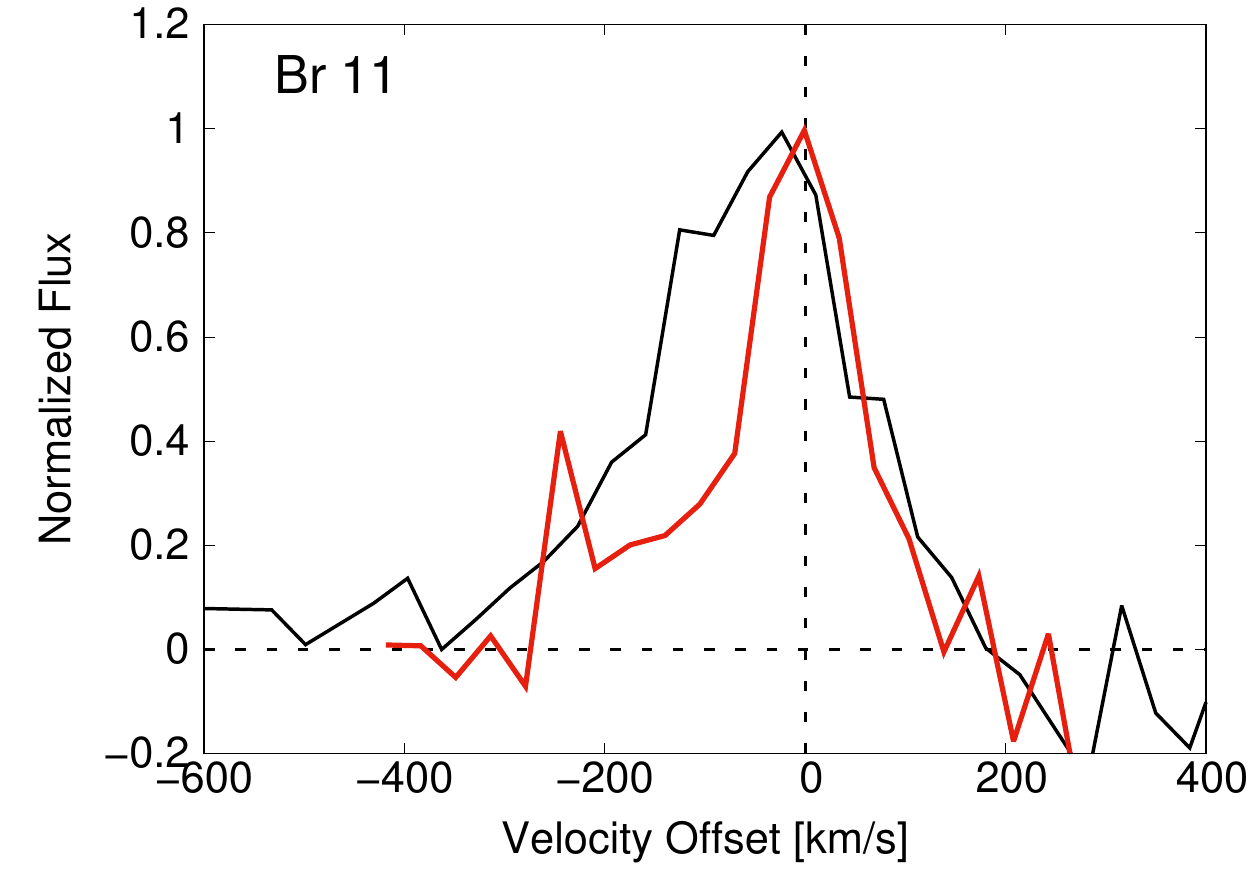}  
     \caption{Normalized spectra for the upper Brackett lines (red) compared with Br$\gamma$ (black) for J182940. }
     \label{ser8-Br}
  \end{figure*}

The observations of a large number of HI lines from the Brackett series allow a detailed study of the decrement of this series from which information on the line excitation can be derived. With the exception of M1701117, all of the proto-BDs in our sample show emission in the upper Brackett lines in addition to the Br$\gamma$ line. J182957 and J163136 show the most number of upper Brackett line detections in Br~10--Br~19 (Figs.~\ref{ser1-Br};~\ref{oph1-Br}). A variety in the line profiles is seen with none showing a broad, symmetric profile similar to Br~$\gamma$. For J163136. Br~17 shows a double-peaked profile, while the rest show nearly symmetric profiles. An extended wing and/or a secondary peak is seen in the Br~10, Br~13, and Br~17 profiles. For J182957, Br~10, Br~11, Br~12 show a broad, double-peaked profile, while the rest are narrower in comparison. For J162648, Br~10 shows a similar broad and double-peaked profile while the rest of the upper Brackett lines are narrow and comparatively symmetric (Fig.~\ref{oph3-Br}). J163152 shows emission in the Br~10, Br~11, Br~12, and Br~13 lines (Fig.~\ref{oph3-Br}), with the narrowest profile seen for Br~12. J182940 shows emission in only the Br~10 and Br~11 lines, with a broader profile seen for Br~10 compared to Br~11 and Br~$\gamma$ (Fig.~\ref{ser8-Br}).




There is a general trend of the peak velocity of the upper Brackett lines shifting from red-shifted to blue-shifted for upper energy level from 10 to 13, and then a shift to blue-shifted velocities for levels of 15, 16 and 17. This is seen in all objects. For J182957, J163136, and J162648 that also show emission in the Br~19 line, the peak velocity shifts again to blue-shifted from n$_{up}$ of 17 to 19. Note that the velocity offsets are well above the uncertainty in the wavelength calibration of $\sim$2-4 \AA. The FWHM of these upper Brackett lines are $\geq$100 km s$^{-1}$. 


The notable differences in the Brackett line profile shapes and kinematics indicate that they arise from different regions. It is expected that different lines trace a different emitting region whose size decreases as the n-number increases. The observed Brackett lines are thick and have different optical depths, thus they trace zones at different physical depths in the emitting region. Since the optical depth decreases with increasing upper quantum number, it is expected that high-n lines trace regions more internal than the Br$\gamma$ line (e.g., Nisini et al. 2004). The line-widths of the Brackett lines are thus more likely to reflect the dynamical properties of the gas in the emitting region than other broadening effects.

The symmetrical profiles seen for Brackett lines are difficult to reproduce either by wind models or by magnetospheric accretion models that generally predict blue- or red-shifted peaks and P-Cygni absorption features (e.g., Hartmann et al. 1990; Muzerolle et al. 1998). Line damping due to different broadening mechanisms and/or occultation by circumstellar material can have the effect of filling in the absorption component causing a much more symmetric and centrally peaked profile.

\begin{figure*}
  \centering              
     \includegraphics[width=4.5in]{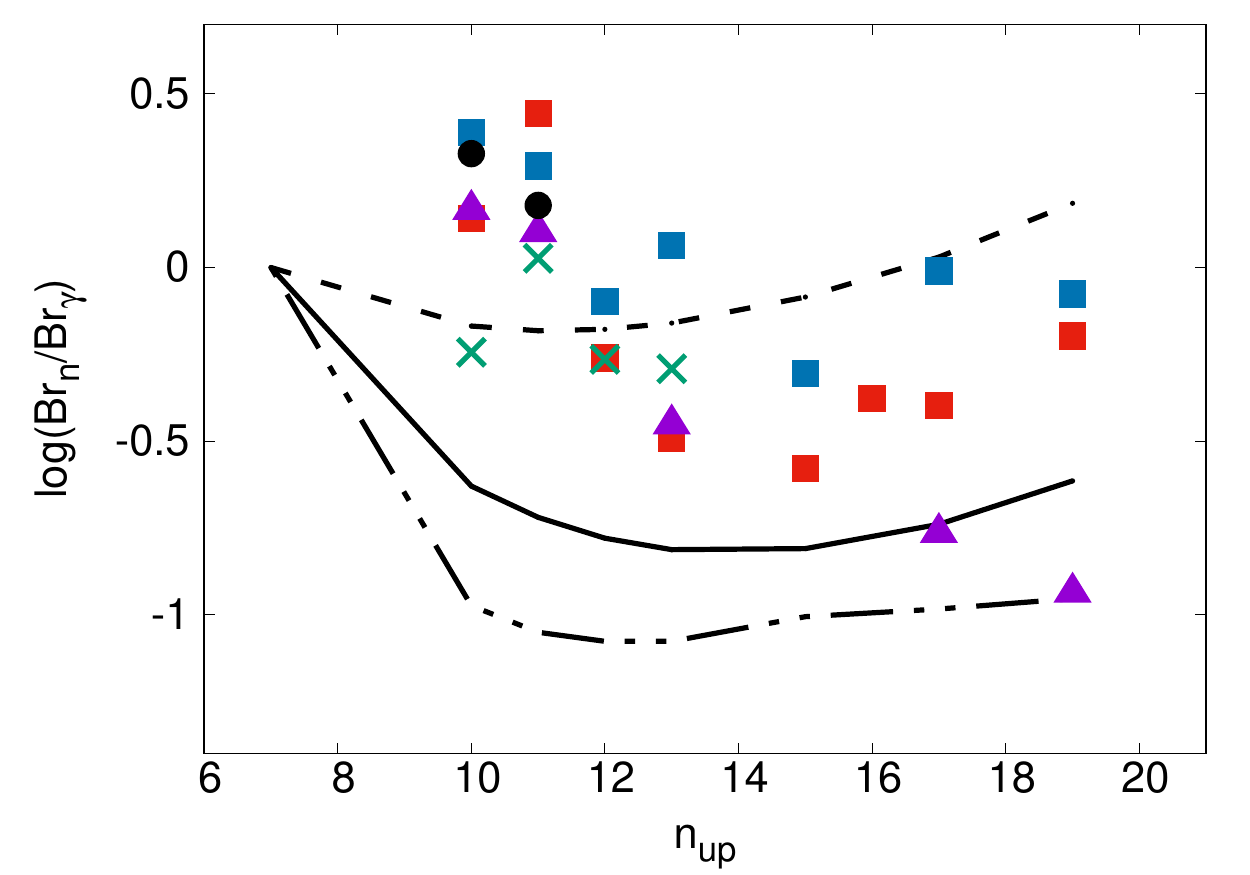}
     \caption{Ratio of the Brackett lines with respect to the Br$\gamma$ plotted as a function of the upper quantum number for J163136 (blue points), J163152 (green points), J162648 (purple points), J182957 (red points), and J182940 (black points). The uncertainty on the observed ratios is $\sim$20\%. The three curves are the ratios expected from Case B recombination assuming (i) T$_{e}$=1000 K, n$_{e}$=10$^{5}$ cm$^{-3}$ (solid line); (ii) T$_{e}$=3000 K, n$_{e}$=10$^{5}$ cm$^{-3}$ (dashed line); (iii) T$_{e}$=500 K, n$_{e}$=10$^{7}$ cm$^{-3}$ (dot-dashed line).  }
     \label{brackett}
  \end{figure*}

In Figure~\ref{brackett}, the ratios of the different Brackett lines with respect to the Br$\gamma$ line intensity are compared with the ratios expected for a Case B recombination, taken from Storey \& Hummer (1995). The ratios have been calculated after correcting the observed lines for the reddening. Nebulae that contain observable amounts of gas generally have quite large optical depths in the Lyman resonance lines of HI, due to which scattering and absorption must be taken into account in calculating the expected line strengths. For these large optical depths, a good approximation is the Case B recombination, which assumes that every Lyman line photon is scattered many times and is converted into lower series photons plus either Lyman alpha or two continuum photons. 


Also plotted in Figure~\ref{brackett} are the ratios expected from Case B recombination assuming three different sets of the electron temperature and density; (i) T$_{e}$=1000 K, n$_{e}$=10$^{5}$ cm$^{-3}$ (solid line); (ii) T$_{e}$=3000 K, n$_{e}$=10$^{5}$ cm$^{-3}$ (dashed line); (iii) T$_{e}$=500 K, n$_{e}$=10$^{7}$ cm$^{-3}$ (dot-dashed line). The first and second sets represent the range in T$_{e}$ and n$_{e}$ estimated for these proto-BDs (Section~\ref{temp_den}).

As seen in Figure~\ref{brackett}, the ratios for the Br~10 and Br~11 lines are higher than the Case B values for all objects, indicating that these Brackett lines are optically thick. The Case C recombination may be more applicable in such an optically thick ($\tau >>$ 1) case where the upper Brackett lines are brighter than Br$\gamma$. For the remaining proto-BDs, the Br$_{n}$/Br$\gamma$ line ratios appear consistent with the ratios expected from Case B recombination (Figure~\ref{brackett}), and lie within the range of T$_{e}$ $\sim$1000 -- 3000 K  and n$_{e} \sim$ 10$^{3}$ -- 10$^{5}$ cm$^{-3}$ estimated for these objects. For J162648 (purple points), a lower T$_{e}$ = 500 K and higher density n$_{e}$ = 10$^{7}$ cm$^{-3}$ Case B scenario is a good fit to the Br~17-19 line ratios, suggesting a denser and cooler environment than the other objects.

We note that there appears to be a systematic offset between the ratios for the proto-BDs and the ratios expected from Case B recombination for T$_{e}$ $\sim$1000, n$_{e} \sim$ 10$^{5}$ cm$^{-3}$ (solid line). The jet and/or source extinction for our targets may have been under-estimated, or the extinction law used is incorrect, or both. In a disk corona, the grains are expected to be large and extinction must be close to grey, and thus ISM extinction curves may not apply (e.g Throop et al. 2001). Also, in some proto-BDs, the H-recombination line emission is too strong, and originates in part from outflow shocks; the post-shock plasma where the bulk of the emission originates may by higher than 3,000 K.  For low-mass YSOs, these collimation shocks are 10s to 100s of AU away from the star (e.g., Bally et al. 2003), while the jet launching region is expected to be $<$10 au in proto-BDs (e.g., Machida et al. 2009). These lines may be obscured by a thick torus or cloud core.



While the sample size is small, the Br~15--19 lines are detected in J182957 (Stage 0), J163136 (Stage I), and J162648 (Stage I), while only the Br~10--11 lines are detected in the more evolved J182940 object (Stage I-T/II). This suggests a possible dependence on the evolutionary stage where the high-n Brackett lines are only detected in the earlier stage objects. However, no clear dependence is seen between the evolutionary stage of the proto-BD and the strength in the upper Brackett lines with respect to Br$\gamma$. J163136 and J162648 are in Stage I while J182957 and J163152 are Stage 0 objects. If higher Brackett line ratios are expected for a deeply embedded source then we would expect the Stage 0 objects to show ratios $>$ 1 in Figure~\ref{brackett}. The uncertainty on the Brackett line ratios is $\sim$20\%. The high observed ratios for Br~10-11 could be due to the uncertainties with continuum-subtraction. Due to the broad features and the different slopes on both sides of the lines, it is difficult to correctly determine the pseudo-continuum which affects the integrated flux measurement. 



 \begin{figure*}
  \centering              
     \includegraphics[width=2.in]{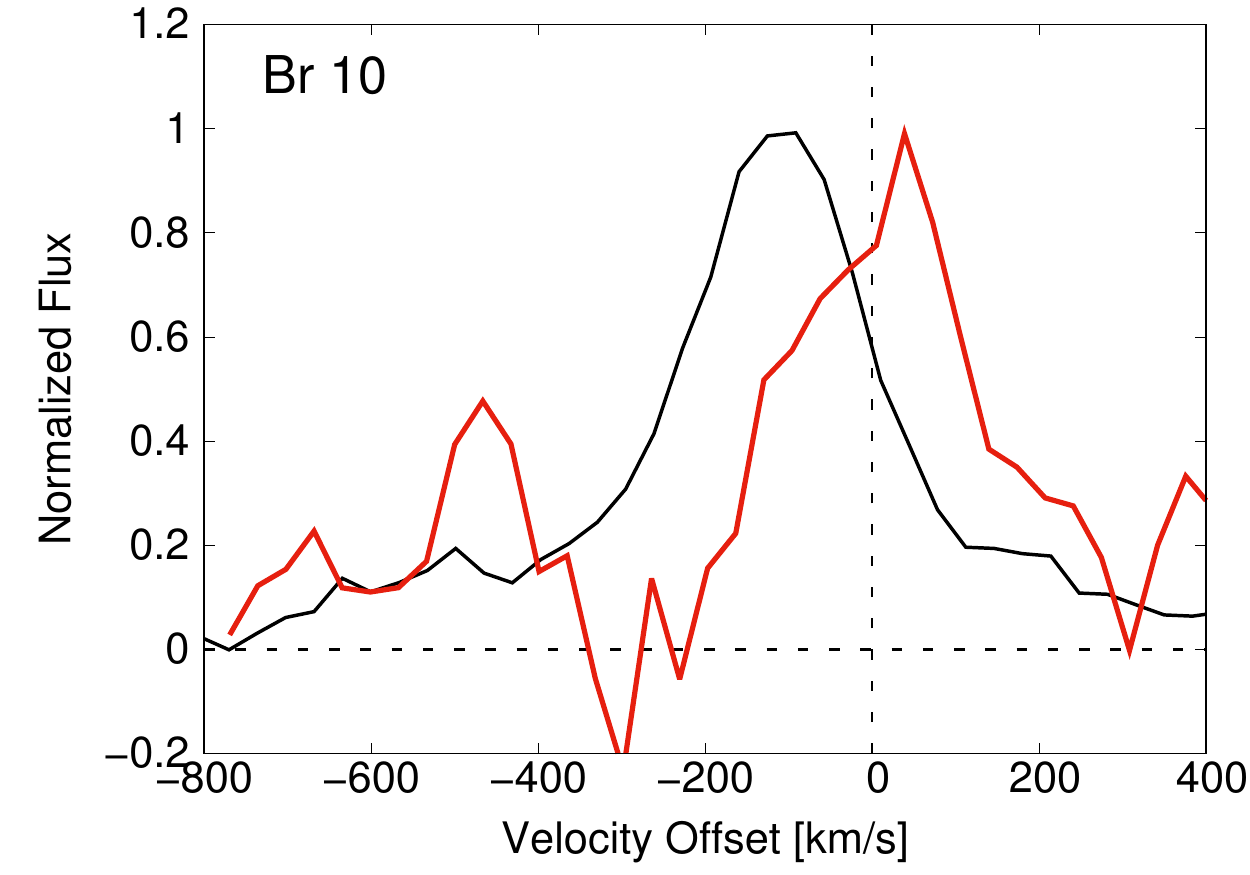}
     \includegraphics[width=2.in]{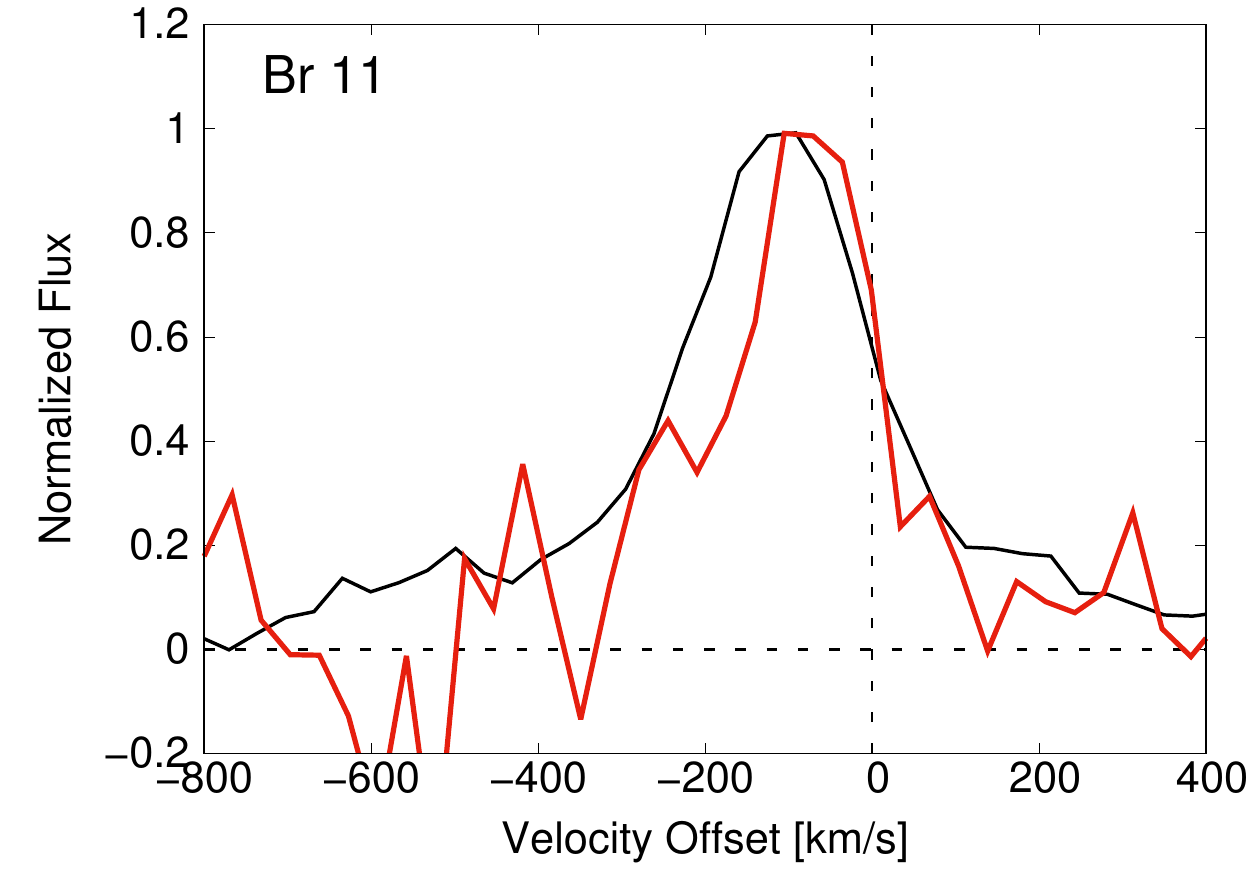}
     \includegraphics[width=2.in]{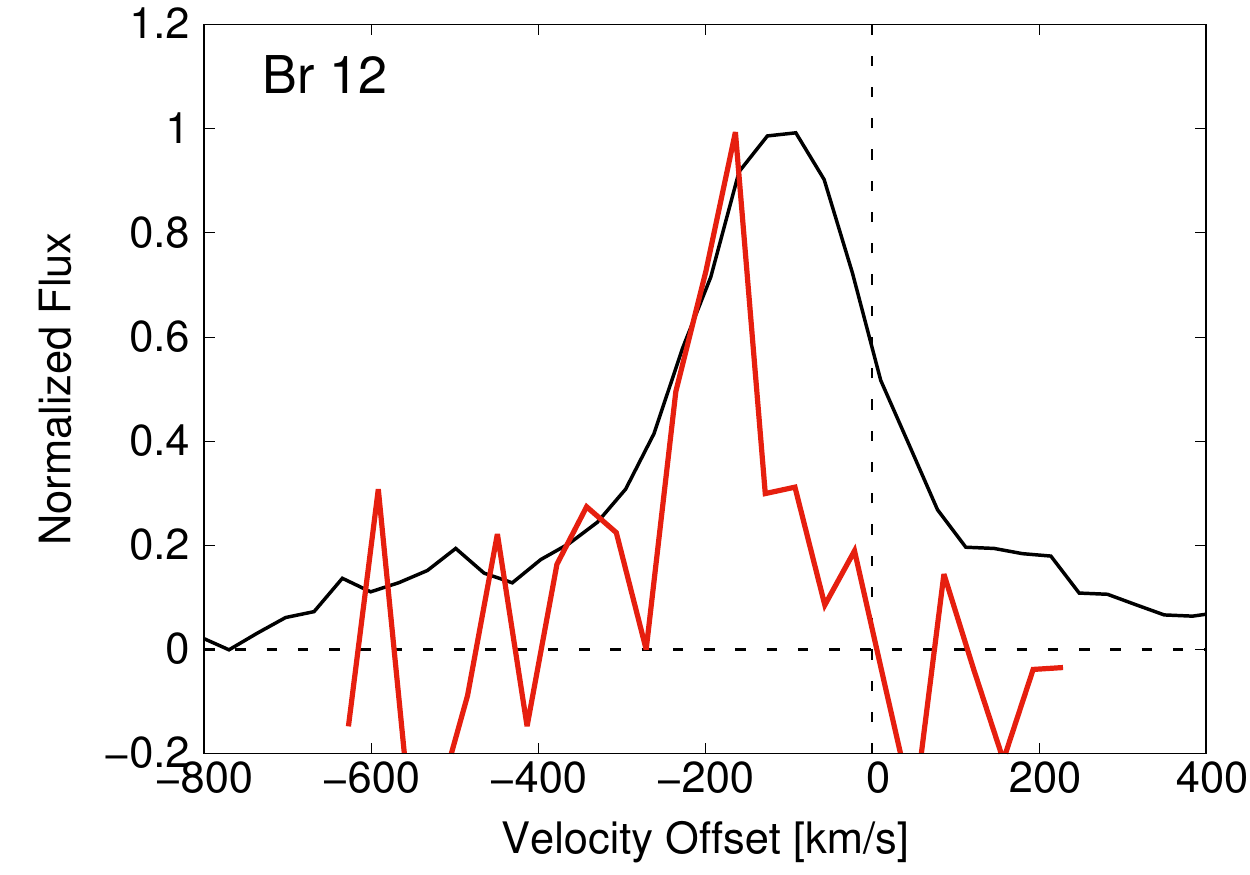}   
     \includegraphics[width=2.in]{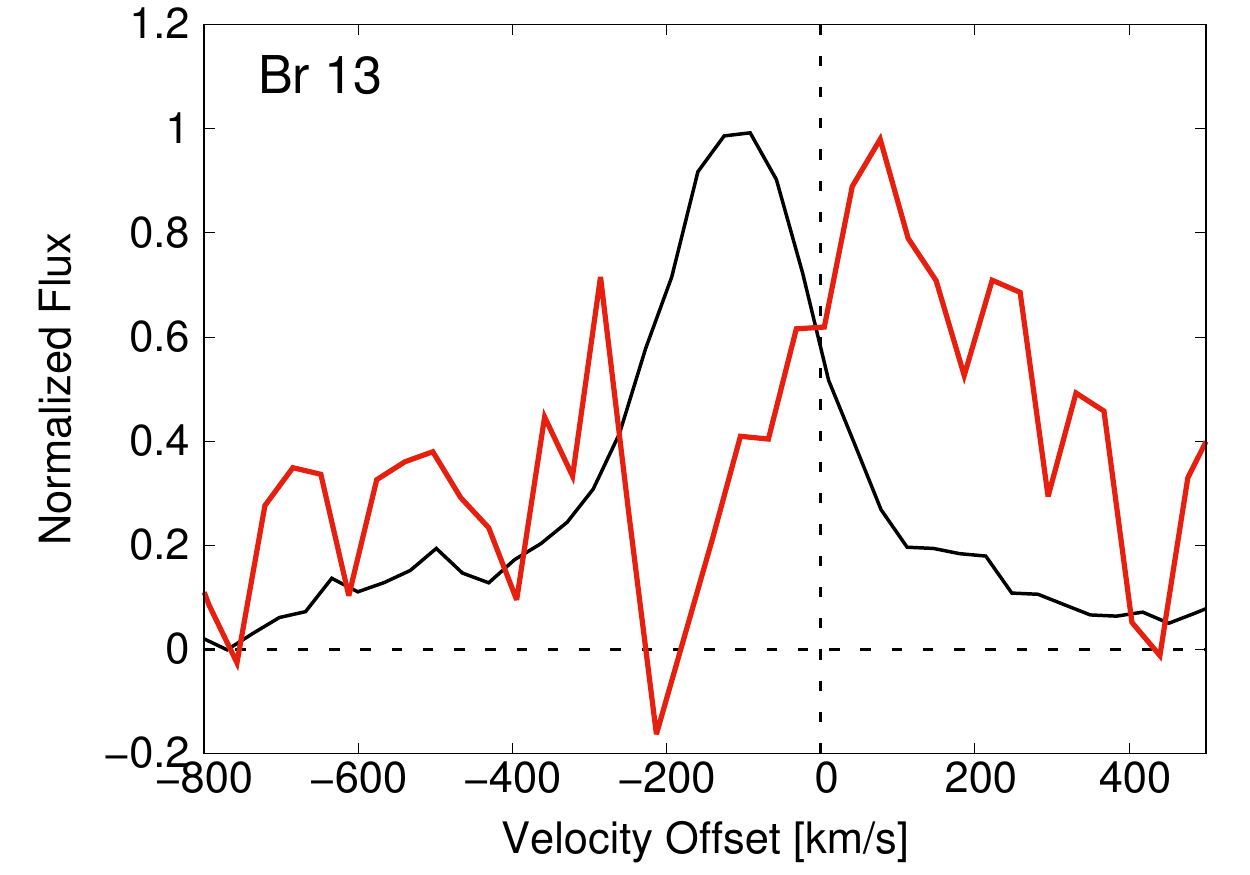}     
     \caption{Normalized spectra for the upper Brackett lines (red) compared with Br$\gamma$ (black) for J163152. }
     \label{oph2-Br}
  \end{figure*}

\section{Conclusions}

We have conducted a near-infrared study of accretion and outflow activity in 6 Class 0/I proto-BDs using VLT/SINFONI spectroscopy and spectro-imaging observations. Several [Fe~II] and H$_{2}$ lines associated with jet/outflow activity are detected for the proto-BDs, including  the accretion diagnostics of Pa~$\beta$ and Br~$\gamma$ lines, and the upper Brackett lines of Br~10-19. The peak velocities of the [Fe~II] lines ($>$100 km s$^{-1}$) are higher than the H$_{2}$ lines, suggesting that [Fe~II] traces the high-velocity jet while H$_{2}$ is likely tracing the low-velocity outflow. The Class 0 (Stage 0) proto-BDs show strong emission in the accretion tracers and the H$_{2}$ lines but the [Fe~II] lines are undetected. In contrast, the Class I (Stage I) objects show emission in both [Fe~II] and H$_{2}$ lines with varied strengths. This suggests an evolutionary trend in the jets from a molecular to an ionic composition. 


The mass accretion rates for the proto-BDs span the range of (2$\times$10$^{-6}$ -- 2$\times$10$^{-8}$) M$_{\sun}$ yr$^{-1}$, while the mass loss rates are in the range of (4$\times$10$^{-8}$ -- 5$\times$10$^{-9}$) M$_{\sun}$ yr$^{-1}$. The mass outflow rate derived using the [Fe~II] lines is at least an order of magnitude higher than H$_{2}$ lines. There is no clear correlation between the evolutionary stage and activity rates for the proto-BDs. 


The main result is that {\it all} of the proto-BDs are driving an outflow, including the relatively evolved Stage I-T/II case. This is expected as outflows are an important component of the early formation stages to conserve the angular momentum. What is also interesting is the range in accretion and outflow rates that are within the range measured for low-mass protostars and higher than Class II brown dwarfs. The overall signatures thus indicate a decline in accretion and outflow activity with evolutionary stage but not with the bolometric luminosity.

We have performed a study of the HI decrement and line profiles from the several Brackett lines of Br~7-19 detected in the proto-BDs. The earlier Stage 0/I objects show emission in the upper Brackett lines of Br~15-19 that are not detected in the more evolved Stage I-T/II object. There is a general trend of the peak velocity of the upper Brackett lines shifting from red-shifted to blue-shifted for upper energy level from 10 to 13, and then a shift to blue-shifted velocities for levels of 15, 16 and 17. The peak velocity shifts again to blue-shifted from n$_{up}$ of 17 to 19. The notable differences in the Brackett line profile shapes and kinematics indicate that they arise from different regions. The ratios of the different Brackett lines with respect to the Br$\gamma$ line intensity appear consistent with the ratios expected from Case B recombination, and lie within the range of T$_{e}$ $\sim$1000 -- 3000 K and n$_{e} \sim$ 10$^{3}$ -- 10$^{5}$ cm$^{-3}$ estimated for these objects.

There is clear extended emission with knots seen in the [Fe~II] and H$_{2}$ spectro-images for 3 proto-BDs, while the rest show compact morphologies with an on-source peak emission. No clear correlation is seen between the evolutionary stage and the presence of extended vs. compact jet emission.

\section*{Acknowledgements}

We thank the referee Fernando Comeron for his valuable comments on the paper. We thank Lowell Tacconi-Garman for help with the data reduction. BR acknowledges funding from the Deutsche Forschungsgemeinschaft (DFG) Projekt number 402837297. Based on observations collected at the European Southern Observatory under ESO programmes 0102.C-0932(A) and 0103.C-0660(A).

\section{Data Availability}

The data underlying this article are available in the ESO archives at http://archive.eso.org/cms.html.



\clearpage

\appendix

\section{Line Fluxes}
\label{fluxes}

The vacuum wavelengths are from Nisini et al. (2005). The line fluxes have not been corrected for extinction. We estimate flux errors of $\sim$20\%--30\%. The line fluxes have been continuum-subtracted. 




\begin{table*}
\centering
\caption{Line fluxes for J182957}
\label{ser1-lines}
\begin{threeparttable}
\begin{tabular}{lccccc} 
\hline
Species & $\lambda_{vac}$ [$\mu$m]  \tnote{a} & $\lambda_{line}$ [$\mu$m]  \tnote{b} & Velocity [km s$^{-1}$] \tnote{c}  & F$_{peak}$ [erg s$^{-1}$ cm$^{-2}$ $\mu$m$^{-1}$] \tnote{d} & Integrated flux [erg s$^{-1}$ cm$^{-2}$] \tnote{d}  	\\
\hline
\relax
[Fe~II] $^{4}D_{7/2}~{\textemdash}~^{6}D_{9/2}$	& 1.25668 	& 1.25725 	& -136  		& 1.35$\times$10$^{-12}$ & 1.22$\times$10$^{-15}$  \\ 
Pa$\beta$							& 1.2818 		& 1.28219 	& -91		& 2.7$\times$10$^{-12}$ & 4.4$\times$10$^{-15}$  \\
\relax
[Fe~II] $^{4}D_{3/2}~{\textemdash}~^{6}D_{1/2}$ ?	& 1.295 	& 1.29886 	& -896 		& 1.95$\times$10$^{-12}$ & 1.20$\times$10$^{-15}$  \\
?										& --		& 1.31583		& --	& 1.9$\times$10$^{-12}$ & 1.33$\times$10$^{-15}$  \\
Mg $\textrm{I}$ 						& 1.505 		& 1.50570		& -139 		& 2.5$\times$10$^{-12}$ & 2.92$\times$10$^{-15}$  \\
H$_{2}$ (4-2) $O$(3) ?					& 1.5094			& 1.50901		& +77		& 1.77$\times$10$^{-12}$ & 8.5$\times$10$^{-16}$  \\
H$_{2}$ (1-7) $P$(8)	 					& 1.52465 	& 1.52422	  & +85	& 2.3$\times$10$^{-12}$ & 1.34$\times$10$^{-15}$ \\
Br~19								& 1.527		& 1.5291	& -413	& 1.9$\times$10$^{-12}$ & 2.56$\times$10$^{-15}$  \\
\relax
[Fe~II] $^{4}D_{5/2}~{\textemdash}~^{4}F_{9/2}$	& 1.53347 	& 1.53339 	& +16 		& 2.5$\times$10$^{-12}$ & 1.5$\times$10$^{-15}$  \\
H$_{2}$ (5-3) $Q$(6) ?					& 1.5424 		& 1.53963	 	& +538		& 2.0$\times$10$^{-12}$ & 1.72$\times$10$^{-15}$ \\
Br~17	 							& 1.544 		& 1.54333 	& +130 		& 2.6$\times$10$^{-12}$ & 1.63$\times$10$^{-15}$ \\
Br~16								& 1.555233	& 1.55425		& +189		& 2.4$\times$10$^{-12}$ & 1.7$\times$10$^{-15}$ \\
H$_{2}$ (5-3) $Q$(7) ?					& 1.5608		& 1.55991		& +171		& 2.4$\times$10$^{-12}$ & 1.35$\times$10$^{-15}$ \\
H$_{2}$ (4-2) $O$(4) ?					& 1.5631		& 1.56361		& -98		& 2.3$\times$10$^{-12}$ & 2.02$\times$10$^{-15}$ \\
?									& --			& 1.56576		& -- 			& 2.4$\times$10$^{-12}$ & 1.04$\times$10$^{-15}$  \\
Br~15								& 1.571		& 1.57044		& +107		& 2.5$\times$10$^{-12}$ & 1.07$\times$10$^{-15}$  \\
H$_{2}$ (5-3) $Q$(8) ?					& 1.5821		& 1.5835		& -265		& 3.3$\times$10$^{-12}$ & 1.46$\times$10$^{-15}$  \\
?									& --			& 1.60008		& --			& 2.8$\times$10$^{-12}$ & 1.6$\times$10$^{-15}$  \\
\relax
[Fe~II] $^{4}D_{3/2}~{\textemdash}~^{4}F_{7/2}$	& 1.601		& 1.60339		& -448	& 3.30$\times$10$^{-12}$ & 1.61$\times$10$^{-15}$  \\
H$_{2}$ (6-4) $Q$(3) ?					& 1.6095		& 1.60846		& +194		& 3.2$\times$10$^{-12}$ & 1.1$\times$10$^{-15}$  \\
Br~13	 							& 1.612 		& 1.61314 	& -212 		& 3.44$\times$10$^{-12}$ & 1.30$\times$10$^{-15}$  \\
H$_{2}$ (6-4) $Q$(4) ?					& 1.6214		& 1.61977		& +302		& 3.1$\times$10$^{-12}$ & 1.57$\times$10$^{-15}$  \\
H$_{2}$ (4-2) $O$(5) ?					& 1.6219		& 1.62387		& -364		& 3.5$\times$10$^{-12}$ & 3.22$\times$10$^{-15}$  \\
? 									& 	--		& 1.63206 	& 	--		& 3.3$\times$10$^{-12}$ & 7.09$\times$10$^{-16}$  \\
? 									& 	--		& 1.63908	 	& 	--		& 3.6$\times$10$^{-12}$ & 1.22$\times$10$^{-15}$  \\
Br~12								& 1.64072		& 1.64142		& -128		& 3.7$\times$10$^{-12}$ & 2.22$\times$10$^{-15}$  \\
\relax
[Fe~II] $^{4}D_{7/2}~{\textemdash}~^{4}F_{9/2}$	& 1.64354		& 1.64434		& -146		& 3.77$\times$10$^{-12}$ & 1.42$\times$10$^{-15}$  \\
H$_{2}$ (1-0) $S$(11)						& 1.650		& 1.64844		& +284		& 3.7$\times$10$^{-12}$ & 2.0$\times$10$^{-15}$  \\
H$_{2}$ (1-0) $S$(10)						& 1.666		& 1.6695		& -630		& 4.7$\times$10$^{-12}$ & 4.2$\times$10$^{-15}$  \\
Br~11									& 1.681		& 1.68139		& -70		& 4.77$\times$10$^{-12}$ & 1.13$\times$10$^{-14}$  \\
?										& --			& 1.69075		& --		&  4.82$\times$10$^{-12}$ & 2.76$\times$10$^{-15}$  \\
H$_{2}$ (6-4) $Q$(8) ?						& 1.7013		& 1.70109		& +37	&  5.0$\times$10$^{-12}$ & 3.0$\times$10$^{-15}$  \\
\relax
[Fe~II] $^{4}D_{3/2}~{\textemdash}~^{4}F_{5/2}$	& 1.71113		& 1.71259		& -256		& 5.0$\times$10$^{-12}$ & 2.55$\times$10$^{-15}$  \\
Br~10									& 1.737		& 1.73638		& +107	& 5.2$\times$10$^{-12}$ & 5.6$\times$10$^{-15}$  \\
H$_{2}$ (1-0) $S$(1)							& 2.12183		& 2.12209		& -37		& 7.0$\times$10$^{-12}$ & 1.97$\times$10$^{-15}$  \\
Br$\gamma$							& 2.16553		& 2.16619		& -91		& 9.5$\times$10$^{-12}$ & 9.43$\times$10$^{-15}$  \\
H$_{2}$ (1-0) $Q$(1)  					& 2.40659		& 2.40678		 & -24	& 1.15$\times$10$^{-11}$	& 2.78$\times$10$^{-15}$ \\

H$_{2}$ (1-0) $Q$(3)  					& 2.42373		& 2.42393		 & -25	& 1.2$\times$10$^{-11}$	& 2.44$\times$10$^{-15}$ \\

\hline
\end{tabular}
\begin{tablenotes}
  \item[a] The vacuum wavelength. 
  \item[b] The line center wavelength.   
  \item[c] The line center velocity. The uncertainty on the velocity measurement is $\sim$20\%. 
  \item[d] The error on flux measurements is 20\%-30\%.  
\end{tablenotes}
\end{threeparttable}
\end{table*}

\begin{table*}
\centering
\caption{Line fluxes for J182940}
\label{ser8-lines}
\begin{threeparttable}
\begin{tabular}{lccccc} 
\hline
Species & $\lambda_{vac}$ [$\mu$m]  \tnote{a} & $\lambda_{line}$ [$\mu$m]  \tnote{b} & Velocity [km s$^{-1}$] \tnote{c}  & F$_{peak}$ [erg s$^{-1}$ cm$^{-2}$ $\mu$m$^{-1}$] \tnote{d} & Integrated flux [erg s$^{-1}$ cm$^{-2}$] \tnote{d}  	\\
\hline

H$_{2}$ (3-1) $S$(7)	 ?					& 1.130 		& 1.12922 & +207 	& 8.2$\times$10$^{-13}$ & 2.58$\times$10$^{-15}$ \\
\relax
[PII]									& 1.189 		& 1.18838 & +156 	& 8.7$\times$10$^{-13}$ & 1.48$\times$10$^{-15}$ \\

H$_{2}$ (4-2) $S$(7)	 ?					& 1.205 		& 1.20505 & -12 	& 1.8$\times$10$^{-12}$ & 8.07$\times$10$^{-16}$ \\

H$_{2}$ (4-2) $S$(6)	 ?					& 1.21573	 	& 1.21535 & +94 	& 1.0$\times$10$^{-12}$ & 5.76$\times$10$^{-16}$ \\
\relax
[Fe~II]	 $^{4}D_{7/2}~{\textemdash}~^{6}D_{9/2}$& 1.25668 	& 1.25696 & -67	& 4.2$\times$10$^{-12}$	& 2.72$\times$10$^{-15}$ \\
\relax
[Fe~II]	 $^{4}D_{1/2}~{\textemdash}~^{6}D_{1/2}$& 1.27034		& 1.27044 & -24	& 1.18$\times$10$^{-12}$	& 6.74$\times$10$^{-16}$ \\
\relax
[Fe~II]	 $^{4}D_{3/2}~{\textemdash}~^{6}D_{3/2}$& 1.27877		& 1.279 & -54	& 1.6$\times$10$^{-12}$	& 1.17$\times$10$^{-15}$ \\

Pa$\beta$							& 1.2818 		& 1.28219	 & -91	& 2.47$\times$10$^{-12}$	& 2.2$\times$10$^{-15}$ \\
\relax
[Fe~II]	 $^{4}D_{5/2}~{\textemdash}~^{6}D_{5/2}$& 1.29427		& 1.29451	 & -56	& 1.77$\times$10$^{-12}$	& 1.18$\times$10$^{-15}$ \\	
\relax
[Fe~II]	 $^{4}D_{3/2}~{\textemdash}~^{6}D_{1/2}$& 1.29777		& 1.29799 & -51	& 1.16$\times$10$^{-12}$	& 8.11$\times$10$^{-16}$ \\

H$_{2}$ (4-2) $S$(1)+(3-1) $Q$(1)					& 1.314		& 1.31409	 & -20	& 1.3$\times$10$^{-12}$	& 4.31$\times$10$^{-16}$ \\
\relax
[Fe~II]	 $^{4}D_{5/2}~{\textemdash}~^{4}F_{9/2}$	& 1.53347		& 1.53378 & -61	& 2.4$\times$10$^{-12}$	& 1.56$\times$10$^{-15}$ \\	
\relax
[Fe~II]	 $^{4}D_{3/2}~{\textemdash}~^{4}F_{7/2}$	& 1.59947		& 1.59988 & -77	& 2.4$\times$10$^{-12}$	& 7.33$\times$10$^{-16}$ \\	
\relax
[Fe~II]	 $^{4}D_{7/2}~{\textemdash}~^{4}F_{9/2}$	& 1.64354		& 1.64395	 & -75	& 7.75$\times$10$^{-12}$	& 1.48$\times$10$^{-15}$ \\	
\relax
[Fe~II]	 $^{4}D_{1/2}~{\textemdash}~^{4}F_{5/2}$	& 1.66376		& 1.66423 & -85	& 2.4$\times$10$^{-12}$	& 7.35$\times$10$^{-16}$ \\
\relax
[Fe~II]	 $^{4}D_{5/2}~{\textemdash}~^{4}F_{7/2}$	& 1.67687		& 1.6773 & -77	& 3.0$\times$10$^{-12}$	& 7.57$\times$10$^{-16}$ \\
Br~11								& 1.681		& 1.681	& 0.0		& 2.2$\times$10$^{-12}$	& 7.43$\times$10$^{-16}$ \\
Br~10								& 1.737		& 1.73619	& +140	& 2.4$\times$10$^{-12}$	& 1.04$\times$10$^{-15}$ \\

H$_{2}$ (1-0) $S$(2)						& 2.03375		& 2.03389 & -21	& 3.6$\times$10$^{-12}$	& 4.62$\times$10$^{-16}$ \\	
\relax
[Fe~II]			& 2.047		& 2.04639		& +89	& 2.5$\times$10$^{-12}$	& 4.8$\times$10$^{-16}$ \\	

H$_{2}$ (1-0) $S$(1)						& 2.12183		& 2.12185	 & -3		& 7.6$\times$10$^{-12}$	& 7.5$\times$10$^{-16}$ \\

Br$\gamma$							& 2.16553		& 2.16570	 & -23	& 3.95$\times$10$^{-12}$	& 1.57$\times$10$^{-15}$ \\	

H$_{2}$ (1-0) $S$(0)						& 2.22330		& 2.22328 & +3		& 4.6$\times$10$^{-12}$	& 5.98$\times$10$^{-16}$ \\

H$_{2}$ (2-1) $S$(1)						& 2.248		& 2.24778 & +30	& 4.1$\times$10$^{-12}$	& 2.1$\times$10$^{-16}$ \\

H$_{2}$ (1-0) $Q$(1)  					& 2.40659		& 2.40654	 & +6	& 1.2$\times$10$^{-11}$	& 6.7$\times$10$^{-16}$ \\

H$_{2}$ (1-0) $Q$(2)  					& 2.41343		& 2.4134 	 & +4	& 7.1$\times$10$^{-12}$	& 3.56$\times$10$^{-16}$ \\

H$_{2}$ (1-0) $Q$(3)  					& 2.42373		& 2.42369  & +5	& 1.16$\times$10$^{-11}$	& 7.38$\times$10$^{-16}$ \\

?									& --			& 2.44647	 & --		& 3.14$\times$10$^{-12}$	& 5.75$\times$10$^{-16}$ \\

?									& --			& 2.45039	 & --		& 2.75$\times$10$^{-12}$	& 5.33$\times$10$^{-16}$ \\

H$_{2}$ (1-0) $Q$(5)  					& 2.455		& 2.4548  & +24	& 3.2$\times$10$^{-12}$	& 4.08$\times$10$^{-16}$ \\
		 	
\hline
\end{tabular}
\begin{tablenotes}
  \item[a] The vacuum wavelength. 
  \item[b] The line center wavelength.   
  \item[c] The line center velocity. The uncertainty on the velocity measurement is $\sim$20\%. 
  \item[d] The error on flux measurements is 20\%-30\%.  
\end{tablenotes}
\end{threeparttable}
\end{table*}

\begin{table*}
\centering
\caption{Line fluxes for J162648}
\label{oph3-lines}
\begin{threeparttable}
\begin{tabular}{lccccc} 
\hline
Species & $\lambda_{vac}$ [$\mu$m]  \tnote{a} & $\lambda_{line}$ [$\mu$m]  \tnote{b} & Velocity [km s$^{-1}$] \tnote{c}  & F$_{peak}$ [erg s$^{-1}$ cm$^{-2}$ $\mu$m$^{-1}$] \tnote{d} & Integrated flux [erg s$^{-1}$ cm$^{-2}$] \tnote{d}  	\\
\hline

Mg $\textrm{I}$  						& 1.505 		& 1.5057 	& -398 	& 1.4$\times$10$^{-12}$ & 4.17$\times$10$^{-16}$ \\
?									& --			& 1.50687	 & --		& 1.12$\times$10$^{-12}$ & 5.28$\times$10$^{-16}$ \\
?									& --			& 1.51896	 & --		& 1.1$\times$10$^{-12}$ & 4.9$\times$10$^{-16}$ \\
H$_{2}$ (1-7) $P$(8)	 ?					& 1.52465 	& 1.52403	& +122	& 1.48$\times$10$^{-12}$ & 7.86$\times$10$^{-16}$ \\
Br~19  								& 1.527		& 1.5289 	& -373	& 1.34$\times$10$^{-12}$ & 7.03$\times$10$^{-16}$ \\	
\relax	
[Fe~II]	 $^{4}D_{5/2}~{\textemdash}~^{4}F_{9/2}$	& 1.53347		& 1.53319 & +55	& 1.5$\times$10$^{-12}$ & 5.95$\times$10$^{-16}$ \\
?									& --			& 1.53943 & --		& 1.28$\times$10$^{-12}$ & 5.54$\times$10$^{-16}$ \\	
Br~17	 							& 1.544		& 1.54333	 & +130	& 1.7$\times$10$^{-12}$ & 1.04$\times$10$^{-15}$ \\
?									& --			& 1.56556 & --		& 1.94$\times$10$^{-12}$ & 4.93$\times$10$^{-16}$ \\	
?									& --			& 1.5835 	 & --		& 2.36$\times$10$^{-12}$ & 1.85$\times$10$^{-15}$ \\	
\relax	
[Fe~II]	 $^{4}D_{3/2}~{\textemdash}~^{4}F_{7/2}$	& 1.59947		& 1.6032 	 & -698	& 2.6$\times$10$^{-12}$ & 1.55$\times$10$^{-15}$ \\	
Br~13	 							& 1.612		& 1.61295	 & -177	& 2.75$\times$10$^{-12}$ & 2.14$\times$10$^{-15}$ \\
?									& --			& 1.62387  & --		& 2.6$\times$10$^{-12}$ & 2.42$\times$10$^{-15}$ \\  	
\relax
[Fe~II]	 $^{4}D_{7/2}~{\textemdash}~^{4}F_{9/2}$	& 1.64354		& 1.64415  & -111	& 3.0$\times$10$^{-12}$ & 9.86$\times$10$^{-16}$ \\
?									& --			& 1.64805	 & --		& 3.04$\times$10$^{-12}$ & 2.41$\times$10$^{-15}$ \\  
H$_{2}$ (1-0) $S$(11)					& 1.650		& 1.65039	 & -71	& 3.3$\times$10$^{-12}$ & 2.13$\times$10$^{-15}$ \\  
?									& --			& 1.6695 	 & --		& 4.14$\times$10$^{-12}$ & 4.52$\times$10$^{-15}$ \\  

Br~11								& 1.681		& 1.681  & 0.0	& 4.07$\times$10$^{-12}$ & 7.68$\times$10$^{-15}$ \\

H$_{2}$ (1-0) $S$(9)	 					& 1.688 		& 1.69056  & -454 	& 4.3$\times$10$^{-12}$ & 1.56$\times$10$^{-15}$ \\  
?									& --			& 1.69563  & --		& 4.1$\times$10$^{-12}$ & 2.64$\times$10$^{-15}$ \\  
?									& --			& 1.70089  & --		& 5.0$\times$10$^{-12}$ & 2.5$\times$10$^{-15}$ \\  
 
Br~10							& 1.737		& 1.73638  & +107	& 5.6$\times$10$^{-12}$ & 8.88$\times$10$^{-15}$ \\
\relax
[Fe~II] ?							& 2.016		& 2.01258	  & +509	& 1.51$\times$10$^{-11}$ & 5.65$\times$10$^{-16}$ \\	 
Mg~I ?							& 2.110		& 2.11278	 & -395	& 1.23$\times$10$^{-11}$ & 4.97$\times$10$^{-15}$ \\	 
H$_{2}$ (1-0) $S$(1)						& 2.12183		& 2.12209  & -37	& 1.25$\times$10$^{-11}$ & 2.2$\times$10$^{-15}$ \\	 
Br$\gamma$							& 2.16553		& 2.16617	  & -116	& 1.89$\times$10$^{-11}$ & 1.82$\times$10$^{-14}$ \\

\hline
\end{tabular}
\begin{tablenotes}
  \item[a] The vacuum wavelength. 
  \item[b] The line center wavelength.   
  \item[c] The line center velocity. The uncertainty on the velocity measurement is $\sim$20\%. 
  \item[d] The error on flux measurements is 20\%-30\%.  
\end{tablenotes}
\end{threeparttable}
\end{table*}

\begin{table*}
\centering
\caption{Line fluxes for J163152}
\label{oph2-lines}
\begin{threeparttable}
\begin{tabular}{lccccc} 
\hline
Species & $\lambda_{vac}$ [$\mu$m]  \tnote{a} & $\lambda_{line}$ [$\mu$m]  \tnote{b} & Velocity [km s$^{-1}$] \tnote{c}  & F$_{peak}$ [erg s$^{-1}$ cm$^{-2}$ $\mu$m$^{-1}$] \tnote{d} & Integrated flux [erg s$^{-1}$ cm$^{-2}$] \tnote{d}  	\\
\hline

H$_{2}$ (2-0) $S$(2)	 ?				& 1.138		& 1.13531 & +709 	& 6.15$\times$10$^{-13}$ & 1.22$\times$10$^{-15}$ \\
\relax
[FeII]$^{4}D_{3/2}{\textemdash}~^{6}D_{3/2}$+Pa$\beta$	& 1.28--1.2818 		& 1.28248	 & --	& 1.23$\times$10$^{-12}$ & 1.46$\times$10$^{-15}$ \\
Br~13 							& 1.612		& 1.61158 & +78	& 9.2$\times$10$^{-13}$ 	& 7.97$\times$10$^{-16}$ \\
Br~12				& 1.64072		& 1.64161	 & -163	& 1.17$\times$10$^{-12}$ & 8.46$\times$10$^{-16}$ \\
Br~11				& 1.681		& 1.68159  & -105	& 1.48$\times$10$^{-12}$ & 1.65$\times$10$^{-15}$ \\
Br~10				& 1.737		& 1.73677  & +40	& 1.86$\times$10$^{-12}$ & 8.88$\times$10$^{-16}$ \\
H$_{2}$ (1-0) $S$(2)		& 2.03375		& 2.03389	  & -21	& 2.7$\times$10$^{-12}$ & 1.3$\times$10$^{-15}$ \\
H$_{2}$ (1-0) $S$(1)		& 2.12183		& 2.12211	  & -40	& 4.26$\times$10$^{-12}$ & 1.23$\times$10$^{-15}$ \\
Br$\gamma$			& 2.16553		& 2.16618   & -118	& 5.76$\times$10$^{-12}$ & 8.40$\times$10$^{-15}$ \\
H$_{2}$ (1-0) $S$(0)		& 2.22330		& 2.22352   & -30	& 4.77$\times$10$^{-12}$ & 1.81$\times$10$^{-15}$ \\
H$_{2}$ (1-0) $Q$(1)  	& 2.40659		& 2.40678	  & -24	& 8.65$\times$10$^{-12}$ & 1.80$\times$10$^{-15}$ \\
H$_{2}$ (1-0) $Q$(2)  	& 2.414		& 2.41658	  & -547	& 9.07$\times$10$^{-12}$ & 1.17$\times$10$^{-15}$ \\
?					& --			& 2.41977	& --		& 8.97$\times$10$^{-12}$ & 1.83$\times$10$^{-15}$ \\
H$_{2}$ (1-0) $Q$(3)  	& 2.42373		& 2.42393	  & -25	& 8.56$\times$10$^{-12}$ & 2.05$\times$10$^{-15}$ \\
?					& --			& 2.42687	  & --		& 7.8$\times$10$^{-12}$ & 1.12$\times$10$^{-15}$ \\
H$_{2}$ (1-0) $Q$(4)  	& 2.437		& 2.43594	 & +130	& 1.16$\times$10$^{-11}$ & 2.40$\times$10$^{-15}$ \\
?					& --			& 2.44353	  & --		& 7.9$\times$10$^{-12}$ & 1.24$\times$10$^{-15}$ \\
?				  	& --			& 2.44647	 & --		& 1.0$\times$10$^{-11}$ & 2.80$\times$10$^{-15}$ \\

\hline
\end{tabular}
\begin{tablenotes}
  \item[a] The vacuum wavelength. 
  \item[b] The line center wavelength.   
  \item[c] The line center velocity. The uncertainty on the velocity measurement is $\sim$20\%. 
  \item[d] The error on flux measurements is 20\%-30\%.  
\end{tablenotes}
\end{threeparttable}
\end{table*}

\begin{table*}
\centering
\caption{Line fluxes for J163136}
\label{oph1-lines}
\begin{threeparttable}
\begin{tabular}{lccccc} 
\hline
Species & $\lambda_{vac}$ [$\mu$m]  \tnote{a} & $\lambda_{line}$ [$\mu$m]  \tnote{b} & Velocity [km s$^{-1}$] \tnote{c}  & F$_{peak}$ [erg s$^{-1}$ cm$^{-2}$ $\mu$m$^{-1}$] \tnote{d} & Integrated flux [erg s$^{-1}$ cm$^{-2}$] \tnote{d}  	\\
\hline

H$_{2}$ (2-0) $S$(2)	 ?					& 1.138		& 1.13516 & +749 	& 1.20$\times$10$^{-12}$ & 4.07$\times$10$^{-16}$ \\
?									& --			& 1.14676 & --		& 1.18$\times$10$^{-12}$ & 7.36$\times$10$^{-16}$ \\
\relax
[FeII]	 $^{4}D_{5/2}~{\textemdash}~^{6}F_{9/2}$ ? & 1.1534	& 1.1543 & -234 	& 9.8$\times$10$^{-13}$ & 8.43$\times$10$^{-16}$ \\

H$_{2}$ (3-1) $S$(3)						& 1.186		& 1.18519 & +205	& 1.05$\times$10$^{-12}$ & 1.58$\times$10$^{-16}$ \\
H$_{2}$ (3-1) $S$(1) ?					& 1.233		& 1.23507 & -503	& 1.14$\times$10$^{-12}$ & 6.52$\times$10$^{-16}$ \\
\relax
[Fe~II]	 $^{4}D_{7/2}~{\textemdash}~^{6}D_{9/2}$& 1.25668 	& 1.25725	 & -136	& 1.5$\times$10$^{-12}$ & 2.08$\times$10$^{-15}$ \\
\relax
[Fe~II]	 $^{4}D_{3/2}~{\textemdash}~^{6}D_{3/2}$& 1.280 	& 1.27914	 & +201	& 1.58$\times$10$^{-12}$ & 1.11$\times$10$^{-15}$ \\
Pa$\beta$							& 1.2818 		& 1.28233	 & -124	& 2.90$\times$10$^{-12}$ & 2.89$\times$10$^{-15}$ \\
\relax
[Fe~II]	 $^{4}D_{5/2}~{\textemdash}~^{6}D_{5/2}$& 1.29427		& 1.2948 	& -123	& 1.5$\times$10$^{-12}$ & 7.76$\times$10$^{-16}$ \\
Mg $\textrm{I}$  						& 1.505 		& 1.5057	& -140	& 2.07$\times$10$^{-12}$ & 6.9$\times$10$^{-16}$ \\
?									& --			& 1.51896	& --		& 1.8$\times$10$^{-12}$ & 1.12$\times$10$^{-15}$ \\

H$_{2}$ (1-7) $P$(8)						& 1.52465		& 1.52442 & +45	& 2.05$\times$10$^{-12}$ & 1.26$\times$10$^{-15}$ \\

Br~19								& 1.527		& 1.5289	& -373	& 1.87$\times$10$^{-12}$ & 2.04$\times$10$^{-15}$ \\
\relax	
[Fe~II]	 $^{4}D_{5/2}~{\textemdash}~^{4}F_{9/2}$& 1.53347		& 1.53339	 & +16	& 2.05$\times$10$^{-12}$ & 2.99$\times$10$^{-15}$ \\	

?									& --			& 1.54002	& --		& 1.8$\times$10$^{-12}$ & 1.22$\times$10$^{-15}$ \\	

Br~17 								& 1.544		& 1.54333 & +130	& 2.10$\times$10$^{-12}$ & 2.38$\times$10$^{-15}$ \\	
Br~15 								& 1.571		& 1.57063 & +71	& 2.42$\times$10$^{-12}$ & 1.20$\times$10$^{-15}$ \\

\relax
[Fe~II]	 $^{4}D_{3/2}~{\textemdash}~^{4}F_{7/2}$& 1.601	& 1.6032	& -412	& 2.6$\times$10$^{-12}$ & 2.05$\times$10$^{-15}$ \\

Br~13 								& 1.612		& 1.61314 & -212	& 2.8$\times$10$^{-12}$ & 2.82$\times$10$^{-15}$ \\

?									& --			& 1.62387	& --		& 2.67$\times$10$^{-12}$ & 2.44$\times$10$^{-15}$ \\

Br~12 								& 1.64072		& 1.64161 & -163	& 3.16$\times$10$^{-12}$ & 1.95$\times$10$^{-15}$ \\
\relax
[Fe~II]	 $^{4}D_{1/2}~{\textemdash}~^{4}F_{5/2}$& 1.64354		& 1.64415 & -111	& 3.72$\times$10$^{-12}$ & 2.01$\times$10$^{-15}$ \\	

H$_{2}$(1-0)$S$(10)+[FeII]$^{4}D_{1/2}{\textemdash}~^{4}F_{5/2}$	& 1.666--1.665	& 1.66969		& -- 	& 3.5$\times$10$^{-12}$ & 3.27$\times$10$^{-15}$ \\

Br~11 								& 1.681		& 1.68139 & -70	& 4.05$\times$10$^{-12}$ & 4.78$\times$10$^{-15}$ \\

?	& --	& 1.69095		& --	& 3.66$\times$10$^{-12}$ 	& 2.7$\times$10$^{-15}$ \\
?	& --	& 1.69602		& --	& 3.63$\times$10$^{-12}$ 	& 2.01$\times$10$^{-15}$ \\
?	& --	& 1.70109		& --	& 4.03$\times$10$^{-12}$ 	& 4.26$\times$10$^{-15}$ \\

\relax
[Fe~II] $^{4}D_{3/2}~{\textemdash}~^{4}F_{5/2}$& 1.71113		& 1.71279	 & -291	& 3.67$\times$10$^{-12}$ & 2.67$\times$10$^{-15}$ \\	
Br~10 								& 1.737		& 1.73677 & +40 	& 4.7$\times$10$^{-12}$ & 5.96$\times$10$^{-15}$ \\
H$_{2}$ (1-0) $S$(2)						& 2.03375		& 2.03389	 &  -21	& 4.23$\times$10$^{-12}$ & 2.04$\times$10$^{-15}$ \\	

Mg~I + Al~I ? 							& 2.110 -- 2.118	& 2.11352	& --	& 4.33$\times$10$^{-12}$ & 2.83$\times$10$^{-15}$ \\	

H$_{2}$ (1-0) $S$(1)						& 2.12183		& 2.12209  & -37	& 6.14$\times$10$^{-12}$ & 3.07$\times$10$^{-15}$ \\	 
Br$\gamma$							& 2.16553		& 2.16619	 & -91	& 6.3$\times$10$^{-12}$ & 7.77$\times$10$^{-15}$ \\
H$_{2}$ (1-0) $S$(0)						& 2.22330		& 2.22352 & -30	& 5.7$\times$10$^{-12}$ & 3.9$\times$10$^{-15}$ \\

H$_{2}$ (2-1) $S$(1) 					& 2.248		& 2.24778	 & +29	& 5.6$\times$10$^{-12}$ & 1.96$\times$10$^{-15}$ \\

?	& --	& 2.27595		& --	& 6.44$\times$10$^{-12}$ 	& 8.3$\times$10$^{-16}$ \\

H$_{2}$ (1-0) $Q$(1)  					& 2.40659		& 2.40678 & -24	& 9.37$\times$10$^{-12}$ & 1.55$\times$10$^{-15}$ \\ 
H$_{2}$ (1-0) $Q$(2)  					& 2.41343		& 2.41364 & -26	& 7.15$\times$10$^{-12}$ & 1.65$\times$10$^{-15}$ \\ 
H$_{2}$ (1-0) $Q$(3)  					& 2.42373		& 2.42393 & -25	& 8.9$\times$10$^{-12}$ & 2.78$\times$10$^{-15}$ \\ 	

\hline
\end{tabular}
\begin{tablenotes}
  \item[a] The vacuum wavelength. 
  \item[b] The line center wavelength.   
  \item[c] The line center velocity. The uncertainty on the velocity measurement is $\sim$20\%. 
  \item[d] The error on flux measurements is 20\%-30\%.  
\end{tablenotes}
\end{threeparttable}
\end{table*}

\begin{table*}
\centering
\caption{Line fluxes for M1701117}
\label{M170-lines}
\begin{threeparttable}
\begin{tabular}{lccccc} 
\hline
Species & $\lambda_{vac}$ [$\mu$m]  \tnote{a} & $\lambda_{line}$ [$\mu$m]  \tnote{b} & Velocity [km s$^{-1}$] \tnote{c}  & F$_{peak}$ [erg s$^{-1}$ cm$^{-2}$ $\mu$m$^{-1}$] \tnote{d} & Integrated flux [erg s$^{-1}$ cm$^{-2}$] \tnote{d}  	\\
\hline
\relax
[Fe~II]	 $^{4}D_{7/2}~{\textemdash}~^{6}D_{9/2}$& 1.25668 	& 1.25739 & -169	& 3.40$\times$10$^{-12}$ & 2.46$\times$10$^{-15}$ \\ 
Pa$\beta$							& 1.2818 		& 1.28262 & -192	& 3.12$\times$10$^{-12}$ & 4.12$\times$10$^{-15}$ \\ 	 
\relax
[Fe~II]	 $^{4}D_{7/2}~{\textemdash}~^{4}F_{9/2}$& 1.64354		& 1.64434 & -146	& 2.90$\times$10$^{-12}$ & 3.83$\times$10$^{-15}$ \\ 
H$_{2}$ (1-0) $S$(2)						& 2.03375		& 2.03414 & -57	& 2.78$\times$10$^{-12}$ & 2.31$\times$10$^{-15}$ \\   
H$_{2}$ (1-0) $S$(2)						& 2.073		& 2.07383 & -120	& 2.65$\times$10$^{-12}$ & 1.59$\times$10$^{-15}$ \\  
H$_{2}$ (1-0) $S$(1)						& 2.12183		& 2.12234 & -72	& 4.89$\times$10$^{-12}$ & 5.39$\times$10$^{-15}$ \\  	
H$_{2}$ (2-1) $S$(2)						& 2.154		& 2.15468 & -95	& 2.79$\times$10$^{-12}$ & 5.85$\times$10$^{-16}$ \\  
Br$\gamma$							& 2.16553		& 2.16644	 & -126	& 3.13$\times$10$^{-12}$ & 5.71$\times$10$^{-15}$ \\  
H$_{2}$ (1-0) $S$(0)						& 2.22330		& 2.22377 & -63	& 3.58$\times$10$^{-12}$ & 2.25$\times$10$^{-15}$ \\  	
H$_{2}$ (2-1) $S$(1)						& 2.248		& 2.24802	 & -3		& 3.53$\times$10$^{-12}$ & 2.67$\times$10$^{-15}$ \\  	
?									& -- 			&  2.30707 & -- 		& 4.92$\times$10$^{-12}$ & 2.31$\times$10$^{-15}$ \\  	
H$_{2}$ (1-0) $Q$(1)  					& 2.40659		& 2.40703  & -55	& 5.53$\times$10$^{-12}$ & 1.16$\times$10$^{-15}$ \\ 
H$_{2}$ (1-0) $Q$(2)  					& 2.41343		& 2.41389  & -57	& 4.52$\times$10$^{-12}$ & 9.96$\times$10$^{-16}$ \\ 
H$_{2}$ (1-0) $Q$(3)  					& 2.42373		& 2.42442  & -85	& 5.07$\times$10$^{-12}$ & 8.93$\times$10$^{-16}$ \\ 
H$_{2}$ (1-0) $Q$(4)  					& 2.437		& 2.4379    & -111	& 5.17$\times$10$^{-12}$ & 4.76$\times$10$^{-16}$ \\ 
H$_{2}$ (1-0) $Q$(5)  					& 2.455		& 2.45529  & -35	& 5.63$\times$10$^{-12}$ & 4.98$\times$10$^{-16}$ \\ 

\hline
\end{tabular}
\begin{tablenotes}
  \item[a] The vacuum wavelength. 
  \item[b] The line center wavelength.   
  \item[c] The line center velocity. The uncertainty on the velocity measurement is $\sim$20\%. 
  \item[d] The error on flux measurements is 20\%-30\%.
\end{tablenotes}
\end{threeparttable}
\end{table*}


\bsp	
\label{lastpage}

\begin{thebibliography}{99}

\bibitem[\protect\citeauthoryear{Allard et al.}{2003}]{allard} Allard, N. F.; Allard, F.; Hauschildt, P. H.; Kielkopf, J. F.; Machin, L. 2003, A\&A, 411, 473
\bibitem[\protect\citeauthoryear{Alcala et al.}{2017}]{alcala} Alcala et al. 2017, A\&A, 600, 20
\bibitem[\protect\citeauthoryear{Antoniucci et al.}{2008}]{ant08} Antoniucci et al. 2008, A\&A, 479, 503
\bibitem[\protect\citeauthoryear{Antoniucci et al.}{2011}]{ant11} Antoniucci et al. 2011, A\&A, 534, 32
\bibitem[\protect\citeauthoryear{Antoniucci et al.}{2014}]{ant14} Antoniucci et al. 2014, A\&A, 566, 129
\bibitem[\protect\citeauthoryear{Antoniucci et al.}{2017}]{ant17} Antoniucci et al. 2017, A\&A, 599, 105
\bibitem[\protect\citeauthoryear{Bally}{2016}]{bally16} Bally, J. 2016, ARA\&A, 54, 491
\bibitem[\protect\citeauthoryear{Bally et al.}{2003}]{bally03} Bally, Feigelson, Reipurth 2003, ApJ, 584, 843
\bibitem[\protect\citeauthoryear{Baraffe et al.}{2003}]{baraffe03} Baraffe, I. et al., 2003, A\&A, 402, 701
\bibitem[\protect\citeauthoryear{Baraffe et al.}{2017}]{baraffe17} Baraffe, I., Elbakyan, V. G., Vorobyov, E. I., \& Chabrier, G. 2017, A\&A, 597, A19
\bibitem[\protect\citeauthoryear{Caratti o Garatti et al.}{2006}]{car06} Caratti o Garatti et al. 2006, A\&A, 449, 1077
\bibitem[\protect\citeauthoryear{Davis et al.}{2001}]{davis01} Davis et al. 2001, MNRAS, 326, 524
\bibitem[\protect\citeauthoryear{Davis et al.}{2003}]{davis03} Davis et al. 2003, A\&A, 397, 693
\bibitem[\protect\citeauthoryear{Davis et al.}{2011}]{davis11} Davis et al. 2011, A\&A, 528, 3
\bibitem[\protect\citeauthoryear{Dzib et al.}{2010}]{dzib} Dzib, Sergio; Loinard, Laurent; Mioduszewski, Amy J.; Boden, Andrew F.; Rodríguez, Luis F.; Torres, Rosa M., 2010, ApJ, 718, 610
\bibitem[\protect\citeauthoryear{Evans et al.}{2009}]{evans} Evans et al. 2009, ApJSS, 181, 321
\bibitem[\protect\citeauthoryear{Frank et al.}{2014}]{frank} Frank et al. 2014, PPVI, 451
\bibitem[\protect\citeauthoryear{Gaia Collaboration}{2018}]{gaia} Gaia Collaboration 2018, A\&A, 616, 1G
\bibitem[\protect\citeauthoryear{Garcia Lopez et al.}{2008}]{garcia08} Garcia Lopez et al. 2008, A\&A, 487, 1019
\bibitem[\protect\citeauthoryear{Garcia Lopez et al.}{2013}]{garcia13} Garcia Lopez et al. 2013, A\&A, 552, L2
\bibitem[\protect\citeauthoryear{Hartmann et al.}{1990}]{hart} Hartmann et al. 1990, ApJ, 349, 168
\bibitem[\protect\citeauthoryear{Machida et al.}{2009}]{mach09} Machida, M., N. et al. 2009, ApJ, 699, L157
\bibitem[\protect\citeauthoryear{Mamajek}{2008}]{mamajek} Mamajek 2008, AN, 329, 10
\bibitem[\protect\citeauthoryear{Muzerolle et al.}{1998}]{muz} Muzerolle et al. 1998, AJ, 116, 2965
\bibitem[\protect\citeauthoryear{Nisini et al.}{2002}]{nis02} Nisini et al. 2002, A\&A, 393, 1035
\bibitem[\protect\citeauthoryear{Nisini et al.}{2005}]{nis05} Nisini et al. 2005, A\&A, 441, 159
\bibitem[\protect\citeauthoryear{Nisini et al.}{2016}]{nis16} Nisini et al. 2016, A\&A, 595, 76
\bibitem[\protect\citeauthoryear{Nisini et al.}{2004}]{nis04} Nisini et al. 2004, A\&A, 421, 187


\bibitem[\protect\citeauthoryear{Ortiz-Le\'{o}n et al.}{2018}]{ortiz} Ortiz-Le\'{o}n, G., N., et al. 2018, ApJ, 869L, 33
\bibitem[\protect\citeauthoryear{Ossenkopf \& Henning}{1994}]{ossen} Ossenkopf, V.; Henning, Th., 1994, A\&A, 291, 943
\bibitem[\protect\citeauthoryear{Reipurth \& Bally}{2001}]{reipurth} Reipurth \& Bally 2001, ARA\&A, 39, 403

\bibitem[\protect\citeauthoryear{Riaz et al.}{2012}]{riaz12} Riaz, B. et al. 2012, MNRAS, 419, 1887

\bibitem[\protect\citeauthoryear{Riaz \& Whelan}{2015}]{r15} Riaz \& Whelan 2015, ApJL, 815, L31
\bibitem[\protect\citeauthoryear{Riaz et al.}{2015}]{riaz15} Riaz, B. et al. 2015, MNRAS, 446, 2550
\bibitem[\protect\citeauthoryear{Riaz et al.}{2016}]{r16} Riaz, B. et al. 2016, ApJ, 831, 189
\bibitem[\protect\citeauthoryear{Riaz et al.}{2017}]{r17} Riaz et al. 2017, ApJ, 844, 47
\bibitem[\protect\citeauthoryear{Riaz et al.}{2019}]{r19} Riaz, B.; Machida, M. N.; Stamatellos, D., 2019, MNRAS, 486, 4114
\bibitem[\protect\citeauthoryear{Savage \& Sembach}{1996}]{savage} Savage, Blair D. \& Sembach, Kenneth R., 1996, ARA\&A, 34, 279
\bibitem[\protect\citeauthoryear{Schlafly et al.}{2014}]{schlafly} Schlafly et al. 2014, ApJ, 786, 29
\bibitem[\protect\citeauthoryear{Storey \& Hummer}{1995}]{storey} Storey, P. J. \& Hummer, D. G., 1995, MNRAS, 272, 41
\bibitem[\protect\citeauthoryear{Takami et al.}{2006}]{takami} Takami et al. 2006, ApJ, 641, 357
\bibitem[\protect\citeauthoryear{Throop et al.}{2001}]{throop} Throop, Bally, Esposito, 2001, Science, 292, 1686
\bibitem[\protect\citeauthoryear{van Kempen et al.}{2009}]{van} van Kempen, T. A., et al. 2009, A\&A, 508, 259
\bibitem[\protect\citeauthoryear{Vorobyov et al.}{2017}]{vor} Vorobyov, E. et al. 2017, A\&A, 600, 36
\bibitem[\protect\citeauthoryear{Whelan et al.}{2018}]{whelan} Whelan, E. T.; Riaz, B.; Rouz\'{e}, B. 2018, A\&A, 610, 19
\bibitem[\protect\citeauthoryear{Whelan et al.}{2014}]{w14} Whelan et al. 2014, AN, 335, 537
\bibitem[\protect\citeauthoryear{Whitney et al.}{2003}]{whitney} Whitney, Barbara A.; Wood, Kenneth; Bjorkman, J. E.; Cohen, Martin, 2003, ApJ, 598, 1079

\end{thebibliography}
\end{document}